\documentclass[acmsmall,screen,dvipsnames]{acmart}
\usepackage{mathpartir}
\usepackage{paracol}
\usepackage{cleveref}
\usepackage{xspace}
\usepackage{xstring}
\usepackage{subcaption}
\usepackage{stmaryrd}
\usepackage{mathtools}

\newcommand\strong{\triangleright}
\newcommand\option{\mathop{?}}
\newcommand\weak{\mathop{:}}

\newcommand\concat{\mathop{+\!\!+}}

\newcommand\force{\mathbf{force}}
\newcommand\uniform[1]{\lfloor #1 \rfloor }
\newcommand\unique[1]{\llfloor #1 \rrfloor } %

\newcommand\apps{\mathop{\mathcolor{App}{●}}}

\newcommand\tconcats{\mathop{\mathcolor{Concat}{⊕}}}

\colorlet{Pattern}{RedOrange}
\colorlet{Store}{LimeGreen}
\colorlet{App}{ForestGreen}
\colorlet{STLC}{Cerulean}
\colorlet{Subtyping}{NavyBlue}
\colorlet{Concat}{Purple}
\newcommand\app[3]{#1
  \,\mathop{\mathcolor{App}{●}} \, #2
  \,\mathop{\mathcolor{App}{↠}} \, #3}
\newcommand\appb[2]{#1
  \,\mathop{\mathcolor{App}{●}} \, #2}
\newcommand\sub{\,\mathop{\mathcolor{Subtyping}{\leqslant}}\,}
\newcommand\tconcat[3]{#1 \,\mathop{\mathcolor{Concat}{⊕}} \, #2 \,
  \mathop{\mathcolor{Concat}{\looparrowright}} ~ #3}
\newcommand\tconcatb[2]{#1 \,\mathop{\mathcolor{Concat}{⊕}} \, #2}
\newcommand\patsep[3]{#1 \mathcolor{Pattern}{:} #2 \mathcolor{Pattern}{\odot} #3}
\newcommand\envbox[2]{#1 \blacktriangleright #2}

\newcommand{\lang}{\texorpdfstring{\ensuremath{\lambda^{\rho\vdash}}}{λρ⊢}\xspace}

\makeatletter
\newcommand{\customlabel}[2]{%
   \phantomsection%
   \edef\@currentlabel{#1}%
   \protected@write\@auxout{}{\string\newlabel{#2}{{#1}{\thepage}{#1}{\@currentHref}{}}}%
}
\makeatother

\newcommand{\ruleref}[1]{\textsc{\ref{#1}}}

\theoremstyle{plain}
\newtheorem{theorem}{Theorem}[section]
\newtheorem{lemma}[theorem]{Lemma}

\usepackage{listings}
\usepackage{wrapfig}
\usepackage{tikz}

\definecolor{listing-number}{RGB}{108,113,196}

\lstdefinestyle{basic}{%
  basicstyle=\small\ttfamily,
  commentstyle=\color{green!60!black},
  stringstyle=\color{orange},
  literate=*{0}{{{\color{listing-number}0}}}1
    {1}{{{\color{listing-number}1}}}1
    {2}{{{\color{listing-number}2}}}1
    {3}{{{\color{listing-number}3}}}1
    {4}{{{\color{listing-number}4}}}1
    {5}{{{\color{listing-number}5}}}1
    {6}{{{\color{listing-number}6}}}1
    {7}{{{\color{listing-number}7}}}1
    {8}{{{\color{listing-number}8}}}1
    {9}{{{\color{listing-number}9}}}1
    {|>}{$\strong$}1
}

\makeatletter
\def\addToLiterate#1{\edef\lst@literate{\unexpanded\expandafter{\lst@literate}\unexpanded{#1}}}
\lst@Key{add to literate}{}{\addToLiterate{#1}}
\makeatother

\lstdefinestyle{python}{
    language=Python,
    style=basic,
    keywordstyle=\color{blue},
    commentstyle=\color{green!60!black},
}

\lstdefinestyle{swift}{
    language=Swift,
    style=basic,
    keywordstyle=\color{blue},
}

\lstdefinestyle{javascript}{
  style=basic,
  keywordstyle=\color{blue},
  morekeywords={function,undefined,typeof},
}

\lstdefinestyle{scala}{
  language=Scala,
  style=basic,
  keywordstyle=\color{blue},
  add to literate={~}{{$\;$}}1
}

\lstdefinestyle{config}{
  style=basic,
  basicstyle=\small\ttfamily,
  morecomment=[l]{\#},
  morekeywords={type,record,opt},
  keywordstyle=\color{Thistle},
  literate=*{+}{{{\color{ForestGreen}\textbf{+}}}}1
}

\lstdefinestyle{presence}{
  style=basic,
  basicstyle=\small\ttfamily,
  morekeywords={pre,abs},
  keywordstyle=\color{Thistle},
}

\lstdefinestyle{nix}{
  style=basic,
  morecomment=[l]{\#},
  keywordstyle=\color{Thistle},
  morestring=[b]",
  morekeywords={let},
  add to literate=
     {\{}{{{\color{Thistle}\{}}}1
     {\}}{{{\color{Thistle}\}}}}1
     {1}{{{1}}}1
     {2}{{{2}}}1
     {3}{{{3}}}1
}

\lstdefinestyle{mylang}{
    style=basic,
    keywordstyle=\color{blue},
    morestring=[b]",
    frame=single,
    rulecolor=\color{gray!50},
    morekeywords={force,if,then,else,let,ref,true,false,in},
    add to literate={\{}{{{\color{magenta}\{}}}1
             {\}}{{{\color{magenta}\}}}}1
             {\\}{{{\color{magenta}$\lambda$}}}1
             {\\-}{{{\color{magenta}\textbackslash}}}1
             {\\/}{{{\color{magenta}$\forall$}}}1
             {->}{$\to$}1
             {~}{{$\;$}}1
             {⩽}{{$\mathcolor{Subtyping}{\leqslant}$}}1
             {<>}{{$\diamond$}}1
}

\usepackage{pifont}
\newcommand{\cmark}{\ding{51}}%
\newcommand{\xmark}{\ding{55}}%

\newcommand{\highlight}[2]{\ensuremath{\mathchoice%
  {\colorbox{#1}{\(\displaystyle#2\)}}%
  {\colorbox{#1}{\(\textstyle#2\)}}%
  {\colorbox{#1}{\(\scriptstyle#2\)}}%
  {\colorbox{#1}{\(\scriptscriptstyle#2\)}}}}%
\newcommand{\ghl}[1]{\highlight{gray!15}{#1}}

\usepackage{enumitem}

\setlist[enumerate]{leftmargin=5.5mm}
\setlist[itemize]{leftmargin=4.5mm}

\Crefformat{figure}{#2Fig.~#1#3}
\Crefformat{section}{\S#2#1#3}
\Crefformat{subsection}{\S#2#1#3}
\Crefformat{subsubsection}{\S#2#1#3}
\usepackage[utf8]{inputenc}
\usepackage[T1]{fontenc} %

\newcommand{\lottalias}[1]{#1}
\newcommand{\lottaliassep}{,}
\newcommand{\lottproduction}[2]{#2}
\newcommand{\lottkw}[1]{\mathbf{#1}}
\newcommand{\lottsym}[1]{{#1}}
\newcommand{\lottsepbycomprehension}[1]{\overline{#1}}

\newcommand{\lottsepbycomprehensionpatcollection}[3]{\overline{#1}^{#2 \in #3}}

\newcommand{\lottjudgementcomprehensionpatcollection}[3]{\overline{#1}^{#2 \in #3}}
\newcommand{\lottproductionsep}{\mid}

\newcommand{\lottnonterminal}[3]{$#2$ & $\Coloneqq$ & $#3$ & \textit{\MakeLowercase{#1}}}
\def\currentlottproductionsep{\relax}
\newcommand{\lottnonterminalforcename}[2]{
\let\oldlottnonterminal\lottnonterminal
\renewcommand{\lottnonterminal}[3]{\oldlottnonterminal{#1}{##2}{##3}}
#2%
}
\newcommand{\lottnonterminalfilter}[2]{
\global\let\oldlottproduction\lottproduction
\global\let\oldlottproductionsep\lottproductionsep
\global\let\oldcurrentlottproductionsep\currentlottproductionsep
\renewcommand{\lottproductionsep}{}
\global\let\lottproductionsep\lottproductionsep
\renewcommand{\currentlottproductionsep}{\dots \oldlottproductionsep}
\global\let\currentlottproductionsep\currentlottproductionsep
\renewcommand{\lottproduction}[2]{
\IfSubStr{,#1,}{,##1,}{
\currentlottproductionsep
\oldlottproduction{##1}{##2}
\renewcommand{\currentlottproductionsep}{\oldlottproductionsep}
}{}
}
\global\let\lottproduction\lottproduction
#2%
\global\let\lottproduction\oldlottproduction
\global\let\lottproductionsep\oldlottproductionsep
\global\let\currentlottproductionsep\oldcurrentlottproductionsep
}
\newcommand{\lottnonterminalfilterout}[2]{
\global\let\oldlottproduction\lottproduction
\global\let\oldlottproductionsep\lottproductionsep
\renewcommand{\lottproductionsep}{}
\global\let\lottproductionsep\lottproductionsep
\renewcommand{\lottproduction}[2]{
\IfSubStr{,#1,}{,##1,}{}{
\currentlottproductionsep
\oldlottproduction{##1}{##2}
\renewcommand{\currentlottproductionsep}{\oldlottproductionsep}
}
}
\global\let\lottproduction\lottproduction
#2%
\global\let\lottproduction\oldlottproduction
\global\let\lottproductionsep\oldlottproductionsep
}
\newcommand{\lotthypothesis}[1]{#1}

\def\lottinferenceruledefaultnarrower{0.5}
\newcommand{\lottinferencerulenarrower}[1]{\ifcsname lottinferencerulenarrower#1\endcsname\csname lottinferencerulenarrower#1\endcsname\else\lottinferenceruledefaultnarrower\fi}
\newcommand{\lottinferenceruletarget}[1]{\ifcsname lottjudgementlabelprefix\endcsname\MakeLinkTarget*{rule:\lottjudgementlabelprefix-#1}\expandafter\def\csname @currentlabel\endcsname{\lottjudgementlabelprefix-#1}\label{rule:\lottjudgementlabelprefix-#1}\fi}
\newcommand{\lottinferencerulelabel}[1]{\ifcsname lottjudgementlabelprefix\endcsname\lottjudgementlabelprefix-#1\else#1\fi}
\newcommand{\lottinferencerule}[3]{\inferrule*[narrower=\lottinferencerulenarrower{#1},lab=\lottinferenceruletarget{#1}\lottinferencerulelabel{#1},leftskip=0.5em,rightskip=0.5em]{#2}{#3}}
\newcommand{\lottinferencerulesep}{\and}
\newcommand{\lottjudgement}[3]{
\noindent\fbox{$#2$}\hfill\emph{(#1)}
\nopagebreak
\begin{mathparpagebreakable}#3\end{mathparpagebreakable}%
}

\newcommand{\lottjudgementforcename}[2]{
\begingroup
\let\oldlottjudgement\lottjudgement
\renewcommand{\lottjudgement}[3]{\oldlottjudgement{#1}{##2}{##3}}
#2%
\endgroup
}
\newcommand{\lottjudgementforceexample}[2]{
\begingroup
\let\oldlottjudgement\lottjudgement
\renewcommand{\lottjudgement}[3]{\oldlottjudgement{##1}{#1}{##3}}
#2%
\endgroup
}
\newcommand{\lottjudgementforcenameandexample}[3]{
\begingroup
\let\oldlottjudgement\lottjudgement
\renewcommand{\lottjudgement}[3]{\oldlottjudgement{#1}{#2}{##3}}
#3%
\endgroup
}
\newcommand{\lottjudgementfilter}[2]{
\begingroup
\let\oldlottinferencerule\lottinferencerule
\let\oldlottinferencerulesep\lottinferencerulesep
\renewcommand{\lottinferencerulesep}{}
\renewcommand{\lottinferencerule}[3]{
\IfSubStr{,#1,}{,##1,}{
\oldlottinferencerule{##1}{##2}{##3}
\renewcommand{\lottinferencerulesep}{\oldlottinferencerulesep}
}{
\renewcommand{\lottinferencerulesep}{}
}
}
#2%
\endgroup
}
\newcommand{\lottjudgementfilterout}[2]{
\begingroup
\let\oldlottinferencerule\lottinferencerule
\let\oldlottinferencerulesep\lottinferencerulesep
\renewcommand{\lottinferencerulesep}{}
\renewcommand{\lottinferencerule}[3]{
\IfSubStr{,#1,}{,##1,}{
\renewcommand{\lottinferencerulesep}{}
}{
\oldlottinferencerule{##1}{##2}{##3}
\renewcommand{\lottinferencerulesep}{\oldlottinferencerulesep}
}
}
#2%
\endgroup
}
\newcommand{\lottjudgementlabel}[2]{
\begingroup
\def\lottjudgementlabelprefix{#1}
#2%
\endgroup
}
\newcommand{\lottref}[2]{\textsc{\ref{rule:#1-#2}}}
\makeatletter
\kv@set@family@handler{lottnarrower}{
\expandafter\def\csname lottinferencerulenarrower#1\endcsname{#2}
}
\makeatother
\newcommand{\lottjudgementnarrower}[3][]{
\begingroup
\def\lottinferenceruledefaultnarrower{#2}
\kvsetkeys{lottnarrower}{#1}
#3%
\endgroup
}
\newenvironment{lottgrammar}{%
\tabularx{\linewidth}{r c >{\raggedright}X r}
}{%
\endtabularx
}
\newwrite\lottembedoutput
\newcommand{\lottembed}[1]{%
\def\lottembedfilebase{\lottembedfileprefix\pdfmdfivesum{#1}}%
\IfFileExists{\lottembedfilebase.tex}{$\input{\lottembedfilebase.tex}$}{%
\immediate\openout\lottembedoutput=\lottembedfilebase.lottinput%
\immediate\write\lottembedoutput{#1}%
\immediate\closeout\lottembedoutput%
\textcolor{red}{$#1$}%
}%
}
\newcommand{\lottignore}[1]{\relax}
 \def\lottembedfileprefix{lott-embed/}
\usepackage{tabularx}
\usepackage{bbding}
\usepackage{scalerel}
\usepackage{unicode-math-input}
\usepackage{stackengine}

\newcommand\doubleplus{+\kern-0.6ex+}

\umiDeclareMathChar{●}{\bullet}
\umiDeclareMathChar{◇}{\diamond}
\umiDeclareMathChar{▹}{\triangleright}
\newcommand{\strength}{\scaleobj{0.6}{\text{\FiveStar}}}
\newcommand{\rstrength}{\mathrlap{\scaleobj{0.6}{\text{\FiveStar}}}{\kern0.125em\raisebox{0.125em}{\scaleobj{0.35}{\mathcolor{white}{\text{\FiveStar}}}}}}

\setcopyright{none}
\renewcommand\footnotetextcopyrightpermission[1]{}

\begin{document}

\title{Record Types Everywhere}
\subtitle{A Record Calculus for Arguments and First-Class Environments}

\author{Matthew Toohey}
\orcid{0009-0006-2253-7193}
\affiliation{%
  \institution{University of Toronto}
  \city{Toronto}
  \country{Canada}
}
\email{mtoohey@cs.toronto.edu}

\author{Ara Jamalzadeh}
\orcid{0009-0007-4728-8100}
\affiliation{%
  \institution{University of Toronto}
  \city{Toronto}
  \country{Canada}
}
\email{a.jamalzadeh@mail.utoronto.ca}

\author{Ningning Xie}
\orcid{0000-0002-5961-1493}
\affiliation{%
  \institution{University of Toronto}
  \city{Toronto}
  \country{Canada}
}
\email{ningningxie@cs.toronto.edu}

\begin{abstract}
  This paper introduces a novel calculus that models first-class environments
  and function application paradigms using standard record types. By proposing
  new extensions to general record types, we achieve a design that is remarkably
  simple and principled. We formalise this design in a core calculus and prove
  several desirable properties, including type soundness, normalization, and a simulation of
  the simply typed lambda calculus. Additionally, we extend the design with mutable
  references. The calculus, its semantics, along with the extension, and all
  stated properties are fully mechanized in the Lean 4 proof assistant,
  alongside an interpreter formally verified to be equivalent to the
  operational semantics.
\end{abstract}

\maketitle

\section{Introduction}

The lambda calculus~\cite{church1941calculi} provides a foundational model for
computation and serves as the theoretical basis for modern functional
programming languages. A lambda abstraction $\lambda x.~e$ denotes a function
binding a variable $x$ within the body expression $e$, with its core evaluation
step defined by the $\beta$-reduction rule via substitution: $(\lambda x.~e_1) ~
e_2 \longrightarrow e_1[e_2/x]$.

However, managing variable bindings remains a classic challenge in language
implementation, reasoning, and mechanization. In particular, ensuring variable
freshness can be quite tricky. While paper formalisms and pen-and-paper proofs often
assume $\alpha$-equivalence,
concrete implementations typically require more complex machinery
such as de Bruijn indices~\cite{de1972lambda} to define capture-avoiding
substitution. De Bruijn indices, however, are famously difficult to formalise
and implement correctly. Moreover, substitution and de Bruijn indices are often inefficient
because reduction requires a traversal of the expression body. Finally,
$\beta$-reduction fails to capture many common argument-passing paradigms such
as \textit{named arguments}, which allow arguments to be supplied by explicit
parameter names.

In this paper, we show how standard record types can be leveraged to model
\textit{first-class environments}~\cite{gelernter1987environments,
  sato2001explicit, tan2024case} and function arguments, establishing an
\textit{environment-based model} that sidesteps traditional binding and
inefficiency issues while capturing diverse function application paradigms.
Our design is remarkably simple and principled; it uncovers subtle interactions
that emerge when combining features and provides a general framework for
function application paradigms by exploiting inherent record properties.

We offer the following contributions:

\begin{itemize}
\item \textbf{A novel design:} We propose a design that models
  first-class environments and function arguments using standard record types
  (\Cref{sec:overview}). Specifically, our design:
  \begin{itemize}
  \item Is simple and principled, yet naturally expresses
    diverse features, including first-class
    environments~(\Cref{sec:overview-env}), along with named
    (\Cref{sec:overview-named}-\Cref{sec:overview-argument-labels}), positional
    (\Cref{sec:positional}), and optional
    (\Cref{sec:overview-optional}) arguments, while supporting a form of
    partial application~(\Cref{sec:partial-app}, \Cref{sec:overview-optional}).

  \item Generalizes traditional function application paradigms to general
    records (\Cref{sec:overview-optional-values},\Cref{sec:overview-overwriting}); and
  \item Unifies disparate approaches to handling duplicate labels in record
    calculi~\cite{leijen2005extensible,remy1993type} (\Cref{sec:overview-overwriting}).
  \end{itemize}

\item \textbf{Formalisation:} We formalise this design as a core record
  calculus, \lang (\Cref{sec:calculus}). We establish key structural
  properties of its record operators
  (Lemma~\ref{lemma:sub-refl}-\ref{lem:app-con-assoc}), determinacy (\Cref{thm:determinacy}), type soundness
  (\Cref{thm:preservation,thm:progress}), and normalization
  (\Cref{thm:normalization}).
  We also prove that it simulates the simply typed lambda calculus
  (\Cref{thm:typ-conserve,thm:sem-correspondence}).

\item \textbf{Imperative extension:}
  We extend the calculus with mutable references, under which the precise
  behaviour of function application becomes important (\Cref{sec:references}). To
  safely handle type generalization in the presence of side-effects, we adopt
  the approach of \citet{dolan2024rethinking} and adapt it to closures
  and applications.

\item \textbf{Mechanization:} The calculus, including its operational semantics
  and all stated lemmas and theorems, is fully mechanized in the Lean 4 proof
  assistant~(\Cref{sec:mech}).
  The complete proof artifact is provided in the supplementary materials.

\item \textbf{Implementation:} We implement an interpreter for \lang and
  formally verify its equivalence with the operational
  semantics~(\Cref{thm:interpreter-correctness}).

\end{itemize}

\noindent
Lastly, we explore the broader design space in
\Cref{sec:discussion}.

\section{Overview}
\label{sec:overview}

This section provides an overview of our design. We frequently draw on examples
from various programming languages, ensuring the specific language is
clear from the context. For further clarity, programs written in our own syntax
are enclosed in gray boxes.

\subsection{The Lambda Calculus with the Substitution Model}

Functions, or lambdas, are the core abstraction in the lambda
calculus~\cite{church1941calculi}. Its two fundamental rules, governing typing
and evaluation respectively, are presented below:

\begin{center}
\small
\begin{minipage}{0.5\textwidth}
\[
  \inferrule*[right=lam]{\Gamma, x:A \vdash e : B}
  {\Gamma \vdash \lambda x.~e : A \to B}
  \customlabel{lam}{rule:lam0}
\]
\end{minipage}
\hfill
\begin{minipage}{0.45\textwidth}
  $ (\lambda x.~e_1) ~ e_2 \longrightarrow e_1[e_2/x]$ \enskip \textsc{\small beta}
  \customlabel{beta}{rule:beta}
\end{minipage}
\end{center}

\noindent
Rule \ruleref{rule:lam0} type-checks a function by extending the typing context
$\Gamma$ with the argument's type to type-check the body $e$, resulting in the
function type $A \to B$. The evaluation rule \ruleref{rule:beta}
substitutes the variable $x$ with the argument $e_2$ within the body $e_1$.

However, while these two rules are standard in presenting the lambda calculus,
they are not without issues. We discuss below some of the challenges of reasoning
about~and~implementing~them.

\paragraph{\textbf{Reasoning about bindings}}
\label{sec:overview-bindings}

Reasoning about variable bindings is notoriously difficult, as substitution
complicates the metatheory. The substitution function $e_1 [e_2/x]$ must itself be
defined with care: when substituting within a lambda $\lambda y.~e$, it must
respect shadowing when $x$ and $y$ coincide and, when reducing open terms, avoid
capturing the free variables of $e_2$. Type preservation then rests on a
substitution lemma.

To manage bindings, it is common to assume the alpha-convention, where
a bound variable can be renamed at any point to ensure its freshness. While this
convention is standard in the literature, it does not directly yield a
straightforward algorithm for implementation or mechanization.

Another well-known alternative is the use of de Bruijn
indices~\cite{de1972lambda}, where variables are represented as natural numbers
pointing to their corresponding binding sites. However, it is widely
acknowledged that implementing and reasoning about shifting, the key operator in
de Bruijn indices used to adjust binding numbers during substitution, can be
error-prone and non-trivial.

\paragraph{\textbf{Inefficiency of the substitution model}}

Another issue is that the substitution model is highly inefficient. Every
substitution requires a full traversal of the expression, a process that becomes
exceptionally expensive for functions with large bodies. Worse, nested
applications trigger this substitution process multiple times, leading to a
compounding cost.

Due to this inherent inefficiency, practical languages do not implement
beta-reduction using the substitution model. Instead, it is much more common to
utilize an \textit{environment-based model}. Rather than eagerly rewriting the
function body every time a variable is bound, the system maintains an
environment that maps variable names to their actual values. Consequently, when
a function is applied to an argument, the language simply adds that
variable-value pairing to the current environment and proceeds to evaluate the
function body. By avoiding substitution, an environment-based model also
simplifies the metatheory~\cite[\S2.2 and \S4.1]{tan2024case}.

In the literature, a body of work has focused on modelling environments, ranging
from calculi with explicit substitutions~\cite{abadi1989explicit,
  curien1991abstract,biernacka2007concrete} to the study of \textit{first-class
  environments}~\cite{gelernter1987environments, sato2001explicit, tan2024case}.

\paragraph{\textbf{Missing function application paradigms}}

Furthermore, the vanilla lambda calculus does not natively support
various function application paradigms common in practical programming languages.
For example, many languages
support \textit{named arguments}, which significantly enhance programming
convenience by allowing arguments to be provided with explicit names. To
illustrate, consider a function written in Python syntax:
\begin{lstlisting}[style=python]
def transfer_funds(amount, from_account, to_account): ...
\end{lstlisting}
When a programmer wants to call this function using standard positional
arguments, they must remember the exact order of the parameters:
\begin{lstlisting}[style=python]
transfer_funds(1000, "Alice", "Bob")
\end{lstlisting}
Without looking up the function's definition, this code is difficult to read and
highly error-prone. If a developer accidentally swaps \lstinline[style=python]{"Alice"} and
\lstinline[style=python]{"Bob"}, the code will still execute, but the funds will
be transferred from and to the wrong person.

Languages that support named arguments, such as Python, allow programmers to
specify the name of the parameter directly alongside the value at the call site:
\begin{lstlisting}[style=Python]
transfer_funds(amount=1000, from_account="Alice", to_account="Bob")
\end{lstlisting}
This way, it becomes immediately clear from the call site which
account is transferring to which.

Named arguments represent one example of function application paradigms; other
important examples include, e.g. optional arguments, which we will discuss
further in subsequent sections.

\subsection{First-Class Environments as Records}
\label{sec:overview-env}

In this paper, we develop an environment-based model for the lambda calculus by
representing environments as records. This perspective is motivated by the close
similarity between a typing context $\Gamma$, which maps variables to their
types, and a record type ${\rho}$, which maps labels to types. Consequently, we
can update the typing rule \ruleref{rule:lam0} to rule \ruleref{rule:lam1} below:

\begin{center}
\begin{minipage}{0.4\textwidth}
\[
  \fcolorbox{white}{blue!10!white}{
  $\begin{array}{llll}
  \text{record type} & \rho & \Coloneqq & \varepsilon \mid \rho, i \strong A
  \end{array}$
  }
\]
\end{minipage}
\begin{minipage}{0.45\textwidth}
  \small
  \[
  \inferrule*[right=T-lam]{ \rho, i \strong A \vdash e : B }
  { \rho \vdash \lambda i. ~ e : A \to B }
  \customlabel{T-lam}{rule:lam1}
  \]
\end{minipage}
\end{center}

Records are typically expected to be commutative, a property usually captured by
a subtyping rule that permutes fields. Under this view, for example,
\lstinline[style=mylang]|{ name:String, age:Int }| is equivalent to
\lstinline[style=mylang]|{ age:Int, name:String }|.
This equivalence naturally reflects the intuition that the typing context is
\begin{wrapfigure}{r}{0.2\textwidth}
  \small
  \vspace{-5pt}
$
  \inferrule*[right=T-var]{ }
  { \rho, i \strong A \vdash i : A }
  \customlabel{T-var}{rule:var}
$
  \vspace{-20pt}
\end{wrapfigure}%
unordered. Therefore, we can formulate a
variable typing rule \ruleref{rule:var} as shown on the right, which, instead of requiring $i \in
\rho$, simply requires $i$ to be the final entry in the context.

\paragraph{\textbf{Closures}}

Similarly, during evaluation, an environment record value $o$ tracks the values
bound to variables. As a result, evaluating lambdas produces \textit{closures}
that capture the variable bindings along with the function:

\begin{center}
\begin{minipage}{0.4\textwidth}
\[
  \fcolorbox{white}{orange!10!white}{$
  \begin{array}{llll}
  \text{record value} & o    & \Coloneqq & \varepsilon \mid o, i = v
  \end{array}
  $}
\]
\end{minipage}
\begin{minipage}{0.45\textwidth}
  \[
  \inferrule*[right={\enskip E-lam}]{}
  { o \vdash \lambda i. ~e \longrightarrow \langle  o \mid \lambda i. ~e \rangle }
  \customlabel{E-lam}{rule:e-lam}
  \]
\end{minipage}
\end{center}

\noindent
When the closure is applied to an argument, the function body $e$ is then
evaluated under the environment $o$ extended with $i$ bound to that argument.

\paragraph{\textbf{First-class environments}}

With first-class environments~\cite{gelernter1987environments, sato2001explicit,
  tan2024case}, we can treat environments as first-class objects that can be
directly manipulated. For example, we can introduce an operator $@$ that reifies
the environment into a record, with the typing rule $\rho \vdash @ : \{ \, \rho \, \}
$ and the evaluation rule $ o \vdash @ \longrightarrow \{ \, o \, \} $. We can then
apply record operators, such as projection, to the environment. In particular, a
variable $i$ can be considered syntactic sugar for $@.i$, which reifies the
environment and then projects out the $i$ field.
As another example, a box operator $\envbox{e_1}{e_2}$
evaluates $e_2$ within the environment $e_1$, allowing us to directly manipulate
the environment in which an expression is evaluated.

As demonstrated by \citet{tan2024case}, first-class environments are useful for
modelling first-class modules~\cite{cardelli1997program}, sandboxed modules, and
object capabilities~\cite{miller2006robust,miller2000capability}. Our
representation of environments as records naturally supports these paradigms,
since record objects are themselves first-class.

\subsection{Named Arguments as Records}
\label{sec:overview-named}

Having observed how function applications with named arguments operate in our
\lstinline|transfer_funds| example, another natural perspective we take is to
represent function arguments as records.

That is, functions are now typed as $\{ \, \rho \, \} \to A$, allowing them to
introduce multiple entries into the context. We introduce record patterns $c$
with the following typing rule \ruleref{rule:lam2}:

\begin{center}
\begin{minipage}{0.4\textwidth}
\[
  \fcolorbox{white}{green!10!white}{
    $
  \begin{array}{llll}
  \text{record pattern} & c    & \Coloneqq & \varepsilon \mid c, i
  \end{array}
$
  }
\]
\end{minipage}
\begin{minipage}{0.45\textwidth}
  \small
  \[
  \inferrule*[right=T-lam2]{ c \mathcolor{Pattern}{:} \rho_2 \\ \rho_1, \rho_2 \vdash e : A }
  { \rho_1 \vdash \lambda \{ \, c \, \}. ~ e : \{ \, \rho_2 \, \} \to A }
  \customlabel{T-lam2}{rule:lam2}
  \]
\end{minipage}
\end{center}

\noindent
Here, the record pattern $c$ defines named function arguments. Rule
\ruleref{rule:lam2} assigns the type $\rho_2$ to the record pattern $c$,
and updates the typing context to $\rho_1, \rho_2$, which simply appends $\rho_2$
to $\rho_1$.

As an example, we can define \lstinline{transfer_funds} as follows:
\begin{lstlisting}[style=mylang]
\{ amount, from_account, to_account }. ...
\end{lstlisting}

Function applications can now provide records as named arguments at the call
site. For example, the previous example can be written as:
\begin{lstlisting}[style=mylang]
transfer_funds { amount=1000, from_account="Alice", to_account="Bob" }
\end{lstlisting}

Thanks to record commutativity, this is equivalent to the following expression, as
is desired:
\begin{lstlisting}[style=mylang]
transfer_funds { to_account="Bob", from_account="Alice", amount=1000 }
\end{lstlisting}

\paragraph{\textbf{Shadowing in first-class environments}}

One may wonder what happens when $\rho_1$ and $\rho_2$ in rule
\ruleref{rule:lam2} share same labels. We support \textit{duplicate labels} (or
\textit{scoped labels}~\cite{leijen2005extensible}), which, as we represent
environments as records, arise naturally in an environment
when \textit{shadowing} occurs.

Duplicate labels are non-commutative with respect to each other, while remaining
commutative with different labels. Specifically, projecting a duplicate label
from a record returns the rightmost corresponding field (unlike
\cite{leijen2005extensible}, which always returns the leftmost one). For
example, \lstinline[style=mylang]|{ i=2, i=1~}.i| returns \lstinline|1|.
To access the second rightmost one, we can simply write
\lstinline[style=mylang]|({ i=2, i=1~}\-i).i|, which removes (\textbackslash) the rightmost
\lstinline{i} field from the record, and then accesses the \lstinline{i} field,
which effectively retrieves the shadowed \lstinline{i} value \lstinline{2}.

The interpretation that projecting a duplicate label retrieves the rightmost
corresponding field directly coincides with the fact that new variables are
always appended to the right of an environment (as in rule \ruleref{rule:lam2}),
thus shadowing prior ones.
Therefore, our previous syntactic sugar $@.i$ for variable lookup
still holds, as it accesses the latest $i$ field in the environment. With
first-class environments, we can also easily access shadowed fields, e.g. by
$(@\textbackslash i).i$.

\subsection{Argument Labels and Parameter Names}
\label{sec:overview-argument-labels}

We can further separate external \textit{argument labels}, used at the call
site, from internal \textit{parameter names}, used within the function
definition. This distinction allows a function call to read like a sentence,
while keeping the function definition itself clear and logical.

The following Swift program, adapted from its official documentation,
illustrates this distinction between argument labels and parameter names (the \lstinline{-> String} annotation specifies the~return~type):
\begin{lstlisting}[style=swift]
func greet(person: String, from hometown: String) -> String  {
  "Hello \(person) from \(hometown)." }
\end{lstlisting}
Here, \lstinline{person} and \lstinline{hometown} bind to the parameters within
the function body, while \lstinline{from} serves as the argument label used at
the call site.
We can then call the function as:
\begin{lstlisting}[style=swift]
greet(person: "Alice", from: "Mars")  // "Hello Alice from Mars"
\end{lstlisting}

To support the distinction between argument labels and parameter names, we can
generalize both record patterns and the typing rule \ruleref{rule:lam2} to rule
\ruleref{rule:lam3}:

\begin{center}
\begin{minipage}{0.45\textwidth}
\[
  \fcolorbox{white}{green!10!white}{
  $\begin{array}{llll}
  \text{record pattern} & c    & \Coloneqq & \varepsilon \mid c, i_1 \diamond i_2
  \end{array}
  $
  }
\]
\end{minipage}
\begin{minipage}{0.45\textwidth}
  \small
  \[
  \inferrule*[right=T-lam3]{ \patsep{c}{\rho_2}{\rho_3} \\ \rho_1, \rho_3 \vdash e : A }
  { \rho_1 \vdash \lambda \{ \, c \, \}. ~ e : \{ \, \rho_2 \, \} \to A }
  \customlabel{T-lam3}{rule:lam3}
  \]
\end{minipage}
\end{center}

\noindent
Here, the record pattern $c$ captures the distinction ($\diamond$) between argument labels and
parameter names, which are separated ($\mathcolor{Pattern}{\odot}$) in \ruleref{rule:lam3} and
assigned to record types $\rho_2$ and $\rho_3$. For example, $i_1 \diamond i_2$ can be
separated into $i_1\,\strong$ \lstinline{Int} and $i_2\,\strong$ \lstinline{Int}. The former,
$\rho_2$, represents the external argument labels and is used as the argument
type in the conclusion, while the latter, $\rho_3$, represents the internal
parameter names and extends the typing context for typing $e$. The
corresponding rule in \Cref{sec:typing} additionally
performs generalization over the function type; we omit this here
for simplicity.

We can now express the example as follows, assuming string interpolation:
\begin{lstlisting}[style=mylang]
\{ person, from <> hometown }. "Hello \(person) from \(hometown)."
\end{lstlisting}
For clarity, we often omit the argument label (e.g. \lstinline{person})
when it is identical to the~parameter~name.

\subsection{Positional Arguments}
\label{sec:positional}

The design so far requires every argument to be named. However,
\textit{positional arguments} remain valuable when parameter roles are intuitive
enough that explicit labels would only add visual noise. For instance, a
function \lstinline{sub(x, y)} is naturally expected to compute
\lstinline{(x - y)} rather than \lstinline{(y - x)}.

Languages that support named arguments often provide mechanisms for positional
arguments as well. In Swift, for example, function arguments are named at the call
site by default.
When an argument label is unnecessary, Swift developers can use an underscore
(\lstinline{_}) as the argument label,
For example, one can define
\lstinline{sub} in Swift as the following:
\begin{lstlisting}[style=swift]
func sub(_ a: Int, _ b: Int) -> Int  { a - b }
\end{lstlisting}
Here the underscores indicate that the function can be called with positional
arguments. This allows the argument labels to be omitted at the call site, e.g.
\lstinline[style=swift]{sub(8, 3)}.

It turns out we can naturally support positional arguments
using
\textit{duplicate labels} (\Cref{sec:overview-named}).
Specifically, by treating the underscore (\lstinline{_}) as a normal field
label, we can define \lstinline{sub} as
\begin{lstlisting}[style=mylang]
sub = \{ _ <> a, _ <> b }. a - b
\end{lstlisting}
The positional application \lstinline[style=mylang]|sub { 8, 3 }| can then be desugared simply
into
\lstinline[style=mylang]+sub { _ = 8, _ = 3 }+,
where the argument \lstinline[style=mylang]|{ _ = 8, _ = 3 }| is a record with duplicate labels
(\lstinline{_}). Since duplicate labels do not commute with respect to each
other, this effectively models positional arguments.

\paragraph{\textbf{Mixing named and positional arguments}}

By interpreting positional arguments as normal records with duplicate labels,
our design interprets positional arguments as simply a special case of named arguments.
As noted, non-duplicate labels are
commutative, while duplicate labels (including the dedicated label for
positional arguments) are non-commutative with respect to each other (thus
preserving the significance of their order), yet remain commutative with other
labels.

As an example, if \lstinline{div} is defined with a named \lstinline{dividend}
 and a positional \lstinline{divisor}:
\begin{lstlisting}[style=mylang]
div = \{ dividend , _ <> divisor }. dividend / divisor
\end{lstlisting}
then both \lstinline[style=mylang]|(div { dividend=4, 2~})| and
\lstinline[style=mylang]|(div { 2, dividend=4~})| are valid and produce the same
result, while \lstinline[style=mylang]|(div { 4, 2 })| is invalid,
as named arguments can never be passed positionally.

Our design preserves simplicity by, like Swift, requiring arguments to be
declared as named or positional (simply using a dedicated label) in the function
definition, making the system easy to reason about. But unlike Swift, which
requires the order of arguments at the call site to strictly match the order
defined in the function declaration, our system ensures commutativity between
distinct labels, allowing for expressive order mixing at the call site. This is
particularly useful when combined with partial application, which we discuss
below.

\subsection{Partial Application}
\label{sec:partial-app}

So far, we have assumed that all arguments are passed together in a single
record. We now generalize this behaviour to support partial application, where
only a subset of the arguments are provided at the call site.
Partial application is particularly flexible when combined with named arguments
and commutativity. For example, consider a function \lstinline{drop} that takes
a \lstinline{count} and a \lstinline{list}, and drops \lstinline{count} elements
from the list. We can fix the list argument and define a
specialized \lstinline{dropN} function simply by
writing \lstinline[style=mylang]+drop { list = ['a', 'b', 'c', 'd'] }+.

To achieve partial application,
we introduce the following typing rule for function application:
\begin{mathpar}
  \small
  \inferrule*[right=T-app]{ \rho \vdash e_1 : \{ \, \rho_1 \, \} \to A \\
    \rho \vdash e_2 : \{ \, \rho_2 \, \} \\
    \app{\rho_1}{\rho_2}{\rho_3}
  }
  { \rho \vdash e_1 ~ e_2 : \{ \, \rho_3 \, \} \to A }
  \customlabel{T-app}{rule:app}
\end{mathpar}
Here, the function $e_1$ expects the arguments specified by
$\rho_1$, while the argument record $e_2$ provides $\rho_2$.
The judgement
$\app{\rho_1}{\rho_2}{\rho_3}$ computes the remaining arguments, $\rho_3$.
In this case, it simply removes $\rho_2$ from $\rho_1$, though the application
process will become more sophisticated later.
Consequently,
the application yields a function that accepts the remaining arguments, typed as
$\{ \, \rho_3 \, \} \to A$.

This rule can be refined in several ways. First, if $\rho_2$ is empty, the
application does not change the function type at all. To eliminate such ``dummy''
applications, we can introduce the side condition $\rho_2 \neq \varepsilon$. Second,
when $\rho_3$ is empty, the resulting type is $\{ \, \} \to A$. In this scenario,
the application is actually saturated, allowing us to simplify the
type to $A$; we will revisit this design choice in
\Cref{sec:overview-optional}.

We now extend the evaluation rules to support partial application. Specifically,
we extend closures to keep track of two record values: one for the environment,
and the other for the arguments accumulated so far. This behaviour is captured by
the following rules:
\begin{mathpar}
  \inferrule*[right={\enskip E-Lam2}]{}
  { o \vdash \lambda \{ c \}. ~e \longrightarrow \langle  o ; \varepsilon \mid \lambda \{c\}. ~e \rangle }
  \customlabel{E-lam2}{rule:e-lam2}
  \and
  \inferrule*[right={\enskip E-App}]{}
  { \langle  o_1 ; o_2  \mid \lambda \{ c \}. ~e \rangle ~ \{ o_3 \}
    \longrightarrow  \langle  o_1 ; o_2, o_3  \mid \lambda \{ c \}. ~e \rangle }
  \customlabel{E-app}{rule:e-app}
\end{mathpar}
When a function abstraction evaluates to a closure, it initializes an empty
record $\varepsilon$ for the arguments (rule \ruleref{rule:e-lam2}). Subsequent
function applications then simply accumulate the provided arguments into this
record (rule \ruleref{rule:e-app}). This dual-record setup will serve us well in
the next section. Once an application is fully saturated, a beta reduction
can occur using all collected arguments~(see~also~\Cref{sec:overview-optional}).

\subsection{Optional Arguments}
\label{sec:overview-optional}

In many programming languages, the developer can specify a default value for
parameters. For example, we may want to define an exponential function that
squares its input by default:

\begin{lstlisting}[style=mylang]
power = \{ base, exponent := 2 }. base ** exponent
\end{lstlisting}

\noindent
Calling \lstinline[style=mylang]|(power {~base=3~})| returns \lstinline{9}, while calling
\lstinline[style=mylang]|(power {~base=3, exponent=3~})| returns \lstinline{27}.
We refer to arguments with default values as \textit{optional arguments}, as
they can be omitted at the call site, while other parameters are considered
\textit{required arguments}.

Integrating optional arguments into the record
pattern is straightforward; we extend $c$ to include an additional case where a
default value $v$ can be provided.
\[
  \fcolorbox{white}{green!10!white}{
  $\begin{array}{llll}
  \text{record pattern} & c    & \Coloneqq & \varepsilon \mid c,i_1 \diamond i_2 \mid c, i_1 \diamond i_2 := v
  \end{array}
  $
  }
\]
We now discuss the typing and evaluation rules for optional arguments.

\paragraph{\textbf{Typing optional arguments}}

Since functions with optional arguments such as \lstinline{power} can be called
with only the required arguments or with optional arguments overwriting the
default values, their type signatures must reflect both arities, so that both
\lstinline[style=mylang]+(power {~base=3, exponent=3~})+
and \lstinline[style=mylang]+(power {~base=3~})+
type-check and saturate
the application.

To this end, we introduce \textit{optional types} in record types:
\[
  \fcolorbox{white}{blue!10!white}{$
  \begin{array}{llllllll}
  \text{record type} & \rho & \Coloneqq & \varepsilon \mid \rho, i \star A \hspace{40pt} &
  \text{strength}    & \star & \Coloneqq & \strong \; \mid \option \\
  \end{array}
  $}
\]
A record field $i$ is now associated with a \textit{strength} $\star$, which is
either \textit{strong} ($\strong$) or \textit{optional} ($\option$).
When used in function argument types,
$(\strong)$ indicates a required argument and $(\option)$ indicates an optional argument.
Then, to type-check a lambda definition with optional arguments, the
separation operation ($\mathcolor{Pattern}{\odot}$) in the typing rule \ruleref{rule:lam3}
can be extended to handle the additional case for $c$ and produce optional
types in $\rho_2$.
For example, the \lstinline{power} function can be typed as
\begin{lstlisting}[style=mylang]
{ base |> Int, exponent ? Int } -> Int
\end{lstlisting}
The type indicates that \lstinline{base} is a required argument while
\lstinline{exponent} is an optional argument.

To type-check function applications
such as
\lstinline[style=mylang]|(power { base=3, exponent=3 })|
and \lstinline[style=mylang]|(power { base=3 })|,
we
introduce subtyping rules between records. Importantly, the subtyping relation
includes the following two rules (omitting other rules such as reflexivity):
\begin{mathpar}
  \small
  \inferrule*[right=S-empty]{
    {\lottsym{⌊} ρ \lottsym{⌋}}^{\lottsym{?}}
  }
  {\lottsym{ε} \, \lottsym{\mathcolor{Subtyping}{⩽}} \, ρ}
  \customlabel{S-empty}{rule:empty}
  \and
  \inferrule*[right=S-ext]{
    ρ₀ \, \lottsym{\mathcolor{Subtyping}{⩽}} \, ρ₁ \lottsym{,} \, ρ₂\\
    i \, \lottsym{∉} \, ρ₁ \\
    A \, \lottsym{\mathcolor{Subtyping}{⩽}} \, B
  }
  {i \, \strong \, A \lottsym{,} \, ρ₀ \, \lottsym{\mathcolor{Subtyping}{⩽}} \, ρ₁ \lottsym{,} \, i \, \strength \, B \lottsym{,} \, ρ₂}
  \customlabel{S-ext}{rule:ext}
\end{mathpar}
In the first subtyping rule \ruleref{rule:empty}, the premise ${\lottsym{⌊} ρ
  \lottsym{⌋}}^{\lottsym{?}}$ ensures that all entries in $ρ$ have the optional
strength, and the conclusion specifies that the empty record is a subtype of
such a record. In the second rule \ruleref{rule:ext},
the premises are standard,
and the conclusion specifies that for the same field $i$, the left type has strength
$(\strong)$, while the right type has strength $(\strong)$ or $(\option)$.

With these subtyping rules, we can derive two subtyping relations on the type
of \lstinline{power}. First,
\lstinline[style=mylang]+{ base |> ~Int, exponent ? Int } ~->~ Int ~ ⩽ ~ { base |> ~Int, exponent |> ~Int } -> Int+.
This relation holds because
\lstinline[style=mylang]+{ base |> ~Int, exponent |> ~Int } ~ ⩽ ~ { base |> ~Int, exponent ? Int }+
by rule \ruleref{rule:ext}. Thus \lstinline{power} can take two
arguments, allowing \lstinline[style=mylang]+(power {~base=3, exponent=3~})+ to
type-check.

Second, we can derive:
\lstinline[style=mylang]+{ base |> ~Int, exponent ? Int } ~->~ Int ~⩽~ { base |> ~Int } -> Int+.
This relation holds because
\lstinline[style=mylang]+{ base |> ~Int } ~ ⩽ ~ { base |> ~Int, exponent ? Int }+ by
\ruleref{rule:ext}
and
\ruleref{rule:empty}. Thus, \lstinline{power} can also take a single
argument, allowing \lstinline[style=mylang]|(power {~base=3~})| to
type-check.

\paragraph{\textbf{Evaluation with optional arguments}}

We now discuss the evaluation rule for function application with optional
arguments. A subtle aspect is that these default values can be
overwritten at the call site. Therefore, we must distinguish between arguments
that can be updated and those that cannot. To achieve this, we update record values as follows:
\begin{center}
\vspace{-4pt}
\begin{minipage}{0.4\textwidth}
\[\arraycolsep=1.9pt
  \fcolorbox{white}{orange!10!white}{$
  \begin{array}{llll}
  \text{record value} & o    & \Coloneqq & \varepsilon \mid o, i = v \mid o, i \weak v
  \end{array}
  $}
\]
\end{minipage}
\hfill
\begin{minipage}{0.57\textwidth}
\begin{mathpar}
  \inferrule*[right={\enskip E-app2}]{}
  { \langle  o_1 ; o_2  \mid \lambda \{ c \}. ~e \rangle ~ \{ o_3 \}
    \longrightarrow  \langle  o_1 ; o_2 \concat o_3  \mid \lambda \{ c \}. ~e \rangle }
  \customlabel{E-app2}{rule:e-app2}
\end{mathpar}
\end{minipage}
\vspace{4pt}
\end{center}
Here, $i=v$ means that the value for the label has been set, while $i\weak v$ means
that $i$ has a default value $v$, which can be overwritten when the caller
provides an explicit value.

Consequently,
the previous evaluation rule \ruleref{rule:e-lam2} can be
straightforwardly updated to
retrieve all default parameters from the record pattern $c$ and use them to form the
initial argument list with $i:v$.
The updated evaluation rule \ruleref{rule:e-app2} above now uses a record
concatenation operation, $o_2 \concat o_3$, instead of simply $(o_2,
o_3)$ as in rule \ruleref{rule:e-app}. Intuitively, this operation allows
arguments in $o_3$ to overwrite the default ones in $o_2$, so that $(i\weak 1) \concat (i=2)$ results in $i=2$.

\paragraph{\textbf{Partial application with optional arguments}}

Now that we have the ability to overwrite optional arguments, a natural question
arises: is the following expression well-typed?
\begin{lstlisting}[style=mylang]
power { base = 3 } { exponent = 3 }
\end{lstlisting}
In other words, is \lstinline[style=mylang]|power {~base=3~}| a partially
applied function that still waits for a potential value to overwrite the optional
argument, or is it a saturated application (using the default value) that
immediately returns the evaluation result \lstinline{9} and therefore cannot
accept more arguments?

Swift does not natively support partial application. On the other hand, OCaml,
for example, requires a positional argument to appear after any optional
arguments, so that it can determine when those optional arguments should be
``erased'' and replaced with their default values. If an optional argument is
placed last, OCaml issues a warning as it cannot determine when the caller has
finished providing arguments. In this case, OCaml would issue this warning for
the definition~of~\lstinline{power}.

Our design prioritizes simplicity by requiring an explicit construct,
denoted as $\force$, to deterministically trigger the evaluation of an
expression once it has received all its arguments, as given by rule
\begin{wrapfigure}{r}{0.25\textwidth}
\vspace{-10pt}
$
  \inferrule*[right=T-force]{\rho \vdash e : \{\,\} \to A}{ \rho \vdash \force ~ e : A }
  \customlabel{T-force}{rule:force}
$
\vspace{-20pt}
\end{wrapfigure}
\ruleref{rule:force}.
The rule states that when an expression $e$ has received all its arguments and
has type $\{\,\} \to A$, $\force~e$ explicitly triggers the computation
to yield a result of type $A$.
The use of an explicit $\force$ is somewhat reminiscent of
\textit{call-by-push-value}~\cite{levy2001cbpv,levy2006call}, which
distinguishes a computation from its thunk, where the thunk is a value that can
be forced at any time. In our setting, a partially applied function can be
viewed as always producing a value (since it merely accumulates arguments
via the evaluation rule \ruleref{rule:e-app2}), while its actual computation is
only triggered by an explicit force.

Now, \lstinline[style=mylang]|(power {~base=3~})| can be considered
either a partially applied function, meaning that
\lstinline[style=mylang]|(power {~base=3~} ~~{~exponent=3~})| is well-typed, or a
saturated application. This works because, in either case, the actual
computation has not yet occurred, and \lstinline[style=mylang]|power {~base=3~}|
simply evaluates to a closure $\langle base=3, exponent \weak 2 \mid \lambda \{ \, base,
exponent \, \} .~ base *\!\!*
~exponent \rangle$. To trigger evaluation, an explicit force is required:
\lstinline[style=mylang]|(force (power {~base=3~}))| triggers the computation and
returns \lstinline{9} by using the default value for the optional argument,
whereas \lstinline[style=mylang]|(force (power {~base=3~} ~~{~exponent=3~}))| first
sets \lstinline{exponent} to \lstinline{3} and then triggers the evaluation to
return \lstinline{27}.

\subsection{Optional Values}
\label{sec:overview-optional-values}

So far, we have developed a system that integrates named, positional, and
optional arguments, introducing both optional types and the necessary subtyping
rules to handle them properly.

However, one aspect of this design remains somewhat unsatisfying: although we
introduced the optional type $(i \option A)$, it is currently only utilized on the
left-hand side of a function type, such as \lstinline[style=mylang]|{~i ? Int~} ~->~Int|, and no
standalone value of type \lstinline[style=mylang]|{~i ? Int~}| actually exists. Yet, how might such a
value be useful?
In the presence of duplicate labels, it turns out there is a subtle interaction
between the subtyping rules and duplicate labels,
which can be resolved by such values.
Specifically, suppose we have a function
\lstinline[style=mylang]|(f : {~i ? Int, i ? Int } ~-> ~Int)| that takes two
optional arguments sharing the same label \lstinline[style=mylang]{(i ? Int)}.
This situation can easily arise when \lstinline{i} is the positional label
\lstinline{(_)}. We assume the arguments
have default values of \lstinline{3} and \lstinline{5}, respectively.
By subtyping, we have:
\begin{lstlisting}[style=mylang]
{ i ? Int, i ? Int } -> Int ⩽ { i |> Int } -> Int
\end{lstlisting}
This holds because \lstinline[style=mylang]+{ i |> Int } ~⩽ ~{ i ? Int, i ? Int }+.
The derivation on the right first applies
\begin{wrapfigure}{r}{0.31\textwidth}
  \vspace{-8pt}
  \small
  $
  \inferrule*[right=\ruleref{rule:ext}]{
    \inferrule*[right=\ruleref{rule:empty}]{ }
    {
      \epsilon  \sub i \option \mathbf{Int}
    }
  }{ \ghl{i \strong \mathbf{Int}} \sub \ghl{i \option \mathbf{Int}}, i \option \mathbf{Int} }
  $
  \vspace{-8pt}
\end{wrapfigure}%
rule \ruleref{rule:ext}
to match ($i \strong \mathbf{Int}$) with the first ($i \option
\mathbf{Int}$), and then applies rule \ruleref{rule:empty} to relate the empty record
with the second ($i \option \mathbf{Int}$).

Now, because
\lstinline[style=mylang]+(f : { i |> Int } ~-> ~Int)+ holds via subtyping, we can
type-check the application
\lstinline[style=mylang]|(f { i = 1 })|
which provides
the value \lstinline{1} for an optional argument. According to the subtyping derivation,
\lstinline[style=mylang]|(f {~i=1~})|
always overwrites the first optional \lstinline{i} rather than the second.
However, if
we wanted to overwrite the second optional \lstinline{i} while leaving the first
to take its default value, the current rules unfortunately offer no
way to achieve this.
The choice matters since it causes different behaviours,
meaning we would expect the closure of
\lstinline{f} to store \lstinline{(i=1, i:5)} in the first case, and
\lstinline{(i:3, i=1)} in the second.

To resolve this issue, we extend record values to introduce \textit{optional values},
written as $i \option$:
\[
  \fcolorbox{white}{orange!10!white}{$
  \begin{array}{llll}
  \text{record value} & o    & \Coloneqq & \varepsilon \mid o, i = v \mid o, i\weak v  \mid o, i
                                           \option \\
  \end{array}
  $}
\]
Notably, an optional value $(i \option)$ does \textit{not} carry a concrete value for
$i$, and it is assigned the record type $\{ \, i \option A \, \}$, which now possesses a
valid inhabitant. Moreover, an optional value is a first-class citizen that
can be passed to or returned from a function just like any other value.

An optional value like $\{ \, i \option \, \}$ is not particularly useful on its own, but it
becomes highly helpful in the presence of duplicate labels. Because it
occupies a specific $i$ position within a record, it can be used to explicitly
specify which $i$ label is being targeted during function application. For
instance, using the function \lstinline{f} from the previous example, we can now
write the application:
\begin{lstlisting}[style=mylang]
f { i?, i = 1 }
\end{lstlisting}
In this case, the application leaves the first $i$ label with its default value
and explicitly sets the second $i$ to \lstinline{1}.
In the evaluation rule \ruleref{rule:e-app2}, the record concatenation operation
$(o_2 \concat o_3)$ handles an optional value by leaving the values in $o_2$
unchanged. For example, $ (i\weak 3, i\weak 5) \concat \, (i \option, i=1) $ yields $(i
\weak 3,i=1)$.
Therefore, the closure of \lstinline{f} has \lstinline[style=mylang]{(i:3, i=1)}.
This use of optional values is analogous to the use of
\lstinline[style=javascript]{undefined} in JavaScript and TypeScript,
where a default parameter is used if an argument is explicitly passed as
\lstinline[style=javascript]{undefined}.

\subsection{Overwriting Arguments}
\label{sec:overview-overwriting}

In the record typing literature, approaches to handling duplicate labels generally
fall into two camps. The first employs scoped
labels~\cite{leijen2005extensible}, where both fields coexist in the record, but
one shadows the other. The second employs overwriting~\cite{remy1993type}, where
extending a record with an existing label overwrites the previous field. In our
current formalism, duplicate labels correspond to the scoped approach.
On the other hand, the behaviour of overwriting an optional argument closely
resembles the latter camp.

Taking this observation further, it turns out that we can unify both approaches
within our formalism by allowing record fields to specify whether they are
overwritable. This design offers the benefit that it subsumes optional arguments
as simply a special case of the unified design. Moreover, to our knowledge, this
is the first system to combine scoped labels and overwriting labels, where all
labels permit duplicates within a record, with overwriting behaviour managed via
fine-grained, field-level specifications.

Specifically, we extend record types one final time to introduce
the \textit{weak} strength:
\[
  \fcolorbox{white}{blue!10!white}{$
  \begin{array}{llllllll}
    \text{record type} & \rho & \Coloneqq & \varepsilon \mid \rho, i \star A \hspace{30pt} &
    \text{strength}    & \star & \Coloneqq & \strong  \mid \weak  \mid  \option \\
  \end{array}
  $}
\]
Here, $(\strong)$ and $(\option)$ are as before, and $(\weak)$ represents a new
strength which we refer to as \textit{weak}, that denotes an overwritable field.

Conveniently, we have already introduced \textit{weak values} $(i \weak
v)$ in \Cref{sec:overview-optional}, which we used there to represent
default values that can be overwritten at runtime. Therefore, the
definition of record values from \Cref{sec:overview-optional-values} (which previously
incorporated optional values) serves as our final formulation.

Altogether, our system in its final form features three record strengths, along with their
corresponding forms of record values.
First, $(i=v)$ of type $(i\strong A)$ denotes a normal field. Second,
$(i\option)$ of $(i\option A)$ denotes an optional value. The type also
indicates an optional argument on the left-hand side of a function; the
intuition is that an optional argument can accept an optional value as it
already carries a default. Finally, $(i \weak v)$ of type $(i\weak A)$
represents an overwritable field.
\Cref{sec:calculus} formally defines the static and dynamic semantics of
records. In particular, core operations, including application and
concatenation, are systematically defined based on record strengths.

Introducing the weak strength offers several benefits. First, it shifts the
decision of whether to overwrite or bind a parameter from the
\textit{parameters} to the \textit{arguments}.
Under this mechanism, functions only require parameters
of strength~$(\weak)$.
Thus, the separation operator
($\patsep{c}{\rho_2}{\rho_3}$) in typing rule \ruleref{rule:lam3} yields
records $\rho_2$ and $\rho_3$ of strength $(\weak)$.

We can also simplify record
patterns by removing patterns with default values, effectively returning to the
definition in \Cref{sec:overview-argument-labels}:
\[
   \fcolorbox{white}{green!10!white}{
  $\begin{array}{llll}
    \text{record pattern} & c    & \Coloneqq & \varepsilon \mid c, i_1 \diamond i_2
  \end{array}
  $
}\tag{from \Cref{sec:overview-argument-labels}}
\]

As an example, we can express the \lstinline{power} function with a default
exponent value of \lstinline[style=mylang]{2} simply as
\begin{lstlisting}[style=mylang]
power' = (\{ base, exponent }. base ** exponent) { exponent : 2 }
\end{lstlisting}
Here, the lambda has
type \lstinline[style=mylang]|{~base : Int, exponent : Int~}~->~Int|.
It is applied to a record containing a weak value.
As such, \lstinline{exponent} gets the value \lstinline[style=mylang]{2} while
remaining open to being overwritten, effectively yielding an optional argument;
namely, \lstinline[style=mylang]|power' : {~base : Int, exponent ? Int~}~->~Int|.

We can then apply, e.g. \lstinline[style=mylang]+(power' { exponent : 3 })+ to
overwrite the default value.
Notably, since the argument is itself an overwritable field, the resulting expression can be further
applied to overwrite the exponent value again.
On the other hand, applying \lstinline[style=mylang]+(power' { exponent = 2~})+
will fix the \lstinline{exponent} parameter.
That is, the argument decides whether it provides a (new) default value
for the parameter, or fully fixes it.

\subsection{Summary}

We have discussed first-class environments represented via records, and
demonstrated how records can also serve as function arguments, covering named,
positional, and optional arguments, supported uniformly through three strengths
over record types and values.
Our design is certainly not the only viable one in the
design space of function application paradigms. Indeed, different languages make
different design choices, as we will discuss in \Cref{sec:design-space}.
However,
while many languages feature forms of named and optional arguments,
their designs often exist only as implementation details rather than core
formalisms.

The motivation for our formulation is
threefold: first, to capture the core essence of these diverse paradigms and
demonstrate how they can be expressed in a principled manner using familiar
record types; second, to uncover the subtle interactions that arise when
different features are combined; and third, to provide a framework that yields
several advantages
in terms of record commutativity, shadowing, and overwriting.
The design also
yields an intriguing record calculus by itself, naturally generalizing function
application paradigms to general records such as optional and weak
fields, which are useful for expressing various programs. Given that record types are
widely utilized across various domains, we anticipate that our formulation can
be inspiring in these areas as well; we expand on this discussion in \Cref{sec:applications}.

\section{A Record Calculus}
\label{sec:calculus}

This section presents the \lang calculus, which formalises the features
described in the prior section.

\subsection{Syntax}

\begin{figure}
  \small
  \begin{tabularx}{\linewidth}{r c l >{\kern-4.5em}r X r c l >{\kern-4.5em}r}
    $e$ & $\Coloneqq$ & \multicolumn{6}{>{\raggedright}p{8cm}}{\renewcommand{\lottnonterminal}[3]{##3}$n\mid\text{$\lottsym{λ} \, \lottsym{\{} \, c \, \lottsym{\}} \lottsym{.} \, e$}\mid\text{$\lottsym{⟨} o₀ \lottsym{;} \, o₁ \, \lottsym{|} \, \lottsym{λ} \, \lottsym{\{} \, c \, \lottsym{\}} \lottsym{.} \, e \lottsym{⟩}$}\mid\lottnonterminalfilterout{nat,op,lam,closure,mproj,diff,let,loc,ref,deref,set}{%
\lottnonterminal{Term}{
\lottalias{e}
}{
\lottproduction{nat}{n}
\lottproductionsep
\lottproduction{op}{e₀ \, b \, e₁}
\lottproductionsep
\lottproduction{lam}{\lottsym{λ} \, \lottsym{\{} \, c \, \lottsym{\}} \lottsym{.} \, e}
\lottproductionsep
\lottproduction{closure}{\lottsym{⟨} r₀ \lottsym{;} \, r₁ \, \lottsym{|} \, \lottsym{λ} \, \lottsym{\{} \, c \, \lottsym{\}} \lottsym{.} \, e \lottsym{⟩}}
\lottproductionsep
\lottproduction{app}{e₀ \, e₁}
\lottproductionsep
\lottproduction{force}{\lottkw{force} \, e}
\lottproductionsep
\lottproduction{box}{e₀ \, \lottsym{▶} \, e₁}
\lottproductionsep
\lottproduction{env}{\lottsym{@}}
\lottproductionsep
\lottproduction{record}{\lottsym{\{} \, r \, \lottsym{\}}}
\lottproductionsep
\lottproduction{concat}{e₀ \, \lottsym{⧺} \, e₁}
\lottproductionsep
\lottproduction{proj}{e \lottsym{.} i}
\lottproductionsep
\lottproduction{mproj}{e \lottsym{.} \lottsym{\{} \, \lottsepbycomprehension{{i}_{i}} \, \lottsym{\}}}
\lottproductionsep
\lottproduction{diff}{e \lottsym{\textbackslash} \lottsym{\{} \, \lottsepbycomprehension{{i}_{i}} \, \lottsym{\}}}
\lottproductionsep
\lottproduction{let}{\lottkw{let} \, t \, \lottsym{=} \, e₀ \, \lottkw{in} \, e₁}
\lottproductionsep
\lottproduction{loc}{l}
\lottproductionsep
\lottproduction{ref}{\lottkw{ref} \, e}
\lottproductionsep
\lottproduction{deref}{\lottsym{!} e}
\lottproductionsep
\lottproduction{set}{e₀ \, \lottsym{≔} \, e₁}
}
}\mid\text{$e \lottsym{.} \lottsym{\{} \, \lottsepbycomprehension{{i}_{j}} \, \lottsym{\}}$}\mid e \lottsym{\textbackslash} \lottsym{\{} \, \lottsepbycomprehension{{i}_{j}} \, \lottsym{\}}\mid\text{$\lottkw{let} \, t \, \lottsym{=} \, e₀ \, \lottkw{in} \, e₁$}$}& \textit{term}\\
    \\
\lottnonterminal{Record Pattern}{
\lottalias{c}
}{
\lottproduction{empty}{\lottsym{ε}}
\lottproductionsep
\lottproduction{ext}{c \lottsym{,} \, i \, \lottsym{◇} \, t}
}
     &&\lottnonterminalfilterout{ref}{%
\lottnonterminal{Type}{
\lottalias{A}
\lottaliassep
\lottalias{B}
}{
\lottproduction{var}{a}
\lottproductionsep
\lottproduction{nat}{\lottsym{ℕ}}
\lottproductionsep
\lottproduction{arr}{\lottsym{\{} \, ρ \, \lottsym{\}} \, \lottsym{→} \, A}
\lottproductionsep
\lottproduction{record}{\lottsym{\{} \, ρ \, \lottsym{\}}}
\lottproductionsep
\lottproduction{quant}{\lottsym{∀} \, a \lottsym{.} \, A}
\lottproductionsep
\lottproduction{ref}{\lottkw{Ref} \, A}
}
}\\
\lottnonterminal{Pattern}{
\lottalias{t}
}{
\lottproduction{label}{i}
\lottproductionsep
\lottproduction{record}{\lottsym{\{} \, c \, \lottsym{\}}}
}
     &&%
\lottnonterminal{Record Type}{
\lottalias{ρ}
}{
\lottproduction{empty}{\lottsym{ε}}
\lottproductionsep
\lottproduction{ext}{ρ \lottsym{,} \, i \, \strength \, A}
}
\\
\lottnonterminal{Record}{
\lottalias{r}
}{
\lottproduction{empty}{\lottsym{ε}}
\lottproductionsep
\lottproduction{ext}{r \lottsym{,} \, i \, \rstrength \, e}
\lottproductionsep
\lottproduction{ext?}{r \lottsym{,} \, i \, \lottsym{?}}
}
     &&%
\lottnonterminal{Strength}{
\lottalias{\strength}
}{
\lottproduction{strong'}{\lottsym{▹}}
\lottproductionsep
\lottproduction{weak}{\lottsym{:}}
\lottproductionsep
\lottproduction{optional}{\lottsym{?}}
}
\\
\lottnonterminal{Required Strength}{
\lottalias{\rstrength}
}{
\lottproduction{strong}{\lottsym{=}}
\lottproductionsep
\lottproduction{weak}{\lottsym{:}}
}
     &&\lottnonterminalfilterout{ref}{%
\lottnonterminal{Monotype}{
\lottalias{τ}
}{
\lottproduction{var}{a}
\lottproductionsep
\lottproduction{nat}{\lottsym{ℕ}}
\lottproductionsep
\lottproduction{arr}{\lottsym{\{} \, ϱ \, \lottsym{\}} \, \lottsym{→} \, τ}
\lottproductionsep
\lottproduction{record}{\lottsym{\{} \, ϱ \, \lottsym{\}}}
\lottproductionsep
\lottproduction{ref}{\lottkw{Ref} \, τ}
}
}\\
    &&&&&%
\lottnonterminal{Record Monotype}{
\lottalias{ϱ}
}{
\lottproduction{empty}{\lottsym{ε}}
\lottproductionsep
\lottproduction{ext}{ϱ \lottsym{,} \, i \, \strength \, τ}
}
\\
    \lottnonterminalfilterout{loc}{%
\lottnonterminal{Value}{
\lottalias{v}
}{
\lottproduction{nat}{n}
\lottproductionsep
\lottproduction{closure}{\lottsym{⟨} o₀ \lottsym{;} \, o₁ \, \lottsym{|} \, \lottsym{λ} \, \lottsym{\{} \, c \, \lottsym{\}} \lottsym{.} \, e \lottsym{⟩}}
\lottproductionsep
\lottproduction{record}{\lottsym{\{} \, o \, \lottsym{\}}}
\lottproductionsep
\lottproduction{loc}{l}
}
} \\
\lottnonterminal{Record Value}{
\lottalias{o}
}{
\lottproduction{empty}{\lottsym{ε}}
\lottproductionsep
\lottproduction{ext}{o \lottsym{,} \, i \, \rstrength \, v}
\lottproductionsep
\lottproduction{ext?}{o \lottsym{,} \, i \, \lottsym{?}}
}
   \end{tabularx}
  \caption{Syntax}
  \label{fig:syntax}
\end{figure}

\Cref{fig:syntax} presents the syntax. Terms $e$ consist
of natural numbers $n$,
lambdas $\lottsym{λ} \, \lottsym{\{} \, c \, \lottsym{\}} \lottsym{.} \, e$ whose bindings are described by record
patterns, closures $\lottsym{⟨} o₀ \lottsym{;} \, o₁ \, \lottsym{|} \, \lottsym{λ} \, \lottsym{\{} \, c \, \lottsym{\}} \lottsym{.} \, e \lottsym{⟩}$ containing record value
components for the captured environment and accumulated arguments, applications
$e₀ \, e₁$, force expressions $\lottkw{force} \, e$, boxes
$e₀ \, \lottsym{▶} \, e₁$, environment reification $\lottsym{@}$, records
$\lottsym{\{} \, r \, \lottsym{\}}$, concatenation $e₀ \, \lottsym{⧺} \, e₁$, projections
$e \lottsym{.} i$, multi-projections
$e \lottsym{.} \lottsym{\{} \, \lottsepbycomprehension{{i}_{j}} \, \lottsym{\}}$, difference $e \lottsym{\textbackslash}
\lottsym{\{} \, \lottsepbycomprehension{{i}_{j}} \, \lottsym{\}}$, and let
expressions $\lottkw{let} \, t \, \lottsym{=} \, e₀ \, \lottkw{in} \, e₁$ which support pattern-matching.
Notably, variables $i$ are not included, as they are simply syntactic sugar for $@.i$.

Record patterns $c$ are either empty
$\lottsym{ε}$, or an extension $c \lottsym{,} \, i \, \lottsym{◇} \, t$, where the label $i$ specifies
the argument label, and $t$ corresponds to parameter names.
These record patterns are more general than those in \Cref{sec:overview};
here, $t$ is either simply a parameter name $i$, or a nested
record pattern $\lottsym{\{} \, c \, \lottsym{\}}$.

Records $r$ include the empty record $\lottsym{ε}$, a
regular extension $r \lottsym{,} \, i \, \rstrength \, e$, or an optional extension
$r \lottsym{,} \, i \, \lottsym{?}$. The strength $\rstrength$ in regular extensions is called a required
strength, either $\lottsym{=}$ or $\lottsym{:}$, which ensures that an expression
for the corresponding field must be present.

Values consist of natural numbers, as well as closures and records whose record entries are themselves
values. Notably, they do not include lambdas, as we will discuss in
\Cref{sec:eval}.

Types $A$ include type variables $a$, natural numbers $\lottsym{ℕ}$, function types
$\lottsym{\{} \, ρ \, \lottsym{\}} \, \lottsym{→} \, A$, record types $\lottsym{\{} \, ρ \, \lottsym{\}}$, and polymorphic types
$\lottsym{∀} \, a \lottsym{.} \, A$. Record types $ρ$ are either empty
$\lottsym{ε}$ or an extension ($ρ \lottsym{,} \, i \, \strength \, A$) that
maps a label to a type and a type-level strength. Type-level strengths
$\strength$ consist of strong $\lottsym{▹}$, weak $\lottsym{:}$, and
optional $\lottsym{?}$.
Monotypes $τ$ and record monotypes $ϱ$
exclude polymorphism.

As a notational convenience, we also allow the $\lottsym{=}$ strength at the
type level, treating it identically to $\lottsym{▹}$. This enables typing
judgements such as $(i = v) : (i = A)$. This means required strengths
$\rstrength$ can also appear at the type level when the optional
strength $\lottsym{?}$ must be excluded, e.g. for projections.

\subsection{Typing}
\label{sec:typing}

\begin{figure}
  \small
  \lottjudgementfilterout{op,loc,ref,deref,set}{\lottjudgementlabel{T}{\lottjudgementnarrower[closure=0.6,app=0.4,box=0.3,concat=0.4,let=0.4]{1}{%
\lottjudgement{Typing}{\ifdefined\store \mathcolor{Store}{ρₛ} \lottsym{;} \, \fi ρ \, \lottsym{⊢} \, e \, \lottsym{:} \, A}{
\lottinferencerule{sub}{
\lotthypothesis{\ifdefined\store \mathcolor{Store}{ρₛ} \lottsym{;} \, \fi ρ \, \lottsym{⊢} \, e \, \lottsym{:} \, A}\\
\lotthypothesis{A \, \lottsym{\mathcolor{Subtyping}{⩽}} \, B}
}{\ifdefined\store \mathcolor{Store}{ρₛ} \lottsym{;} \, \fi ρ \, \lottsym{⊢} \, e \, \lottsym{:} \, B}
\lottinferencerulesep
\lottinferencerule{nat}{
}{\ifdefined\store \mathcolor{Store}{ρₛ} \lottsym{;} \, \fi ρ \, \lottsym{⊢} \, n \, \lottsym{:} \, \lottsym{ℕ}}
\lottinferencerulesep
\lottinferencerule{op}{
\lotthypothesis{\ifdefined\store \mathcolor{Store}{ρₛ} \lottsym{;} \, \fi ρ \, \lottsym{⊢} \, e₀ \, \lottsym{:} \, \lottsym{ℕ}}\\
\lotthypothesis{\ifdefined\store \mathcolor{Store}{ρₛ} \lottsym{;} \, \fi ρ \, \lottsym{⊢} \, e₁ \, \lottsym{:} \, \lottsym{ℕ}}
}{\ifdefined\store \mathcolor{Store}{ρₛ} \lottsym{;} \, \fi ρ \, \lottsym{⊢} \, e₀ \, b \, e₁ \, \lottsym{:} \, \lottsym{ℕ}}
\lottinferencerulesep
\lottinferencerule{lam}{\ifdefined\store\overline{a_j} \, \lottsym{∉} \, \lottkw{ftv} \lottsym{(} \mathcolor{Store}{ρₛ} \lottsym{)}\\\fi\overline{a_j} \, \lottsym{∉} \, \lottkw{ftv} \lottsym{(} ρ₀ \lottsym{)}\\
\lotthypothesis{c \, \lottsym{\mathcolor{Pattern}{:}} \, ρ₁ \, \lottsym{\mathcolor{Pattern}{⊙}} \, ρ₂}\\
\lotthypothesis{\ifdefined\store \mathcolor{Store}{ρₛ} \lottsym{;} \, \fi ρ₀ \lottsym{,} \, ρ₂ \, \lottsym{⊢} \, e \, \lottsym{:} \, A}
}{\ifdefined\store \mathcolor{Store}{ρₛ} \lottsym{;} \, \fi ρ₀ \, \lottsym{⊢} \, \lottsym{λ} \, \lottsym{\{} \, c \, \lottsym{\}} \lottsym{.} \, e \, \lottsym{:} \, \lottsym{∀} \, \lottsepbycomprehension{{a}_{j}} \lottsym{.} \, \lottsym{\{} \, ρ₁ \, \lottsym{\}} \, \lottsym{→} \, A}
\lottinferencerulesep
\lottinferencerule{closure}{\ifdefined\store\overline{a_j} \, \lottsym{∉} \, \lottkw{ftv} \lottsym{(} \mathcolor{Store}{ρₛ} \lottsym{)}\\\fi
\lotthypothesis{\ifdefined\store \mathcolor{Store}{ρₛ} \lottsym{;} \, \fi \lottsym{ε} \, \lottsym{⊢} \, \lottsym{\{} \, o₀ \, \lottsym{\}} \, \lottsym{:} \, \lottsym{\{} \, ρ₁ \, \lottsym{\}}}\\
\lotthypothesis{\lottsym{⌊} ρ₁ \lottsym{⌋}}\\\overline{a_j} \, \lottsym{∉} \, \lottkw{ftv} \lottsym{(} ρ₁ \lottsym{)}\\
\lotthypothesis{\ifdefined\store \mathcolor{Store}{ρₛ} \lottsym{;} \, \fi \lottsym{ε} \, \lottsym{⊢} \, \lottsym{\{} \, o₁ \, \lottsym{\}} \, \lottsym{:} \, \lottsym{∀} \, \lottsepbycomprehension{{a}_{j}} \lottsym{.} \, \lottsym{\{} \, ρ₂ \, \lottsym{\}}}\\
\lotthypothesis{c \, \lottsym{\mathcolor{Pattern}{:}} \, ρ₃ \, \lottsym{\mathcolor{Pattern}{⊙}} \, ρ₄}\\
\lotthypothesis{ρ₃ \, \lottsym{\mathcolor{App}{●}} \, ρ₂ \, \lottsym{\mathcolor{App}{↠}} \, ρ₅}\\
\lotthypothesis{\ifdefined\store \mathcolor{Store}{ρₛ} \lottsym{;} \, \fi ρ₁ \lottsym{,} \, ρ₄ \, \lottsym{⊢} \, e \, \lottsym{:} \, A}
}{\ifdefined\store \mathcolor{Store}{ρₛ} \lottsym{;} \, \fi ρ₀ \, \lottsym{⊢} \, \lottsym{⟨} o₀ \lottsym{;} \, o₁ \, \lottsym{|} \, \lottsym{λ} \, \lottsym{\{} \, c \, \lottsym{\}} \lottsym{.} \, e \lottsym{⟩} \, \lottsym{:} \, \lottsym{∀} \, \lottsepbycomprehension{{a}_{j}} \lottsym{.} \, \lottsym{\{} \, ρ₅ \, \lottsym{\}} \, \lottsym{→} \, A}
\lottinferencerulesep
\lottinferencerule{app}{
\lotthypothesis{\ifdefined\store \mathcolor{Store}{ρₛ} \lottsym{;} \, \fi ρ₀ \, \lottsym{⊢} \, e₀ \, \lottsym{:} \, \lottsym{∀} \, \lottsepbycomprehension{{a}_{j}} \lottsym{.} \, \lottsym{\{} \, ρ₁ \, \lottsym{\}} \, \lottsym{→} \, A}\\
\lotthypothesis{\ifdefined\store \mathcolor{Store}{ρₛ} \lottsym{;} \, \fi ρ₀ \, \lottsym{⊢} \, e₁ \, \lottsym{:} \, \lottsym{∀} \, \lottsepbycomprehension{{a}_{j}} \lottsym{.} \, \lottsym{\{} \, ρ₂ \, \lottsym{\}}}\\
\lotthypothesis{ρ₁ \, \lottsym{\mathcolor{App}{●}} \, ρ₂ \, \lottsym{\mathcolor{App}{↠}} \, ρ₃}
}{\ifdefined\store \mathcolor{Store}{ρₛ} \lottsym{;} \, \fi ρ₀ \, \lottsym{⊢} \, e₀ \, e₁ \, \lottsym{:} \, \lottsym{∀} \, \lottsepbycomprehension{{a}_{j}} \lottsym{.} \, \lottsym{\{} \, ρ₃ \, \lottsym{\}} \, \lottsym{→} \, A}
\lottinferencerulesep
\lottinferencerule{force}{
\lotthypothesis{\ifdefined\store \mathcolor{Store}{ρₛ} \lottsym{;} \, \fi ρ \, \lottsym{⊢} \, e \, \lottsym{:} \, \lottsym{\{} \, \lottsym{\}} \, \lottsym{→} \, A}
}{\ifdefined\store \mathcolor{Store}{ρₛ} \lottsym{;} \, \fi ρ \, \lottsym{⊢} \, \lottkw{force} \, e \, \lottsym{:} \, A}
\lottinferencerulesep
\lottinferencerule{box}{
\lotthypothesis{\ifdefined\store \mathcolor{Store}{ρₛ} \lottsym{;} \, \fi ρ₀ \, \lottsym{⊢} \, e₀ \, \lottsym{:} \, \lottsym{\{} \, ρ₁ \, \lottsym{\}}}\\
\lotthypothesis{\lottsym{⌊} ρ₁ \lottsym{⌋}}\\
\lotthypothesis{\ifdefined\store \mathcolor{Store}{ρₛ} \lottsym{;} \, \fi ρ₁ \, \lottsym{⊢} \, e₁ \, \lottsym{:} \, A}
}{\ifdefined\store \mathcolor{Store}{ρₛ} \lottsym{;} \, \fi ρ₀ \, \lottsym{⊢} \, e₀ \, \lottsym{▶} \, e₁ \, \lottsym{:} \, A}
\lottinferencerulesep
\lottinferencerule{env}{
}{\ifdefined\store \mathcolor{Store}{ρₛ} \lottsym{;} \, \fi ρ \, \lottsym{⊢} \, \lottsym{@} \, \lottsym{:} \, \lottsym{\{} \, ρ \, \lottsym{\}}}
\lottinferencerulesep
\lottinferencerule{record}{
\lotthypothesis{\ifdefined\store \mathcolor{Store}{ρₛ} \lottsym{;} \, \fi ρ₀ \, \lottsym{⊢} \, r \, \lottsym{:} \, ρ₁}
}{\ifdefined\store \mathcolor{Store}{ρₛ} \lottsym{;} \, \fi ρ₀ \, \lottsym{⊢} \, \lottsym{\{} \, r \, \lottsym{\}} \, \lottsym{:} \, \lottsym{\{} \, ρ₁ \, \lottsym{\}}}
\lottinferencerulesep
\lottinferencerule{concat}{
\lotthypothesis{\ifdefined\store \mathcolor{Store}{ρₛ} \lottsym{;} \, \fi ρ₀ \, \lottsym{⊢} \, e₀ \, \lottsym{:} \, \lottsym{\{} \, ρ₁ \, \lottsym{\}}}\\
\lotthypothesis{\ifdefined\store \mathcolor{Store}{ρₛ} \lottsym{;} \, \fi ρ₀ \, \lottsym{⊢} \, e₁ \, \lottsym{:} \, \lottsym{\{} \, ρ₂ \, \lottsym{\}}}\\
\lotthypothesis{ρ₁ \, \lottsym{\mathcolor{Concat}{⊕}} \, ρ₂ \, \lottsym{\mathcolor{Concat}{↬}} \, ρ₃}
}{\ifdefined\store \mathcolor{Store}{ρₛ} \lottsym{;} \, \fi ρ₀ \, \lottsym{⊢} \, e₀ \, \lottsym{⧺} \, e₁ \, \lottsym{:} \, \lottsym{\{} \, ρ₃ \, \lottsym{\}}}
\lottinferencerulesep
\lottinferencerule{proj}{
\lotthypothesis{\ifdefined\store \mathcolor{Store}{ρₛ} \lottsym{;} \, \fi ρ₀ \, \lottsym{⊢} \, e \, \lottsym{:} \, \lottsym{\{} \, ρ₁ \lottsym{,} \, i \, \rstrength \, A \, \lottsym{\}}}
}{\ifdefined\store \mathcolor{Store}{ρₛ} \lottsym{;} \, \fi ρ₀ \, \lottsym{⊢} \, e \lottsym{.} i \, \lottsym{:} \, A}
\lottinferencerulesep
\lottinferencerule{mproj}{
\lotthypothesis{\ifdefined\store \mathcolor{Store}{ρₛ} \lottsym{;} \, \fi ρ₀ \, \lottsym{⊢} \, e \, \lottsym{:} \, \lottsym{\{} \, ρ₁ \lottsym{,} \, \lottsepbycomprehension{{i}_{j} \, {\rstrength}_{j} \, {A}_{j}} \, \lottsym{\}}}
}{\ifdefined\store \mathcolor{Store}{ρₛ} \lottsym{;} \, \fi ρ₀ \, \lottsym{⊢} \, e \lottsym{.} \lottsym{\{} \, \lottsepbycomprehension{{i}_{j}} \, \lottsym{\}} \, \lottsym{:} \, \lottsym{\{} \, \lottsepbycomprehension{{i}_{j} \, {\rstrength}_{j} \, {A}_{j}} \, \lottsym{\}}}
\lottinferencerulesep
\lottinferencerule{diff}{
\lotthypothesis{\ifdefined\store \mathcolor{Store}{ρₛ} \lottsym{;} \, \fi ρ₀ \, \lottsym{⊢} \, e \, \lottsym{:} \, \lottsym{\{} \, ρ₁ \lottsym{,} \, \lottsepbycomprehension{{i}_{j} \, {\rstrength}_{j} \, {A}_{j}} \, \lottsym{\}}}
}{\ifdefined\store \mathcolor{Store}{ρₛ} \lottsym{;} \, \fi ρ₀ \, \lottsym{⊢} \, e \lottsym{\textbackslash} \lottsym{\{} \, \lottsepbycomprehension{{i}_{j}} \, \lottsym{\}} \, \lottsym{:} \, \lottsym{\{} \, ρ₁ \, \lottsym{\}}}
\lottinferencerulesep
\lottinferencerule{let}{
\lotthypothesis{\ifdefined\store \mathcolor{Store}{ρₛ} \lottsym{;} \, \fi ρ₀ \, \lottsym{⊢} \, e₀ \, \lottsym{:} \, A}\\
\lotthypothesis{t \, \lottsym{\mathcolor{Pattern}{:}} \, A \, \lottsym{\mathcolor{Pattern}{⊙}} \, ρ₁}\\
\lotthypothesis{\ifdefined\store \mathcolor{Store}{ρₛ} \lottsym{;} \, \fi ρ₀ \lottsym{,} \, ρ₁ \, \lottsym{⊢} \, e₁ \, \lottsym{:} \, B}
}{\ifdefined\store \mathcolor{Store}{ρₛ} \lottsym{;} \, \fi ρ₀ \, \lottsym{⊢} \, \lottkw{let} \, t \, \lottsym{=} \, e₀ \, \lottkw{in} \, e₁ \, \lottsym{:} \, B}
\lottinferencerulesep
\lottinferencerule{loc}{
}{\ifdefined\store \mathcolor{Store}{ρₛ₀ \lottsym{,} \, l \, \lottsym{:} \, A \lottsym{,} \, ρₛ₁} \lottsym{;} \, \fi ρ \, \lottsym{⊢} \, l \, \lottsym{:} \, \lottkw{Ref} \, A}
\lottinferencerulesep
\lottinferencerule{ref}{
\lotthypothesis{\ifdefined\store \mathcolor{Store}{ρₛ} \lottsym{;} \, \fi ρ \, \lottsym{⊢} \, e \, \lottsym{:} \, A}
}{\ifdefined\store \mathcolor{Store}{ρₛ} \lottsym{;} \, \fi ρ \, \lottsym{⊢} \, \lottkw{ref} \, e \, \lottsym{:} \, \lottkw{Ref} \, A}
\lottinferencerulesep
\lottinferencerule{deref}{
\lotthypothesis{\ifdefined\store \mathcolor{Store}{ρₛ} \lottsym{;} \, \fi ρ \, \lottsym{⊢} \, e \, \lottsym{:} \, \lottkw{Ref} \, A}
}{\ifdefined\store \mathcolor{Store}{ρₛ} \lottsym{;} \, \fi ρ \, \lottsym{⊢} \, \lottsym{!} e \, \lottsym{:} \, A}
\lottinferencerulesep
\lottinferencerule{set}{
\lotthypothesis{\ifdefined\store \mathcolor{Store}{ρₛ} \lottsym{;} \, \fi ρ \, \lottsym{⊢} \, e₀ \, \lottsym{:} \, \lottkw{Ref} \, A}\\
\lotthypothesis{\ifdefined\store \mathcolor{Store}{ρₛ} \lottsym{;} \, \fi ρ \, \lottsym{⊢} \, e₁ \, \lottsym{:} \, A}
}{\ifdefined\store \mathcolor{Store}{ρₛ} \lottsym{;} \, \fi ρ \, \lottsym{⊢} \, e₀ \, \lottsym{≔} \, e₁ \, \lottsym{:} \, \lottsym{\{} \, \lottsym{\}}}
}
}}}

  \medskip

  \lottjudgementforcename{Record Typing}{\lottjudgementlabel{T}{%
\lottjudgement{Typing}{\ifdefined\store \mathcolor{Store}{ρₛ} \lottsym{;} \, \fi ρ₀ \, \lottsym{⊢} \, r \, \lottsym{:} \, ρ₁}{
\lottinferencerule{empty}{
}{\ifdefined\store \mathcolor{Store}{ρₛ} \lottsym{;} \, \fi ρ \, \lottsym{⊢} \, \lottsym{ε} \, \lottsym{:} \, \lottsym{ε}}
\lottinferencerulesep
\lottinferencerule{ext}{
\lotthypothesis{\ifdefined\store \mathcolor{Store}{ρₛ} \lottsym{;} \, \fi ρ₀ \, \lottsym{⊢} \, r \, \lottsym{:} \, ρ₁}\\
\lotthypothesis{\ifdefined\store \mathcolor{Store}{ρₛ} \lottsym{;} \, \fi ρ₀ \, \lottsym{⊢} \, e \, \lottsym{:} \, A}
}{\ifdefined\store \mathcolor{Store}{ρₛ} \lottsym{;} \, \fi ρ₀ \, \lottsym{⊢} \, r \lottsym{,} \, i \, \rstrength \, e \, \lottsym{:} \, ρ₁ \lottsym{,} \, i \, \rstrength \, A}
\lottinferencerulesep
\lottinferencerule{ext?}{
\lotthypothesis{\ifdefined\store \mathcolor{Store}{ρₛ} \lottsym{;} \, \fi ρ₀ \, \lottsym{⊢} \, r \, \lottsym{:} \, ρ₁}
}{\ifdefined\store \mathcolor{Store}{ρₛ} \lottsym{;} \, \fi ρ₀ \, \lottsym{⊢} \, r \lottsym{,} \, i \, \lottsym{?} \, \lottsym{:} \, ρ₁ \lottsym{,} \, i \, \lottsym{?} \, A}
}
}}

  \caption{Typing}
  \label{fig:typing}

  \smallskip
  \lottjudgementfilterout{ref}{\lottjudgementlabel{S}{%
\lottjudgement{Subtyping}{A \, \lottsym{\mathcolor{Subtyping}{⩽}} \, B}{
\lottinferencerule{var}{
}{a \, \lottsym{\mathcolor{Subtyping}{⩽}} \, a}
\lottinferencerulesep
\lottinferencerule{nat}{
}{\lottsym{ℕ} \, \lottsym{\mathcolor{Subtyping}{⩽}} \, \lottsym{ℕ}}
\lottinferencerulesep
\lottinferencerule{arr}{
\lotthypothesis{ρ₁ \, \lottsym{\mathcolor{Subtyping}{⩽}} \, ρ₀}\\
\lotthypothesis{A \, \lottsym{\mathcolor{Subtyping}{⩽}} \, B}
}{\lottsym{\{} \, ρ₀ \, \lottsym{\}} \, \lottsym{→} \, A \, \lottsym{\mathcolor{Subtyping}{⩽}} \, \lottsym{\{} \, ρ₁ \, \lottsym{\}} \, \lottsym{→} \, B}
\lottinferencerulesep
\lottinferencerule{record}{
\lotthypothesis{ρ₀ \, \lottsym{\mathcolor{Subtyping}{⩽}} \, ρ₁}
}{\lottsym{\{} \, ρ₀ \, \lottsym{\}} \, \lottsym{\mathcolor{Subtyping}{⩽}} \, \lottsym{\{} \, ρ₁ \, \lottsym{\}}}
\lottinferencerulesep
\lottinferencerule{quantL}{
\lotthypothesis{A\left[τ/a\right] \, \lottsym{\mathcolor{Subtyping}{⩽}} \, B}
}{\lottsym{∀} \, a \lottsym{.} \, A \, \lottsym{\mathcolor{Subtyping}{⩽}} \, B}
\lottinferencerulesep
\lottinferencerule{quantR}{a \, \lottsym{∉} \, \lottkw{ftv} \lottsym{(} A \lottsym{)}\\
\lotthypothesis{A \, \lottsym{\mathcolor{Subtyping}{⩽}} \, B}
}{A \, \lottsym{\mathcolor{Subtyping}{⩽}} \, \lottsym{∀} \, a \lottsym{.} \, B}
\lottinferencerulesep
\lottinferencerule{ref}{
}{\lottkw{Ref} \, A \, \lottsym{\mathcolor{Subtyping}{⩽}} \, \lottkw{Ref} \, A}
}
}}

  \begingroup
  \renewcommand{\lottinferencerule}[3]{\relax}
  \renewcommand{\lottinferencerulesep}{\relax}
  \renewcommand{\lottjudgement}[3]{\fbox{$#2$}}
  \begin{subfigure}[t]{0.625\textwidth}
    \noindent
\lottjudgement{Left Subtyping}{ρ₀ \, \lottsym{\mathcolor{Subtyping}{⩽}} \, ρ₁}{
\lottinferencerule{empty}{
\lotthypothesis{{\lottsym{⌊} ρ \lottsym{⌋}}^{\lottsym{?}}}
}{\lottsym{ε} \, \lottsym{\mathcolor{Subtyping}{⩽}} \, ρ}
\lottinferencerulesep
\lottinferencerule{ext}{
\lotthypothesis{ρ₀ \, \lottsym{\mathcolor{Subtyping}{⩽}} \, ρ₁ \lottsym{,} \, ρ₂}\\
\lotthypothesis{i \, \lottsym{∉} \, ρ₁}\\
\lotthypothesis{\strength₀ \, \lottsym{\mathcolor{Subtyping}{⩽}} \, \strength₁}\\
\lotthypothesis{A \, \lottsym{\mathcolor{Subtyping}{⩽}} \, B}
}{i \, \strength₀ \, A \lottsym{,} \, ρ₀ \, \lottsym{\mathcolor{Subtyping}{⩽}} \, ρ₁ \lottsym{,} \, i \, \strength₁ \, B \lottsym{,} \, ρ₂}
}
\hfill\emph{(Record Subtyping)}
  \end{subfigure}
  \hfill
  \begin{subfigure}[t]{0.325\textwidth}
    \noindent
\lottjudgement{Subtyping}{\strength₀ \, \lottsym{\mathcolor{Subtyping}{⩽}} \, \strength₁}{
\lottinferencerule{refl}{
}{\strength \, \lottsym{\mathcolor{Subtyping}{⩽}} \, \strength}
\lottinferencerulesep
\lottinferencerule{lt}{
}{\lottsym{:} \, \lottsym{\mathcolor{Subtyping}{⩽}} \, \lottsym{?}}
}
\hfill\emph{(Strength Subtyping)}
  \end{subfigure}
  \endgroup

  \begingroup
  \renewcommand{\lottjudgement}[3]{\begin{mathparpagebreakable}#3\end{mathparpagebreakable}}
  \begin{subfigure}[b]{0.625\textwidth}
    \centering
    \smallskip
    \lottjudgementforcename{Record Subtyping}{\lottjudgementlabel{S}{\lottjudgementnarrower{0.7}{%
\lottjudgement{Left Subtyping}{ρ₀ \, \lottsym{\mathcolor{Subtyping}{⩽}} \, ρ₁}{
\lottinferencerule{empty}{
\lotthypothesis{{\lottsym{⌊} ρ \lottsym{⌋}}^{\lottsym{?}}}
}{\lottsym{ε} \, \lottsym{\mathcolor{Subtyping}{⩽}} \, ρ}
\lottinferencerulesep
\lottinferencerule{ext}{
\lotthypothesis{ρ₀ \, \lottsym{\mathcolor{Subtyping}{⩽}} \, ρ₁ \lottsym{,} \, ρ₂}\\
\lotthypothesis{i \, \lottsym{∉} \, ρ₁}\\
\lotthypothesis{\strength₀ \, \lottsym{\mathcolor{Subtyping}{⩽}} \, \strength₁}\\
\lotthypothesis{A \, \lottsym{\mathcolor{Subtyping}{⩽}} \, B}
}{i \, \strength₀ \, A \lottsym{,} \, ρ₀ \, \lottsym{\mathcolor{Subtyping}{⩽}} \, ρ₁ \lottsym{,} \, i \, \strength₁ \, B \lottsym{,} \, ρ₂}
}
}}}
  \end{subfigure}
  \hfill
  \begin{subfigure}[b]{0.325\textwidth}
    \centering
    \smallskip
    \lottjudgementforcename{Strength Subtyping}{\lottjudgementlabel{S}{%
\lottjudgement{Subtyping}{\strength₀ \, \lottsym{\mathcolor{Subtyping}{⩽}} \, \strength₁}{
\lottinferencerule{refl}{
}{\strength \, \lottsym{\mathcolor{Subtyping}{⩽}} \, \strength}
\lottinferencerulesep
\lottinferencerule{lt}{
}{\lottsym{:} \, \lottsym{\mathcolor{Subtyping}{⩽}} \, \lottsym{?}}
}
}}
  \end{subfigure}
  \endgroup

  \caption{Subtyping}
  \label{fig:subtyping}
\end{figure}

\Cref{fig:typing} presents the typing rules of our calculus. The typing judgement
$\ifdefined\store \mathcolor{Store}{ρₛ} \lottsym{;} \, \fi ρ \, \lottsym{⊢} \, e \, \lottsym{:} \, A$ states that under $ρ$, expression
$e$ has type $A$. %
Typing relies on the subtyping relation in \Cref{fig:subtyping} as well as
auxiliary judgements in \Cref{fig:pattern-sep}-\ref{fig:concatenation}, which we
elaborate on after explaining the typing rules.
During typing, we preserve the invariant that $ρ$ always has strength
($\lottsym{:}$), written $\lottsym{⌊} ρ \lottsym{⌋}$. This invariant is maintained since when the
environment is extended with a new entry, a term of the corresponding type always exists at runtime.
Moreover, as noted in \Cref{sec:overview-overwriting}, functions only require
parameters of strength ($\lottsym{:}$).

Rule \lottref{T}{sub} is the standard subsumption rule. Rule \lottref{T}{lam}
uses record pattern separation ($c \, \lottsym{\mathcolor{Pattern}{:}} \, ρ₁ \,
\lottsym{\mathcolor{Pattern}{⊙}} \, ρ₂$) to extract argument labels
$ρ₁$ and parameter names $ρ₂$ from the record pattern
$c$. The type is then generalized over the type variables
$\overline{a_j}$. Notably, we perform generalization at function abstractions
(as well as closures) rather than let-expressions, a design choice we justify in
\Cref{sec:references}.
Rule \lottref{T}{closure} type-checks the captured environment $o₀$
and accumulated arguments $o₁$ under the empty context.
$\uniform{\rho_1}$
ensures that all entries in $ρ₁$ have
weak strength, and
the relation $ρ₃ \, \lottsym{\mathcolor{App}{●}} \, ρ₂ \,
\lottsym{\mathcolor{App}{↠}} \, ρ₅$ computes the function's remaining
arguments. The rule also performs generalization.
Aside from its handling of quantifiers, which we also defer to
\Cref{sec:references}, rule \lottref{T}{app} is straightforward. Rule
\lottref{T}{force} eliminates a function type that expects an empty argument.

Rules \lottref{T}{box} and \lottref{T}{env} handle first-class environments,
allowing records to be converted into environments and vice versa. Rule
\lottref{T}{box} also ensures that the entries of $\rho_1$ have weak strength.

To type-check a record, rule \lottref{T}{record} delegates to the record typing
judgement $ρ₀ \, \lottsym{⊢} \, r \, \lottsym{:} \, ρ₁$, at the bottom of
\Cref{fig:typing}, which simply examines entries inductively. Rule
\lottref{T}{concat} types record concatenation, where $ρ₁ \,
\lottsym{\mathcolor{Concat}{⊕}} \, ρ₂ \, \lottsym{\mathcolor{Concat}{↬}} \, ρ₃ $
calculates the concatenated type. Rules \lottref{T}{proj}, \lottref{T}{mproj},
and \lottref{T}{diff} follow a similar structure. Notably, all of them use
required strengths, ensuring we can only project or subtract fields that actually
exist.

Finally, \lottref{T}{let} first types $e_0$, and then uses pattern separation to
resolve the pattern $t$.

\paragraph{\textbf{Subtyping} \textnormal{(\Cref{fig:subtyping})}}

The subtyping rules for types $A \, \lottsym{\mathcolor{Subtyping}{⩽}} \, B$ are
largely standard, with rule \lottref{S}{quantL} handling type instantiation.
Record subtyping $ρ₀ \, \lottsym{\mathcolor{Subtyping}{⩽}} \, ρ₁$ introduces
more interesting design choices.
Rule \lottref{S}{empty} allows a record type to be extended with
entries at the end (as in \Cref{sec:overview-optional}). The
hypothesis ensures that all such entries must have optional strength. When applied
to function arguments, this behaviour allows us to forget optional arguments,
such that their defaults are used.
This is a limited, inverted form of width subtyping: we omit the usual
rule, which would allow arbitrary entries to be dropped (see
\Cref{sec:applications}).
Rule \lottref{S}{empty} has two important interactions with typing rules. First, it
enables \lottref{T}{force} to expect a strictly empty argument list, instead of
one containing an arbitrary amount of optional arguments.
Second, it justifies why rules \lottref{T}{mproj} and \lottref{T}{diff} are restricted
to fields with required strengths $\rstrength$; otherwise, rule
\lottref{S}{empty} could be applied to introduce ambiguity,
where a label could refer to either an existing field or an
optional one introduced by subtyping.

In rule \lottref{S}{ext}, the strengths $\star_0$ and $\star_1$ of entry $i$ on
the two sides are related by strength subtyping. The rule subsumes record permutation, allowing $i$ to commute with all
entries in $ρ₁$,
as long as duplicate labels are not swapped.
This is enforced by the premise $i \notin ρ₁$.\footnote{%
Notably, \lottref{S}{ext}
traverses the first record from left to right. This keeps the rule simple;
a right-to-left formulation is possible but, since
\lottref{S}{empty} appends optional entries at the end, needs a slightly more
involved condition on the occurrences of $i$.}
The typing rules \lottref{T}{proj}, \lottref{T}{mproj}, and
\lottref{T}{diff} utilize this permutation behaviour (through subsumption) to
ensure the entries they manipulate are positioned at the tail of the record
types within their premises.

Lastly, strength subtyping $\strength₀ \, \lottsym{\mathcolor{Subtyping}{⩽}} \, \strength₁$
allows the weak strength $\lottsym{:}$ to be converted to optional $\lottsym{?}$ (rule
\lottref{S}{lt}). When considered from the perspective of function arguments,
this allows us to discard the default value of an optional argument, thereby
converting it into a required one. Strength subtyping does not relate
$\lottsym{▹}$ to anything other than itself, since doing so would prevent
strong arguments from always having the intended effect of eliminating the
corresponding function argument.

\begin{figure}
  \small
    \begin{subfigure}[t]{0.525\textwidth}
      \centering
      \lottjudgementforcename{Record Pattern Separation}{\lottjudgementlabel{P}{%
\lottjudgement{Sep}{c \, \lottsym{\mathcolor{Pattern}{:}} \, ρ₀ \, \lottsym{\mathcolor{Pattern}{⊙}} \, ρ₁}{
\lottinferencerule{empty}{
}{\lottsym{ε} \, \lottsym{\mathcolor{Pattern}{:}} \, \lottsym{ε} \, \lottsym{\mathcolor{Pattern}{⊙}} \, \lottsym{ε}}
\lottinferencerulesep
\lottinferencerule{ext}{
\lotthypothesis{c \, \lottsym{\mathcolor{Pattern}{:}} \, ρ₀ \, \lottsym{\mathcolor{Pattern}{⊙}} \, ρ₁}\\
\lotthypothesis{t \, \lottsym{\mathcolor{Pattern}{:}} \, A \, \lottsym{\mathcolor{Pattern}{⊙}} \, ρ₂}
}{\lottsym{(} c \lottsym{,} \, i \, \lottsym{◇} \, t \lottsym{)} \, \lottsym{\mathcolor{Pattern}{:}} \, \lottsym{(} ρ₀ \lottsym{,} \, i \, \lottsym{:} \, A \lottsym{)} \, \lottsym{\mathcolor{Pattern}{⊙}} \, ρ₁ \lottsym{,} \, ρ₂}
}
}}
    \end{subfigure}
    \hfill
    \begin{subfigure}[t]{0.425\textwidth}
      \centering
      \lottjudgementforcename{Pattern Separation}{\lottjudgementlabel{P}{%
\lottjudgement{Sep}{t \, \lottsym{\mathcolor{Pattern}{:}} \, A \, \lottsym{\mathcolor{Pattern}{⊙}} \, ρ}{
\lottinferencerule{label}{
}{i \, \lottsym{\mathcolor{Pattern}{:}} \, A \, \lottsym{\mathcolor{Pattern}{⊙}} \, \lottsym{(} i \, \lottsym{:} \, A \lottsym{)}}
\lottinferencerulesep
\lottinferencerule{record}{
\lotthypothesis{c \, \lottsym{\mathcolor{Pattern}{:}} \, ρ₀ \, \lottsym{\mathcolor{Pattern}{⊙}} \, ρ₁}
}{\lottsym{\{} \, c \, \lottsym{\}} \, \lottsym{\mathcolor{Pattern}{:}} \, \lottsym{\{} \, ρ₀ \, \lottsym{\}} \, \lottsym{\mathcolor{Pattern}{⊙}} \, ρ₁}
}
}}
    \end{subfigure}

    \caption{Pattern separation}
    \label{fig:pattern-sep}

    \bigskip
  \lottjudgementforcename{Record Type App}{\lottjudgementlabel{A}{%
\lottjudgement{App}{ρ₀ \, \lottsym{\mathcolor{App}{●}} \, ρ₁ \, \lottsym{\mathcolor{App}{↠}} \, ρ₂}{
\lottinferencerule{empty}{
}{\lottsym{ε} \, \lottsym{\mathcolor{App}{●}} \, \lottsym{ε} \, \lottsym{\mathcolor{App}{↠}} \, \lottsym{ε}}
\lottinferencerulesep
\lottinferencerule{weakStrong}{
\lotthypothesis{ρ₀ \, \lottsym{\mathcolor{App}{●}} \, ρ₁ \, \lottsym{\mathcolor{App}{↠}} \, ρ₂}
}{ρ₀ \lottsym{,} \, i \, \lottsym{:} \, A \, \lottsym{\mathcolor{App}{●}} \, ρ₁ \lottsym{,} \, i \, \lottsym{▹} \, A \, \lottsym{\mathcolor{App}{↠}} \, ρ₂}
\lottinferencerulesep
\lottinferencerule{optionalStrong}{
\lotthypothesis{ρ₀ \, \lottsym{\mathcolor{App}{●}} \, ρ₁ \, \lottsym{\mathcolor{App}{↠}} \, ρ₂}
}{ρ₀ \lottsym{,} \, i \, \lottsym{?} \, A \, \lottsym{\mathcolor{App}{●}} \, ρ₁ \lottsym{,} \, i \, \lottsym{▹} \, A \, \lottsym{\mathcolor{App}{↠}} \, ρ₂}
\lottinferencerulesep
\lottinferencerule{ext}{
\lotthypothesis{ρ₀ \, \lottsym{\mathcolor{App}{●}} \, ρ₁ \, \lottsym{\mathcolor{App}{↠}} \, ρ₂}\\
\lotthypothesis{\strength₀ \, \lottsym{\mathcolor{App}{●}} \, \strength₁ \, \lottsym{\mathcolor{App}{↠}} \, \strength₂}
}{ρ₀ \lottsym{,} \, i \, \strength₀ \, A \, \lottsym{\mathcolor{App}{●}} \, ρ₁ \lottsym{,} \, i \, \strength₁ \, A \, \lottsym{\mathcolor{App}{↠}} \, ρ₂ \lottsym{,} \, i \, \strength₂ \, A}
}
}}

  \lottjudgementforcename{Strength App}{\lottjudgementlabel{A}{%
\lottjudgement{App}{\strength₀ \, \lottsym{\mathcolor{App}{●}} \, \strength₁ \, \lottsym{\mathcolor{App}{↠}} \, \strength₂}{
\lottinferencerule{later}{
}{\lottsym{:} \, \lottsym{\mathcolor{App}{●}} \, \lottsym{?} \, \lottsym{\mathcolor{App}{↠}} \, \lottsym{:}}
\lottinferencerulesep
\lottinferencerule{set}{
}{\lottsym{:} \, \lottsym{\mathcolor{App}{●}} \, \lottsym{:} \, \lottsym{\mathcolor{App}{↠}} \, \lottsym{?}}
\lottinferencerulesep
\lottinferencerule{leave}{
}{\lottsym{?} \, \lottsym{\mathcolor{App}{●}} \, \lottsym{?} \, \lottsym{\mathcolor{App}{↠}} \, \lottsym{?}}
\lottinferencerulesep
\lottinferencerule{replace}{
}{\lottsym{?} \, \lottsym{\mathcolor{App}{●}} \, \lottsym{:} \, \lottsym{\mathcolor{App}{↠}} \, \lottsym{?}}
}
}}

  \caption{Application}
  \label{fig:application}

  \bigskip

  \lottjudgementforcename{Record Type Concat}{\lottjudgementlabel{C}{%
\lottjudgement{Concat}{ρ₀ \, \lottsym{\mathcolor{Concat}{⊕}} \, ρ₁ \, \lottsym{\mathcolor{Concat}{↬}} \, ρ₂}{
\lottinferencerule{empty}{
}{\lottsym{ε} \, \lottsym{\mathcolor{Concat}{⊕}} \, \lottsym{ε} \, \lottsym{\mathcolor{Concat}{↬}} \, \lottsym{ε}}
\lottinferencerulesep
\lottinferencerule{strongL}{
\lotthypothesis{ρ₀ \, \lottsym{\mathcolor{Concat}{⊕}} \, ρ₁ \, \lottsym{\mathcolor{Concat}{↬}} \, ρ₂}
}{ρ₀ \lottsym{,} \, i \, \lottsym{▹} \, A \, \lottsym{\mathcolor{Concat}{⊕}} \, ρ₁ \, \lottsym{\mathcolor{Concat}{↬}} \, ρ₂ \lottsym{,} \, i \, \lottsym{▹} \, A}
\lottinferencerulesep
\lottinferencerule{ext}{
\lotthypothesis{ρ₀ \, \lottsym{\mathcolor{Concat}{⊕}} \, ρ₁ \, \lottsym{\mathcolor{Concat}{↬}} \, ρ₂}\\
\lotthypothesis{\strength₀ \, \lottsym{\mathcolor{Concat}{⊕}} \, \strength₁ \, \lottsym{\mathcolor{Concat}{↬}} \, \strength₂}
}{ρ₀ \lottsym{,} \, i \, \strength₀ \, A \, \lottsym{\mathcolor{Concat}{⊕}} \, ρ₁ \lottsym{,} \, i \, \strength₁ \, A \, \lottsym{\mathcolor{Concat}{↬}} \, ρ₂ \lottsym{,} \, i \, \strength₂ \, A}
}
}}

  \lottjudgementforcename{Strength Concat}{\lottjudgementlabel{C}{%
\lottjudgement{Concat}{\strength₀ \, \lottsym{\mathcolor{Concat}{⊕}} \, \strength₁ \, \lottsym{\mathcolor{Concat}{↬}} \, \strength₂}{
\lottinferencerule{strongR}{
}{\lottsym{:} \, \lottsym{\mathcolor{Concat}{⊕}} \, \lottsym{▹} \, \lottsym{\mathcolor{Concat}{↬}} \, \lottsym{▹}}
\lottinferencerulesep
\lottinferencerule{later}{
}{\lottsym{?} \, \lottsym{\mathcolor{Concat}{⊕}} \, \strength \, \lottsym{\mathcolor{Concat}{↬}} \, \strength}
\lottinferencerulesep
\lottinferencerule{leave}{
}{\lottsym{:} \, \lottsym{\mathcolor{Concat}{⊕}} \, \lottsym{?} \, \lottsym{\mathcolor{Concat}{↬}} \, \lottsym{:}}
\lottinferencerulesep
\lottinferencerule{replace}{
}{\lottsym{:} \, \lottsym{\mathcolor{Concat}{⊕}} \, \lottsym{:} \, \lottsym{\mathcolor{Concat}{↬}} \, \lottsym{:}}
}
}}

  \caption{Concatenation}
  \label{fig:concatenation}

\end{figure}

\paragraph{\textbf{Pattern separation} \textnormal{(\Cref{fig:pattern-sep})}}

The judgement $c \, \lottsym{\mathcolor{Pattern}{:}} \, ρ₀ \,
\lottsym{\mathcolor{Pattern}{⊙}} \, ρ₁$ separates argument labels $ρ₀$ from
parameter names $ρ₁$ in the record pattern $c$. Rule \lottref{P}{ext} is more
interesting: it recurses on the record pattern $c$, and also uses the other
judgement $t \, \lottsym{\mathcolor{Pattern}{:}} \, A \,
\lottsym{\mathcolor{Pattern}{⊙}} \, \rho_2$, which states that a parameter
pattern $t$ produces an argument type $A$ and its corresponding parameter
names $\rho_2$. The result then extends $ρ₀$ with argument label $i:A$, and
extends $ρ₁$ with $\rho_2$. Note that the weak strength $\lottsym{:}$ is used
for the new argument entry, since all arguments start out as required. For the parameter pattern $t$, rule
\lottref{P}{label} guesses a type $A$ and uses the given parameter name $(i:A)$. Rule
\lottref{P}{record} handles nested record patterns. %
As a simple example, $i \diamond \{ \, age \diamond n \, \} \,
\lottsym{\mathcolor{Pattern}{:}} \, (i : \{ \, age : \lottkw{Nat} \, \}) \,
\lottsym{\mathcolor{Pattern}{⊙}} \, (n : \lottkw{Nat})$.

\paragraph{\textbf{Application} \textnormal{(\Cref{fig:application})}}
\label{sec:typing-application}

The judgement $\app{\rho_0}{\rho_1}{\rho_2}$ models function application: when a
function expecting parameters $\rho_0$ receives an argument of type $\rho_1$, it
computes the remaining parameter type $\rho_2$. This judgement assumes that
entries in $\rho_0$ and $\rho_1$ appear in a convenient order, as record
subtyping takes care of commutativity and optional entry insertion.

These rules directly reflect the behaviour of strengths outlined in
\Cref{sec:overview-overwriting}. Rule \lottref{A}{empty} is straightforward.
Rules \lottref{A}{weakStrong} and \lottref{A}{optionalStrong} handle cases where
a strong argument is supplied, thereby eliminating the parameter from the
resulting type. Otherwise, the parameter remains in some form via rule
\lottref{A}{ext}, where its resulting strength is determined by a
strength-specific variant of the judgement, $\app{\star_0}{\star_1}{\star_2}$.
For a required parameter, it remains required if it is provided an optional
argument (rule \lottref{A}{later}), and becomes optional when provided with a
weak argument (rule \lottref{A}{set}). Finally, an optional parameter remains
optional regardless of the argument's strength in rules \lottref{A}{leave} and
\lottref{A}{replace}.

\paragraph{\textbf{Concatenation} \textnormal{(\Cref{fig:concatenation})}}
\label{sec:typing-concatenation}

The record concatenation judgement $\tconcat{\rho_0}{\rho_1}{\rho_2}$ computes the
resulting record type $\rho_2$. As before, we assume that entries appear in a
convenient order within $\rho_0$ and $\rho_1$. Concatenation depends on record
strength and is thus not commutative, as label order
matters for shadowing and overwriting.
Rule \lottref{C}{strongL} always preserves a strong entry on the left;
otherwise \lottref{C}{ext} combines the two strengths via
$\tconcat{\star_0}{\star_1}{\star_2}$: a strong entry on the right overwrites a
weak one (\mbox{\lottref{C}{strongR}}), an optional entry is overwritten
(\lottref{C}{later}) or ignored when the left entry is weak
(\lottref{C}{leave}), and a weak entry replaces a matching weak entry
(\lottref{C}{replace}).

\subsection{Example}
\label{sec:example}

To illustrate how application interacts with subtyping, consider the partial
application \[{(λ \{ \_ ◇ i₀, \_ ◇ i₁ \}. @.i₀ - @.i₁) ~\{ \_ = 2 \}}\] which
supplies only the first of two positional arguments (we assume simple arithmetic
operators and omit uninteresting derivation details). To type it, subtyping first
adds an optional entry for the missing second argument through
\lottref{S}{empty}:
\vspace{-6pt}
\begin{mathpar}
  \vspace{-15pt}
  \scriptsize
  \renewcommand{\TirNameStyle}[1]{\scriptsize\textsc{#1}}
  \inferrule*[left=\fbox{\textbf{T}}\textrm{ =},right=\lottref{T}{sub}]{
    \inferrule*[right=\lottref{T}{record}]{ }{
      \text{$\ifdefined\store \mathcolor{Store}{\lottsym{ε}} \lottsym{;} \, \fi \lottsym{ε} \, \lottsym{⊢} \, \lottsym{\{} \, \_ \, \lottsym{=} \, 2 \, \lottsym{\}} \, \lottsym{:} \, \lottsym{\{} \, \_ \, \lottsym{▹} \, \lottsym{ℕ} \, \lottsym{\}}$}
    }
    \and
    \inferrule*[right=\lottref{S}{record}]{
      \inferrule*[right=\lottref{S}{ext}]{
        \inferrule*[right=\lottref{S}{empty}]{
          \text{${\lottsym{⌊} \_ \, \lottsym{?} \, \lottsym{ℕ} \lottsym{⌋}}^{\lottsym{?}}$}
        }{
          \text{$\lottsym{ε} \, \lottsym{\mathcolor{Subtyping}{⩽}} \, \_ \, \lottsym{?} \, \lottsym{ℕ}$}
        }
        \and
        \cdots
      }{
        \text{$\_ \, \lottsym{▹} \, \lottsym{ℕ} \, \lottsym{\mathcolor{Subtyping}{⩽}} \, \_ \, \lottsym{▹} \, \lottsym{ℕ} \lottsym{,} \, \_ \, \lottsym{?} \, \lottsym{ℕ}$}
      }
    }{
      \text{$\lottsym{\{} \, \_ \, \lottsym{▹} \, \lottsym{ℕ} \, \lottsym{\}} \, \lottsym{\mathcolor{Subtyping}{⩽}} \, \lottsym{\{} \, \_ \, \lottsym{▹} \, \lottsym{ℕ} \lottsym{,} \, \_ \, \lottsym{?} \, \lottsym{ℕ} \, \lottsym{\}}$}
    }
  }{
    \text{$\ifdefined\store \mathcolor{Store}{\lottsym{ε}} \lottsym{;} \, \fi \lottsym{ε} \, \lottsym{⊢} \, \lottsym{\{} \, \_ \, \lottsym{=} \, 2 \, \lottsym{\}} \, \lottsym{:} \, \lottsym{\{} \, \_ \, \lottsym{▹} \, \lottsym{ℕ} \lottsym{,} \, \_ \, \lottsym{?} \, \lottsym{ℕ} \, \lottsym{\}}$}
  }
\end{mathpar}
The application judgement then eliminates the first parameter but leaves the
second one:
\begin{mathpar}
  \scriptsize
  \renewcommand{\TirNameStyle}[1]{\scriptsize\textsc{#1}}
  \inferrule*[left=\fbox{\textbf{A}}\textrm{ =},right=\lottref{A}{ext}]{
    \inferrule*[right=\lottref{A}{weakStrong}]{
      \inferrule*[right=\lottref{A}{empty}]{
      }{
        \text{$\lottsym{ε} \, \lottsym{\mathcolor{App}{●}} \, \lottsym{ε} \, \lottsym{\mathcolor{App}{↠}} \, \lottsym{ε}$}
      }
    }{
      \text{$\_ \, \lottsym{:} \, \lottsym{ℕ} \, \lottsym{\mathcolor{App}{●}} \, \_ \, \lottsym{▹} \, \lottsym{ℕ} \, \lottsym{\mathcolor{App}{↠}} \, \lottsym{ε}$}
    }
    \and
    \inferrule*[right=\lottref{A}{later}]{
    }{
      \text{$\lottsym{:} \, \lottsym{\mathcolor{App}{●}} \, \lottsym{?} \, \lottsym{\mathcolor{App}{↠}} \, \lottsym{:}$}
    }
  }{
    \text{$\_ \, \lottsym{:} \, \lottsym{ℕ} \lottsym{,} \, \_ \, \lottsym{:} \, \lottsym{ℕ} \, \lottsym{\mathcolor{App}{●}} \, \_ \, \lottsym{▹} \, \lottsym{ℕ} \lottsym{,} \, \_ \, \lottsym{?} \, \lottsym{ℕ} \, \lottsym{\mathcolor{App}{↠}} \, \_ \, \lottsym{:} \, \lottsym{ℕ}$}
  }
\end{mathpar}
Finally, rule \lottref{T}{app} combines the two, yielding a function that still
expects the second argument:
\begin{mathpar}
  \scriptsize
  \renewcommand{\TirNameStyle}[1]{\scriptsize\textsc{#1}}
  \inferrule*[right=\lottref{T}{app}]{
    \inferrule*[right=\lottref{T}{lam}]{ }{
      \text{$\ifdefined\store \mathcolor{Store}{\lottsym{ε}} \lottsym{;} \, \fi \lottsym{ε} \, \lottsym{⊢} \, \lottsym{λ} \, \lottsym{\{} \, \_ \, \lottsym{◇} \, i₀ \lottsym{,} \, \_ \, \lottsym{◇} \, i₁ \, \lottsym{\}} \lottsym{.} \, \lottsym{@} \lottsym{.} i₀ \, \lottsym{-} \, \lottsym{@} \lottsym{.} i₁ \, \lottsym{:} \, \lottsym{\{} \, \_ \, \lottsym{:} \, \lottsym{ℕ} \lottsym{,} \, \_ \, \lottsym{:} \, \lottsym{ℕ} \, \lottsym{\}} \, \lottsym{→} \, \lottsym{ℕ}$}
    }
    \and
    \fbox{\textbf{T}}
    \and
    \fbox{\textbf{A}}
  }{
    \text{$\ifdefined\store \mathcolor{Store}{\lottsym{ε}} \lottsym{;} \, \fi \lottsym{ε} \, \lottsym{⊢} \, \lottsym{(} \lottsym{λ} \, \lottsym{\{} \, \_ \, \lottsym{◇} \, i₀ \lottsym{,} \, \_ \, \lottsym{◇} \, i₁ \, \lottsym{\}} \lottsym{.} \, \lottsym{@} \lottsym{.} i₀ \, \lottsym{-} \, \lottsym{@} \lottsym{.} i₁ \lottsym{)} \, \lottsym{\{} \, \_ \, \lottsym{=} \, 2 \, \lottsym{\}} \, \lottsym{:} \, \lottsym{\{} \, \_ \, \lottsym{:} \, \lottsym{ℕ} \, \lottsym{\}} \, \lottsym{→} \, \lottsym{ℕ}$}
  }
\end{mathpar}
The appendix contains further examples involving pattern separation and
concatenation.

\subsection{Evaluation}
\label{sec:eval}

\Cref{fig:small-step} presents the operational semantics:
$o \, \lottsym{⊢} \, e \ifdefined\store \lottsym{;} \, \mathcolor{Store}{oₛ} \fi \, \lottsym{\longrightarrow} \, e' \ifdefined\store \lottsym{;} \, \mathcolor{Store}{oₛ'} \fi$ evaluates $e$ to $e'$ under the evaluation
environment $o$. Many of the rules are routine; here we focus on the most
interesting ones.

\begin{figure}
  \small
  \lottjudgementforcenameandexample{Operational Semantics}{o \, \lottsym{⊢} \, e \, \lottsym{\longrightarrow} \, e'}{\lottjudgementfilterout{opl,opr,op,ref,refv,deref,derefLoc,setl,setr,set}{\lottjudgementlabel{E}{%
\lottjudgement{Small Step}{r \, \lottsym{⊢} \, e \ifdefined\store \lottsym{;} \, \mathcolor{Store}{rₛ} \fi \, \lottsym{\longrightarrow} \, e' \ifdefined\store \lottsym{;} \, \mathcolor{Store}{rₛ'} \fi}{
\lottinferencerule{opl}{
\lotthypothesis{o \, \lottsym{⊢} \, e₀ \ifdefined\store \lottsym{;} \, \mathcolor{Store}{oₛ} \fi \, \lottsym{\longrightarrow} \, e₀' \ifdefined\store \lottsym{;} \, \mathcolor{Store}{oₛ'} \fi}
}{o \, \lottsym{⊢} \, e₀ \, b \, e₁ \ifdefined\store \lottsym{;} \, \mathcolor{Store}{oₛ} \fi \, \lottsym{\longrightarrow} \, e₀' \, b \, e₁ \ifdefined\store \lottsym{;} \, \mathcolor{Store}{oₛ'} \fi}
\lottinferencerulesep
\lottinferencerule{opr}{
\lotthypothesis{o \, \lottsym{⊢} \, e \ifdefined\store \lottsym{;} \, \mathcolor{Store}{oₛ} \fi \, \lottsym{\longrightarrow} \, e' \ifdefined\store \lottsym{;} \, \mathcolor{Store}{oₛ'} \fi}
}{o \, \lottsym{⊢} \, v \, b \, e \ifdefined\store \lottsym{;} \, \mathcolor{Store}{oₛ} \fi \, \lottsym{\longrightarrow} \, v \, b \, e' \ifdefined\store \lottsym{;} \, \mathcolor{Store}{oₛ'} \fi}
\lottinferencerulesep
\lottinferencerule{op}{
\lotthypothesis{n₀ \, b \, n₁ \, \lottsym{=} \, n₂}
}{o \, \lottsym{⊢} \, n₀ \, b \, n₁ \ifdefined\store \lottsym{;} \, \mathcolor{Store}{oₛ} \fi \, \lottsym{\longrightarrow} \, n₂ \ifdefined\store \lottsym{;} \, \mathcolor{Store}{oₛ} \fi}
\lottinferencerulesep
\lottinferencerule{lam}{
}{o \, \lottsym{⊢} \, \lottsym{λ} \, \lottsym{\{} \, c \, \lottsym{\}} \lottsym{.} \, e \ifdefined\store \lottsym{;} \, \mathcolor{Store}{oₛ} \fi \, \lottsym{\longrightarrow} \, \lottsym{⟨} o \lottsym{;} \, \lottsym{ε} \, \lottsym{|} \, \lottsym{λ} \, \lottsym{\{} \, c \, \lottsym{\}} \lottsym{.} \, e \lottsym{⟩} \ifdefined\store \lottsym{;} \, \mathcolor{Store}{oₛ} \fi}
\lottinferencerulesep
\lottinferencerule{appl}{
\lotthypothesis{o \, \lottsym{⊢} \, e₀ \ifdefined\store \lottsym{;} \, \mathcolor{Store}{oₛ} \fi \, \lottsym{\longrightarrow} \, e₀' \ifdefined\store \lottsym{;} \, \mathcolor{Store}{oₛ'} \fi}
}{o \, \lottsym{⊢} \, e₀ \, e₁ \ifdefined\store \lottsym{;} \, \mathcolor{Store}{oₛ} \fi \, \lottsym{\longrightarrow} \, e₀' \, e₁ \ifdefined\store \lottsym{;} \, \mathcolor{Store}{oₛ'} \fi}
\lottinferencerulesep
\lottinferencerule{appr}{
\lotthypothesis{o \, \lottsym{⊢} \, e \ifdefined\store \lottsym{;} \, \mathcolor{Store}{oₛ} \fi \, \lottsym{\longrightarrow} \, e' \ifdefined\store \lottsym{;} \, \mathcolor{Store}{oₛ'} \fi}
}{o \, \lottsym{⊢} \, v \, e \ifdefined\store \lottsym{;} \, \mathcolor{Store}{oₛ} \fi \, \lottsym{\longrightarrow} \, v \, e' \ifdefined\store \lottsym{;} \, \mathcolor{Store}{oₛ'} \fi}
\lottinferencerulesep
\lottinferencerule{app}{
}{o₀ \, \lottsym{⊢} \, \lottsym{⟨} o₁ \lottsym{;} \, o₂ \, \lottsym{|} \, \lottsym{λ} \, \lottsym{\{} \, c \, \lottsym{\}} \lottsym{.} \, e \lottsym{⟩} \, \lottsym{\{} \, o₃ \, \lottsym{\}} \ifdefined\store \lottsym{;} \, \mathcolor{Store}{oₛ} \fi \, \lottsym{\longrightarrow} \, \lottsym{⟨} o₁ \lottsym{;} \, o₂ \, \lottsym{⧺} \, o₃ \, \lottsym{|} \, \lottsym{λ} \, \lottsym{\{} \, c \, \lottsym{\}} \lottsym{.} \, e \lottsym{⟩} \ifdefined\store \lottsym{;} \, \mathcolor{Store}{oₛ} \fi}
\lottinferencerulesep
\lottinferencerule{force}{
\lotthypothesis{o \, \lottsym{⊢} \, e \ifdefined\store \lottsym{;} \, \mathcolor{Store}{oₛ} \fi \, \lottsym{\longrightarrow} \, e' \ifdefined\store \lottsym{;} \, \mathcolor{Store}{oₛ'} \fi}
}{o \, \lottsym{⊢} \, \lottkw{force} \, e \ifdefined\store \lottsym{;} \, \mathcolor{Store}{oₛ} \fi \, \lottsym{\longrightarrow} \, \lottkw{force} \, e' \ifdefined\store \lottsym{;} \, \mathcolor{Store}{oₛ'} \fi}
\lottinferencerulesep
\lottinferencerule{forceClosure}{
\lotthypothesis{\lottkw{expand} \lottsym{(} c \lottsym{;} \, o₂ \lottsym{)} \, \lottsym{=} \, o₃}
}{o₀ \, \lottsym{⊢} \, \lottkw{force} \, \lottsym{⟨} o₁ \lottsym{;} \, o₂ \, \lottsym{|} \, \lottsym{λ} \, \lottsym{\{} \, c \, \lottsym{\}} \lottsym{.} \, e \lottsym{⟩} \ifdefined\store \lottsym{;} \, \mathcolor{Store}{oₛ} \fi \, \lottsym{\longrightarrow} \, \lottsym{\{} \, o₁ \lottsym{,} \, o₃ \, \lottsym{\}} \, \lottsym{▶} \, e \ifdefined\store \lottsym{;} \, \mathcolor{Store}{oₛ} \fi}
\lottinferencerulesep
\lottinferencerule{boxl}{
\lotthypothesis{o \, \lottsym{⊢} \, e₀ \ifdefined\store \lottsym{;} \, \mathcolor{Store}{oₛ} \fi \, \lottsym{\longrightarrow} \, e₀' \ifdefined\store \lottsym{;} \, \mathcolor{Store}{oₛ'} \fi}
}{o \, \lottsym{⊢} \, e₀ \, \lottsym{▶} \, e₁ \ifdefined\store \lottsym{;} \, \mathcolor{Store}{oₛ} \fi \, \lottsym{\longrightarrow} \, e₀' \, \lottsym{▶} \, e₁ \ifdefined\store \lottsym{;} \, \mathcolor{Store}{oₛ'} \fi}
\lottinferencerulesep
\lottinferencerule{boxr}{
\lotthypothesis{o₁ \, \lottsym{⊢} \, e \ifdefined\store \lottsym{;} \, \mathcolor{Store}{oₛ} \fi \, \lottsym{\longrightarrow} \, e' \ifdefined\store \lottsym{;} \, \mathcolor{Store}{oₛ'} \fi}
}{o₀ \, \lottsym{⊢} \, \lottsym{\{} \, o₁ \, \lottsym{\}} \, \lottsym{▶} \, e \ifdefined\store \lottsym{;} \, \mathcolor{Store}{oₛ} \fi \, \lottsym{\longrightarrow} \, \lottsym{\{} \, o₁ \, \lottsym{\}} \, \lottsym{▶} \, e' \ifdefined\store \lottsym{;} \, \mathcolor{Store}{oₛ'} \fi}
\lottinferencerulesep
\lottinferencerule{box}{
}{o₀ \, \lottsym{⊢} \, \lottsym{\{} \, o₁ \, \lottsym{\}} \, \lottsym{▶} \, v \ifdefined\store \lottsym{;} \, \mathcolor{Store}{oₛ} \fi \, \lottsym{\longrightarrow} \, v \ifdefined\store \lottsym{;} \, \mathcolor{Store}{oₛ} \fi}
\lottinferencerulesep
\lottinferencerule{env}{
}{o \, \lottsym{⊢} \, \lottsym{@} \ifdefined\store \lottsym{;} \, \mathcolor{Store}{oₛ} \fi \, \lottsym{\longrightarrow} \, \lottsym{\{} \, o \, \lottsym{\}} \ifdefined\store \lottsym{;} \, \mathcolor{Store}{oₛ} \fi}
\lottinferencerulesep
\lottinferencerule{extl}{
\lotthypothesis{o \, \lottsym{⊢} \, \lottsym{\{} \, r \, \lottsym{\}} \ifdefined\store \lottsym{;} \, \mathcolor{Store}{oₛ} \fi \, \lottsym{\longrightarrow} \, \lottsym{\{} \, r' \, \lottsym{\}} \ifdefined\store \lottsym{;} \, \mathcolor{Store}{oₛ'} \fi}
}{o \, \lottsym{⊢} \, \lottsym{\{} \, r \lottsym{,} \, i \, \rstrength \, e \, \lottsym{\}} \ifdefined\store \lottsym{;} \, \mathcolor{Store}{oₛ} \fi \, \lottsym{\longrightarrow} \, \lottsym{\{} \, r' \lottsym{,} \, i \, \rstrength \, e \, \lottsym{\}} \ifdefined\store \lottsym{;} \, \mathcolor{Store}{oₛ'} \fi}
\lottinferencerulesep
\lottinferencerule{extl?}{
\lotthypothesis{o \, \lottsym{⊢} \, \lottsym{\{} \, r \, \lottsym{\}} \ifdefined\store \lottsym{;} \, \mathcolor{Store}{oₛ} \fi \, \lottsym{\longrightarrow} \, \lottsym{\{} \, r' \, \lottsym{\}} \ifdefined\store \lottsym{;} \, \mathcolor{Store}{oₛ'} \fi}
}{o \, \lottsym{⊢} \, \lottsym{\{} \, r \lottsym{,} \, i \, \lottsym{?} \, \lottsym{\}} \ifdefined\store \lottsym{;} \, \mathcolor{Store}{oₛ} \fi \, \lottsym{\longrightarrow} \, \lottsym{\{} \, r' \lottsym{,} \, i \, \lottsym{?} \, \lottsym{\}} \ifdefined\store \lottsym{;} \, \mathcolor{Store}{oₛ'} \fi}
\lottinferencerulesep
\lottinferencerule{extr}{
\lotthypothesis{o₀ \, \lottsym{⊢} \, e \ifdefined\store \lottsym{;} \, \mathcolor{Store}{oₛ} \fi \, \lottsym{\longrightarrow} \, e' \ifdefined\store \lottsym{;} \, \mathcolor{Store}{oₛ'} \fi}
}{o₀ \, \lottsym{⊢} \, \lottsym{\{} \, o₁ \lottsym{,} \, i \, \rstrength \, e \, \lottsym{\}} \ifdefined\store \lottsym{;} \, \mathcolor{Store}{oₛ} \fi \, \lottsym{\longrightarrow} \, \lottsym{\{} \, o₁ \lottsym{,} \, i \, \rstrength \, e' \, \lottsym{\}} \ifdefined\store \lottsym{;} \, \mathcolor{Store}{oₛ'} \fi}
\lottinferencerulesep
\lottinferencerule{concatl}{
\lotthypothesis{o \, \lottsym{⊢} \, e₀ \ifdefined\store \lottsym{;} \, \mathcolor{Store}{oₛ} \fi \, \lottsym{\longrightarrow} \, e₀' \ifdefined\store \lottsym{;} \, \mathcolor{Store}{oₛ'} \fi}
}{o \, \lottsym{⊢} \, e₀ \, \lottsym{⧺} \, e₁ \ifdefined\store \lottsym{;} \, \mathcolor{Store}{oₛ} \fi \, \lottsym{\longrightarrow} \, e₀' \, \lottsym{⧺} \, e₁ \ifdefined\store \lottsym{;} \, \mathcolor{Store}{oₛ'} \fi}
\lottinferencerulesep
\lottinferencerule{concatr}{
\lotthypothesis{o \, \lottsym{⊢} \, e \ifdefined\store \lottsym{;} \, \mathcolor{Store}{oₛ} \fi \, \lottsym{\longrightarrow} \, e' \ifdefined\store \lottsym{;} \, \mathcolor{Store}{oₛ'} \fi}
}{o \, \lottsym{⊢} \, v \, \lottsym{⧺} \, e \ifdefined\store \lottsym{;} \, \mathcolor{Store}{oₛ} \fi \, \lottsym{\longrightarrow} \, v \, \lottsym{⧺} \, e' \ifdefined\store \lottsym{;} \, \mathcolor{Store}{oₛ'} \fi}
\lottinferencerulesep
\lottinferencerule{concat}{
}{o₀ \, \lottsym{⊢} \, \lottsym{\{} \, o₁ \, \lottsym{\}} \, \lottsym{⧺} \, \lottsym{\{} \, o₂ \, \lottsym{\}} \ifdefined\store \lottsym{;} \, \mathcolor{Store}{oₛ} \fi \, \lottsym{\longrightarrow} \, \lottsym{\{} \, o₁ \, \lottsym{⧺} \, o₂ \, \lottsym{\}} \ifdefined\store \lottsym{;} \, \mathcolor{Store}{oₛ} \fi}
\lottinferencerulesep
\lottinferencerule{proj}{
\lotthypothesis{o \, \lottsym{⊢} \, e \ifdefined\store \lottsym{;} \, \mathcolor{Store}{oₛ} \fi \, \lottsym{\longrightarrow} \, e' \ifdefined\store \lottsym{;} \, \mathcolor{Store}{oₛ'} \fi}
}{o \, \lottsym{⊢} \, e \lottsym{.} i \ifdefined\store \lottsym{;} \, \mathcolor{Store}{oₛ} \fi \, \lottsym{\longrightarrow} \, e' \lottsym{.} i \ifdefined\store \lottsym{;} \, \mathcolor{Store}{oₛ'} \fi}
\lottinferencerulesep
\lottinferencerule{projRecord}{
\lotthypothesis{\lottkw{proj} \lottsym{(} o₁ \lottsym{;} \, i \lottsym{)} \, \lottsym{=} \, v}
}{o₀ \, \lottsym{⊢} \, \lottsym{\{} \, o₁ \, \lottsym{\}} \lottsym{.} i \ifdefined\store \lottsym{;} \, \mathcolor{Store}{oₛ} \fi \, \lottsym{\longrightarrow} \, v \ifdefined\store \lottsym{;} \, \mathcolor{Store}{oₛ} \fi}
\lottinferencerulesep
\lottinferencerule{mproj}{
\lotthypothesis{o \, \lottsym{⊢} \, e \ifdefined\store \lottsym{;} \, \mathcolor{Store}{oₛ} \fi \, \lottsym{\longrightarrow} \, e' \ifdefined\store \lottsym{;} \, \mathcolor{Store}{oₛ'} \fi}
}{o \, \lottsym{⊢} \, e \lottsym{.} \lottsym{\{} \, \lottsepbycomprehension{{i}_{j}} \, \lottsym{\}} \ifdefined\store \lottsym{;} \, \mathcolor{Store}{oₛ} \fi \, \lottsym{\longrightarrow} \, e' \lottsym{.} \lottsym{\{} \, \lottsepbycomprehension{{i}_{j}} \, \lottsym{\}} \ifdefined\store \lottsym{;} \, \mathcolor{Store}{oₛ'} \fi}
\lottinferencerulesep
\lottinferencerule{mprojRecord}{
\lotthypothesis{\lottkw{mproj} \lottsym{(} o₁ \lottsym{;} \, \lottsepbycomprehension{{i}_{j}} \lottsym{)} \, \lottsym{=} \, o₂}
}{o₀ \, \lottsym{⊢} \, \lottsym{\{} \, o₁ \, \lottsym{\}} \lottsym{.} \lottsym{\{} \, \lottsepbycomprehension{{i}_{j}} \, \lottsym{\}} \ifdefined\store \lottsym{;} \, \mathcolor{Store}{oₛ} \fi \, \lottsym{\longrightarrow} \, \lottsym{\{} \, o₂ \, \lottsym{\}} \ifdefined\store \lottsym{;} \, \mathcolor{Store}{oₛ} \fi}
\lottinferencerulesep
\lottinferencerule{diff}{
\lotthypothesis{o \, \lottsym{⊢} \, e \ifdefined\store \lottsym{;} \, \mathcolor{Store}{oₛ} \fi \, \lottsym{\longrightarrow} \, e' \ifdefined\store \lottsym{;} \, \mathcolor{Store}{oₛ'} \fi}
}{o \, \lottsym{⊢} \, e \lottsym{\textbackslash} \lottsym{\{} \, \lottsepbycomprehension{{i}_{j}} \, \lottsym{\}} \ifdefined\store \lottsym{;} \, \mathcolor{Store}{oₛ} \fi \, \lottsym{\longrightarrow} \, e' \lottsym{\textbackslash} \lottsym{\{} \, \lottsepbycomprehension{{i}_{j}} \, \lottsym{\}} \ifdefined\store \lottsym{;} \, \mathcolor{Store}{oₛ'} \fi}
\lottinferencerulesep
\lottinferencerule{diffRecord}{
\lotthypothesis{\lottkw{diff} \lottsym{(} o₁ \lottsym{;} \, \lottsepbycomprehension{{i}_{j}} \lottsym{)} \, \lottsym{=} \, o₂}
}{o₀ \, \lottsym{⊢} \, \lottsym{\{} \, o₁ \, \lottsym{\}} \lottsym{\textbackslash} \lottsym{\{} \, \lottsepbycomprehension{{i}_{j}} \, \lottsym{\}} \ifdefined\store \lottsym{;} \, \mathcolor{Store}{oₛ} \fi \, \lottsym{\longrightarrow} \, \lottsym{\{} \, o₂ \, \lottsym{\}} \ifdefined\store \lottsym{;} \, \mathcolor{Store}{oₛ} \fi}
\lottinferencerulesep
\lottinferencerule{let}{
\lotthypothesis{o \, \lottsym{⊢} \, e₀ \ifdefined\store \lottsym{;} \, \mathcolor{Store}{oₛ} \fi \, \lottsym{\longrightarrow} \, e₀' \ifdefined\store \lottsym{;} \, \mathcolor{Store}{oₛ'} \fi}
}{o \, \lottsym{⊢} \, \lottkw{let} \, t \, \lottsym{=} \, e₀ \, \lottkw{in} \, e₁ \ifdefined\store \lottsym{;} \, \mathcolor{Store}{oₛ} \fi \, \lottsym{\longrightarrow} \, \lottkw{let} \, t \, \lottsym{=} \, e₀' \, \lottkw{in} \, e₁ \ifdefined\store \lottsym{;} \, \mathcolor{Store}{oₛ'} \fi}
\lottinferencerulesep
\lottinferencerule{letv}{
\lotthypothesis{\lottkw{expand} \lottsym{(} t \lottsym{;} \, v \lottsym{)} \, \lottsym{=} \, o₁}
}{o₀ \, \lottsym{⊢} \, \lottkw{let} \, t \, \lottsym{=} \, v \, \lottkw{in} \, e \ifdefined\store \lottsym{;} \, \mathcolor{Store}{oₛ} \fi \, \lottsym{\longrightarrow} \, \lottsym{\{} \, o₀ \lottsym{,} \, o₁ \, \lottsym{\}} \, \lottsym{▶} \, e \ifdefined\store \lottsym{;} \, \mathcolor{Store}{oₛ} \fi}
\lottinferencerulesep
\lottinferencerule{ref}{
\lotthypothesis{o \, \lottsym{⊢} \, e \ifdefined\store \lottsym{;} \, \mathcolor{Store}{oₛ} \fi \, \lottsym{\longrightarrow} \, e' \ifdefined\store \lottsym{;} \, \mathcolor{Store}{oₛ'} \fi}
}{o \, \lottsym{⊢} \, \lottkw{ref} \, e \ifdefined\store \lottsym{;} \, \mathcolor{Store}{oₛ} \fi \, \lottsym{\longrightarrow} \, \lottkw{ref} \, e' \ifdefined\store \lottsym{;} \, \mathcolor{Store}{oₛ'} \fi}
\lottinferencerulesep
\lottinferencerule{refv}{
\lotthypothesis{l \, \lottsym{∉} \, oₛ}
}{o \, \lottsym{⊢} \, \lottkw{ref} \, v \ifdefined\store \lottsym{;} \, \mathcolor{Store}{oₛ} \fi \, \lottsym{\longrightarrow} \, l \ifdefined\store \lottsym{;} \, \mathcolor{Store}{oₛ \lottsym{,} \, l \, \lottsym{:} \, v} \fi}
\lottinferencerulesep
\lottinferencerule{deref}{
\lotthypothesis{o \, \lottsym{⊢} \, e \ifdefined\store \lottsym{;} \, \mathcolor{Store}{oₛ} \fi \, \lottsym{\longrightarrow} \, e' \ifdefined\store \lottsym{;} \, \mathcolor{Store}{oₛ'} \fi}
}{o \, \lottsym{⊢} \, \lottsym{!} e \ifdefined\store \lottsym{;} \, \mathcolor{Store}{oₛ} \fi \, \lottsym{\longrightarrow} \, \lottsym{!} e' \ifdefined\store \lottsym{;} \, \mathcolor{Store}{oₛ'} \fi}
\lottinferencerulesep
\lottinferencerule{derefLoc}{
}{o \, \lottsym{⊢} \, \lottsym{!} l \ifdefined\store \lottsym{;} \, \mathcolor{Store}{oₛ} \fi \, \lottsym{\longrightarrow} \, \lottsym{\{} \, oₛ \, \lottsym{\}} \lottsym{.} l \ifdefined\store \lottsym{;} \, \mathcolor{Store}{oₛ} \fi}
\lottinferencerulesep
\lottinferencerule{setl}{
\lotthypothesis{o \, \lottsym{⊢} \, e₀ \ifdefined\store \lottsym{;} \, \mathcolor{Store}{oₛ} \fi \, \lottsym{\longrightarrow} \, e₀' \ifdefined\store \lottsym{;} \, \mathcolor{Store}{oₛ'} \fi}
}{o \, \lottsym{⊢} \, e₀ \, \lottsym{≔} \, e₁ \ifdefined\store \lottsym{;} \, \mathcolor{Store}{oₛ} \fi \, \lottsym{\longrightarrow} \, e₀' \, \lottsym{≔} \, e₁ \ifdefined\store \lottsym{;} \, \mathcolor{Store}{oₛ'} \fi}
\lottinferencerulesep
\lottinferencerule{setr}{
\lotthypothesis{o \, \lottsym{⊢} \, e \ifdefined\store \lottsym{;} \, \mathcolor{Store}{oₛ} \fi \, \lottsym{\longrightarrow} \, e' \ifdefined\store \lottsym{;} \, \mathcolor{Store}{oₛ'} \fi}
}{o \, \lottsym{⊢} \, v \, \lottsym{≔} \, e \ifdefined\store \lottsym{;} \, \mathcolor{Store}{oₛ} \fi \, \lottsym{\longrightarrow} \, v \, \lottsym{≔} \, e' \ifdefined\store \lottsym{;} \, \mathcolor{Store}{oₛ'} \fi}
\lottinferencerulesep
\lottinferencerule{set}{
}{o \, \lottsym{⊢} \, l \, \lottsym{≔} \, v \ifdefined\store \lottsym{;} \, \mathcolor{Store}{oₛ} \fi \, \lottsym{\longrightarrow} \, \lottsym{\{} \, \lottsym{\}} \ifdefined\store \lottsym{;} \, \mathcolor{Store}{oₛ \, \lottsym{⧺} \, l \, \lottsym{:} \, v} \fi}
}
}}}

  \caption{Operational semantics}
  \label{fig:small-step}
\end{figure}

First, rule \lottref{E}{lam} evaluates lambdas into closures with an empty
record of accumulated arguments. The current environment is captured inside the
closure to preserve an important invariant: values~(e.g. closures) are
self-contained and do not depend on the surrounding scope. In rule
\lottref{E}{app}, the accumulated arguments in $o₂$ are concatenated
with the new arguments $o₃$. The meta-notation $\doubleplus$~(also
used in rule \lottref{E}{concat}) is analogous to the type-level concatenation
relation; its complete definition is provided in the appendix. The resulting
term remains well-typed, a property guaranteed in part by
\Cref{lem:app-con-assoc}.

Closures are eliminated via rule \lottref{E}{forceClosure}, which converts them
into boxes. Here, the $\lottkw{expand}(c; o_2)$ operation destructures $o_2$ and
renames entries according to the record pattern $c$. The rule then evaluates $e$
under the captured environment $o_1$ extended with the arguments $o_3$.
Similarly, rule \lottref{E}{letv} performs a corresponding conversion to
evaluate let-bindings.

Rule \lottref{E}{boxr} is the only rule that updates the evaluation environment.
Specifically, the rule ignores the current environment $o_0$, and evaluates $e$
under the environment $o_1$. Rule \lottref{E}{box} discards the environment for
values, since values are self-contained. Rule \lottref{E}{env} is the only rule
where the evaluation environment is accessed, serving a dual role to that of
rule \lottref{E}{boxr}.

Finally, rules \lottref{E}{projRecord}, \lottref{E}{mprojRecord}, and
\lottref{E}{diffRecord} handle the basic record operations. Auxiliary meta-operations
$\lottkw{proj}$, $\lottkw{mproj}$, and $\lottkw{diff}$ behave straightforwardly; their
full definitions are omitted for brevity and included in the appendix.
Mechanically, these notations traverse $o₁$ from right to left to
select or discard the designated entries.

\subsection{Desirable Properties and Type Soundness}
\label{sec:properties}

In this section, we demonstrate that our calculus enjoys several desirable
properties, which facilitate reasoning about system behaviour, and we then prove
type soundness and normalization.

\paragraph{\textbf{Subtyping}}

First, we prove that subtyping is reflexive and transitive:

\begin{lemma}[Subtyping reflexivity]
  \label{lemma:sub-refl}
  $A \, \lottsym{\mathcolor{Subtyping}{⩽}} \, A$ for all $A$.
\end{lemma}

\begin{lemma}[Subtyping transitivity]
  \label[lemma]{lem:sub-trans}
  If $A \, \lottsym{\mathcolor{Subtyping}{⩽}} \, B$ and $B \, \lottsym{\mathcolor{Subtyping}{⩽}} \, C$, then $A \, \lottsym{\mathcolor{Subtyping}{⩽}} \, C$.
\end{lemma}

Record subtyping is preserved when entries are appended;
the $\lottsym{⌊} ρ₁ \lottsym{⌋}$ condition ensures
that \lottref{S}{empty} has not added optional entries on the left-hand side
for a name that appears in $ρ₂$:

\begin{lemma}[Record subtyping appending]
  \label[lemma]{lem:sub-append}
  If $ρ₀ \, \lottsym{\mathcolor{Subtyping}{⩽}} \, ρ₁$, $ρ₂ \, \lottsym{\mathcolor{Subtyping}{⩽}} \, ρ₃$, and $\lottsym{⌊} ρ₁ \lottsym{⌋}$, then
  $ρ₀ \lottsym{,} \, ρ₂ \, \lottsym{\mathcolor{Subtyping}{⩽}} \, ρ₁ \lottsym{,} \, ρ₃$.
\end{lemma}

Moreover, subtyping on the environment is admissible:

\begin{lemma}[Environment subtyping]
  If $\ifdefined\store \mathcolor{Store}{ρₛ} \lottsym{;} \, \fi ρ₀ \, \lottsym{⊢} \, e \, \lottsym{:} \, A$ with $\lottsym{⌊} ρ₀ \lottsym{⌋}$, and $ρ₁ \, \lottsym{\mathcolor{Subtyping}{⩽}} \, ρ₀$,
  then $\ifdefined\store \mathcolor{Store}{ρₛ} \lottsym{;} \, \fi ρ₁ \, \lottsym{⊢} \, e \, \lottsym{:} \, A$.
\end{lemma}

\noindent
Intuitively, this holds since the only way to access the environment is via
$\lottsym{@}$ through rule \lottref{T}{env}, and we can always apply the subsumption
rule (\lottref{T}{sub}) and the record subtyping rule (\lottref{S}{record})
immediately after.
The second hypothesis ensures the subtyping relationship can be maintained using
\Cref{lem:sub-append} even when new bindings are introduced.

\paragraph{\textbf{Concatenation}}

The following lemma establishes the associativity property of concatenation. For
concision, we often embed record operators directly within other relations. For
instance, we write $\tconcat{(\tconcatb{\rho_0}{\rho_1})}{\rho_2}{\rho_3}$,
which implies there exists $\rho$ such that $\tconcat{\rho_0}{\rho_1}{\rho}$
and $\tconcat{\rho}{\rho_2}{\rho_3}$.

\begin{lemma}[Associativity of $\tconcats$]
    $\tconcat{(\tconcatb{\rho_0}{\rho_1})}{\rho_2}{\rho_3}$ if and only if
    $\tconcat{\rho_0}{(\tconcatb{\rho_1}{\rho_2})}{\rho_3}$.
\end{lemma}

\paragraph{\textbf{Application and record concatenation}}

The next lemma establishes an associativity-like property between application
and concatenation. Intuitively, it shows that given an initial $ρ₀$,
consecutively applying it first to $ρ_1$ and then to $ρ_2$ yields the same result as
first concatenating $ρ_1$ and $ρ_2$ and then applying $ρ_0$ to the result.

\begin{lemma}[Associativity of $\apps$ over $\tconcats$]
  \label[lemma]{lem:app-con-assoc}
  $\app{(\appb{\rho_0}{\rho_1})}{\rho_2}{\rho_3}$ if and only if
  $\app{\rho_0}{(\tconcatb{\rho_1}{\rho_2})}{\rho_3}$.
\end{lemma}

\paragraph{\textbf{Type soundness}}

First, we prove that the operational semantics enjoy determinacy:

\begin{theorem}[Determinacy]
  \label{thm:determinacy}
  If $o \, \lottsym{⊢} \, e \ifdefined\store \lottsym{;} \, \mathcolor{Store}{oₛ} \fi \, \lottsym{\longrightarrow} \, e₀ \ifdefined\store \lottsym{;} \, \mathcolor{Store}{\lottsym{ε}} \fi$ and $o \, \lottsym{⊢} \, e \ifdefined\store \lottsym{;} \, \mathcolor{Store}{oₛ} \fi \, \lottsym{\longrightarrow} \, e₁ \ifdefined\store \lottsym{;} \, \mathcolor{Store}{\lottsym{ε}} \fi$ then
  $e₀ = e₁$.
\end{theorem}

We then establish type soundness for \lang through progress and preservation.

\begin{theorem}[Preservation]
  \label{thm:preservation}
  Given $\ifdefined\store \mathcolor{Store}{ρₛ} \lottsym{;} \, \fi \lottsym{ε} \, \lottsym{⊢} \, \lottsym{\{} \, o \, \lottsym{\}} \, \lottsym{:} \, \lottsym{\{} \, ρ \, \lottsym{\}}$ with $\uniform{\rho}$,
  if $\ifdefined\store \mathcolor{Store}{ρₛ} \lottsym{;} \, \fi ρ \, \lottsym{⊢} \, e \, \lottsym{:} \, A$,  and
  $o \, \lottsym{⊢} \, e \ifdefined\store \lottsym{;} \, \mathcolor{Store}{oₛ} \fi \, \lottsym{\longrightarrow} \, e' \ifdefined\store \lottsym{;} \, \mathcolor{Store}{oₛ'} \fi$,~then~$\ifdefined\store \mathcolor{Store}{ρₛ \lottsym{,} \, ρₛ'} \lottsym{;} \, \fi ρ \, \lottsym{⊢} \, e' \, \lottsym{:} \, A$.
\end{theorem}

\begin{theorem}[Progress]
  \label{thm:progress}
  Given $\ifdefined\store \mathcolor{Store}{ρₛ} \lottsym{;} \, \fi \lottsym{ε} \, \lottsym{⊢} \, \lottsym{\{} \, o \, \lottsym{\}} \, \lottsym{:} \, \lottsym{\{} \, ρ \, \lottsym{\}}$ with $\uniform{\rho}$,
  if $\ifdefined\store \mathcolor{Store}{ρₛ} \lottsym{;} \, \fi ρ \, \lottsym{⊢} \, e \, \lottsym{:} \, A$,
  then either $e$ is a value, or there exists
  $e'$ such that $o \, \lottsym{⊢} \, e \ifdefined\store \lottsym{;} \, \mathcolor{Store}{oₛ} \fi \, \lottsym{\longrightarrow} \, e' \ifdefined\store \lottsym{;} \, \mathcolor{Store}{oₛ'} \fi$.
\end{theorem}

\paragraph{\textbf{Normalization}}

Additionally, since our reduction is non-standard, we also formally prove its
normalization property. The proof proceeds via a logical
relation~\cite{tait1967intensional}:
\vspace{-10pt}

\begin{gather*}
  \begin{aligned}
  \text{$ρ \lottsym{;} \, o \, \lottsym{⊢} \, \lottkw{SN}_{\lottsym{\{} \, ρ₀ \, \lottsym{\}} \, \lottsym{→} \, A} \lottsym{(} e \lottsym{)}$} &= \text{$\ifdefined\store \mathcolor{Store}{\lottsym{ε}} \lottsym{;} \, \fi ρ \, \lottsym{⊢} \, e \, \lottsym{:} \, \lottsym{\{} \, ρ₀ \, \lottsym{\}} \, \lottsym{→} \, A$} ∧ \text{$o \, \lottsym{⊢} \, \lottkw{SN} \lottsym{(} e \lottsym{)}$} ∧ ∀ ρ₁ \, ρ₂, \text{$ρ₀ \, \lottsym{\mathcolor{App}{●}} \, ρ₁ \, \lottsym{\mathcolor{App}{↠}} \, ρ₂$} ⇒ \text{${\lottsym{⌊} ρ₂ \lottsym{⌋}}^{\lottsym{?}}$} ⇒ \\&\hspace{-4em} ∀ v, (∀ ρ' o', \text{$\ifdefined\store \mathcolor{Store}{\lottsym{ε}} \lottsym{;} \, \fi \lottsym{ε} \, \lottsym{⊢} \, \lottsym{\{} \, o' \, \lottsym{\}} \, \lottsym{:} \, \lottsym{\{} \, ρ' \, \lottsym{\}}$} ⇒ \text{$\lottsym{⌊} ρ' \lottsym{⌋}$} ⇒ \text{$ρ' \lottsym{;} \, o' \, \lottsym{⊢} \, \lottkw{SN}_{\lottsym{\{} \, ρ₁ \, \lottsym{\}}} \lottsym{(} v \lottsym{)}$}) ⇒ \text{$ρ \lottsym{;} \, o \, \lottsym{⊢} \, \lottkw{SN}_{A} \lottsym{(} \lottkw{force} \, \lottsym{(} e \, v \lottsym{)} \lottsym{)}$}\\
  \text{$ρ \lottsym{;} \, o \, \lottsym{⊢} \, \lottkw{SN}_{\lottsym{\{} \, ρ' \, \lottsym{\}}} \lottsym{(} e \lottsym{)}$} &= \text{$\ifdefined\store \mathcolor{Store}{\lottsym{ε}} \lottsym{;} \, \fi ρ \, \lottsym{⊢} \, e \, \lottsym{:} \, \lottsym{\{} \, ρ' \, \lottsym{\}}$} ∧ ∃ \, r, \text{$o \, \lottsym{⊢} \, e \ifdefined\store \lottsym{;} \, \mathcolor{Store}{\lottsym{ε}} \fi \, \lottsym{\overset{*}{\longrightarrow}} \, \lottsym{\{} \, r \, \lottsym{\}} \ifdefined\store \lottsym{;} \, \mathcolor{Store}{\lottsym{ε}} \fi$} ∧ ∀ (e', A) ∈ (r, ρ'), \text{$ρ \lottsym{;} \, o \, \lottsym{⊢} \, \lottkw{SN}_{A} \lottsym{(} e' \lottsym{)}$}
  \end{aligned}\\
  \text{$ρ \lottsym{;} \, o \, \lottsym{⊢} \, \lottkw{SN}_{\lottsym{ℕ}} \lottsym{(} e \lottsym{)}$} = \text{$\ifdefined\store \mathcolor{Store}{\lottsym{ε}} \lottsym{;} \, \fi ρ \, \lottsym{⊢} \, e \, \lottsym{:} \, \lottsym{ℕ}$} ∧ \text{$o \, \lottsym{⊢} \, \lottkw{SN} \lottsym{(} e \lottsym{)}$}\qquad\text{$ρ \lottsym{;} \, o \, \lottsym{⊢} \, \lottkw{SN}_{\lottsym{∀} \, a \lottsym{.} \, A} \lottsym{(} e \lottsym{)}$} = \text{$\ifdefined\store \mathcolor{Store}{\lottsym{ε}} \lottsym{;} \, \fi ρ \, \lottsym{⊢} \, e \, \lottsym{:} \, \lottsym{∀} \, a \lottsym{.} \, A$} ∧ ∀ \, τ, ρ \lottsym{;} \, o \, \lottsym{⊢} \, \lottkw{SN}_{A [τ/a]} \lottsym{(} e \lottsym{)}
\end{gather*}
\vspace{-12pt}

\noindent
In the function case, we quantify over record types whose application leaves
only entries of optional strength ($\uniform{\rho_2}^?$), and over an argument
value which satisfies the relation under any well-typed environment. Terms of
record types are required to reduce to concrete records whose entries all satisfy
the relation. The quantifier case is straightforward with predicative
polymorphism. The most interesting aspect of these proofs is that we avoid
substitutions by assuming the first-class environment satisfies the relation.
We prove the fundamental theorem and then normalization:

\begin{theorem}[Fundamental]
  Given $\ifdefined\store \mathcolor{Store}{\lottsym{ε}} \lottsym{;} \, \fi ρ \, \lottsym{⊢} \, e \, \lottsym{:} \, A$ with $\uniform{\rho}$, if
  $\lottsym{ε} \lottsym{;} \, \lottsym{ε} \, \lottsym{⊢} \, \lottkw{SN}_{\lottsym{\{} \, ρ \, \lottsym{\}}} \lottsym{(} \lottsym{\{} \, o \, \lottsym{\}} \lottsym{)}$ then $ρ \lottsym{;} \, o \, \lottsym{⊢} \, \lottkw{SN}_{A} \lottsym{(} e \lottsym{)}$.
\end{theorem}

\vspace{-5pt}

\begin{theorem}[Normalization]
  \label{thm:normalization}
  If $ρ \lottsym{;} \, o \, \lottsym{⊢} \, \lottkw{SN}_{A} \lottsym{(} e \lottsym{)}$ then $e$ terminates.
\end{theorem}

\subsection{Syntax-Directed Typing}

The declarative typing rules are not syntax-directed: subsumption
(\lottref{T}{sub}) may be applied anywhere, and quantifiers are instantiated
implicitly through subtyping. In this section we present a syntax-directed
variant that is sound and complete with
respect to the declarative system. Full algorithmic type inference is outside
the scope of this paper, but the syntax-directed rules show how our features
can be handled without guessing where to apply subsumption. Selected rules are
given in \Cref{fig:syntax-directed}; the full set of typing and updated
auxiliary judgement rules can be found in the appendix.

\begin{figure}[H]
  \vspace{-7pt}
  \small
  \begingroup
  \renewcommand{\lottinferencerule}[3]{\relax}
  \renewcommand{\lottinferencerulesep}{\relax}
  \renewcommand{\lottjudgement}[3]{\fbox{$#2$\rule[-0.2\baselineskip]{0pt}{0.1\baselineskip}}}
\lottjudgement{Syntax Directed}{\ifdefined\store \mathcolor{Store}{ρₛ} \lottsym{;} \, \fi ρ \, \lottsym{⊢}_s \, e \, \lottsym{:} \, A}{
\lottinferencerule{nat}{
}{\ifdefined\store \mathcolor{Store}{ρₛ} \lottsym{;} \, \fi ρ \, \lottsym{⊢}_s \, n \, \lottsym{:} \, \lottsym{ℕ}}
\lottinferencerulesep
\lottinferencerule{op}{
\lotthypothesis{\ifdefined\store \mathcolor{Store}{ρₛ} \lottsym{;} \, \fi ρ \, \lottsym{⊢}_s \, e₀ \, \lottsym{:} \, A}\\
\lotthypothesis{A \, \lottsym{\mathcolor{Subtyping}{⩽}} \, \lottsym{ℕ}}\\
\lotthypothesis{\ifdefined\store \mathcolor{Store}{ρₛ} \lottsym{;} \, \fi ρ \, \lottsym{⊢}_s \, e₁ \, \lottsym{:} \, B}\\
\lotthypothesis{B \, \lottsym{\mathcolor{Subtyping}{⩽}} \, \lottsym{ℕ}}
}{\ifdefined\store \mathcolor{Store}{ρₛ} \lottsym{;} \, \fi ρ \, \lottsym{⊢}_s \, e₀ \, b \, e₁ \, \lottsym{:} \, \lottsym{ℕ}}
\lottinferencerulesep
\lottinferencerule{lam}{\ifdefined\store\overline{a_j} \, \lottsym{∉} \, \lottkw{ftv} \lottsym{(} \mathcolor{Store}{ρₛ} \lottsym{)}\\\fi\overline{a_j} \, \lottsym{∉} \, \lottkw{ftv} \lottsym{(} ρ₀ \lottsym{)}\\
\lotthypothesis{c \, \lottsym{\mathcolor{Pattern}{:}} \, ρ₁ \, \lottsym{\mathcolor{Pattern}{⊙}} \, ρ₂}\\
\lotthypothesis{\ifdefined\store \mathcolor{Store}{ρₛ} \lottsym{;} \, \fi ρ₀ \lottsym{,} \, ρ₂ \, \lottsym{⊢}_s \, e \, \lottsym{:} \, A}
}{\ifdefined\store \mathcolor{Store}{ρₛ} \lottsym{;} \, \fi ρ₀ \, \lottsym{⊢}_s \, \lottsym{λ} \, \lottsym{\{} \, c \, \lottsym{\}} \lottsym{.} \, e \, \lottsym{:} \, \lottsym{∀} \, \lottsepbycomprehension{{a}_{j}} \lottsym{.} \, \lottsym{\{} \, ρ₁ \, \lottsym{\}} \, \lottsym{→} \, A}
\lottinferencerulesep
\lottinferencerule{closure}{\ifdefined\store\overline{a_j} \, \lottsym{∉} \, \lottkw{ftv} \lottsym{(} \mathcolor{Store}{ρₛ} \lottsym{)}\\\fi
\lotthypothesis{\ifdefined\store \mathcolor{Store}{ρₛ} \lottsym{;} \, \fi \lottsym{ε} \, \lottsym{⊢}_s \, \lottsym{\{} \, o₀ \, \lottsym{\}} \, \lottsym{:} \, A}\\
\lotthypothesis{A \, \lottsym{\mathcolor{Subtyping}{⩽}} \, \lottsym{\{} \, ρ₁ \, \lottsym{\}}}\\
\lotthypothesis{\lottsym{⌊} ρ₁ \lottsym{⌋}}\\\overline{a_j} \, \lottsym{∉} \, \lottkw{ftv} \lottsym{(} ρ₁ \lottsym{)}\\
\lotthypothesis{\ifdefined\store \mathcolor{Store}{ρₛ} \lottsym{;} \, \fi \lottsym{ε} \, \lottsym{⊢}_s \, \lottsym{\{} \, o₁ \, \lottsym{\}} \, \lottsym{:} \, B}\\
\lotthypothesis{B \, \lottsym{\mathcolor{Subtyping}{⩽}} \, \lottsym{∀} \, \lottsepbycomprehension{{a}_{j}} \lottsym{.} \, \lottsym{\{} \, ρ₂ \, \lottsym{\}}}\\
\lotthypothesis{c \, \lottsym{\mathcolor{Pattern}{:}} \, ρ₃ \, \lottsym{\mathcolor{Pattern}{⊙}} \, ρ₄}\\
\lotthypothesis{ρ₃ \, \lottsym{\mathcolor{App}{●}} \, ρ₂ \, \lottsym{\mathcolor{App}{↠}} \, ρ₅}\\
\lotthypothesis{\ifdefined\store \mathcolor{Store}{ρₛ} \lottsym{;} \, \fi ρ₁ \lottsym{,} \, ρ₄ \, \lottsym{⊢}_s \, e \, \lottsym{:} \, C}
}{\ifdefined\store \mathcolor{Store}{ρₛ} \lottsym{;} \, \fi ρ₀ \, \lottsym{⊢}_s \, \lottsym{⟨} o₀ \lottsym{;} \, o₁ \, \lottsym{|} \, \lottsym{λ} \, \lottsym{\{} \, c \, \lottsym{\}} \lottsym{.} \, e \lottsym{⟩} \, \lottsym{:} \, \lottsym{∀} \, \lottsepbycomprehension{{a}_{j}} \lottsym{.} \, \lottsym{\{} \, ρ₅ \, \lottsym{\}} \, \lottsym{→} \, C}
\lottinferencerulesep
\lottinferencerule{app}{\overline{a₂_l} \, \lottsym{∉} \, \lottkw{ftv} \lottsym{(} ρ₁ \lottsym{)}\\\overline{a₂_l} \, \lottsym{∉} \, \lottkw{ftv} \lottsym{(} A \lottsym{)}\\\overline{a₂_l} \, \lottsym{∉} \, \lottkw{ftv} \lottsym{(} ρ₂ \lottsym{)}\\
\lotthypothesis{\ifdefined\store \mathcolor{Store}{ρₛ} \lottsym{;} \, \fi ρ₀ \, \lottsym{⊢}_s \, e₀ \, \lottsym{:} \, \lottsym{∀} \, \lottsepbycomprehension{{a₀}_{j}} \lottsym{.} \, \lottsym{\{} \, ρ₁ \, \lottsym{\}} \, \lottsym{→} \, A}\\
\lotthypothesis{\ifdefined\store \mathcolor{Store}{ρₛ} \lottsym{;} \, \fi ρ₀ \, \lottsym{⊢}_s \, e₁ \, \lottsym{:} \, \lottsym{∀} \, \lottsepbycomprehension{{a₁}_{k}} \lottsym{.} \, \lottsym{\{} \, ρ₂ \, \lottsym{\}}}\\
\lotthypothesis{ρ₁\left[\lottsepbycomprehension{{τ₀}_j/{a₀}_{j}}\right] \, \lottsym{\mathcolor{App}{●}}_s \, ρ₂\left[\lottsepbycomprehension{{τ₁}_j/{a₁}_{k}}\right] \, \lottsym{\mathcolor{App}{↠}} \, ρ₃}
}{\ifdefined\store \mathcolor{Store}{ρₛ} \lottsym{;} \, \fi ρ₀ \, \lottsym{⊢}_s \, e₀ \, e₁ \, \lottsym{:} \, \lottsym{∀} \, \lottsepbycomprehension{{a₂}_{l}} \lottsym{.} \, \lottsym{\{} \, ρ₃ \, \lottsym{\}} \, \lottsym{→} \, A\left[\lottsepbycomprehension{{τ₀}_j/{a₀}_{j}}\right]}
\lottinferencerulesep
\lottinferencerule{appAnyL}{\overline{a₂_l} \, \lottsym{∉} \, \lottkw{ftv} \lottsym{(} ρ₂ \lottsym{)}\\
\lotthypothesis{\ifdefined\store \mathcolor{Store}{ρₛ} \lottsym{;} \, \fi ρ₀ \, \lottsym{⊢}_s \, e₀ \, \lottsym{:} \, \lottsym{∀} \, \lottsepbycomprehension{{a₀}_{j}} \lottsym{.} \, {a}_{m}}\\
\lotthypothesis{\ifdefined\store \mathcolor{Store}{ρₛ} \lottsym{;} \, \fi ρ₀ \, \lottsym{⊢}_s \, e₁ \, \lottsym{:} \, \lottsym{∀} \, \lottsepbycomprehension{{a₁}_{k}} \lottsym{.} \, \lottsym{\{} \, ρ₂ \, \lottsym{\}}}\\
\lotthypothesis{ϱ \, \lottsym{\mathcolor{App}{●}}_s \, ρ₂\left[\lottsepbycomprehension{{τ₁}_j/{a₁}_{k}}\right] \, \lottsym{\mathcolor{App}{↠}} \, ρ₃}
}{\ifdefined\store \mathcolor{Store}{ρₛ} \lottsym{;} \, \fi ρ₀ \, \lottsym{⊢}_s \, e₀ \, e₁ \, \lottsym{:} \, \lottsym{∀} \, \lottsepbycomprehension{{a₂}_{l}} \lottsym{.} \, \lottsym{\{} \, ρ₃ \, \lottsym{\}} \, \lottsym{→} \, τ₀}
\lottinferencerulesep
\lottinferencerule{appAnyR}{\overline{a₂_l} \, \lottsym{∉} \, \lottkw{ftv} \lottsym{(} ρ₁ \lottsym{)}\\\overline{a₂_l} \, \lottsym{∉} \, \lottkw{ftv} \lottsym{(} A \lottsym{)}\\
\lotthypothesis{\ifdefined\store \mathcolor{Store}{ρₛ} \lottsym{;} \, \fi ρ₀ \, \lottsym{⊢}_s \, e₀ \, \lottsym{:} \, \lottsym{∀} \, \lottsepbycomprehension{{a₀}_{j}} \lottsym{.} \, \lottsym{\{} \, ρ₁ \, \lottsym{\}} \, \lottsym{→} \, A}\\
\lotthypothesis{\ifdefined\store \mathcolor{Store}{ρₛ} \lottsym{;} \, \fi ρ₀ \, \lottsym{⊢}_s \, e₁ \, \lottsym{:} \, \lottsym{∀} \, \lottsepbycomprehension{{a₁}_{k}} \lottsym{.} \, {a}_{m}}\\
\lotthypothesis{ρ₁\left[\lottsepbycomprehension{{τ}_j/{a₀}_{j}}\right] \, \lottsym{\mathcolor{App}{●}}_s \, ϱ \, \lottsym{\mathcolor{App}{↠}} \, ρ₃}
}{\ifdefined\store \mathcolor{Store}{ρₛ} \lottsym{;} \, \fi ρ₀ \, \lottsym{⊢}_s \, e₀ \, e₁ \, \lottsym{:} \, \lottsym{∀} \, \lottsepbycomprehension{{a₂}_{l}} \lottsym{.} \, \lottsym{\{} \, ρ₃ \, \lottsym{\}} \, \lottsym{→} \, A\left[\lottsepbycomprehension{{τ}_j/{a₀}_{j}}\right]}
\lottinferencerulesep
\lottinferencerule{appAny}{
\lotthypothesis{\ifdefined\store \mathcolor{Store}{ρₛ} \lottsym{;} \, \fi ρ₀ \, \lottsym{⊢}_s \, e₀ \, \lottsym{:} \, \lottsym{∀} \, \lottsepbycomprehension{{a₀}_{j}} \lottsym{.} \, {a}_{m₀}}\\
\lotthypothesis{\ifdefined\store \mathcolor{Store}{ρₛ} \lottsym{;} \, \fi ρ₀ \, \lottsym{⊢}_s \, e₁ \, \lottsym{:} \, \lottsym{∀} \, \lottsepbycomprehension{{a₁}_{k}} \lottsym{.} \, {a}_{m₁}}\\
\lotthypothesis{ϱ₁ \, \lottsym{\mathcolor{App}{●}}_s \, ϱ₂ \, \lottsym{\mathcolor{App}{↠}} \, ρ₃}
}{\ifdefined\store \mathcolor{Store}{ρₛ} \lottsym{;} \, \fi ρ₀ \, \lottsym{⊢}_s \, e₀ \, e₁ \, \lottsym{:} \, \lottsym{∀} \, \lottsepbycomprehension{{a₂}_{l}} \lottsym{.} \, \lottsym{\{} \, ρ₃ \, \lottsym{\}} \, \lottsym{→} \, τ}
\lottinferencerulesep
\lottinferencerule{force}{
\lotthypothesis{\ifdefined\store \mathcolor{Store}{ρₛ} \lottsym{;} \, \fi ρ₀ \, \lottsym{⊢}_s \, e \, \lottsym{::} \, \lottsym{∀} \, \lottsepbycomprehension{{a}_{j}} \lottsym{.} \, \lottsym{\{} \, ρ₁ \, \lottsym{\}} \, \lottsym{→} \, A}\\
\lotthypothesis{{\lottsym{⌊} ρ₁ \lottsym{⌋}}^{\lottsym{?}}}
}{\ifdefined\store \mathcolor{Store}{ρₛ} \lottsym{;} \, \fi ρ₀ \, \lottsym{⊢}_s \, \lottkw{force} \, e \, \lottsym{:} \, A\left[\lottsepbycomprehension{{τ}_j/{a}_{j}}\right]}
\lottinferencerulesep
\lottinferencerule{box}{
\lotthypothesis{\ifdefined\store \mathcolor{Store}{ρₛ} \lottsym{;} \, \fi ρ₀ \, \lottsym{⊢}_s \, e₀ \, \lottsym{::} \, \lottsym{∀} \, \lottsepbycomprehension{{a}_{j}} \lottsym{.} \, \lottsym{\{} \, ρ₁ \, \lottsym{\}}}\\
\lotthypothesis{\lottsym{⌊} ρ₁ \lottsym{⌋}}\\
\lotthypothesis{\ifdefined\store \mathcolor{Store}{ρₛ} \lottsym{;} \, \fi ρ₁\left[\lottsepbycomprehension{{τ}_j/{a}_{j}}\right] \, \lottsym{⊢}_s \, e₁ \, \lottsym{:} \, A}
}{\ifdefined\store \mathcolor{Store}{ρₛ} \lottsym{;} \, \fi ρ₀ \, \lottsym{⊢}_s \, e₀ \, \lottsym{▶} \, e₁ \, \lottsym{:} \, A}
\lottinferencerulesep
\lottinferencerule{env}{
}{\ifdefined\store \mathcolor{Store}{ρₛ} \lottsym{;} \, \fi ρ \, \lottsym{⊢}_s \, \lottsym{@} \, \lottsym{:} \, \lottsym{\{} \, ρ \, \lottsym{\}}}
\lottinferencerulesep
\lottinferencerule{record}{
\lotthypothesis{\ifdefined\store \mathcolor{Store}{ρₛ} \lottsym{;} \, \fi ρ₀ \, \lottsym{⊢}_s \, r \, \lottsym{:} \, ρ₁}
}{\ifdefined\store \mathcolor{Store}{ρₛ} \lottsym{;} \, \fi ρ₀ \, \lottsym{⊢}_s \, \lottsym{\{} \, r \, \lottsym{\}} \, \lottsym{:} \, \lottsym{\{} \, ρ₁ \, \lottsym{\}}}
\lottinferencerulesep
\lottinferencerule{concat}{
\lotthypothesis{\ifdefined\store \mathcolor{Store}{ρₛ} \lottsym{;} \, \fi ρ₀ \, \lottsym{⊢}_s \, e₀ \, \lottsym{::} \, \lottsym{∀} \, \lottsepbycomprehension{{a₁}_{j}} \lottsym{.} \, \lottsym{\{} \, ρ₁ \, \lottsym{\}}}\\
\lotthypothesis{\ifdefined\store \mathcolor{Store}{ρₛ} \lottsym{;} \, \fi ρ₀ \, \lottsym{⊢}_s \, e₁ \, \lottsym{::} \, \lottsym{∀} \, \lottsepbycomprehension{{a₂}_{k}} \lottsym{.} \, \lottsym{\{} \, ρ₂ \, \lottsym{\}}}\\
\lotthypothesis{ρ₁\left[\lottsepbycomprehension{{τ₁}_j/{a₁}_{j}}\right] \, \lottsym{\mathcolor{Concat}{⊕}}_s \, ρ₂\left[\lottsepbycomprehension{{τ₂}_j/{a₂}_{k}}\right] \, \lottsym{\mathcolor{Concat}{↬}} \, ρ₃}
}{\ifdefined\store \mathcolor{Store}{ρₛ} \lottsym{;} \, \fi ρ₀ \, \lottsym{⊢}_s \, e₀ \, \lottsym{⧺} \, e₁ \, \lottsym{:} \, \lottsym{\{} \, ρ₃ \, \lottsym{\}}}
\lottinferencerulesep
\lottinferencerule{proj}{
\lotthypothesis{\ifdefined\store \mathcolor{Store}{ρₛ} \lottsym{;} \, \fi ρ₀ \, \lottsym{⊢}_s \, e \, \lottsym{::} \, \lottsym{∀} \, \lottsepbycomprehension{{a}_{j}} \lottsym{.} \, \lottsym{\{} \, ρ₁ \lottsym{,} \, i \, \rstrength \, A \lottsym{,} \, ρ₂ \, \lottsym{\}}}\\
\lotthypothesis{i \, \lottsym{∉} \, ρ₂}
}{\ifdefined\store \mathcolor{Store}{ρₛ} \lottsym{;} \, \fi ρ₀ \, \lottsym{⊢}_s \, e \lottsym{.} i \, \lottsym{:} \, A\left[\lottsepbycomprehension{{τ}_j/{a}_{j}}\right]}
\lottinferencerulesep
\lottinferencerule{mproj}{
\lotthypothesis{\ifdefined\store \mathcolor{Store}{ρₛ} \lottsym{;} \, \fi ρ₀ \, \lottsym{⊢}_s \, e \, \lottsym{::} \, \lottsym{∀} \, \lottsepbycomprehension{{a}_{j}} \lottsym{.} \, \lottsym{\{} \, ρ₁ \lottsym{,} \, \lottsepbycomprehensionpatcollection{{i₀}_{k} \, {\rstrength₀}_{k} \, {A}_{k} \lottsym{,} \, {ρ₂}_{k}}{k}{[:n]} \, \lottsym{\}}}\\
\lotthypothesis{\lottjudgementcomprehensionpatcollection{\lottjudgementcomprehensionpatcollection{{i₀}_{k} \, \lottsym{∉} \, {ρ₂}_{l}}{l}{[:k+1]}}{k}{[:n]}}\\
\lotthypothesis{\lottsepbycomprehensionpatcollection{{i₀}_{k} \, {\rstrength₀}_{k} \, {A}_{k}}{k}{[:n]} \, \lottkw{permutes} \, \lottsepbycomprehensionpatcollection{{i₁}_{k} \, {\rstrength₁}_{k} \, {B}_{k}}{k}{[:n]}}
}{\ifdefined\store \mathcolor{Store}{ρₛ} \lottsym{;} \, \fi ρ₀ \, \lottsym{⊢}_s \, e \lottsym{.} \lottsym{\{} \, \lottsepbycomprehensionpatcollection{{i₁}_{k}}{k}{[:n]} \, \lottsym{\}} \, \lottsym{:} \, \lottsym{\{} \, \lottsepbycomprehensionpatcollection{{i₁}_{k} \, {\rstrength₁}_{k} \, {B}_{k}\left[\lottsepbycomprehension{{τ}_j/{a}_{j}}\right]}{k}{[:n]} \, \lottsym{\}}}
\lottinferencerulesep
\lottinferencerule{diff}{
\lotthypothesis{\ifdefined\store \mathcolor{Store}{ρₛ} \lottsym{;} \, \fi ρ₀ \, \lottsym{⊢}_s \, e \, \lottsym{::} \, \lottsym{∀} \, \lottsepbycomprehension{{a}_{j}} \lottsym{.} \, \lottsym{\{} \, ρ₁ \lottsym{,} \, \lottsepbycomprehensionpatcollection{{i₀}_{k} \, {\rstrength₀}_{k} \, {A}_{k} \lottsym{,} \, {ρ₂}_{k}}{k}{[:n]} \, \lottsym{\}}}\\
\lotthypothesis{\lottjudgementcomprehensionpatcollection{\lottjudgementcomprehensionpatcollection{{i₀}_{k} \, \lottsym{∉} \, {ρ₂}_{l}}{l}{[:k+1]}}{k}{[:n]}}\\
\lotthypothesis{\lottsepbycomprehensionpatcollection{{i₀}_{k} \, {\rstrength₀}_{k} \, {A}_{k}}{k}{[:n]} \, \lottkw{permutes} \, \lottsepbycomprehensionpatcollection{{i₁}_{k} \, {\rstrength₁}_{k} \, {B}_{k}}{k}{[:n]}}
}{\ifdefined\store \mathcolor{Store}{ρₛ} \lottsym{;} \, \fi ρ₀ \, \lottsym{⊢}_s \, e \lottsym{\textbackslash} \lottsym{\{} \, \lottsepbycomprehensionpatcollection{{i₁}_{k}}{k}{[:n]} \, \lottsym{\}} \, \lottsym{:} \, \lottsym{\{} \, ρ₁\left[\lottsepbycomprehension{{τ}_j/{a}_{j}}\right] \lottsym{,} \, \lottsepbycomprehensionpatcollection{{ρ₂}_{k}\left[\lottsepbycomprehension{{τ}_j/{a}_{j}}\right]}{k}{[:n]} \, \lottsym{\}}}
\lottinferencerulesep
\lottinferencerule{let}{
\lotthypothesis{\ifdefined\store \mathcolor{Store}{ρₛ} \lottsym{;} \, \fi ρ₀ \, \lottsym{⊢}_s \, e₀ \, \lottsym{:} \, A}\\
\lotthypothesis{t \, \lottsym{\mathcolor{Pattern}{:}}_s \, A \, \lottsym{\mathcolor{Pattern}{⊙}} \, ρ₁}\\
\lotthypothesis{\ifdefined\store \mathcolor{Store}{ρₛ} \lottsym{;} \, \fi ρ₀ \lottsym{,} \, ρ₁ \, \lottsym{⊢}_s \, e₁ \, \lottsym{:} \, B}
}{\ifdefined\store \mathcolor{Store}{ρₛ} \lottsym{;} \, \fi ρ₀ \, \lottsym{⊢}_s \, \lottkw{let} \, t \, \lottsym{=} \, e₀ \, \lottkw{in} \, e₁ \, \lottsym{:} \, B}
\lottinferencerulesep
\lottinferencerule{loc}{
}{\ifdefined\store \mathcolor{Store}{ρₛ₀ \lottsym{,} \, l \, \lottsym{:} \, A \lottsym{,} \, ρₛ₁} \lottsym{;} \, \fi ρ \, \lottsym{⊢}_s \, l \, \lottsym{:} \, \lottkw{Ref} \, A}
\lottinferencerulesep
\lottinferencerule{ref}{
\lotthypothesis{\ifdefined\store \mathcolor{Store}{ρₛ} \lottsym{;} \, \fi ρ \, \lottsym{⊢}_s \, e \, \lottsym{:} \, A}\\
\lotthypothesis{A \, \lottsym{\mathcolor{Subtyping}{⩽}} \, B}
}{\ifdefined\store \mathcolor{Store}{ρₛ} \lottsym{;} \, \fi ρ \, \lottsym{⊢}_s \, \lottkw{ref} \, e \, \lottsym{:} \, \lottkw{Ref} \, B}
\lottinferencerulesep
\lottinferencerule{deref}{
\lotthypothesis{\ifdefined\store \mathcolor{Store}{ρₛ} \lottsym{;} \, \fi ρ \, \lottsym{⊢}_s \, e \, \lottsym{::} \, \lottsym{∀} \, \lottsepbycomprehension{{a}_{j}} \lottsym{.} \, \lottkw{Ref} \, A}
}{\ifdefined\store \mathcolor{Store}{ρₛ} \lottsym{;} \, \fi ρ \, \lottsym{⊢}_s \, \lottsym{!} e \, \lottsym{:} \, A\left[\lottsepbycomprehension{{τ}_j/{a}_{j}}\right]}
\lottinferencerulesep
\lottinferencerule{set}{
\lotthypothesis{\ifdefined\store \mathcolor{Store}{ρₛ} \lottsym{;} \, \fi ρ \, \lottsym{⊢}_s \, e₀ \, \lottsym{::} \, \lottsym{∀} \, \lottsepbycomprehension{{a}_{j}} \lottsym{.} \, \lottkw{Ref} \, A}\\
\lotthypothesis{\ifdefined\store \mathcolor{Store}{ρₛ} \lottsym{;} \, \fi ρ \, \lottsym{⊢}_s \, e₁ \, \lottsym{:} \, B}\\
\lotthypothesis{B \, \lottsym{\mathcolor{Subtyping}{⩽}} \, A\left[\lottsepbycomprehension{{τ}_j/{a}_{j}}\right]}
}{\ifdefined\store \mathcolor{Store}{ρₛ} \lottsym{;} \, \fi ρ \, \lottsym{⊢}_s \, e₀ \, \lottsym{≔} \, e₁ \, \lottsym{:} \, \lottsym{\{} \, \lottsym{\}}}
}
  \space
\lottjudgement{Instantiate Any}{\ifdefined\store \mathcolor{Store}{ρₛ} \lottsym{;} \, \fi ρ \, \lottsym{⊢}_s \, e \, \lottsym{::} \, A}{
\lottinferencerule{instantiate}{
\lotthypothesis{\ifdefined\store \mathcolor{Store}{ρₛ} \lottsym{;} \, \fi ρ \, \lottsym{⊢}_s \, e \, \lottsym{:} \, \lottsym{∀} \, \lottsepbycomprehension{{a}_{j}} \lottsym{.} \, {a}_{k}}
}{\ifdefined\store \mathcolor{Store}{ρₛ} \lottsym{;} \, \fi ρ \, \lottsym{⊢}_s \, e \, \lottsym{::} \, τ}
\lottinferencerulesep
\lottinferencerule{ignore}{
\lotthypothesis{\ifdefined\store \mathcolor{Store}{ρₛ} \lottsym{;} \, \fi ρ \, \lottsym{⊢}_s \, e \, \lottsym{:} \, A}\\\lottsym{∀} a \, k, A \, \lottsym{≠} \, \lottsym{∀} \, \lottsepbycomprehension{{a}_{j}} \lottsym{.} \, a_k
}{\ifdefined\store \mathcolor{Store}{ρₛ} \lottsym{;} \, \fi ρ \, \lottsym{⊢}_s \, e \, \lottsym{::} \, A}
}
\hfill\emph{(Syntax-Directed Typing and Instantiation)}
  \endgroup
  \begingroup
  \renewcommand{\lottjudgement}[3]{#3}
  \begin{mathparpagebreakable}
    \lottjudgementfilter{force,let}{\lottjudgementlabel{ST}{%
\lottjudgement{Syntax Directed}{\ifdefined\store \mathcolor{Store}{ρₛ} \lottsym{;} \, \fi ρ \, \lottsym{⊢}_s \, e \, \lottsym{:} \, A}{
\lottinferencerule{nat}{
}{\ifdefined\store \mathcolor{Store}{ρₛ} \lottsym{;} \, \fi ρ \, \lottsym{⊢}_s \, n \, \lottsym{:} \, \lottsym{ℕ}}
\lottinferencerulesep
\lottinferencerule{op}{
\lotthypothesis{\ifdefined\store \mathcolor{Store}{ρₛ} \lottsym{;} \, \fi ρ \, \lottsym{⊢}_s \, e₀ \, \lottsym{:} \, A}\\
\lotthypothesis{A \, \lottsym{\mathcolor{Subtyping}{⩽}} \, \lottsym{ℕ}}\\
\lotthypothesis{\ifdefined\store \mathcolor{Store}{ρₛ} \lottsym{;} \, \fi ρ \, \lottsym{⊢}_s \, e₁ \, \lottsym{:} \, B}\\
\lotthypothesis{B \, \lottsym{\mathcolor{Subtyping}{⩽}} \, \lottsym{ℕ}}
}{\ifdefined\store \mathcolor{Store}{ρₛ} \lottsym{;} \, \fi ρ \, \lottsym{⊢}_s \, e₀ \, b \, e₁ \, \lottsym{:} \, \lottsym{ℕ}}
\lottinferencerulesep
\lottinferencerule{lam}{\ifdefined\store\overline{a_j} \, \lottsym{∉} \, \lottkw{ftv} \lottsym{(} \mathcolor{Store}{ρₛ} \lottsym{)}\\\fi\overline{a_j} \, \lottsym{∉} \, \lottkw{ftv} \lottsym{(} ρ₀ \lottsym{)}\\
\lotthypothesis{c \, \lottsym{\mathcolor{Pattern}{:}} \, ρ₁ \, \lottsym{\mathcolor{Pattern}{⊙}} \, ρ₂}\\
\lotthypothesis{\ifdefined\store \mathcolor{Store}{ρₛ} \lottsym{;} \, \fi ρ₀ \lottsym{,} \, ρ₂ \, \lottsym{⊢}_s \, e \, \lottsym{:} \, A}
}{\ifdefined\store \mathcolor{Store}{ρₛ} \lottsym{;} \, \fi ρ₀ \, \lottsym{⊢}_s \, \lottsym{λ} \, \lottsym{\{} \, c \, \lottsym{\}} \lottsym{.} \, e \, \lottsym{:} \, \lottsym{∀} \, \lottsepbycomprehension{{a}_{j}} \lottsym{.} \, \lottsym{\{} \, ρ₁ \, \lottsym{\}} \, \lottsym{→} \, A}
\lottinferencerulesep
\lottinferencerule{closure}{\ifdefined\store\overline{a_j} \, \lottsym{∉} \, \lottkw{ftv} \lottsym{(} \mathcolor{Store}{ρₛ} \lottsym{)}\\\fi
\lotthypothesis{\ifdefined\store \mathcolor{Store}{ρₛ} \lottsym{;} \, \fi \lottsym{ε} \, \lottsym{⊢}_s \, \lottsym{\{} \, o₀ \, \lottsym{\}} \, \lottsym{:} \, A}\\
\lotthypothesis{A \, \lottsym{\mathcolor{Subtyping}{⩽}} \, \lottsym{\{} \, ρ₁ \, \lottsym{\}}}\\
\lotthypothesis{\lottsym{⌊} ρ₁ \lottsym{⌋}}\\\overline{a_j} \, \lottsym{∉} \, \lottkw{ftv} \lottsym{(} ρ₁ \lottsym{)}\\
\lotthypothesis{\ifdefined\store \mathcolor{Store}{ρₛ} \lottsym{;} \, \fi \lottsym{ε} \, \lottsym{⊢}_s \, \lottsym{\{} \, o₁ \, \lottsym{\}} \, \lottsym{:} \, B}\\
\lotthypothesis{B \, \lottsym{\mathcolor{Subtyping}{⩽}} \, \lottsym{∀} \, \lottsepbycomprehension{{a}_{j}} \lottsym{.} \, \lottsym{\{} \, ρ₂ \, \lottsym{\}}}\\
\lotthypothesis{c \, \lottsym{\mathcolor{Pattern}{:}} \, ρ₃ \, \lottsym{\mathcolor{Pattern}{⊙}} \, ρ₄}\\
\lotthypothesis{ρ₃ \, \lottsym{\mathcolor{App}{●}} \, ρ₂ \, \lottsym{\mathcolor{App}{↠}} \, ρ₅}\\
\lotthypothesis{\ifdefined\store \mathcolor{Store}{ρₛ} \lottsym{;} \, \fi ρ₁ \lottsym{,} \, ρ₄ \, \lottsym{⊢}_s \, e \, \lottsym{:} \, C}
}{\ifdefined\store \mathcolor{Store}{ρₛ} \lottsym{;} \, \fi ρ₀ \, \lottsym{⊢}_s \, \lottsym{⟨} o₀ \lottsym{;} \, o₁ \, \lottsym{|} \, \lottsym{λ} \, \lottsym{\{} \, c \, \lottsym{\}} \lottsym{.} \, e \lottsym{⟩} \, \lottsym{:} \, \lottsym{∀} \, \lottsepbycomprehension{{a}_{j}} \lottsym{.} \, \lottsym{\{} \, ρ₅ \, \lottsym{\}} \, \lottsym{→} \, C}
\lottinferencerulesep
\lottinferencerule{app}{\overline{a₂_l} \, \lottsym{∉} \, \lottkw{ftv} \lottsym{(} ρ₁ \lottsym{)}\\\overline{a₂_l} \, \lottsym{∉} \, \lottkw{ftv} \lottsym{(} A \lottsym{)}\\\overline{a₂_l} \, \lottsym{∉} \, \lottkw{ftv} \lottsym{(} ρ₂ \lottsym{)}\\
\lotthypothesis{\ifdefined\store \mathcolor{Store}{ρₛ} \lottsym{;} \, \fi ρ₀ \, \lottsym{⊢}_s \, e₀ \, \lottsym{:} \, \lottsym{∀} \, \lottsepbycomprehension{{a₀}_{j}} \lottsym{.} \, \lottsym{\{} \, ρ₁ \, \lottsym{\}} \, \lottsym{→} \, A}\\
\lotthypothesis{\ifdefined\store \mathcolor{Store}{ρₛ} \lottsym{;} \, \fi ρ₀ \, \lottsym{⊢}_s \, e₁ \, \lottsym{:} \, \lottsym{∀} \, \lottsepbycomprehension{{a₁}_{k}} \lottsym{.} \, \lottsym{\{} \, ρ₂ \, \lottsym{\}}}\\
\lotthypothesis{ρ₁\left[\lottsepbycomprehension{{τ₀}_j/{a₀}_{j}}\right] \, \lottsym{\mathcolor{App}{●}}_s \, ρ₂\left[\lottsepbycomprehension{{τ₁}_j/{a₁}_{k}}\right] \, \lottsym{\mathcolor{App}{↠}} \, ρ₃}
}{\ifdefined\store \mathcolor{Store}{ρₛ} \lottsym{;} \, \fi ρ₀ \, \lottsym{⊢}_s \, e₀ \, e₁ \, \lottsym{:} \, \lottsym{∀} \, \lottsepbycomprehension{{a₂}_{l}} \lottsym{.} \, \lottsym{\{} \, ρ₃ \, \lottsym{\}} \, \lottsym{→} \, A\left[\lottsepbycomprehension{{τ₀}_j/{a₀}_{j}}\right]}
\lottinferencerulesep
\lottinferencerule{appAnyL}{\overline{a₂_l} \, \lottsym{∉} \, \lottkw{ftv} \lottsym{(} ρ₂ \lottsym{)}\\
\lotthypothesis{\ifdefined\store \mathcolor{Store}{ρₛ} \lottsym{;} \, \fi ρ₀ \, \lottsym{⊢}_s \, e₀ \, \lottsym{:} \, \lottsym{∀} \, \lottsepbycomprehension{{a₀}_{j}} \lottsym{.} \, {a}_{m}}\\
\lotthypothesis{\ifdefined\store \mathcolor{Store}{ρₛ} \lottsym{;} \, \fi ρ₀ \, \lottsym{⊢}_s \, e₁ \, \lottsym{:} \, \lottsym{∀} \, \lottsepbycomprehension{{a₁}_{k}} \lottsym{.} \, \lottsym{\{} \, ρ₂ \, \lottsym{\}}}\\
\lotthypothesis{ϱ \, \lottsym{\mathcolor{App}{●}}_s \, ρ₂\left[\lottsepbycomprehension{{τ₁}_j/{a₁}_{k}}\right] \, \lottsym{\mathcolor{App}{↠}} \, ρ₃}
}{\ifdefined\store \mathcolor{Store}{ρₛ} \lottsym{;} \, \fi ρ₀ \, \lottsym{⊢}_s \, e₀ \, e₁ \, \lottsym{:} \, \lottsym{∀} \, \lottsepbycomprehension{{a₂}_{l}} \lottsym{.} \, \lottsym{\{} \, ρ₃ \, \lottsym{\}} \, \lottsym{→} \, τ₀}
\lottinferencerulesep
\lottinferencerule{appAnyR}{\overline{a₂_l} \, \lottsym{∉} \, \lottkw{ftv} \lottsym{(} ρ₁ \lottsym{)}\\\overline{a₂_l} \, \lottsym{∉} \, \lottkw{ftv} \lottsym{(} A \lottsym{)}\\
\lotthypothesis{\ifdefined\store \mathcolor{Store}{ρₛ} \lottsym{;} \, \fi ρ₀ \, \lottsym{⊢}_s \, e₀ \, \lottsym{:} \, \lottsym{∀} \, \lottsepbycomprehension{{a₀}_{j}} \lottsym{.} \, \lottsym{\{} \, ρ₁ \, \lottsym{\}} \, \lottsym{→} \, A}\\
\lotthypothesis{\ifdefined\store \mathcolor{Store}{ρₛ} \lottsym{;} \, \fi ρ₀ \, \lottsym{⊢}_s \, e₁ \, \lottsym{:} \, \lottsym{∀} \, \lottsepbycomprehension{{a₁}_{k}} \lottsym{.} \, {a}_{m}}\\
\lotthypothesis{ρ₁\left[\lottsepbycomprehension{{τ}_j/{a₀}_{j}}\right] \, \lottsym{\mathcolor{App}{●}}_s \, ϱ \, \lottsym{\mathcolor{App}{↠}} \, ρ₃}
}{\ifdefined\store \mathcolor{Store}{ρₛ} \lottsym{;} \, \fi ρ₀ \, \lottsym{⊢}_s \, e₀ \, e₁ \, \lottsym{:} \, \lottsym{∀} \, \lottsepbycomprehension{{a₂}_{l}} \lottsym{.} \, \lottsym{\{} \, ρ₃ \, \lottsym{\}} \, \lottsym{→} \, A\left[\lottsepbycomprehension{{τ}_j/{a₀}_{j}}\right]}
\lottinferencerulesep
\lottinferencerule{appAny}{
\lotthypothesis{\ifdefined\store \mathcolor{Store}{ρₛ} \lottsym{;} \, \fi ρ₀ \, \lottsym{⊢}_s \, e₀ \, \lottsym{:} \, \lottsym{∀} \, \lottsepbycomprehension{{a₀}_{j}} \lottsym{.} \, {a}_{m₀}}\\
\lotthypothesis{\ifdefined\store \mathcolor{Store}{ρₛ} \lottsym{;} \, \fi ρ₀ \, \lottsym{⊢}_s \, e₁ \, \lottsym{:} \, \lottsym{∀} \, \lottsepbycomprehension{{a₁}_{k}} \lottsym{.} \, {a}_{m₁}}\\
\lotthypothesis{ϱ₁ \, \lottsym{\mathcolor{App}{●}}_s \, ϱ₂ \, \lottsym{\mathcolor{App}{↠}} \, ρ₃}
}{\ifdefined\store \mathcolor{Store}{ρₛ} \lottsym{;} \, \fi ρ₀ \, \lottsym{⊢}_s \, e₀ \, e₁ \, \lottsym{:} \, \lottsym{∀} \, \lottsepbycomprehension{{a₂}_{l}} \lottsym{.} \, \lottsym{\{} \, ρ₃ \, \lottsym{\}} \, \lottsym{→} \, τ}
\lottinferencerulesep
\lottinferencerule{force}{
\lotthypothesis{\ifdefined\store \mathcolor{Store}{ρₛ} \lottsym{;} \, \fi ρ₀ \, \lottsym{⊢}_s \, e \, \lottsym{::} \, \lottsym{∀} \, \lottsepbycomprehension{{a}_{j}} \lottsym{.} \, \lottsym{\{} \, ρ₁ \, \lottsym{\}} \, \lottsym{→} \, A}\\
\lotthypothesis{{\lottsym{⌊} ρ₁ \lottsym{⌋}}^{\lottsym{?}}}
}{\ifdefined\store \mathcolor{Store}{ρₛ} \lottsym{;} \, \fi ρ₀ \, \lottsym{⊢}_s \, \lottkw{force} \, e \, \lottsym{:} \, A\left[\lottsepbycomprehension{{τ}_j/{a}_{j}}\right]}
\lottinferencerulesep
\lottinferencerule{box}{
\lotthypothesis{\ifdefined\store \mathcolor{Store}{ρₛ} \lottsym{;} \, \fi ρ₀ \, \lottsym{⊢}_s \, e₀ \, \lottsym{::} \, \lottsym{∀} \, \lottsepbycomprehension{{a}_{j}} \lottsym{.} \, \lottsym{\{} \, ρ₁ \, \lottsym{\}}}\\
\lotthypothesis{\lottsym{⌊} ρ₁ \lottsym{⌋}}\\
\lotthypothesis{\ifdefined\store \mathcolor{Store}{ρₛ} \lottsym{;} \, \fi ρ₁\left[\lottsepbycomprehension{{τ}_j/{a}_{j}}\right] \, \lottsym{⊢}_s \, e₁ \, \lottsym{:} \, A}
}{\ifdefined\store \mathcolor{Store}{ρₛ} \lottsym{;} \, \fi ρ₀ \, \lottsym{⊢}_s \, e₀ \, \lottsym{▶} \, e₁ \, \lottsym{:} \, A}
\lottinferencerulesep
\lottinferencerule{env}{
}{\ifdefined\store \mathcolor{Store}{ρₛ} \lottsym{;} \, \fi ρ \, \lottsym{⊢}_s \, \lottsym{@} \, \lottsym{:} \, \lottsym{\{} \, ρ \, \lottsym{\}}}
\lottinferencerulesep
\lottinferencerule{record}{
\lotthypothesis{\ifdefined\store \mathcolor{Store}{ρₛ} \lottsym{;} \, \fi ρ₀ \, \lottsym{⊢}_s \, r \, \lottsym{:} \, ρ₁}
}{\ifdefined\store \mathcolor{Store}{ρₛ} \lottsym{;} \, \fi ρ₀ \, \lottsym{⊢}_s \, \lottsym{\{} \, r \, \lottsym{\}} \, \lottsym{:} \, \lottsym{\{} \, ρ₁ \, \lottsym{\}}}
\lottinferencerulesep
\lottinferencerule{concat}{
\lotthypothesis{\ifdefined\store \mathcolor{Store}{ρₛ} \lottsym{;} \, \fi ρ₀ \, \lottsym{⊢}_s \, e₀ \, \lottsym{::} \, \lottsym{∀} \, \lottsepbycomprehension{{a₁}_{j}} \lottsym{.} \, \lottsym{\{} \, ρ₁ \, \lottsym{\}}}\\
\lotthypothesis{\ifdefined\store \mathcolor{Store}{ρₛ} \lottsym{;} \, \fi ρ₀ \, \lottsym{⊢}_s \, e₁ \, \lottsym{::} \, \lottsym{∀} \, \lottsepbycomprehension{{a₂}_{k}} \lottsym{.} \, \lottsym{\{} \, ρ₂ \, \lottsym{\}}}\\
\lotthypothesis{ρ₁\left[\lottsepbycomprehension{{τ₁}_j/{a₁}_{j}}\right] \, \lottsym{\mathcolor{Concat}{⊕}}_s \, ρ₂\left[\lottsepbycomprehension{{τ₂}_j/{a₂}_{k}}\right] \, \lottsym{\mathcolor{Concat}{↬}} \, ρ₃}
}{\ifdefined\store \mathcolor{Store}{ρₛ} \lottsym{;} \, \fi ρ₀ \, \lottsym{⊢}_s \, e₀ \, \lottsym{⧺} \, e₁ \, \lottsym{:} \, \lottsym{\{} \, ρ₃ \, \lottsym{\}}}
\lottinferencerulesep
\lottinferencerule{proj}{
\lotthypothesis{\ifdefined\store \mathcolor{Store}{ρₛ} \lottsym{;} \, \fi ρ₀ \, \lottsym{⊢}_s \, e \, \lottsym{::} \, \lottsym{∀} \, \lottsepbycomprehension{{a}_{j}} \lottsym{.} \, \lottsym{\{} \, ρ₁ \lottsym{,} \, i \, \rstrength \, A \lottsym{,} \, ρ₂ \, \lottsym{\}}}\\
\lotthypothesis{i \, \lottsym{∉} \, ρ₂}
}{\ifdefined\store \mathcolor{Store}{ρₛ} \lottsym{;} \, \fi ρ₀ \, \lottsym{⊢}_s \, e \lottsym{.} i \, \lottsym{:} \, A\left[\lottsepbycomprehension{{τ}_j/{a}_{j}}\right]}
\lottinferencerulesep
\lottinferencerule{mproj}{
\lotthypothesis{\ifdefined\store \mathcolor{Store}{ρₛ} \lottsym{;} \, \fi ρ₀ \, \lottsym{⊢}_s \, e \, \lottsym{::} \, \lottsym{∀} \, \lottsepbycomprehension{{a}_{j}} \lottsym{.} \, \lottsym{\{} \, ρ₁ \lottsym{,} \, \lottsepbycomprehensionpatcollection{{i₀}_{k} \, {\rstrength₀}_{k} \, {A}_{k} \lottsym{,} \, {ρ₂}_{k}}{k}{[:n]} \, \lottsym{\}}}\\
\lotthypothesis{\lottjudgementcomprehensionpatcollection{\lottjudgementcomprehensionpatcollection{{i₀}_{k} \, \lottsym{∉} \, {ρ₂}_{l}}{l}{[:k+1]}}{k}{[:n]}}\\
\lotthypothesis{\lottsepbycomprehensionpatcollection{{i₀}_{k} \, {\rstrength₀}_{k} \, {A}_{k}}{k}{[:n]} \, \lottkw{permutes} \, \lottsepbycomprehensionpatcollection{{i₁}_{k} \, {\rstrength₁}_{k} \, {B}_{k}}{k}{[:n]}}
}{\ifdefined\store \mathcolor{Store}{ρₛ} \lottsym{;} \, \fi ρ₀ \, \lottsym{⊢}_s \, e \lottsym{.} \lottsym{\{} \, \lottsepbycomprehensionpatcollection{{i₁}_{k}}{k}{[:n]} \, \lottsym{\}} \, \lottsym{:} \, \lottsym{\{} \, \lottsepbycomprehensionpatcollection{{i₁}_{k} \, {\rstrength₁}_{k} \, {B}_{k}\left[\lottsepbycomprehension{{τ}_j/{a}_{j}}\right]}{k}{[:n]} \, \lottsym{\}}}
\lottinferencerulesep
\lottinferencerule{diff}{
\lotthypothesis{\ifdefined\store \mathcolor{Store}{ρₛ} \lottsym{;} \, \fi ρ₀ \, \lottsym{⊢}_s \, e \, \lottsym{::} \, \lottsym{∀} \, \lottsepbycomprehension{{a}_{j}} \lottsym{.} \, \lottsym{\{} \, ρ₁ \lottsym{,} \, \lottsepbycomprehensionpatcollection{{i₀}_{k} \, {\rstrength₀}_{k} \, {A}_{k} \lottsym{,} \, {ρ₂}_{k}}{k}{[:n]} \, \lottsym{\}}}\\
\lotthypothesis{\lottjudgementcomprehensionpatcollection{\lottjudgementcomprehensionpatcollection{{i₀}_{k} \, \lottsym{∉} \, {ρ₂}_{l}}{l}{[:k+1]}}{k}{[:n]}}\\
\lotthypothesis{\lottsepbycomprehensionpatcollection{{i₀}_{k} \, {\rstrength₀}_{k} \, {A}_{k}}{k}{[:n]} \, \lottkw{permutes} \, \lottsepbycomprehensionpatcollection{{i₁}_{k} \, {\rstrength₁}_{k} \, {B}_{k}}{k}{[:n]}}
}{\ifdefined\store \mathcolor{Store}{ρₛ} \lottsym{;} \, \fi ρ₀ \, \lottsym{⊢}_s \, e \lottsym{\textbackslash} \lottsym{\{} \, \lottsepbycomprehensionpatcollection{{i₁}_{k}}{k}{[:n]} \, \lottsym{\}} \, \lottsym{:} \, \lottsym{\{} \, ρ₁\left[\lottsepbycomprehension{{τ}_j/{a}_{j}}\right] \lottsym{,} \, \lottsepbycomprehensionpatcollection{{ρ₂}_{k}\left[\lottsepbycomprehension{{τ}_j/{a}_{j}}\right]}{k}{[:n]} \, \lottsym{\}}}
\lottinferencerulesep
\lottinferencerule{let}{
\lotthypothesis{\ifdefined\store \mathcolor{Store}{ρₛ} \lottsym{;} \, \fi ρ₀ \, \lottsym{⊢}_s \, e₀ \, \lottsym{:} \, A}\\
\lotthypothesis{t \, \lottsym{\mathcolor{Pattern}{:}}_s \, A \, \lottsym{\mathcolor{Pattern}{⊙}} \, ρ₁}\\
\lotthypothesis{\ifdefined\store \mathcolor{Store}{ρₛ} \lottsym{;} \, \fi ρ₀ \lottsym{,} \, ρ₁ \, \lottsym{⊢}_s \, e₁ \, \lottsym{:} \, B}
}{\ifdefined\store \mathcolor{Store}{ρₛ} \lottsym{;} \, \fi ρ₀ \, \lottsym{⊢}_s \, \lottkw{let} \, t \, \lottsym{=} \, e₀ \, \lottkw{in} \, e₁ \, \lottsym{:} \, B}
\lottinferencerulesep
\lottinferencerule{loc}{
}{\ifdefined\store \mathcolor{Store}{ρₛ₀ \lottsym{,} \, l \, \lottsym{:} \, A \lottsym{,} \, ρₛ₁} \lottsym{;} \, \fi ρ \, \lottsym{⊢}_s \, l \, \lottsym{:} \, \lottkw{Ref} \, A}
\lottinferencerulesep
\lottinferencerule{ref}{
\lotthypothesis{\ifdefined\store \mathcolor{Store}{ρₛ} \lottsym{;} \, \fi ρ \, \lottsym{⊢}_s \, e \, \lottsym{:} \, A}\\
\lotthypothesis{A \, \lottsym{\mathcolor{Subtyping}{⩽}} \, B}
}{\ifdefined\store \mathcolor{Store}{ρₛ} \lottsym{;} \, \fi ρ \, \lottsym{⊢}_s \, \lottkw{ref} \, e \, \lottsym{:} \, \lottkw{Ref} \, B}
\lottinferencerulesep
\lottinferencerule{deref}{
\lotthypothesis{\ifdefined\store \mathcolor{Store}{ρₛ} \lottsym{;} \, \fi ρ \, \lottsym{⊢}_s \, e \, \lottsym{::} \, \lottsym{∀} \, \lottsepbycomprehension{{a}_{j}} \lottsym{.} \, \lottkw{Ref} \, A}
}{\ifdefined\store \mathcolor{Store}{ρₛ} \lottsym{;} \, \fi ρ \, \lottsym{⊢}_s \, \lottsym{!} e \, \lottsym{:} \, A\left[\lottsepbycomprehension{{τ}_j/{a}_{j}}\right]}
\lottinferencerulesep
\lottinferencerule{set}{
\lotthypothesis{\ifdefined\store \mathcolor{Store}{ρₛ} \lottsym{;} \, \fi ρ \, \lottsym{⊢}_s \, e₀ \, \lottsym{::} \, \lottsym{∀} \, \lottsepbycomprehension{{a}_{j}} \lottsym{.} \, \lottkw{Ref} \, A}\\
\lotthypothesis{\ifdefined\store \mathcolor{Store}{ρₛ} \lottsym{;} \, \fi ρ \, \lottsym{⊢}_s \, e₁ \, \lottsym{:} \, B}\\
\lotthypothesis{B \, \lottsym{\mathcolor{Subtyping}{⩽}} \, A\left[\lottsepbycomprehension{{τ}_j/{a}_{j}}\right]}
}{\ifdefined\store \mathcolor{Store}{ρₛ} \lottsym{;} \, \fi ρ \, \lottsym{⊢}_s \, e₀ \, \lottsym{≔} \, e₁ \, \lottsym{:} \, \lottsym{\{} \, \lottsym{\}}}
}
}}
    \lottinferencerulesep
    \lottjudgementfilter{instantiate}{\lottjudgementlabel{ST}{%
\lottjudgement{Instantiate Any}{\ifdefined\store \mathcolor{Store}{ρₛ} \lottsym{;} \, \fi ρ \, \lottsym{⊢}_s \, e \, \lottsym{::} \, A}{
\lottinferencerule{instantiate}{
\lotthypothesis{\ifdefined\store \mathcolor{Store}{ρₛ} \lottsym{;} \, \fi ρ \, \lottsym{⊢}_s \, e \, \lottsym{:} \, \lottsym{∀} \, \lottsepbycomprehension{{a}_{j}} \lottsym{.} \, {a}_{k}}
}{\ifdefined\store \mathcolor{Store}{ρₛ} \lottsym{;} \, \fi ρ \, \lottsym{⊢}_s \, e \, \lottsym{::} \, τ}
\lottinferencerulesep
\lottinferencerule{ignore}{
\lotthypothesis{\ifdefined\store \mathcolor{Store}{ρₛ} \lottsym{;} \, \fi ρ \, \lottsym{⊢}_s \, e \, \lottsym{:} \, A}\\\lottsym{∀} a \, k, A \, \lottsym{≠} \, \lottsym{∀} \, \lottsepbycomprehension{{a}_{j}} \lottsym{.} \, a_k
}{\ifdefined\store \mathcolor{Store}{ρₛ} \lottsym{;} \, \fi ρ \, \lottsym{⊢}_s \, e \, \lottsym{::} \, A}
}
}}
  \end{mathparpagebreakable}
  \endgroup

  \caption{Syntax-directed typing}
  \label{fig:syntax-directed}
  \vspace{1pt}
\end{figure}

The usual approach of instantiating quantifiers eagerly at variables and
re-generalizing at $\lottkw{let}$s does not fit our language: variables are
projections out of a first-class environment, types are higher-rank, and we
generalize at abstractions rather than $\lottkw{let}$s. We therefore
instantiate \textit{lazily}, in the elimination rules.
In rule \lottref{ST}{force}, for instance, the first hypothesis accepts any
number of quantifiers, which are instantiated in the conclusion. Without
subtyping to discard optional entries (\lottref{S}{empty}), $ρ₁$ may
be non-empty, so the second hypothesis checks its strengths instead. This
hypothesis also suggests an elaboration that inserts $\lottkw{force}$
automatically: whenever a term of type $\{\rho\} \to A$ with
$\uniform{\rho}^?$ is used at type $A$, the elaboration wraps it in
$\lottkw{force}$, so that programmers need not write it explicitly.
Such an elaboration fits naturally with bidirectional
typing~\cite{dunfield2021bidirectional},
where the expected type is available in checking mode;
we leave this to future work.
The double-colon hypothesis invokes a helper judgement
(\lottref{ST}{instantiate}) for types such as $\forall a.~a$, in which case we
instantiate it with any type. Such types can exist as function arguments.
Finally, auxiliary judgements such as pattern separation, application, and
concatenation also have syntax-directed versions, as used in
\lottref{ST}{let}.

We prove that our syntax-directed typing is sound and complete:

\begin{theorem}[Soundness]
  If $\ifdefined\store \mathcolor{Store}{ρₛ} \lottsym{;} \, \fi ρ \, \lottsym{⊢}_s \, e \, \lottsym{:} \, A$ then $\ifdefined\store \mathcolor{Store}{ρₛ} \lottsym{;} \, \fi ρ \, \lottsym{⊢} \, e \, \lottsym{:} \, A$.
\end{theorem}

\begin{theorem}[Completeness]
  If $\ifdefined\store \mathcolor{Store}{ρₛ} \lottsym{;} \, \fi ρ \, \lottsym{⊢} \, e \, \lottsym{:} \, A$ with $\uniform{\rho}$, then there exists $B$ such
  that $B \, \lottsym{\mathcolor{Subtyping}{⩽}} \, A$ and $\ifdefined\store \mathcolor{Store}{ρₛ} \lottsym{;} \, \fi ρ \, \lottsym{⊢}_s \, e \, \lottsym{:} \, B$.
\end{theorem}

\subsection{Translating the Lambda Calculus}
\label{sec:lambda}

\let\store\undefined

To demonstrate that the use of explicit environments preserves the semantics of
the substitution model, we present a translation from the simply typed lambda calculus (STLC) to \lang.
The syntax and semantics of STLC are standard and given in the appendix.

\paragraph{\textbf{Translations}}

The translation of types, environments, and terms from STLC to \lang is
defined below. We assume STLC uses its own mechanism to
ensure variable freshness, thus we utilize a translation environment
$ι$ to track the variable mapping between
the two calculi. (Our mechanization formalises STLC using the locally nameless
representation~\cite{charguraud2011ln}.)

\smallskip
\noindent
\resizebox{\linewidth}{!}{\small
$\boxed{\lottsym{⟦} \mathcolor{STLC}{\lottsym{τ}} \lottsym{⟧}}
\;\begin{aligned}[t]
  \text{$\lottsym{⟦} \mathcolor{STLC}{ \mathcolor{STLC}{ τ₀ } \, \lottsym{→} \, \mathcolor{STLC}{ τ₁ } } \lottsym{⟧}$} &= \text{$\lottsym{\{} \, \_ \, \lottsym{:} \, \lottsym{⟦} \mathcolor{STLC}{ τ₀ } \lottsym{⟧} \, \lottsym{\}} \, \lottsym{→} \, \lottsym{⟦} \mathcolor{STLC}{ τ₁ } \lottsym{⟧}$} \\
  \text{$\lottsym{⟦} \mathcolor{STLC}{ \lottsym{ℕ} } \lottsym{⟧}$} &= ℕ
\end{aligned}
\quad
\boxed{\lottsym{⟦} \lottsym{ι} \, \lottsym{⊢} \, \mathcolor{STLC}{Γ}⟧}
\;\begin{aligned}[t]
  \lottsym{⟦} ε \, \lottsym{⊢} \, \mathcolor{STLC}{ε} \lottsym{⟧} &= ε \\
  \lottsym{⟦} ι \lottsym{,} \, \mathcolor{STLC}{x} \, \lottsym{∼} \, i \, \lottsym{⊢} \, \mathcolor{STLC}{Γ \lottsym{,} \, x \, \lottsym{:} \, τ} \lottsym{⟧} &= \lottsym{⟦} ι \, \lottsym{⊢} \, \mathcolor{STLC}{Γ} \lottsym{⟧} \lottsym{,} \, i \, \lottsym{:} \, \lottsym{⟦} \mathcolor{STLC}{τ} \lottsym{⟧}
\end{aligned}
\quad
\boxed{\lottsym{⟦} ι \, \lottsym{⊢} \, \mathcolor{STLC}{e} \lottsym{⟧}}
\;\begin{aligned}[t]
  \lottsym{⟦} ι \, \lottsym{⊢} \, \mathcolor{STLC}{x} \lottsym{⟧} &= \lottsym{@} \lottsym{.} i \quad \text{if} ~ \mathcolor{STLC}{x} \, \lottsym{∼} \, i \, \lottsym{∈} \, ι \\
  \lottsym{⟦} ι \, \lottsym{⊢} \, \mathcolor{STLC}{\lottsym{λ} \, x \lottsym{.} \, e} \lottsym{⟧} &= \lottsym{λ} \, \lottsym{\{} \, \_ \, \lottsym{◇} \, i \, \lottsym{\}} \lottsym{.} \, \lottsym{⟦} ι \lottsym{,} \, \mathcolor{STLC}{x} \, \lottsym{∼} \, i \, \lottsym{⊢} \, \mathcolor{STLC}{e} \lottsym{⟧} \\
  \lottsym{⟦} ι \, \lottsym{⊢} \, \mathcolor{STLC}{e₀ \, e₁} \lottsym{⟧} &= \lottkw{force} \, \lottsym{⟦} ι \, \lottsym{⊢} \, \mathcolor{STLC}{e₀} \lottsym{⟧} \, \lottsym{\{} \, \_ \, \lottsym{:} \, \lottsym{⟦} ι \, \lottsym{⊢} \, \mathcolor{STLC}{e₁} \lottsym{⟧} \, \lottsym{\}} \\
  \lottsym{⟦} ι \, \lottsym{⊢} \, \mathcolor{STLC}{n} \lottsym{⟧} &= n
\end{aligned}$}
\par\smallskip

The translation rules are then largely straightforward. STLC lambdas are mapped
to \lang lambdas that have a record pattern with a single positional argument,
and STLC applications are mapped to force expressions of applications with the
argument passed positionally.

\paragraph{\textbf{Simulation}}

First, we prove that typing is preserved under translation:

\begin{theorem}[Preservation of typing]
  \label{thm:typ-conserve}
  If\;\;$\mathcolor{STLC}{ Γ } \, \lottsym{⊢} \, \mathcolor{STLC}{ e } \, \lottsym{:} \, \mathcolor{STLC}{ τ }$, then $\lottsym{⟦} ι \, \lottsym{⊢} \,
  \mathcolor{STLC}{Γ} \lottsym{⟧} \, \lottsym{⊢} \, \lottsym{⟦} ι \,
  \lottsym{⊢} \, \mathcolor{STLC}{e} \lottsym{⟧} \, \lottsym{:} \, \lottsym{⟦}
  \mathcolor{STLC}{τ} \lottsym{⟧}$.
\end{theorem}

Relating the evaluation behaviour of the two calculi is slightly more involved.
We utilize a \textit{term approximation} relation $δ \,
\lottsym{⊢} \, \mathcolor{STLC}{ eₛ } \, \lottsym{≈} \, e$, which states that
an STLC term $\mathcolor{STLC}{ eₛ }$ is approximated by a \lang term $e$ under an
approximation
environment $δ$. We defer the details to the appendix.
We establish the correspondence of semantics between the two calculi:

\begin{theorem}[Correspondence of semantics]
  \label{thm:sem-correspondence}
  $\mathcolor{STLC}{ e } \, \lottsym{\overset{*}{\longrightarrow}} \, \mathcolor{STLC}{ n }$ if and only if $\lottsym{ε} \, \lottsym{⊢} \,
  \lottsym{⟦} ε \, \lottsym{⊢} \, \mathcolor{STLC}{e} \lottsym{⟧} \,
  \lottsym{\overset{*}{\longrightarrow}} \, n$.
\end{theorem}

\def\store{\relax}

\section{Extension: References}
\label{sec:references}

In this section, we extend \lang with mutable references. This particular
extension serves two primary purposes. First, it introduces side-effects, under
which it is critical to have deterministic evaluation behaviour in the presence of
partial application, optional arguments, and overwriting. Second,
consistent with our paradigm of using records to formalise various language
constructs, we leverage records to model stores that map reference locations to their values.

\begin{figure}[b]
  \small
  \begingroup
  \renewcommand{\lottnonterminal}[3]{#2 \Coloneqq #3}
  \noindent
  $\lottnonterminalfilter{ref}{%
\lottnonterminal{Type}{
\lottalias{A}
\lottaliassep
\lottalias{B}
}{
\lottproduction{var}{a}
\lottproductionsep
\lottproduction{nat}{\lottsym{ℕ}}
\lottproductionsep
\lottproduction{arr}{\lottsym{\{} \, ρ \, \lottsym{\}} \, \lottsym{→} \, A}
\lottproductionsep
\lottproduction{record}{\lottsym{\{} \, ρ \, \lottsym{\}}}
\lottproductionsep
\lottproduction{quant}{\lottsym{∀} \, a \lottsym{.} \, A}
\lottproductionsep
\lottproduction{ref}{\lottkw{Ref} \, A}
}
}$\hfill
  $\lottnonterminalfilter{ref}{%
\lottnonterminal{Monotype}{
\lottalias{τ}
}{
\lottproduction{var}{a}
\lottproductionsep
\lottproduction{nat}{\lottsym{ℕ}}
\lottproductionsep
\lottproduction{arr}{\lottsym{\{} \, ϱ \, \lottsym{\}} \, \lottsym{→} \, τ}
\lottproductionsep
\lottproduction{record}{\lottsym{\{} \, ϱ \, \lottsym{\}}}
\lottproductionsep
\lottproduction{ref}{\lottkw{Ref} \, τ}
}
}$\hfill
  $\lottnonterminalfilter{loc,ref,deref,set}{%
\lottnonterminal{Term}{
\lottalias{e}
}{
\lottproduction{nat}{n}
\lottproductionsep
\lottproduction{op}{e₀ \, b \, e₁}
\lottproductionsep
\lottproduction{lam}{\lottsym{λ} \, \lottsym{\{} \, c \, \lottsym{\}} \lottsym{.} \, e}
\lottproductionsep
\lottproduction{closure}{\lottsym{⟨} r₀ \lottsym{;} \, r₁ \, \lottsym{|} \, \lottsym{λ} \, \lottsym{\{} \, c \, \lottsym{\}} \lottsym{.} \, e \lottsym{⟩}}
\lottproductionsep
\lottproduction{app}{e₀ \, e₁}
\lottproductionsep
\lottproduction{force}{\lottkw{force} \, e}
\lottproductionsep
\lottproduction{box}{e₀ \, \lottsym{▶} \, e₁}
\lottproductionsep
\lottproduction{env}{\lottsym{@}}
\lottproductionsep
\lottproduction{record}{\lottsym{\{} \, r \, \lottsym{\}}}
\lottproductionsep
\lottproduction{concat}{e₀ \, \lottsym{⧺} \, e₁}
\lottproductionsep
\lottproduction{proj}{e \lottsym{.} i}
\lottproductionsep
\lottproduction{mproj}{e \lottsym{.} \lottsym{\{} \, \lottsepbycomprehension{{i}_{i}} \, \lottsym{\}}}
\lottproductionsep
\lottproduction{diff}{e \lottsym{\textbackslash} \lottsym{\{} \, \lottsepbycomprehension{{i}_{i}} \, \lottsym{\}}}
\lottproductionsep
\lottproduction{let}{\lottkw{let} \, t \, \lottsym{=} \, e₀ \, \lottkw{in} \, e₁}
\lottproductionsep
\lottproduction{loc}{l}
\lottproductionsep
\lottproduction{ref}{\lottkw{ref} \, e}
\lottproductionsep
\lottproduction{deref}{\lottsym{!} e}
\lottproductionsep
\lottproduction{set}{e₀ \, \lottsym{≔} \, e₁}
}
}$\hfill
  $\lottnonterminalfilter{loc}{%
\lottnonterminal{Value}{
\lottalias{v}
}{
\lottproduction{nat}{n}
\lottproductionsep
\lottproduction{closure}{\lottsym{⟨} o₀ \lottsym{;} \, o₁ \, \lottsym{|} \, \lottsym{λ} \, \lottsym{\{} \, c \, \lottsym{\}} \lottsym{.} \, e \lottsym{⟩}}
\lottproductionsep
\lottproduction{record}{\lottsym{\{} \, o \, \lottsym{\}}}
\lottproductionsep
\lottproduction{loc}{l}
}
}$
  \endgroup

  \medskip

  \begingroup
  \renewcommand{\lottinferencerule}[3]{\relax}
  \renewcommand{\lottinferencerulesep}{\relax}
  \renewcommand{\lottjudgement}[3]{\fbox{$#2$\rule[-0.2\baselineskip]{0pt}{0.1\baselineskip}}}
\lottjudgement{Subtyping}{A \, \lottsym{\mathcolor{Subtyping}{⩽}} \, B}{
\lottinferencerule{var}{
}{a \, \lottsym{\mathcolor{Subtyping}{⩽}} \, a}
\lottinferencerulesep
\lottinferencerule{nat}{
}{\lottsym{ℕ} \, \lottsym{\mathcolor{Subtyping}{⩽}} \, \lottsym{ℕ}}
\lottinferencerulesep
\lottinferencerule{arr}{
\lotthypothesis{ρ₁ \, \lottsym{\mathcolor{Subtyping}{⩽}} \, ρ₀}\\
\lotthypothesis{A \, \lottsym{\mathcolor{Subtyping}{⩽}} \, B}
}{\lottsym{\{} \, ρ₀ \, \lottsym{\}} \, \lottsym{→} \, A \, \lottsym{\mathcolor{Subtyping}{⩽}} \, \lottsym{\{} \, ρ₁ \, \lottsym{\}} \, \lottsym{→} \, B}
\lottinferencerulesep
\lottinferencerule{record}{
\lotthypothesis{ρ₀ \, \lottsym{\mathcolor{Subtyping}{⩽}} \, ρ₁}
}{\lottsym{\{} \, ρ₀ \, \lottsym{\}} \, \lottsym{\mathcolor{Subtyping}{⩽}} \, \lottsym{\{} \, ρ₁ \, \lottsym{\}}}
\lottinferencerulesep
\lottinferencerule{quantL}{
\lotthypothesis{A\left[τ/a\right] \, \lottsym{\mathcolor{Subtyping}{⩽}} \, B}
}{\lottsym{∀} \, a \lottsym{.} \, A \, \lottsym{\mathcolor{Subtyping}{⩽}} \, B}
\lottinferencerulesep
\lottinferencerule{quantR}{a \, \lottsym{∉} \, \lottkw{ftv} \lottsym{(} A \lottsym{)}\\
\lotthypothesis{A \, \lottsym{\mathcolor{Subtyping}{⩽}} \, B}
}{A \, \lottsym{\mathcolor{Subtyping}{⩽}} \, \lottsym{∀} \, a \lottsym{.} \, B}
\lottinferencerulesep
\lottinferencerule{ref}{
}{\lottkw{Ref} \, A \, \lottsym{\mathcolor{Subtyping}{⩽}} \, \lottkw{Ref} \, A}
}
  \space
\lottjudgement{Typing}{\ifdefined\store \mathcolor{Store}{ρₛ} \lottsym{;} \, \fi ρ \, \lottsym{⊢} \, e \, \lottsym{:} \, A}{
\lottinferencerule{sub}{
\lotthypothesis{\ifdefined\store \mathcolor{Store}{ρₛ} \lottsym{;} \, \fi ρ \, \lottsym{⊢} \, e \, \lottsym{:} \, A}\\
\lotthypothesis{A \, \lottsym{\mathcolor{Subtyping}{⩽}} \, B}
}{\ifdefined\store \mathcolor{Store}{ρₛ} \lottsym{;} \, \fi ρ \, \lottsym{⊢} \, e \, \lottsym{:} \, B}
\lottinferencerulesep
\lottinferencerule{nat}{
}{\ifdefined\store \mathcolor{Store}{ρₛ} \lottsym{;} \, \fi ρ \, \lottsym{⊢} \, n \, \lottsym{:} \, \lottsym{ℕ}}
\lottinferencerulesep
\lottinferencerule{op}{
\lotthypothesis{\ifdefined\store \mathcolor{Store}{ρₛ} \lottsym{;} \, \fi ρ \, \lottsym{⊢} \, e₀ \, \lottsym{:} \, \lottsym{ℕ}}\\
\lotthypothesis{\ifdefined\store \mathcolor{Store}{ρₛ} \lottsym{;} \, \fi ρ \, \lottsym{⊢} \, e₁ \, \lottsym{:} \, \lottsym{ℕ}}
}{\ifdefined\store \mathcolor{Store}{ρₛ} \lottsym{;} \, \fi ρ \, \lottsym{⊢} \, e₀ \, b \, e₁ \, \lottsym{:} \, \lottsym{ℕ}}
\lottinferencerulesep
\lottinferencerule{lam}{\ifdefined\store\overline{a_j} \, \lottsym{∉} \, \lottkw{ftv} \lottsym{(} \mathcolor{Store}{ρₛ} \lottsym{)}\\\fi\overline{a_j} \, \lottsym{∉} \, \lottkw{ftv} \lottsym{(} ρ₀ \lottsym{)}\\
\lotthypothesis{c \, \lottsym{\mathcolor{Pattern}{:}} \, ρ₁ \, \lottsym{\mathcolor{Pattern}{⊙}} \, ρ₂}\\
\lotthypothesis{\ifdefined\store \mathcolor{Store}{ρₛ} \lottsym{;} \, \fi ρ₀ \lottsym{,} \, ρ₂ \, \lottsym{⊢} \, e \, \lottsym{:} \, A}
}{\ifdefined\store \mathcolor{Store}{ρₛ} \lottsym{;} \, \fi ρ₀ \, \lottsym{⊢} \, \lottsym{λ} \, \lottsym{\{} \, c \, \lottsym{\}} \lottsym{.} \, e \, \lottsym{:} \, \lottsym{∀} \, \lottsepbycomprehension{{a}_{j}} \lottsym{.} \, \lottsym{\{} \, ρ₁ \, \lottsym{\}} \, \lottsym{→} \, A}
\lottinferencerulesep
\lottinferencerule{closure}{\ifdefined\store\overline{a_j} \, \lottsym{∉} \, \lottkw{ftv} \lottsym{(} \mathcolor{Store}{ρₛ} \lottsym{)}\\\fi
\lotthypothesis{\ifdefined\store \mathcolor{Store}{ρₛ} \lottsym{;} \, \fi \lottsym{ε} \, \lottsym{⊢} \, \lottsym{\{} \, o₀ \, \lottsym{\}} \, \lottsym{:} \, \lottsym{\{} \, ρ₁ \, \lottsym{\}}}\\
\lotthypothesis{\lottsym{⌊} ρ₁ \lottsym{⌋}}\\\overline{a_j} \, \lottsym{∉} \, \lottkw{ftv} \lottsym{(} ρ₁ \lottsym{)}\\
\lotthypothesis{\ifdefined\store \mathcolor{Store}{ρₛ} \lottsym{;} \, \fi \lottsym{ε} \, \lottsym{⊢} \, \lottsym{\{} \, o₁ \, \lottsym{\}} \, \lottsym{:} \, \lottsym{∀} \, \lottsepbycomprehension{{a}_{j}} \lottsym{.} \, \lottsym{\{} \, ρ₂ \, \lottsym{\}}}\\
\lotthypothesis{c \, \lottsym{\mathcolor{Pattern}{:}} \, ρ₃ \, \lottsym{\mathcolor{Pattern}{⊙}} \, ρ₄}\\
\lotthypothesis{ρ₃ \, \lottsym{\mathcolor{App}{●}} \, ρ₂ \, \lottsym{\mathcolor{App}{↠}} \, ρ₅}\\
\lotthypothesis{\ifdefined\store \mathcolor{Store}{ρₛ} \lottsym{;} \, \fi ρ₁ \lottsym{,} \, ρ₄ \, \lottsym{⊢} \, e \, \lottsym{:} \, A}
}{\ifdefined\store \mathcolor{Store}{ρₛ} \lottsym{;} \, \fi ρ₀ \, \lottsym{⊢} \, \lottsym{⟨} o₀ \lottsym{;} \, o₁ \, \lottsym{|} \, \lottsym{λ} \, \lottsym{\{} \, c \, \lottsym{\}} \lottsym{.} \, e \lottsym{⟩} \, \lottsym{:} \, \lottsym{∀} \, \lottsepbycomprehension{{a}_{j}} \lottsym{.} \, \lottsym{\{} \, ρ₅ \, \lottsym{\}} \, \lottsym{→} \, A}
\lottinferencerulesep
\lottinferencerule{app}{
\lotthypothesis{\ifdefined\store \mathcolor{Store}{ρₛ} \lottsym{;} \, \fi ρ₀ \, \lottsym{⊢} \, e₀ \, \lottsym{:} \, \lottsym{∀} \, \lottsepbycomprehension{{a}_{j}} \lottsym{.} \, \lottsym{\{} \, ρ₁ \, \lottsym{\}} \, \lottsym{→} \, A}\\
\lotthypothesis{\ifdefined\store \mathcolor{Store}{ρₛ} \lottsym{;} \, \fi ρ₀ \, \lottsym{⊢} \, e₁ \, \lottsym{:} \, \lottsym{∀} \, \lottsepbycomprehension{{a}_{j}} \lottsym{.} \, \lottsym{\{} \, ρ₂ \, \lottsym{\}}}\\
\lotthypothesis{ρ₁ \, \lottsym{\mathcolor{App}{●}} \, ρ₂ \, \lottsym{\mathcolor{App}{↠}} \, ρ₃}
}{\ifdefined\store \mathcolor{Store}{ρₛ} \lottsym{;} \, \fi ρ₀ \, \lottsym{⊢} \, e₀ \, e₁ \, \lottsym{:} \, \lottsym{∀} \, \lottsepbycomprehension{{a}_{j}} \lottsym{.} \, \lottsym{\{} \, ρ₃ \, \lottsym{\}} \, \lottsym{→} \, A}
\lottinferencerulesep
\lottinferencerule{force}{
\lotthypothesis{\ifdefined\store \mathcolor{Store}{ρₛ} \lottsym{;} \, \fi ρ \, \lottsym{⊢} \, e \, \lottsym{:} \, \lottsym{\{} \, \lottsym{\}} \, \lottsym{→} \, A}
}{\ifdefined\store \mathcolor{Store}{ρₛ} \lottsym{;} \, \fi ρ \, \lottsym{⊢} \, \lottkw{force} \, e \, \lottsym{:} \, A}
\lottinferencerulesep
\lottinferencerule{box}{
\lotthypothesis{\ifdefined\store \mathcolor{Store}{ρₛ} \lottsym{;} \, \fi ρ₀ \, \lottsym{⊢} \, e₀ \, \lottsym{:} \, \lottsym{\{} \, ρ₁ \, \lottsym{\}}}\\
\lotthypothesis{\lottsym{⌊} ρ₁ \lottsym{⌋}}\\
\lotthypothesis{\ifdefined\store \mathcolor{Store}{ρₛ} \lottsym{;} \, \fi ρ₁ \, \lottsym{⊢} \, e₁ \, \lottsym{:} \, A}
}{\ifdefined\store \mathcolor{Store}{ρₛ} \lottsym{;} \, \fi ρ₀ \, \lottsym{⊢} \, e₀ \, \lottsym{▶} \, e₁ \, \lottsym{:} \, A}
\lottinferencerulesep
\lottinferencerule{env}{
}{\ifdefined\store \mathcolor{Store}{ρₛ} \lottsym{;} \, \fi ρ \, \lottsym{⊢} \, \lottsym{@} \, \lottsym{:} \, \lottsym{\{} \, ρ \, \lottsym{\}}}
\lottinferencerulesep
\lottinferencerule{record}{
\lotthypothesis{\ifdefined\store \mathcolor{Store}{ρₛ} \lottsym{;} \, \fi ρ₀ \, \lottsym{⊢} \, r \, \lottsym{:} \, ρ₁}
}{\ifdefined\store \mathcolor{Store}{ρₛ} \lottsym{;} \, \fi ρ₀ \, \lottsym{⊢} \, \lottsym{\{} \, r \, \lottsym{\}} \, \lottsym{:} \, \lottsym{\{} \, ρ₁ \, \lottsym{\}}}
\lottinferencerulesep
\lottinferencerule{concat}{
\lotthypothesis{\ifdefined\store \mathcolor{Store}{ρₛ} \lottsym{;} \, \fi ρ₀ \, \lottsym{⊢} \, e₀ \, \lottsym{:} \, \lottsym{\{} \, ρ₁ \, \lottsym{\}}}\\
\lotthypothesis{\ifdefined\store \mathcolor{Store}{ρₛ} \lottsym{;} \, \fi ρ₀ \, \lottsym{⊢} \, e₁ \, \lottsym{:} \, \lottsym{\{} \, ρ₂ \, \lottsym{\}}}\\
\lotthypothesis{ρ₁ \, \lottsym{\mathcolor{Concat}{⊕}} \, ρ₂ \, \lottsym{\mathcolor{Concat}{↬}} \, ρ₃}
}{\ifdefined\store \mathcolor{Store}{ρₛ} \lottsym{;} \, \fi ρ₀ \, \lottsym{⊢} \, e₀ \, \lottsym{⧺} \, e₁ \, \lottsym{:} \, \lottsym{\{} \, ρ₃ \, \lottsym{\}}}
\lottinferencerulesep
\lottinferencerule{proj}{
\lotthypothesis{\ifdefined\store \mathcolor{Store}{ρₛ} \lottsym{;} \, \fi ρ₀ \, \lottsym{⊢} \, e \, \lottsym{:} \, \lottsym{\{} \, ρ₁ \lottsym{,} \, i \, \rstrength \, A \, \lottsym{\}}}
}{\ifdefined\store \mathcolor{Store}{ρₛ} \lottsym{;} \, \fi ρ₀ \, \lottsym{⊢} \, e \lottsym{.} i \, \lottsym{:} \, A}
\lottinferencerulesep
\lottinferencerule{mproj}{
\lotthypothesis{\ifdefined\store \mathcolor{Store}{ρₛ} \lottsym{;} \, \fi ρ₀ \, \lottsym{⊢} \, e \, \lottsym{:} \, \lottsym{\{} \, ρ₁ \lottsym{,} \, \lottsepbycomprehension{{i}_{j} \, {\rstrength}_{j} \, {A}_{j}} \, \lottsym{\}}}
}{\ifdefined\store \mathcolor{Store}{ρₛ} \lottsym{;} \, \fi ρ₀ \, \lottsym{⊢} \, e \lottsym{.} \lottsym{\{} \, \lottsepbycomprehension{{i}_{j}} \, \lottsym{\}} \, \lottsym{:} \, \lottsym{\{} \, \lottsepbycomprehension{{i}_{j} \, {\rstrength}_{j} \, {A}_{j}} \, \lottsym{\}}}
\lottinferencerulesep
\lottinferencerule{diff}{
\lotthypothesis{\ifdefined\store \mathcolor{Store}{ρₛ} \lottsym{;} \, \fi ρ₀ \, \lottsym{⊢} \, e \, \lottsym{:} \, \lottsym{\{} \, ρ₁ \lottsym{,} \, \lottsepbycomprehension{{i}_{j} \, {\rstrength}_{j} \, {A}_{j}} \, \lottsym{\}}}
}{\ifdefined\store \mathcolor{Store}{ρₛ} \lottsym{;} \, \fi ρ₀ \, \lottsym{⊢} \, e \lottsym{\textbackslash} \lottsym{\{} \, \lottsepbycomprehension{{i}_{j}} \, \lottsym{\}} \, \lottsym{:} \, \lottsym{\{} \, ρ₁ \, \lottsym{\}}}
\lottinferencerulesep
\lottinferencerule{let}{
\lotthypothesis{\ifdefined\store \mathcolor{Store}{ρₛ} \lottsym{;} \, \fi ρ₀ \, \lottsym{⊢} \, e₀ \, \lottsym{:} \, A}\\
\lotthypothesis{t \, \lottsym{\mathcolor{Pattern}{:}} \, A \, \lottsym{\mathcolor{Pattern}{⊙}} \, ρ₁}\\
\lotthypothesis{\ifdefined\store \mathcolor{Store}{ρₛ} \lottsym{;} \, \fi ρ₀ \lottsym{,} \, ρ₁ \, \lottsym{⊢} \, e₁ \, \lottsym{:} \, B}
}{\ifdefined\store \mathcolor{Store}{ρₛ} \lottsym{;} \, \fi ρ₀ \, \lottsym{⊢} \, \lottkw{let} \, t \, \lottsym{=} \, e₀ \, \lottkw{in} \, e₁ \, \lottsym{:} \, B}
\lottinferencerulesep
\lottinferencerule{loc}{
}{\ifdefined\store \mathcolor{Store}{ρₛ₀ \lottsym{,} \, l \, \lottsym{:} \, A \lottsym{,} \, ρₛ₁} \lottsym{;} \, \fi ρ \, \lottsym{⊢} \, l \, \lottsym{:} \, \lottkw{Ref} \, A}
\lottinferencerulesep
\lottinferencerule{ref}{
\lotthypothesis{\ifdefined\store \mathcolor{Store}{ρₛ} \lottsym{;} \, \fi ρ \, \lottsym{⊢} \, e \, \lottsym{:} \, A}
}{\ifdefined\store \mathcolor{Store}{ρₛ} \lottsym{;} \, \fi ρ \, \lottsym{⊢} \, \lottkw{ref} \, e \, \lottsym{:} \, \lottkw{Ref} \, A}
\lottinferencerulesep
\lottinferencerule{deref}{
\lotthypothesis{\ifdefined\store \mathcolor{Store}{ρₛ} \lottsym{;} \, \fi ρ \, \lottsym{⊢} \, e \, \lottsym{:} \, \lottkw{Ref} \, A}
}{\ifdefined\store \mathcolor{Store}{ρₛ} \lottsym{;} \, \fi ρ \, \lottsym{⊢} \, \lottsym{!} e \, \lottsym{:} \, A}
\lottinferencerulesep
\lottinferencerule{set}{
\lotthypothesis{\ifdefined\store \mathcolor{Store}{ρₛ} \lottsym{;} \, \fi ρ \, \lottsym{⊢} \, e₀ \, \lottsym{:} \, \lottkw{Ref} \, A}\\
\lotthypothesis{\ifdefined\store \mathcolor{Store}{ρₛ} \lottsym{;} \, \fi ρ \, \lottsym{⊢} \, e₁ \, \lottsym{:} \, A}
}{\ifdefined\store \mathcolor{Store}{ρₛ} \lottsym{;} \, \fi ρ \, \lottsym{⊢} \, e₀ \, \lottsym{≔} \, e₁ \, \lottsym{:} \, \lottsym{\{} \, \lottsym{\}}}
}
\hfill\emph{(Subtyping and Typing)}
  \endgroup
  \begingroup
  \renewcommand{\lottjudgement}[3]{#3}
  \begin{mathparpagebreakable}
    \lottjudgementfilter{ref}{\lottjudgementlabel{S}{%
\lottjudgement{Subtyping}{A \, \lottsym{\mathcolor{Subtyping}{⩽}} \, B}{
\lottinferencerule{var}{
}{a \, \lottsym{\mathcolor{Subtyping}{⩽}} \, a}
\lottinferencerulesep
\lottinferencerule{nat}{
}{\lottsym{ℕ} \, \lottsym{\mathcolor{Subtyping}{⩽}} \, \lottsym{ℕ}}
\lottinferencerulesep
\lottinferencerule{arr}{
\lotthypothesis{ρ₁ \, \lottsym{\mathcolor{Subtyping}{⩽}} \, ρ₀}\\
\lotthypothesis{A \, \lottsym{\mathcolor{Subtyping}{⩽}} \, B}
}{\lottsym{\{} \, ρ₀ \, \lottsym{\}} \, \lottsym{→} \, A \, \lottsym{\mathcolor{Subtyping}{⩽}} \, \lottsym{\{} \, ρ₁ \, \lottsym{\}} \, \lottsym{→} \, B}
\lottinferencerulesep
\lottinferencerule{record}{
\lotthypothesis{ρ₀ \, \lottsym{\mathcolor{Subtyping}{⩽}} \, ρ₁}
}{\lottsym{\{} \, ρ₀ \, \lottsym{\}} \, \lottsym{\mathcolor{Subtyping}{⩽}} \, \lottsym{\{} \, ρ₁ \, \lottsym{\}}}
\lottinferencerulesep
\lottinferencerule{quantL}{
\lotthypothesis{A\left[τ/a\right] \, \lottsym{\mathcolor{Subtyping}{⩽}} \, B}
}{\lottsym{∀} \, a \lottsym{.} \, A \, \lottsym{\mathcolor{Subtyping}{⩽}} \, B}
\lottinferencerulesep
\lottinferencerule{quantR}{a \, \lottsym{∉} \, \lottkw{ftv} \lottsym{(} A \lottsym{)}\\
\lotthypothesis{A \, \lottsym{\mathcolor{Subtyping}{⩽}} \, B}
}{A \, \lottsym{\mathcolor{Subtyping}{⩽}} \, \lottsym{∀} \, a \lottsym{.} \, B}
\lottinferencerulesep
\lottinferencerule{ref}{
}{\lottkw{Ref} \, A \, \lottsym{\mathcolor{Subtyping}{⩽}} \, \lottkw{Ref} \, A}
}
}}
    \lottinferencerulesep
    \lottjudgementfilter{loc,ref,deref,set}{\lottjudgementnarrower[set=0.2]{1}{\lottjudgementlabel{T}{%
\lottjudgement{Typing}{\ifdefined\store \mathcolor{Store}{ρₛ} \lottsym{;} \, \fi ρ \, \lottsym{⊢} \, e \, \lottsym{:} \, A}{
\lottinferencerule{sub}{
\lotthypothesis{\ifdefined\store \mathcolor{Store}{ρₛ} \lottsym{;} \, \fi ρ \, \lottsym{⊢} \, e \, \lottsym{:} \, A}\\
\lotthypothesis{A \, \lottsym{\mathcolor{Subtyping}{⩽}} \, B}
}{\ifdefined\store \mathcolor{Store}{ρₛ} \lottsym{;} \, \fi ρ \, \lottsym{⊢} \, e \, \lottsym{:} \, B}
\lottinferencerulesep
\lottinferencerule{nat}{
}{\ifdefined\store \mathcolor{Store}{ρₛ} \lottsym{;} \, \fi ρ \, \lottsym{⊢} \, n \, \lottsym{:} \, \lottsym{ℕ}}
\lottinferencerulesep
\lottinferencerule{op}{
\lotthypothesis{\ifdefined\store \mathcolor{Store}{ρₛ} \lottsym{;} \, \fi ρ \, \lottsym{⊢} \, e₀ \, \lottsym{:} \, \lottsym{ℕ}}\\
\lotthypothesis{\ifdefined\store \mathcolor{Store}{ρₛ} \lottsym{;} \, \fi ρ \, \lottsym{⊢} \, e₁ \, \lottsym{:} \, \lottsym{ℕ}}
}{\ifdefined\store \mathcolor{Store}{ρₛ} \lottsym{;} \, \fi ρ \, \lottsym{⊢} \, e₀ \, b \, e₁ \, \lottsym{:} \, \lottsym{ℕ}}
\lottinferencerulesep
\lottinferencerule{lam}{\ifdefined\store\overline{a_j} \, \lottsym{∉} \, \lottkw{ftv} \lottsym{(} \mathcolor{Store}{ρₛ} \lottsym{)}\\\fi\overline{a_j} \, \lottsym{∉} \, \lottkw{ftv} \lottsym{(} ρ₀ \lottsym{)}\\
\lotthypothesis{c \, \lottsym{\mathcolor{Pattern}{:}} \, ρ₁ \, \lottsym{\mathcolor{Pattern}{⊙}} \, ρ₂}\\
\lotthypothesis{\ifdefined\store \mathcolor{Store}{ρₛ} \lottsym{;} \, \fi ρ₀ \lottsym{,} \, ρ₂ \, \lottsym{⊢} \, e \, \lottsym{:} \, A}
}{\ifdefined\store \mathcolor{Store}{ρₛ} \lottsym{;} \, \fi ρ₀ \, \lottsym{⊢} \, \lottsym{λ} \, \lottsym{\{} \, c \, \lottsym{\}} \lottsym{.} \, e \, \lottsym{:} \, \lottsym{∀} \, \lottsepbycomprehension{{a}_{j}} \lottsym{.} \, \lottsym{\{} \, ρ₁ \, \lottsym{\}} \, \lottsym{→} \, A}
\lottinferencerulesep
\lottinferencerule{closure}{\ifdefined\store\overline{a_j} \, \lottsym{∉} \, \lottkw{ftv} \lottsym{(} \mathcolor{Store}{ρₛ} \lottsym{)}\\\fi
\lotthypothesis{\ifdefined\store \mathcolor{Store}{ρₛ} \lottsym{;} \, \fi \lottsym{ε} \, \lottsym{⊢} \, \lottsym{\{} \, o₀ \, \lottsym{\}} \, \lottsym{:} \, \lottsym{\{} \, ρ₁ \, \lottsym{\}}}\\
\lotthypothesis{\lottsym{⌊} ρ₁ \lottsym{⌋}}\\\overline{a_j} \, \lottsym{∉} \, \lottkw{ftv} \lottsym{(} ρ₁ \lottsym{)}\\
\lotthypothesis{\ifdefined\store \mathcolor{Store}{ρₛ} \lottsym{;} \, \fi \lottsym{ε} \, \lottsym{⊢} \, \lottsym{\{} \, o₁ \, \lottsym{\}} \, \lottsym{:} \, \lottsym{∀} \, \lottsepbycomprehension{{a}_{j}} \lottsym{.} \, \lottsym{\{} \, ρ₂ \, \lottsym{\}}}\\
\lotthypothesis{c \, \lottsym{\mathcolor{Pattern}{:}} \, ρ₃ \, \lottsym{\mathcolor{Pattern}{⊙}} \, ρ₄}\\
\lotthypothesis{ρ₃ \, \lottsym{\mathcolor{App}{●}} \, ρ₂ \, \lottsym{\mathcolor{App}{↠}} \, ρ₅}\\
\lotthypothesis{\ifdefined\store \mathcolor{Store}{ρₛ} \lottsym{;} \, \fi ρ₁ \lottsym{,} \, ρ₄ \, \lottsym{⊢} \, e \, \lottsym{:} \, A}
}{\ifdefined\store \mathcolor{Store}{ρₛ} \lottsym{;} \, \fi ρ₀ \, \lottsym{⊢} \, \lottsym{⟨} o₀ \lottsym{;} \, o₁ \, \lottsym{|} \, \lottsym{λ} \, \lottsym{\{} \, c \, \lottsym{\}} \lottsym{.} \, e \lottsym{⟩} \, \lottsym{:} \, \lottsym{∀} \, \lottsepbycomprehension{{a}_{j}} \lottsym{.} \, \lottsym{\{} \, ρ₅ \, \lottsym{\}} \, \lottsym{→} \, A}
\lottinferencerulesep
\lottinferencerule{app}{
\lotthypothesis{\ifdefined\store \mathcolor{Store}{ρₛ} \lottsym{;} \, \fi ρ₀ \, \lottsym{⊢} \, e₀ \, \lottsym{:} \, \lottsym{∀} \, \lottsepbycomprehension{{a}_{j}} \lottsym{.} \, \lottsym{\{} \, ρ₁ \, \lottsym{\}} \, \lottsym{→} \, A}\\
\lotthypothesis{\ifdefined\store \mathcolor{Store}{ρₛ} \lottsym{;} \, \fi ρ₀ \, \lottsym{⊢} \, e₁ \, \lottsym{:} \, \lottsym{∀} \, \lottsepbycomprehension{{a}_{j}} \lottsym{.} \, \lottsym{\{} \, ρ₂ \, \lottsym{\}}}\\
\lotthypothesis{ρ₁ \, \lottsym{\mathcolor{App}{●}} \, ρ₂ \, \lottsym{\mathcolor{App}{↠}} \, ρ₃}
}{\ifdefined\store \mathcolor{Store}{ρₛ} \lottsym{;} \, \fi ρ₀ \, \lottsym{⊢} \, e₀ \, e₁ \, \lottsym{:} \, \lottsym{∀} \, \lottsepbycomprehension{{a}_{j}} \lottsym{.} \, \lottsym{\{} \, ρ₃ \, \lottsym{\}} \, \lottsym{→} \, A}
\lottinferencerulesep
\lottinferencerule{force}{
\lotthypothesis{\ifdefined\store \mathcolor{Store}{ρₛ} \lottsym{;} \, \fi ρ \, \lottsym{⊢} \, e \, \lottsym{:} \, \lottsym{\{} \, \lottsym{\}} \, \lottsym{→} \, A}
}{\ifdefined\store \mathcolor{Store}{ρₛ} \lottsym{;} \, \fi ρ \, \lottsym{⊢} \, \lottkw{force} \, e \, \lottsym{:} \, A}
\lottinferencerulesep
\lottinferencerule{box}{
\lotthypothesis{\ifdefined\store \mathcolor{Store}{ρₛ} \lottsym{;} \, \fi ρ₀ \, \lottsym{⊢} \, e₀ \, \lottsym{:} \, \lottsym{\{} \, ρ₁ \, \lottsym{\}}}\\
\lotthypothesis{\lottsym{⌊} ρ₁ \lottsym{⌋}}\\
\lotthypothesis{\ifdefined\store \mathcolor{Store}{ρₛ} \lottsym{;} \, \fi ρ₁ \, \lottsym{⊢} \, e₁ \, \lottsym{:} \, A}
}{\ifdefined\store \mathcolor{Store}{ρₛ} \lottsym{;} \, \fi ρ₀ \, \lottsym{⊢} \, e₀ \, \lottsym{▶} \, e₁ \, \lottsym{:} \, A}
\lottinferencerulesep
\lottinferencerule{env}{
}{\ifdefined\store \mathcolor{Store}{ρₛ} \lottsym{;} \, \fi ρ \, \lottsym{⊢} \, \lottsym{@} \, \lottsym{:} \, \lottsym{\{} \, ρ \, \lottsym{\}}}
\lottinferencerulesep
\lottinferencerule{record}{
\lotthypothesis{\ifdefined\store \mathcolor{Store}{ρₛ} \lottsym{;} \, \fi ρ₀ \, \lottsym{⊢} \, r \, \lottsym{:} \, ρ₁}
}{\ifdefined\store \mathcolor{Store}{ρₛ} \lottsym{;} \, \fi ρ₀ \, \lottsym{⊢} \, \lottsym{\{} \, r \, \lottsym{\}} \, \lottsym{:} \, \lottsym{\{} \, ρ₁ \, \lottsym{\}}}
\lottinferencerulesep
\lottinferencerule{concat}{
\lotthypothesis{\ifdefined\store \mathcolor{Store}{ρₛ} \lottsym{;} \, \fi ρ₀ \, \lottsym{⊢} \, e₀ \, \lottsym{:} \, \lottsym{\{} \, ρ₁ \, \lottsym{\}}}\\
\lotthypothesis{\ifdefined\store \mathcolor{Store}{ρₛ} \lottsym{;} \, \fi ρ₀ \, \lottsym{⊢} \, e₁ \, \lottsym{:} \, \lottsym{\{} \, ρ₂ \, \lottsym{\}}}\\
\lotthypothesis{ρ₁ \, \lottsym{\mathcolor{Concat}{⊕}} \, ρ₂ \, \lottsym{\mathcolor{Concat}{↬}} \, ρ₃}
}{\ifdefined\store \mathcolor{Store}{ρₛ} \lottsym{;} \, \fi ρ₀ \, \lottsym{⊢} \, e₀ \, \lottsym{⧺} \, e₁ \, \lottsym{:} \, \lottsym{\{} \, ρ₃ \, \lottsym{\}}}
\lottinferencerulesep
\lottinferencerule{proj}{
\lotthypothesis{\ifdefined\store \mathcolor{Store}{ρₛ} \lottsym{;} \, \fi ρ₀ \, \lottsym{⊢} \, e \, \lottsym{:} \, \lottsym{\{} \, ρ₁ \lottsym{,} \, i \, \rstrength \, A \, \lottsym{\}}}
}{\ifdefined\store \mathcolor{Store}{ρₛ} \lottsym{;} \, \fi ρ₀ \, \lottsym{⊢} \, e \lottsym{.} i \, \lottsym{:} \, A}
\lottinferencerulesep
\lottinferencerule{mproj}{
\lotthypothesis{\ifdefined\store \mathcolor{Store}{ρₛ} \lottsym{;} \, \fi ρ₀ \, \lottsym{⊢} \, e \, \lottsym{:} \, \lottsym{\{} \, ρ₁ \lottsym{,} \, \lottsepbycomprehension{{i}_{j} \, {\rstrength}_{j} \, {A}_{j}} \, \lottsym{\}}}
}{\ifdefined\store \mathcolor{Store}{ρₛ} \lottsym{;} \, \fi ρ₀ \, \lottsym{⊢} \, e \lottsym{.} \lottsym{\{} \, \lottsepbycomprehension{{i}_{j}} \, \lottsym{\}} \, \lottsym{:} \, \lottsym{\{} \, \lottsepbycomprehension{{i}_{j} \, {\rstrength}_{j} \, {A}_{j}} \, \lottsym{\}}}
\lottinferencerulesep
\lottinferencerule{diff}{
\lotthypothesis{\ifdefined\store \mathcolor{Store}{ρₛ} \lottsym{;} \, \fi ρ₀ \, \lottsym{⊢} \, e \, \lottsym{:} \, \lottsym{\{} \, ρ₁ \lottsym{,} \, \lottsepbycomprehension{{i}_{j} \, {\rstrength}_{j} \, {A}_{j}} \, \lottsym{\}}}
}{\ifdefined\store \mathcolor{Store}{ρₛ} \lottsym{;} \, \fi ρ₀ \, \lottsym{⊢} \, e \lottsym{\textbackslash} \lottsym{\{} \, \lottsepbycomprehension{{i}_{j}} \, \lottsym{\}} \, \lottsym{:} \, \lottsym{\{} \, ρ₁ \, \lottsym{\}}}
\lottinferencerulesep
\lottinferencerule{let}{
\lotthypothesis{\ifdefined\store \mathcolor{Store}{ρₛ} \lottsym{;} \, \fi ρ₀ \, \lottsym{⊢} \, e₀ \, \lottsym{:} \, A}\\
\lotthypothesis{t \, \lottsym{\mathcolor{Pattern}{:}} \, A \, \lottsym{\mathcolor{Pattern}{⊙}} \, ρ₁}\\
\lotthypothesis{\ifdefined\store \mathcolor{Store}{ρₛ} \lottsym{;} \, \fi ρ₀ \lottsym{,} \, ρ₁ \, \lottsym{⊢} \, e₁ \, \lottsym{:} \, B}
}{\ifdefined\store \mathcolor{Store}{ρₛ} \lottsym{;} \, \fi ρ₀ \, \lottsym{⊢} \, \lottkw{let} \, t \, \lottsym{=} \, e₀ \, \lottkw{in} \, e₁ \, \lottsym{:} \, B}
\lottinferencerulesep
\lottinferencerule{loc}{
}{\ifdefined\store \mathcolor{Store}{ρₛ₀ \lottsym{,} \, l \, \lottsym{:} \, A \lottsym{,} \, ρₛ₁} \lottsym{;} \, \fi ρ \, \lottsym{⊢} \, l \, \lottsym{:} \, \lottkw{Ref} \, A}
\lottinferencerulesep
\lottinferencerule{ref}{
\lotthypothesis{\ifdefined\store \mathcolor{Store}{ρₛ} \lottsym{;} \, \fi ρ \, \lottsym{⊢} \, e \, \lottsym{:} \, A}
}{\ifdefined\store \mathcolor{Store}{ρₛ} \lottsym{;} \, \fi ρ \, \lottsym{⊢} \, \lottkw{ref} \, e \, \lottsym{:} \, \lottkw{Ref} \, A}
\lottinferencerulesep
\lottinferencerule{deref}{
\lotthypothesis{\ifdefined\store \mathcolor{Store}{ρₛ} \lottsym{;} \, \fi ρ \, \lottsym{⊢} \, e \, \lottsym{:} \, \lottkw{Ref} \, A}
}{\ifdefined\store \mathcolor{Store}{ρₛ} \lottsym{;} \, \fi ρ \, \lottsym{⊢} \, \lottsym{!} e \, \lottsym{:} \, A}
\lottinferencerulesep
\lottinferencerule{set}{
\lotthypothesis{\ifdefined\store \mathcolor{Store}{ρₛ} \lottsym{;} \, \fi ρ \, \lottsym{⊢} \, e₀ \, \lottsym{:} \, \lottkw{Ref} \, A}\\
\lotthypothesis{\ifdefined\store \mathcolor{Store}{ρₛ} \lottsym{;} \, \fi ρ \, \lottsym{⊢} \, e₁ \, \lottsym{:} \, A}
}{\ifdefined\store \mathcolor{Store}{ρₛ} \lottsym{;} \, \fi ρ \, \lottsym{⊢} \, e₀ \, \lottsym{≔} \, e₁ \, \lottsym{:} \, \lottsym{\{} \, \lottsym{\}}}
}
}}}
  \end{mathparpagebreakable}
  \endgroup

  \lottjudgementforcenameandexample{Operational Semantics}{o \, \lottsym{⊢} \, e \lottsym{;} \, \mathcolor{Store}{oₛ} \, \lottsym{\longrightarrow} \, e' \lottsym{;} \, \mathcolor{Store}{oₛ'}}{\lottjudgementfilter{refv,derefLoc,set}{\lottjudgementlabel{E}{%
\lottjudgement{Small Step}{r \, \lottsym{⊢} \, e \ifdefined\store \lottsym{;} \, \mathcolor{Store}{rₛ} \fi \, \lottsym{\longrightarrow} \, e' \ifdefined\store \lottsym{;} \, \mathcolor{Store}{rₛ'} \fi}{
\lottinferencerule{opl}{
\lotthypothesis{o \, \lottsym{⊢} \, e₀ \ifdefined\store \lottsym{;} \, \mathcolor{Store}{oₛ} \fi \, \lottsym{\longrightarrow} \, e₀' \ifdefined\store \lottsym{;} \, \mathcolor{Store}{oₛ'} \fi}
}{o \, \lottsym{⊢} \, e₀ \, b \, e₁ \ifdefined\store \lottsym{;} \, \mathcolor{Store}{oₛ} \fi \, \lottsym{\longrightarrow} \, e₀' \, b \, e₁ \ifdefined\store \lottsym{;} \, \mathcolor{Store}{oₛ'} \fi}
\lottinferencerulesep
\lottinferencerule{opr}{
\lotthypothesis{o \, \lottsym{⊢} \, e \ifdefined\store \lottsym{;} \, \mathcolor{Store}{oₛ} \fi \, \lottsym{\longrightarrow} \, e' \ifdefined\store \lottsym{;} \, \mathcolor{Store}{oₛ'} \fi}
}{o \, \lottsym{⊢} \, v \, b \, e \ifdefined\store \lottsym{;} \, \mathcolor{Store}{oₛ} \fi \, \lottsym{\longrightarrow} \, v \, b \, e' \ifdefined\store \lottsym{;} \, \mathcolor{Store}{oₛ'} \fi}
\lottinferencerulesep
\lottinferencerule{op}{
\lotthypothesis{n₀ \, b \, n₁ \, \lottsym{=} \, n₂}
}{o \, \lottsym{⊢} \, n₀ \, b \, n₁ \ifdefined\store \lottsym{;} \, \mathcolor{Store}{oₛ} \fi \, \lottsym{\longrightarrow} \, n₂ \ifdefined\store \lottsym{;} \, \mathcolor{Store}{oₛ} \fi}
\lottinferencerulesep
\lottinferencerule{lam}{
}{o \, \lottsym{⊢} \, \lottsym{λ} \, \lottsym{\{} \, c \, \lottsym{\}} \lottsym{.} \, e \ifdefined\store \lottsym{;} \, \mathcolor{Store}{oₛ} \fi \, \lottsym{\longrightarrow} \, \lottsym{⟨} o \lottsym{;} \, \lottsym{ε} \, \lottsym{|} \, \lottsym{λ} \, \lottsym{\{} \, c \, \lottsym{\}} \lottsym{.} \, e \lottsym{⟩} \ifdefined\store \lottsym{;} \, \mathcolor{Store}{oₛ} \fi}
\lottinferencerulesep
\lottinferencerule{appl}{
\lotthypothesis{o \, \lottsym{⊢} \, e₀ \ifdefined\store \lottsym{;} \, \mathcolor{Store}{oₛ} \fi \, \lottsym{\longrightarrow} \, e₀' \ifdefined\store \lottsym{;} \, \mathcolor{Store}{oₛ'} \fi}
}{o \, \lottsym{⊢} \, e₀ \, e₁ \ifdefined\store \lottsym{;} \, \mathcolor{Store}{oₛ} \fi \, \lottsym{\longrightarrow} \, e₀' \, e₁ \ifdefined\store \lottsym{;} \, \mathcolor{Store}{oₛ'} \fi}
\lottinferencerulesep
\lottinferencerule{appr}{
\lotthypothesis{o \, \lottsym{⊢} \, e \ifdefined\store \lottsym{;} \, \mathcolor{Store}{oₛ} \fi \, \lottsym{\longrightarrow} \, e' \ifdefined\store \lottsym{;} \, \mathcolor{Store}{oₛ'} \fi}
}{o \, \lottsym{⊢} \, v \, e \ifdefined\store \lottsym{;} \, \mathcolor{Store}{oₛ} \fi \, \lottsym{\longrightarrow} \, v \, e' \ifdefined\store \lottsym{;} \, \mathcolor{Store}{oₛ'} \fi}
\lottinferencerulesep
\lottinferencerule{app}{
}{o₀ \, \lottsym{⊢} \, \lottsym{⟨} o₁ \lottsym{;} \, o₂ \, \lottsym{|} \, \lottsym{λ} \, \lottsym{\{} \, c \, \lottsym{\}} \lottsym{.} \, e \lottsym{⟩} \, \lottsym{\{} \, o₃ \, \lottsym{\}} \ifdefined\store \lottsym{;} \, \mathcolor{Store}{oₛ} \fi \, \lottsym{\longrightarrow} \, \lottsym{⟨} o₁ \lottsym{;} \, o₂ \, \lottsym{⧺} \, o₃ \, \lottsym{|} \, \lottsym{λ} \, \lottsym{\{} \, c \, \lottsym{\}} \lottsym{.} \, e \lottsym{⟩} \ifdefined\store \lottsym{;} \, \mathcolor{Store}{oₛ} \fi}
\lottinferencerulesep
\lottinferencerule{force}{
\lotthypothesis{o \, \lottsym{⊢} \, e \ifdefined\store \lottsym{;} \, \mathcolor{Store}{oₛ} \fi \, \lottsym{\longrightarrow} \, e' \ifdefined\store \lottsym{;} \, \mathcolor{Store}{oₛ'} \fi}
}{o \, \lottsym{⊢} \, \lottkw{force} \, e \ifdefined\store \lottsym{;} \, \mathcolor{Store}{oₛ} \fi \, \lottsym{\longrightarrow} \, \lottkw{force} \, e' \ifdefined\store \lottsym{;} \, \mathcolor{Store}{oₛ'} \fi}
\lottinferencerulesep
\lottinferencerule{forceClosure}{
\lotthypothesis{\lottkw{expand} \lottsym{(} c \lottsym{;} \, o₂ \lottsym{)} \, \lottsym{=} \, o₃}
}{o₀ \, \lottsym{⊢} \, \lottkw{force} \, \lottsym{⟨} o₁ \lottsym{;} \, o₂ \, \lottsym{|} \, \lottsym{λ} \, \lottsym{\{} \, c \, \lottsym{\}} \lottsym{.} \, e \lottsym{⟩} \ifdefined\store \lottsym{;} \, \mathcolor{Store}{oₛ} \fi \, \lottsym{\longrightarrow} \, \lottsym{\{} \, o₁ \lottsym{,} \, o₃ \, \lottsym{\}} \, \lottsym{▶} \, e \ifdefined\store \lottsym{;} \, \mathcolor{Store}{oₛ} \fi}
\lottinferencerulesep
\lottinferencerule{boxl}{
\lotthypothesis{o \, \lottsym{⊢} \, e₀ \ifdefined\store \lottsym{;} \, \mathcolor{Store}{oₛ} \fi \, \lottsym{\longrightarrow} \, e₀' \ifdefined\store \lottsym{;} \, \mathcolor{Store}{oₛ'} \fi}
}{o \, \lottsym{⊢} \, e₀ \, \lottsym{▶} \, e₁ \ifdefined\store \lottsym{;} \, \mathcolor{Store}{oₛ} \fi \, \lottsym{\longrightarrow} \, e₀' \, \lottsym{▶} \, e₁ \ifdefined\store \lottsym{;} \, \mathcolor{Store}{oₛ'} \fi}
\lottinferencerulesep
\lottinferencerule{boxr}{
\lotthypothesis{o₁ \, \lottsym{⊢} \, e \ifdefined\store \lottsym{;} \, \mathcolor{Store}{oₛ} \fi \, \lottsym{\longrightarrow} \, e' \ifdefined\store \lottsym{;} \, \mathcolor{Store}{oₛ'} \fi}
}{o₀ \, \lottsym{⊢} \, \lottsym{\{} \, o₁ \, \lottsym{\}} \, \lottsym{▶} \, e \ifdefined\store \lottsym{;} \, \mathcolor{Store}{oₛ} \fi \, \lottsym{\longrightarrow} \, \lottsym{\{} \, o₁ \, \lottsym{\}} \, \lottsym{▶} \, e' \ifdefined\store \lottsym{;} \, \mathcolor{Store}{oₛ'} \fi}
\lottinferencerulesep
\lottinferencerule{box}{
}{o₀ \, \lottsym{⊢} \, \lottsym{\{} \, o₁ \, \lottsym{\}} \, \lottsym{▶} \, v \ifdefined\store \lottsym{;} \, \mathcolor{Store}{oₛ} \fi \, \lottsym{\longrightarrow} \, v \ifdefined\store \lottsym{;} \, \mathcolor{Store}{oₛ} \fi}
\lottinferencerulesep
\lottinferencerule{env}{
}{o \, \lottsym{⊢} \, \lottsym{@} \ifdefined\store \lottsym{;} \, \mathcolor{Store}{oₛ} \fi \, \lottsym{\longrightarrow} \, \lottsym{\{} \, o \, \lottsym{\}} \ifdefined\store \lottsym{;} \, \mathcolor{Store}{oₛ} \fi}
\lottinferencerulesep
\lottinferencerule{extl}{
\lotthypothesis{o \, \lottsym{⊢} \, \lottsym{\{} \, r \, \lottsym{\}} \ifdefined\store \lottsym{;} \, \mathcolor{Store}{oₛ} \fi \, \lottsym{\longrightarrow} \, \lottsym{\{} \, r' \, \lottsym{\}} \ifdefined\store \lottsym{;} \, \mathcolor{Store}{oₛ'} \fi}
}{o \, \lottsym{⊢} \, \lottsym{\{} \, r \lottsym{,} \, i \, \rstrength \, e \, \lottsym{\}} \ifdefined\store \lottsym{;} \, \mathcolor{Store}{oₛ} \fi \, \lottsym{\longrightarrow} \, \lottsym{\{} \, r' \lottsym{,} \, i \, \rstrength \, e \, \lottsym{\}} \ifdefined\store \lottsym{;} \, \mathcolor{Store}{oₛ'} \fi}
\lottinferencerulesep
\lottinferencerule{extl?}{
\lotthypothesis{o \, \lottsym{⊢} \, \lottsym{\{} \, r \, \lottsym{\}} \ifdefined\store \lottsym{;} \, \mathcolor{Store}{oₛ} \fi \, \lottsym{\longrightarrow} \, \lottsym{\{} \, r' \, \lottsym{\}} \ifdefined\store \lottsym{;} \, \mathcolor{Store}{oₛ'} \fi}
}{o \, \lottsym{⊢} \, \lottsym{\{} \, r \lottsym{,} \, i \, \lottsym{?} \, \lottsym{\}} \ifdefined\store \lottsym{;} \, \mathcolor{Store}{oₛ} \fi \, \lottsym{\longrightarrow} \, \lottsym{\{} \, r' \lottsym{,} \, i \, \lottsym{?} \, \lottsym{\}} \ifdefined\store \lottsym{;} \, \mathcolor{Store}{oₛ'} \fi}
\lottinferencerulesep
\lottinferencerule{extr}{
\lotthypothesis{o₀ \, \lottsym{⊢} \, e \ifdefined\store \lottsym{;} \, \mathcolor{Store}{oₛ} \fi \, \lottsym{\longrightarrow} \, e' \ifdefined\store \lottsym{;} \, \mathcolor{Store}{oₛ'} \fi}
}{o₀ \, \lottsym{⊢} \, \lottsym{\{} \, o₁ \lottsym{,} \, i \, \rstrength \, e \, \lottsym{\}} \ifdefined\store \lottsym{;} \, \mathcolor{Store}{oₛ} \fi \, \lottsym{\longrightarrow} \, \lottsym{\{} \, o₁ \lottsym{,} \, i \, \rstrength \, e' \, \lottsym{\}} \ifdefined\store \lottsym{;} \, \mathcolor{Store}{oₛ'} \fi}
\lottinferencerulesep
\lottinferencerule{concatl}{
\lotthypothesis{o \, \lottsym{⊢} \, e₀ \ifdefined\store \lottsym{;} \, \mathcolor{Store}{oₛ} \fi \, \lottsym{\longrightarrow} \, e₀' \ifdefined\store \lottsym{;} \, \mathcolor{Store}{oₛ'} \fi}
}{o \, \lottsym{⊢} \, e₀ \, \lottsym{⧺} \, e₁ \ifdefined\store \lottsym{;} \, \mathcolor{Store}{oₛ} \fi \, \lottsym{\longrightarrow} \, e₀' \, \lottsym{⧺} \, e₁ \ifdefined\store \lottsym{;} \, \mathcolor{Store}{oₛ'} \fi}
\lottinferencerulesep
\lottinferencerule{concatr}{
\lotthypothesis{o \, \lottsym{⊢} \, e \ifdefined\store \lottsym{;} \, \mathcolor{Store}{oₛ} \fi \, \lottsym{\longrightarrow} \, e' \ifdefined\store \lottsym{;} \, \mathcolor{Store}{oₛ'} \fi}
}{o \, \lottsym{⊢} \, v \, \lottsym{⧺} \, e \ifdefined\store \lottsym{;} \, \mathcolor{Store}{oₛ} \fi \, \lottsym{\longrightarrow} \, v \, \lottsym{⧺} \, e' \ifdefined\store \lottsym{;} \, \mathcolor{Store}{oₛ'} \fi}
\lottinferencerulesep
\lottinferencerule{concat}{
}{o₀ \, \lottsym{⊢} \, \lottsym{\{} \, o₁ \, \lottsym{\}} \, \lottsym{⧺} \, \lottsym{\{} \, o₂ \, \lottsym{\}} \ifdefined\store \lottsym{;} \, \mathcolor{Store}{oₛ} \fi \, \lottsym{\longrightarrow} \, \lottsym{\{} \, o₁ \, \lottsym{⧺} \, o₂ \, \lottsym{\}} \ifdefined\store \lottsym{;} \, \mathcolor{Store}{oₛ} \fi}
\lottinferencerulesep
\lottinferencerule{proj}{
\lotthypothesis{o \, \lottsym{⊢} \, e \ifdefined\store \lottsym{;} \, \mathcolor{Store}{oₛ} \fi \, \lottsym{\longrightarrow} \, e' \ifdefined\store \lottsym{;} \, \mathcolor{Store}{oₛ'} \fi}
}{o \, \lottsym{⊢} \, e \lottsym{.} i \ifdefined\store \lottsym{;} \, \mathcolor{Store}{oₛ} \fi \, \lottsym{\longrightarrow} \, e' \lottsym{.} i \ifdefined\store \lottsym{;} \, \mathcolor{Store}{oₛ'} \fi}
\lottinferencerulesep
\lottinferencerule{projRecord}{
\lotthypothesis{\lottkw{proj} \lottsym{(} o₁ \lottsym{;} \, i \lottsym{)} \, \lottsym{=} \, v}
}{o₀ \, \lottsym{⊢} \, \lottsym{\{} \, o₁ \, \lottsym{\}} \lottsym{.} i \ifdefined\store \lottsym{;} \, \mathcolor{Store}{oₛ} \fi \, \lottsym{\longrightarrow} \, v \ifdefined\store \lottsym{;} \, \mathcolor{Store}{oₛ} \fi}
\lottinferencerulesep
\lottinferencerule{mproj}{
\lotthypothesis{o \, \lottsym{⊢} \, e \ifdefined\store \lottsym{;} \, \mathcolor{Store}{oₛ} \fi \, \lottsym{\longrightarrow} \, e' \ifdefined\store \lottsym{;} \, \mathcolor{Store}{oₛ'} \fi}
}{o \, \lottsym{⊢} \, e \lottsym{.} \lottsym{\{} \, \lottsepbycomprehension{{i}_{j}} \, \lottsym{\}} \ifdefined\store \lottsym{;} \, \mathcolor{Store}{oₛ} \fi \, \lottsym{\longrightarrow} \, e' \lottsym{.} \lottsym{\{} \, \lottsepbycomprehension{{i}_{j}} \, \lottsym{\}} \ifdefined\store \lottsym{;} \, \mathcolor{Store}{oₛ'} \fi}
\lottinferencerulesep
\lottinferencerule{mprojRecord}{
\lotthypothesis{\lottkw{mproj} \lottsym{(} o₁ \lottsym{;} \, \lottsepbycomprehension{{i}_{j}} \lottsym{)} \, \lottsym{=} \, o₂}
}{o₀ \, \lottsym{⊢} \, \lottsym{\{} \, o₁ \, \lottsym{\}} \lottsym{.} \lottsym{\{} \, \lottsepbycomprehension{{i}_{j}} \, \lottsym{\}} \ifdefined\store \lottsym{;} \, \mathcolor{Store}{oₛ} \fi \, \lottsym{\longrightarrow} \, \lottsym{\{} \, o₂ \, \lottsym{\}} \ifdefined\store \lottsym{;} \, \mathcolor{Store}{oₛ} \fi}
\lottinferencerulesep
\lottinferencerule{diff}{
\lotthypothesis{o \, \lottsym{⊢} \, e \ifdefined\store \lottsym{;} \, \mathcolor{Store}{oₛ} \fi \, \lottsym{\longrightarrow} \, e' \ifdefined\store \lottsym{;} \, \mathcolor{Store}{oₛ'} \fi}
}{o \, \lottsym{⊢} \, e \lottsym{\textbackslash} \lottsym{\{} \, \lottsepbycomprehension{{i}_{j}} \, \lottsym{\}} \ifdefined\store \lottsym{;} \, \mathcolor{Store}{oₛ} \fi \, \lottsym{\longrightarrow} \, e' \lottsym{\textbackslash} \lottsym{\{} \, \lottsepbycomprehension{{i}_{j}} \, \lottsym{\}} \ifdefined\store \lottsym{;} \, \mathcolor{Store}{oₛ'} \fi}
\lottinferencerulesep
\lottinferencerule{diffRecord}{
\lotthypothesis{\lottkw{diff} \lottsym{(} o₁ \lottsym{;} \, \lottsepbycomprehension{{i}_{j}} \lottsym{)} \, \lottsym{=} \, o₂}
}{o₀ \, \lottsym{⊢} \, \lottsym{\{} \, o₁ \, \lottsym{\}} \lottsym{\textbackslash} \lottsym{\{} \, \lottsepbycomprehension{{i}_{j}} \, \lottsym{\}} \ifdefined\store \lottsym{;} \, \mathcolor{Store}{oₛ} \fi \, \lottsym{\longrightarrow} \, \lottsym{\{} \, o₂ \, \lottsym{\}} \ifdefined\store \lottsym{;} \, \mathcolor{Store}{oₛ} \fi}
\lottinferencerulesep
\lottinferencerule{let}{
\lotthypothesis{o \, \lottsym{⊢} \, e₀ \ifdefined\store \lottsym{;} \, \mathcolor{Store}{oₛ} \fi \, \lottsym{\longrightarrow} \, e₀' \ifdefined\store \lottsym{;} \, \mathcolor{Store}{oₛ'} \fi}
}{o \, \lottsym{⊢} \, \lottkw{let} \, t \, \lottsym{=} \, e₀ \, \lottkw{in} \, e₁ \ifdefined\store \lottsym{;} \, \mathcolor{Store}{oₛ} \fi \, \lottsym{\longrightarrow} \, \lottkw{let} \, t \, \lottsym{=} \, e₀' \, \lottkw{in} \, e₁ \ifdefined\store \lottsym{;} \, \mathcolor{Store}{oₛ'} \fi}
\lottinferencerulesep
\lottinferencerule{letv}{
\lotthypothesis{\lottkw{expand} \lottsym{(} t \lottsym{;} \, v \lottsym{)} \, \lottsym{=} \, o₁}
}{o₀ \, \lottsym{⊢} \, \lottkw{let} \, t \, \lottsym{=} \, v \, \lottkw{in} \, e \ifdefined\store \lottsym{;} \, \mathcolor{Store}{oₛ} \fi \, \lottsym{\longrightarrow} \, \lottsym{\{} \, o₀ \lottsym{,} \, o₁ \, \lottsym{\}} \, \lottsym{▶} \, e \ifdefined\store \lottsym{;} \, \mathcolor{Store}{oₛ} \fi}
\lottinferencerulesep
\lottinferencerule{ref}{
\lotthypothesis{o \, \lottsym{⊢} \, e \ifdefined\store \lottsym{;} \, \mathcolor{Store}{oₛ} \fi \, \lottsym{\longrightarrow} \, e' \ifdefined\store \lottsym{;} \, \mathcolor{Store}{oₛ'} \fi}
}{o \, \lottsym{⊢} \, \lottkw{ref} \, e \ifdefined\store \lottsym{;} \, \mathcolor{Store}{oₛ} \fi \, \lottsym{\longrightarrow} \, \lottkw{ref} \, e' \ifdefined\store \lottsym{;} \, \mathcolor{Store}{oₛ'} \fi}
\lottinferencerulesep
\lottinferencerule{refv}{
\lotthypothesis{l \, \lottsym{∉} \, oₛ}
}{o \, \lottsym{⊢} \, \lottkw{ref} \, v \ifdefined\store \lottsym{;} \, \mathcolor{Store}{oₛ} \fi \, \lottsym{\longrightarrow} \, l \ifdefined\store \lottsym{;} \, \mathcolor{Store}{oₛ \lottsym{,} \, l \, \lottsym{:} \, v} \fi}
\lottinferencerulesep
\lottinferencerule{deref}{
\lotthypothesis{o \, \lottsym{⊢} \, e \ifdefined\store \lottsym{;} \, \mathcolor{Store}{oₛ} \fi \, \lottsym{\longrightarrow} \, e' \ifdefined\store \lottsym{;} \, \mathcolor{Store}{oₛ'} \fi}
}{o \, \lottsym{⊢} \, \lottsym{!} e \ifdefined\store \lottsym{;} \, \mathcolor{Store}{oₛ} \fi \, \lottsym{\longrightarrow} \, \lottsym{!} e' \ifdefined\store \lottsym{;} \, \mathcolor{Store}{oₛ'} \fi}
\lottinferencerulesep
\lottinferencerule{derefLoc}{
}{o \, \lottsym{⊢} \, \lottsym{!} l \ifdefined\store \lottsym{;} \, \mathcolor{Store}{oₛ} \fi \, \lottsym{\longrightarrow} \, \lottsym{\{} \, oₛ \, \lottsym{\}} \lottsym{.} l \ifdefined\store \lottsym{;} \, \mathcolor{Store}{oₛ} \fi}
\lottinferencerulesep
\lottinferencerule{setl}{
\lotthypothesis{o \, \lottsym{⊢} \, e₀ \ifdefined\store \lottsym{;} \, \mathcolor{Store}{oₛ} \fi \, \lottsym{\longrightarrow} \, e₀' \ifdefined\store \lottsym{;} \, \mathcolor{Store}{oₛ'} \fi}
}{o \, \lottsym{⊢} \, e₀ \, \lottsym{≔} \, e₁ \ifdefined\store \lottsym{;} \, \mathcolor{Store}{oₛ} \fi \, \lottsym{\longrightarrow} \, e₀' \, \lottsym{≔} \, e₁ \ifdefined\store \lottsym{;} \, \mathcolor{Store}{oₛ'} \fi}
\lottinferencerulesep
\lottinferencerule{setr}{
\lotthypothesis{o \, \lottsym{⊢} \, e \ifdefined\store \lottsym{;} \, \mathcolor{Store}{oₛ} \fi \, \lottsym{\longrightarrow} \, e' \ifdefined\store \lottsym{;} \, \mathcolor{Store}{oₛ'} \fi}
}{o \, \lottsym{⊢} \, v \, \lottsym{≔} \, e \ifdefined\store \lottsym{;} \, \mathcolor{Store}{oₛ} \fi \, \lottsym{\longrightarrow} \, v \, \lottsym{≔} \, e' \ifdefined\store \lottsym{;} \, \mathcolor{Store}{oₛ'} \fi}
\lottinferencerulesep
\lottinferencerule{set}{
}{o \, \lottsym{⊢} \, l \, \lottsym{≔} \, v \ifdefined\store \lottsym{;} \, \mathcolor{Store}{oₛ} \fi \, \lottsym{\longrightarrow} \, \lottsym{\{} \, \lottsym{\}} \ifdefined\store \lottsym{;} \, \mathcolor{Store}{oₛ \, \lottsym{⧺} \, l \, \lottsym{:} \, v} \fi}
}
}}}

  \caption{Extension with references (selected rules)}
  \label{fig:ref}
\end{figure}

\paragraph{\textbf{Syntax and semantics}}

\Cref{fig:ref} presents the extended syntax, typing, and operational
semantics. Congruence reduction rules and syntax-directed typing are omitted for
space but can be found in the appendix.
Types now include the reference type $\lottkw{Ref} \, A$. Terms include
locations $l$ (treated as a form of label),
reference allocation $\lottkw{ref} \, e$,
dereferencing $\lottsym{!} e$, and assignment
$e₀ \, \lottsym{≔} \, e₁$. Accordingly, values are also extended to include locations.
Moreover, we use the record value $o$ as a store that maps locations to values,
and its corresponding type $ρₛ$ as the store typing. A store type is
well-formed, written $\unique{ρₛ}$, when all its entries have
strength ($\lottsym{:}$), and all locations are unique.

The extended subtyping and typing are standard. Subtyping of references is
invariant. The typing judgement $\mathcolor{Store}{ρₛ} \lottsym{;} \, ρ \,
\lottsym{⊢} \, e \, \lottsym{:} \, A$ now incorporates a store type
$ρₛ$ using a record type. Rule \lottref{T}{loc} types locations as
references to the type specified in the store.

The evaluation judgement is also extended to thread the store through each
step. Rule \lottref{E}{refv} allocates a fresh name in the
store. (In the mechanization,
the operational semantics maintain determinacy by picking
the fresh name to be the maximal variable index in the store plus one.)
Rule \lottref{E}{derefLoc} projects the corresponding entry out of the
store. Finally, rule \lottref{E}{set} leverages the overwriting behaviour of
record concatenation to update the corresponding store entry.

\paragraph{\textbf{Generalization and the value restriction}}
\label{sec:generalization}

It may seem surprising that we did not modify any of the original typing rules.
In systems with polymorphism and generalization, it is well-known that adding
mutable state can easily yield type \textit{unsoundness}. As an example,
consider the program on
\begin{wrapfigure}{r}{0.43\textwidth}
  \vspace{-10pt}
\begin{lstlisting}[style=mylang,basicstyle=\footnotesize\ttfamily]
let x = ref (\{ i }. i) in
let { } = (x := (\{ i }. i + 1)) in
force !x { i = true }
\end{lstlisting}
 \vspace{-10pt}
\end{wrapfigure}%
the right. Naive let-generalization will assign $x$ a polymorphic type
$\lottsym{∀} \, a. \, \lottkw{Ref} \, (\{ \, i: a \,\} \to a)$. The type variable $a$ is then
instantiated to $ℕ$ on line 2, and to $𝔹$ on line 3,
allowing the program to type-check. However, the program fails at runtime, as
it applies a function expecting a natural number to a boolean.

To preserve safety, a standard remedy is the \textit{value
  restriction}~\cite{wright1995simple}, which limits let-generalization to
syntactic values. Applying this restriction would prevent $x$ from being
generalized, and thus the program fails to type-check.
While this approach guarantees soundness, it
doesn't integrate very nicely with our calculus: let-generalization introduces
universal quantifiers at the topmost level of a binding, while we often want to
localize polymorphism within individual record fields, rather than enforcing
top-level quantifiers over an entire record.

We therefore adapt the approach recently proposed by \citet{dolan2024rethinking},
which performs generalization at functions (specifically rules \lottref{T}{lam} and
\lottref{T}{closure} in our case), rather than let-bindings.
This approach can be considered a more fine-grained variant of the value
restriction. It restores type soundness; for example, in the program above
$x$ can be typed either with $\lottkw{Ref} \, (\lottsym{∀} \, a. \, \{ \, i : a \, \} \to a)$
or with a monotype, %
causing a type error in either case.
Moreover, the approach also increases typeability, as it allows the individual components of a non-value
compound expression to generalize independently. In our calculus, this design
naturally facilitates the construction of records with polymorphic fields,
e.g.
\lstinline[style=mylang]|let x = { id = \{ i }. i, incr = \{ j }. j + 1 }~in e| assigns $x$ type $\{ \, id \strong \forall \, a. \, \{ \, i : a \, \} \to a, incr \strong \{ \, j : ℕ \, \} \to ℕ \, \}$. %
We also extend this approach to \lottref{T}{app}: when the argument is
polymorphic over the same type variables as the function, the application
keeps those quantifiers rather than instantiating them.
This is useful for partial applications that do not require particular
instantiations for some quantifiers, and it is sound here because reduction (and thus
reference allocation) does not occur until a $\lottkw{force}$ is encountered.

\paragraph{\textbf{Type soundness}}

We have proven the determinacy, progress, and preservation theorems for the
extended semantics, where the theorems are now adapted to account for the
presence of stores; we omit their statements for space.
The syntax-directed system likewise extends to references and remains
sound and complete.

\section{Mechanization and Implementation}
\label{sec:mech}

\paragraph{\textbf{Mechanization}}

We have mechanized \lang using the Lean 4 proof assistant \cite{moura2021lean},
covering all semantics and theorems. \Cref{tab:code-stats} summarizes
the statistics.
By using explicit environments, we avoid the term-level binding complications
described in \Cref{sec:overview-bindings}, allowing labels and locations to
be represented simply as natural numbers. However, without
explicit type environments, we adopt the locally nameless representation
\cite{charguraud2011ln} for type variables.

\begin{wraptable}{r}{0.36\textwidth}
  \footnotesize
  \caption{Code stats of mechanization and implementation}
  \label{tab:code-stats}
\begin{tabular}{l r r}
  \textbf{Component} & \textbf{LOC} & \textbf{Decls} \\
  \hline
  \lang Type Soundness & 16260 & 773 \\
  STLC Correspondence & 1366 & 65 \\
  Big-Step Equivalence & 396 & 10 \\
  Normalization & 5932 & 93 \\
  Syntax-Directed Typing & 4695 & 11 \\
  Implementation & 934 & 48 \\
  \hline
  \textbf{Total} & 29619 & 1000
\end{tabular}
 \end{wraptable}

Most of the proofs proceeded without significant difficulty. One particular
challenge in a few cases, such as \Cref{lem:sub-trans},
was to show that the induction used to prove them
was well-founded. This arose as Lean has proof irrelevance: since Lean
treats all propositions of the same type as equal, the depth of a derivation
cannot be used directly in termination measures. We circumvented this limitation
by defining versions of subtyping and multi-step evaluation that
are explicitly indexed by their derivation depth, and using these numerical indices
as well-founded measures.
\paragraph{\textbf{Implementation}}
\label{sec:impl}

Since one of our motivations for using first-class environments is to close the
gap between specification and implementation identified by
\citet{tan2024case},
we have implemented an interpreter and verified it against our
operational semantics. Because the extended \lang supports
general references, and it's known that references can encode recursion (via
Landin's knot~\cite{landin1964mechanical}), well-typed programs are not
guaranteed to terminate. We thus leverage Lean's \texttt{partial\_fixpoint}\footnote{\url{https://lean-lang.org/doc/reference/4.33.0/Definitions/Recursive-Definitions/\#partial-fixpoint}}
feature and the option monad to define our evaluation function \texttt{eval}.
This frees us from providing a termination proof for \texttt{eval}, while still
allowing us to formally reason about it. For proof purposes, Lean assigns
\texttt{none} to non-terminating inputs.

We then prove the equivalence of the operational semantics
and the evaluation function \texttt{eval}.
The first clause of the following theorem expresses that for terminating inputs,
the operational semantics and \texttt{eval} agree. The second clause states that
for well-typed inputs, \texttt{eval} yields \texttt{none} precisely when there
are infinite reduction steps (denoted by $\overset{ω}{\longrightarrow}$).

\begin{theorem}[Interpreter correctness]\leavevmode
  \label{thm:interpreter-correctness}
  \begin{itemize}
    \item $o \, \lottsym{⊢} \, e \ifdefined\store \lottsym{;} \, \mathcolor{Store}{oₛ} \fi \, \lottsym{\overset{*}{\longrightarrow}} \, v \ifdefined\store \lottsym{;} \, \mathcolor{Store}{oₛ'} \fi$ if and only if
      \texttt{eval o e }\textcolor{Store}{\texttt{o}\textsubscript{\texttt{s}}}\texttt{ = some (v, }\textcolor{Store}{\texttt{o}\textsubscript{\texttt{s}}\texttt{'}}\texttt{)}.
    \item Given $\ifdefined\store \mathcolor{Store}{ρₛ} \lottsym{;} \, \fi \lottsym{ε} \, \lottsym{⊢} \, \lottsym{\{} \, o \, \lottsym{\}} \, \lottsym{:} \, \lottsym{\{} \, ρ \, \lottsym{\}}$ with $\uniform{\rho}$,
      and $\ifdefined\store \mathcolor{Store}{ρₛ} \lottsym{;} \, \fi \lottsym{ε} \, \lottsym{⊢} \, \lottsym{\{} \, oₛ \, \lottsym{\}} \, \lottsym{:} \, \lottsym{\{} \, ρₛ \, \lottsym{\}}$ with $\unique{\rho_s}$,
      if $\ifdefined\store \mathcolor{Store}{ρₛ} \lottsym{;} \, \fi ρ \, \lottsym{⊢} \, e \, \lottsym{:} \, A$, then $o \, \lottsym{⊢} \, e \ifdefined\store \lottsym{;} \, \mathcolor{Store}{oₛ} \fi \, \lottsym{\overset{ω}{\longrightarrow}}$ if and only
      if \texttt{eval o e }\textcolor{Store}{\texttt{o}\textsubscript{\texttt{s}}}\texttt{ = none}.
  \end{itemize}
\end{theorem}

\noindent
We prove this theorem by using the transitivity of equivalence, introducing an
intermediate big-step operational semantics for \lang (whose full rules are
provided in the appendix), which is proven
equivalent to both the interpreter's \texttt{eval} function and the small-step operational
semantics.
\section{Discussion}
\label{sec:discussion}

\subsection{The Design Space}
\label{sec:design-space}

Function application paradigms encompass a broad design space. In this section,
we examine how these various features are supported across programming
languages.
The selected languages are (1) Python (dynamically typed), (2) TypeScript
(gradually typed), (3) Swift, (4) OCaml, (5) Scala~(all statically typed), and
(6) Lean 4 (dependently typed). While many more languages exist, these languages
offer a diverse and representative spectrum of modern language design decisions.
We summarize the comparison in \Cref{tab:design-space}:\footnote{While
  language-specific workarounds can often simulate missing capabilities, our
  evaluation mostly focuses on native support, acknowledging that a simple
  binary classification can occasionally obscure these underlying nuances.}

\newcommand{\tblnote}[1]{\textsuperscript{\textdagger#1}}

\begin{table}[H]
\vspace{-8pt}
\small
\begin{tabular}{c | c c c c c c | c}

  & Python
  & Typescript
  & Swift
  & OCaml
  & Scala
  & Lean 4
  & $\lang$
  \\
  \hline

  named arguments
  & \lstinline[basicstyle=\fontsize{7}{7}\ttfamily]{f(a=x)} %
  & \lstinline[basicstyle=\fontsize{7}{7}\ttfamily]|f({a: x})| %
  & \lstinline[basicstyle=\fontsize{7}{7}\ttfamily]{f(a: x)} %
  & \lstinline[basicstyle=\fontsize{7}{7}\ttfamily]{f ~a:x} %
  & \lstinline[basicstyle=\fontsize{7}{7}\ttfamily]{f(a = x)} %
  & \lstinline[basicstyle=\fontsize{7}{7}\ttfamily]{f (a := x)} %
  & \lstinline[basicstyle=\fontsize{7}{7}\ttfamily]|f {a=x}| %
  \\
  \hline

  first-class arguments
  & \cmark %
  & \cmark\rlap{\tblnote{1}} %
  & \xmark %
  & \xmark %
  & \xmark %
  & \xmark %
  & \cmark %
  \\

  optional arguments
  & \cmark %
  & \cmark %
  & \cmark %
  & \cmark %
  & \cmark %
  & \cmark %
  & \cmark %
  \\

  distinct args and params
  & \xmark %
  & \cmark %
  & \cmark %
  & \cmark %
  & \xmark %
  & \cmark\rlap{\tblnote{2}} %
  & \cmark %
  \\

  commutative arguments
  & \cmark\rlap{\tblnote{3}} %
  & \cmark %
  & \xmark %
  & \cmark\rlap{\tblnote{4}} %
  & \cmark %
  & \cmark\rlap{\tblnote{5}} %
  & \cmark %
  \\

  partial application
  & \xmark %
  & \xmark %
  & \xmark %
  & \cmark %
  & \cmark %
  & \cmark %
  & \cmark %
  \\

\end{tabular}
\vspace{12pt}
\caption{A comparison table of languages with named arguments (with \tblnote{} explained in \Cref{sec:design-space})}
\vspace{-12pt}
\label{tab:design-space}
\end{table}

All languages listed in the table offer some degree of support for \textbf{named
  arguments}. Although TypeScript\tblnote{1} lacks native support, simulating this behaviour
by passing a single object is a widespread idiom in its community; we thus
record ticks for both named and first-class arguments.
Statically typed languages often support named arguments through elaboration,
while dynamically typed ones typically resolve them by matching labels against
function parameters at runtime.

We use \textbf{first-class arguments} to mean that a function's arguments can be stored
within a data structure and subsequently supplied to a function call. For
example, in Python, the example in \Cref{sec:overview-named} can be rewritten
using the dictionary unpacking operator \verb|**|:
\begin{lstlisting}[style=python,basicstyle=\footnotesize\ttfamily]
transaction = { amount: 1000, "from_account": "Alice", "to_account": "Bob" }
transfer_funds(**transaction)
\end{lstlisting}
Optional type checkers for Python can type such keyword arguments with a record
type~\cite{pep692}.
$\lang$ inherently supports first-class arguments by modelling arguments as
records. On the other hand, this feature is rarely supported among statically typed
languages.

All listed languages support certain forms of \textbf{optional arguments} that
can be omitted at call sites. This unanimous adoption stems partly from our
focus on languages with named arguments, which provide the parameter
disambiguation necessary to make optional arguments viable.
A few languages do not require default values for optional arguments. In
TypeScript, for instance, an omitted argument defaults to
\lstinline[style=javascript]{undefined}, and its presence must be
checked by~e.g.~\lstinline[style=javascript]{c === undefined}.

In both Swift and OCaml,
\textbf{distinct argument labels and parameter
  names} are specified within the
function definition. TypeScript mimics this behaviour via property-renaming
syntax during object destructuring. In Lean 4\tblnote{2}, this feature emerges
as a consequence of type equivalence, where the desired argument label can be set by
annotating the function with a new type.

\textbf{Commutative arguments} are typical among named arguments, though
exceptions like Swift exist. Positional arguments cannot commute with
themselves, as their order determines their identity. It remains to ask whether
positional arguments can commute with named ones. In Python\tblnote{3},
positional arguments must always precede named arguments. Scala allows a
positional argument to follow a named one, if the entire sequence up to that
point aligns perfectly with the definition-site order. OCaml\tblnote{4} is more
flexible and allows positional arguments to commute with named arguments, as
long as a positional argument appears last to mark the end of application.
Lastly, as a dependently typed system, Lean 4\tblnote{5} permits argument
commutation provided that the permutation does not violate any underlying type
dependencies.
However,
while these languages allow named arguments to be passed commutatively at the
call site, they do not support structural commutativity.
For example, if a higher-order function \lstinline{g} expects an argument of type
\lstinline[style=mylang]|{ name:String, age:Int }~->~Int|,
passing a function \lstinline[style=mylang]|f : { age: Int, name: String }~->~Int|
results in a type error. Conversely, it type-checks in
$\lang$, which natively supports commutativity.

With \textbf{partial application}, applying a function to a subset of its
arguments results in a new function taking the remaining parameters. Unlike
traditional currying, which is constrained by strict sequential evaluation,
general partial application allows a function to be partially applied to an
arbitrary subset of arguments. Scala achieves this via wildcard placeholder
syntax:
\lstinline[style=scala,basicstyle=\footnotesize\ttfamily]{transfer_from_alice =~ transfer_funds(_ : Int, "Alice", _ : String)}.
Nevertheless, Scala requires programmers to explicitly mark each unapplied
parameter with \lstinline{_}. This mechanism also replaces the original parameter
names with compiler-generated names
(\lstinline[style=nix]{v1}, \lstinline[style=nix]{v2}, \ldots), obscuring the purpose of the remaining arguments.
As discussed in \Cref{sec:overview-optional},
partial application interacts with optional arguments.
While OCaml requires a subsequent positional argument after optional ones, Scala and Lean 4 automatically fill in missing optional arguments with their default values during compilation.

\subsection{Discussion: Broader Applications}
\label{sec:applications}

In this section, we broaden our scope to examine how various components of our
design can be helpful in alternative contexts,
highlighting the underlying principles of our design; the appendix discusses
further applications.

\paragraph{\textbf{Safe interface upgrades with optional subtyping}}

Record-like data structures are widely used to describe software interfaces,
which naturally evolve over time. As a simple example, consider the user profile
interface (a) in \Cref{fig:interface}, which specifies a profile sent from a client
to a server:

\begin{figure}[H]
\vspace{-11pt}
\noindent
\begin{subfigure}[t]{0.33\textwidth}
\begin{lstlisting}[style=config,basicstyle=\footnotesize\ttfamily]
# (a) a profile interface
type profile = {
  bio : text
  theme_color : num }
\end{lstlisting}
\end{subfigure}
\begin{subfigure}[t]{0.31\textwidth}
\begin{lstlisting}[style=config,basicstyle=\footnotesize\ttfamily]
# (b) upgraded interface
type profile = {
  bio : text
  theme_color : num
+ dark_mode : bool }
\end{lstlisting}
\end{subfigure}
\begin{subfigure}[t]{0.34\textwidth}
  \begin{lstlisting}[style=config,basicstyle=\footnotesize\ttfamily]
# (c) safe upgraded interface
type profile = {
  bio : text
  theme_color : num
+ dark_mode : opt bool }
  \end{lstlisting}
\end{subfigure}
\caption{Interface upgrades (specified with \textcolor{ForestGreen}{\textbf{+}})}
\label{fig:interface}
\end{figure}

A new software release may introduce additional features and consequently update the
interface. For example, interface (b) upgrades the \lstinline{profile}
interface with a \lstinline{dark_mode} field to specify whether the server
should render a dark theme. However, legacy clients that have not upgraded will
now send profiles lacking this field, creating a type mismatch and causing
runtime errors. In this context, an upgrade is considered \textit{safe} if a
client sending data according to the old format does not cause a failure when
the server expects the new format (and vice versa).

To resolve this challenge, the interface description language
Candid~\cite{candid} requires that an interface for inbound data (i.e. data sent
from the client to the server) may only be replaced by a supertype (e.g. fewer
fields); therefore, the upgrade (b)
is rejected. On the other hand,
Candid permits a record subtyping rule that is
``unusual but necessary in practice''~\cite{candid2024}, where new fields are
allowed in the supertype as long as they are optional. Under this rule,
interface (a) is considered a subtype of (c), where the newly added \lstinline{dark_mode} field has
type \lstinline[style=config]{opt bool}. Thus
interface (c) is accepted as a safe upgraded interface.
We note that this use of optional types is similar to ours, and the subtyping
rule corresponds to the extra right entries in our \lottref{S}{empty} rule.

Interestingly, Candid combines the standard record width subtyping
with the above subtyping rule for optionals, which leads to surprising
consequences. Specifically, via record width subtyping, interface (a) is a subtype
of \lstinline[style=config]|{ bio : text }|, which removed
the \lstinline[style=config]|theme_color| field.
Via subtyping for optionals, %
\lstinline[style=config]|{ bio : text }| is a subtype of
\lstinline[style=config]|{ bio : text, theme_color : opt string }|.
It follows,
via subtyping transitivity, that
(a) is a subtype of %
\lstinline[style=config]|{ bio : text, theme_color : opt string }|; effectively,
this changes the type of \lstinline{theme_color} from \lstinline{nat} to
\lstinline[style=config]|opt string|.
However, if one were to compare \lstinline{nat} and
\lstinline[style=config]|opt string| directly, subtyping would fail.
Since subtyping transitivity is crucial for supporting sequences of upgrades,
Candid's subtyping tolerates
such type mismatches at runtime, yielding \lstinline{null} values for the
affected fields.
On the other hand,
our system omits the record width subtyping rule entirely, and thus
this specific issue does not arise.
Specifically,
width subtyping effectively
allows a function to accept extraneous arguments (e.g.
\lstinline[style=mylang]|sub { name = "Alice" }|) that serve no purpose. Such applications are disallowed;
instead, a programmer can use record projection to extract the desired
fields from an argument before passing it to a function.

\paragraph{\textbf{Packaging with overwritable fields}}

The Nix language is a functional domain-specific language that powers the Nix
packaging system~\cite{nix}. Nix evaluates
expressions to create \textit{derivations}, which describe how to build
reproducible packages.
For example, consider a package recipe defined as a function of
its dependencies, including an \lstinline{openssl} argument:
\begin{lstlisting}[style=nix,basicstyle=\footnotesize\ttfamily]
let app = { stdenv, openssl }:
      stdenv.mkDerivation { name = "app"; buildInputs = [ openssl ]; };
\end{lstlisting}
We can instantiate this recipe using the \lstinline{callPackage} function,
providing an attribute set that explicitly binds \lstinline{stdenv}, as well as
the \lstinline{openssl} argument to the \lstinline[style=nix]{openssl_1_1}
package:
\begin{lstlisting}[style=nix,basicstyle=\footnotesize\ttfamily]
myApp = pkgs.callPackage app { stdenv = pkgs.stdenv; openssl = pkgs.openssl_1_1; };
\end{lstlisting}
More interestingly, \lstinline{callPackage} allows arguments to be customized
\textit{after the fact} using a derivation's \lstinline{override} function. For
example, even after \lstinline{myApp} has been instantiated, we can override its
\lstinline{openssl} argument with a new value to update the package to OpenSSL
3.0:
\begin{lstlisting}[style=nix,basicstyle=\footnotesize\ttfamily]
mySecuredApp = myApp.override { openssl = pkgs.openssl_3; };
\end{lstlisting}
\noindent
Here, \lstinline{myApp.override} overrides the \lstinline{openssl} argument
originally supplied by \lstinline{callPackage}, producing a new derivation with
the exact same arguments as \lstinline{myApp} except for the updated
\lstinline{openssl} dependency. Note that since Nix is functional, this does not
mutate the original package. This style of overriding is made possible by Nix's
evaluation strategy, which evaluates expressions to construct a derivation graph
but defers the actual compilation and building of binaries until a build is
explicitly requested (e.g. via \lstinline{nix-build}). This behaviour aligns with
our overwritable arguments feature, which allows a function's arguments to be
overwritten even after the function has been applied, and the evaluation of a
function is deferred until it is explicitly \lstinline[style=mylang]{force}d.

\section{Related Work}
\label{sec:related}

\newcommand{\lambdae}{\ensuremath{\lambda_\mathrm{E}}\xspace}

We have discussed comparisons to various formalisms and languages throughout the
paper.
While many production languages feature forms of named and optional arguments,
their designs are mostly described
informally~\cite{rytz2010named,flatt2009keyword}.
Formally, \citet{garrigue2001labeled}
presented typing rules for labelled and optional arguments for OCaml, noting that
the application rule becomes complex because it ``integrates all the
commuting and discarding machinery''.
More recently, \citet{sun2025named}
modelled named arguments using intersection types and optional arguments using
union types; they identified a tension with record width subtyping that closely
mirrors the Candid issue discussed in \Cref{sec:applications}.

One recent exploration of first-class environments is by \citet{tan2024case}.
Our approach differs from theirs in several significant ways. First, their
calculus \lambdae uses de Bruijn indices, and thus
environments are not commutative. Moreover, they explicitly disallow
lookups from environments with duplicate labels. Further, \lambdae lacks
support for polymorphism or function application paradigms.

Row calculi have a rich history in the literature.
Most relevant to our setting
are \citet{leijen2005extensible}, which supports duplicate labels, and
\citet{remy1993type}, which allows existing labels to be overwritten.
Furthermore, \citet{osinski2006polymorphic} introduced a record concatenation
operator (\lstinline{&}) that permits duplicate labels only when they have the
same type, in which case the value of the field in the second operand is
selected (similarly to our shadowing behaviour).
CDuce~\cite{cduce} supports optional fields in record types and distinguishes
open record types, which admit extra fields, from closed ones. Read this
way, \lang records are closed: a supertype may only add optional fields.
For a recent and comprehensive
overview of the design space of record calculi, we refer the reader to
\citet{morris2019abstracting}.

\section*{Data Availability Statement}

The Lean 4 development accompanying this paper is provided in the supplementary
materials. It contains the mechanized proofs of all stated lemmas and theorems,
as well as the code of the interpreter and the proof of its equivalence with the
operational semantics.

\bibliography{main.bib}

\newpage

\appendix

\begingroup

\renewcommand{\lottinferenceruletarget}[1]{\ifcsname lottjudgementlabelprefix\endcsname\MakeLinkTarget*{rule:appendix-\lottjudgementlabelprefix-#1}\expandafter\def\csname @currentlabel\endcsname{\lottjudgementlabelprefix-#1}\label{rule:appendix-\lottjudgementlabelprefix-#1}\fi}
\renewcommand{\lottref}[2]{\textsc{\ref{rule:appendix-#1-#2}}}

\section{More Discussions}

\subsection{Broader Applications}

\paragraph{\textbf{Typed XML documents with optional values}}

CDuce~\cite{cduce} is a typed functional language targeted at XML applications,
in which the attributes of an XML element form a record. Since XML attributes
are often optional, a record type can mark a field as optional, e.g.
\lstinline+{ kind =? String }+, as with our optional values.
CDuce distinguishes \textit{open} record types, which admit new fields, from
\textit{closed} ones, which do not.
Since subtyping in CDuce is defined as set inclusion, width subtyping requires
an open supertype, while a closed type may only add optional fields. Read this
way, \lang records are closed. Under
the open reading, the type \lstinline[style=config]|{ bio : text }| from the
Candid example in \Cref{sec:applications} admits a
\lstinline[style=config]|theme_color| field of any type, and is therefore
\textit{not} a subtype of \lstinline[style=config]|{ bio : text, theme_color : opt string }|.
CDuce records have unique labels, however; extending \lang with open records
would require care, as the additional fields of an open record must not shadow
existing ones.

\paragraph{\textbf{Masking effects with optionals}}
Optional values emerge naturally in the presence of scoped labels. In
particular, our formulation of optional values is closely related to the
\textit{mask} (also called \textit{inject} or \textit{lift}) operation found in
\textit{algebraic effect
  handlers}~\cite{plotkin2001adequacy,plotkin2013handling} when effects are
represented as scoped labels~\cite{biernacki2017handle,leijen2014koka}. In such
systems, the effect type system tracks effect operations performed within an
expression, and handlers eliminate these effects, with operations being handled
dynamically by default with the innermost enclosing handler. If a developer
intends for an effect to be handled by the second innermost handler instead,
they can explicitly mask the innermost handler. For instance, $\mathsf{mask} \langle \mathsf{print}
\rangle ~ (\mathsf{print}~())$ carries the duplicate effects $\{\mathsf{print},
\mathsf{print}\}$, though the first $\mathsf{print}$ label does not correspond
to an actual operation, and $\mathsf{print}$ is then handled by the second
innermost handler.

\paragraph{\textbf{Presence types with strength subtyping}}

Consider a function %
\lstinline[style=mylang]+eq : \/ a. {~ _ : a, _ : a }~->~bool+, which takes two
arguments of the same type and returns a boolean result. Assume we have
two terms:
\lstinline[style=mylang]+alice : {~name : String, age : Int }+
and
\lstinline[style=mylang]+bob : {~name : String, age ? Int }+.
In a system without subtyping, the application
\lstinline{(eq alice bob)} would be rejected as
the type variable \lstinline{a} cannot be instantiated to match both arguments.
Fortunately,
via the subtyping rule \lottref{S}{lt}, we can also type
\lstinline[style=mylang]+alice : {~name : String, age ? Int }+.
Consequently, \lstinline[style=mylang]+ (eq { alice, bob })+ type-checks successfully by instantiating \lstinline{a} with
\lstinline[style=mylang]+{~name : String, age ? Int }+.

This approach is reminiscent of record calculi where labels have
\textit{presence types}~\cite{remy1993type}. In such a system, one can
write
\lstinline[style=presence]+alice : { name : pre(String), age : pre(Int); abs }+
and \lstinline{bob :} \lstinline[style=presence]+{ name : pre(String); abs }+,
where \lstinline[style=presence]{"abs"} after \lstinline{";"} is a "template"
type for all absent fields.
However,
\citet{remy1993type} noted a
limitation of the calculus: the expression \lstinline[style=mylang]+ (eq { alice, bob })+ fails to type-check
as
\lstinline[style=presence]+pre(Int)+
cannot be unified with
\lstinline[style=presence]+abs+.
(The issue is also known as
the \textit{poisoning problem}~\cite{once1999},
where type information from one expression pollutes another,
here, via the presence of the \lstinline[style=presence]+abs+ field.)
To address this, \citet{remy1993type} discussed an extension featuring a subtyping relation
\lstinline[style=presence]+pre+ $\sub$ \lstinline[style=presence]+abs+, which directly mirrors our subtyping rule \lottref{S}{lt}.

\subsection{Extensions}

This section explores several potential extensions to our design.

\paragraph{\textbf{Variadic arguments}}
\label{sec:variadic-arguments}

Another common paradigm for function application is \textit{variadic
  arguments}~(or variable-length arguments), If the arguments are homogeneous,
this feature can be considered as passing a list. More
interestingly, in dynamically typed languages like Python, we can pass a
variable number of arguments of different types, as demonstrated in the
program on
\begin{wrapfigure}{r}{0.42\textwidth}
\vspace{-6pt}
\begin{lstlisting}[style=python,basicstyle=\footnotesize\ttfamily]
def display_info(name, **kwargs):
  for key, value in kwargs.items():
    print(f"{key}: {value}")
\end{lstlisting}
\vspace{-15pt}
\end{wrapfigure}
the right. Here, \lstinline{display_info} first takes
\lstinline{name}, and then a variable number of
named arguments via \lstinline{kwargs} and simply prints each key-value pair. We
can then call it like this: %
\lstinline[style=python]{display_info(name="Alice", age=2, is_logged_in=True)}.
In Python, varying named arguments can only be at the tail of a function.

Intuitively, this feature aligns well with \textit{row polymorphism} featuring a
polymorphic tail~\cite{wand1987type}. Specifically, we can define a function
with type %
\lstinline[style=mylang]+ \/ t. {name |> ~String | t } ~ -> ~ Unit +, where
\lstinline{t} represents a row variable, capturing all
remaining arguments passed to the function.
\citet{laurent2026revisiting} realize this idea for R with set-theoretic types,
encoding argument lists as records whose tail is closed, open, or a row variable
that captures the arguments of the ellipsis parameter $(\ldots)$ and can be forwarded to another call.

While promising, in the presence of duplicated labels, this design is most effective when paired with
\textit{qualified types}~\cite{jones1994qualified} to enforce constraints on the
tail. For example, consider using the \lstinline{name} variable within the body
of the function above. If the tail \lstinline{t} does not contain
\lstinline{name}, the variable clearly refers to the first parameter. However,
if \lstinline{t} is instantiated with a record that also contains a
\lstinline{name} label, it shadows the first parameter. As a result, the
behaviour of the function depends on the instantiation of \lstinline{t}. To
prevent this ambiguity, a constraint such as \lstinline{(name} $\notin$
\lstinline{t)} can be introduced into the function type
signature~\cite{gaster1996polymorphic}, which restricts \lstinline{t} from being
instantiated with any record containing a \lstinline{name} label, thereby ensuring
deterministic behaviour (see \citet{morris2019abstracting} for
more general row theories). We leave the exploration of this approach to future
work.

\paragraph{\textbf{Dependent default parameters}}

\textit{Dependent default parameters}
are evaluated at call-time and can refer to previously declared
parameters. For example, we can define a JavaScript function:
\lstinline[style=javascript]|function add(x, y, z = x + y) { return z; }|.
This feature is, however, not widely supported among programming languages,
especially those (e.g. OCaml) that do not guarantee a strict
left-to-right evaluation order for function arguments.
In Scala, this behaviour can be encoded via currying: %
\lstinline[style=scala]|def add(x: Int, y: Int)(z: Int = x + y): Int = z|.
We can achieve a similar encoding in our calculus:
\lstinline[style=mylang]|\{ x, y }. ((\{ z }. z) { z : x + y })|, and calling
this requires two \lstinline[style=mylang]{force}s.

\paragraph{\textbf{Type components}}
\label{sec:type-components}

Another potential extension is to include type components within records,
thereby enabling System F-style type abstractions and applications. This
extension on its own should be straightforward. On the other hand, we may want
to allow subsequent fields to depend on earlier type components in the record,
which approaches OCaml-style modules. In either case, it may be more practical
to erase types before evaluation. We leave this investigation to future work.

\section{Additional Typing Examples}
\label{sec:appendix-examples}

This section presents two further typing derivations.
As before, we omit uninteresting derivation details.

\paragraph{\textbf{Let-binding with pattern separation}}

Consider the following let-binding, whose pattern lists its labels in a
different order from the record it destructures:
\[
  \lottkw{let} \, \{ \, iⱼ, iₖ \, \} = \{ \, iₖ : 3, iⱼ : 5 \, \} \, \lottkw{in} \, @.iⱼ + @.iₖ
\]
To type this example, subtyping first permutes the entries of the record type
through \lottref{S}{ext}, so that they follow the order of the pattern:
\begin{mathpar}
  \footnotesize
  \renewcommand{\TirNameStyle}[1]{\scriptsize\textsc{#1}}
  \inferrule*[left=\fbox{\textbf{B}}\textrm{ =},right=\lottref{T}{sub}]{
    \inferrule*[right=\lottref{T}{record}]{ }{
      \text{$\ifdefined\store \mathcolor{Store}{\lottsym{ε}} \lottsym{;} \, \fi \lottsym{ε} \, \lottsym{⊢} \, \lottsym{\{} \, iₖ \, \lottsym{:} \, 3 \lottsym{,} \, iⱼ \, \lottsym{:} \, 5 \, \lottsym{\}} \, \lottsym{:} \, \lottsym{\{} \, iₖ \, \lottsym{:} \, \lottsym{ℕ} \lottsym{,} \, iⱼ \, \lottsym{:} \, \lottsym{ℕ} \, \lottsym{\}}$}
    }
    \and
    \inferrule*[right=\lottref{S}{record}]{
      \inferrule*[right=\lottref{S}{ext}]{
        \inferrule*[right=\lottref{S}{ext}]{
          \inferrule*[right=\lottref{S}{empty}]{
            \text{${\lottsym{⌊} \lottsym{ε} \lottsym{⌋}}^{\lottsym{?}}$}
          }{
            \text{$\lottsym{ε} \, \lottsym{\mathcolor{Subtyping}{⩽}} \, \lottsym{ε}$}
          }
          \and
          \cdots
        }{
          \text{$iⱼ \, \lottsym{:} \, \lottsym{ℕ} \, \lottsym{\mathcolor{Subtyping}{⩽}} \, iⱼ \, \lottsym{:} \, \lottsym{ℕ}$}
        }
        \and
        \cdots
      }{
        \text{$iₖ \, \lottsym{:} \, \lottsym{ℕ} \lottsym{,} \, iⱼ \, \lottsym{:} \, \lottsym{ℕ} \, \lottsym{\mathcolor{Subtyping}{⩽}} \, iⱼ \, \lottsym{:} \, \lottsym{ℕ} \lottsym{,} \, iₖ \, \lottsym{:} \, \lottsym{ℕ}$}
      }
    }{
      \text{$\lottsym{\{} \, iₖ \, \lottsym{:} \, \lottsym{ℕ} \lottsym{,} \, iⱼ \, \lottsym{:} \, \lottsym{ℕ} \, \lottsym{\}} \, \lottsym{\mathcolor{Subtyping}{⩽}} \, \lottsym{\{} \, iⱼ \, \lottsym{:} \, \lottsym{ℕ} \lottsym{,} \, iₖ \, \lottsym{:} \, \lottsym{ℕ} \, \lottsym{\}}$}
    }
  }{
    \text{$\ifdefined\store \mathcolor{Store}{\lottsym{ε}} \lottsym{;} \, \fi \lottsym{ε} \, \lottsym{⊢} \, \lottsym{\{} \, iₖ \, \lottsym{:} \, 3 \lottsym{,} \, iⱼ \, \lottsym{:} \, 5 \, \lottsym{\}} \, \lottsym{:} \, \lottsym{\{} \, iⱼ \, \lottsym{:} \, \lottsym{ℕ} \lottsym{,} \, iₖ \, \lottsym{:} \, \lottsym{ℕ} \, \lottsym{\}}$}
  }
\end{mathpar}
Pattern separation then splits the pattern into the type of the record it
destructures and the parameter names it binds:
\begin{mathpar}
  \footnotesize
  \renewcommand{\TirNameStyle}[1]{\scriptsize\textsc{#1}}
  \inferrule*[left=\fbox{\textbf{P}}\textrm{ =},right=\lottref{P}{record}]{
    \inferrule*[right=\lottref{P}{ext}]{
      \inferrule*[right=\lottref{P}{ext}]{
        \inferrule*[right=\lottref{P}{empty}]{
        }{
          \text{$\lottsym{ε} \, \lottsym{\mathcolor{Pattern}{:}} \, \lottsym{ε} \, \lottsym{\mathcolor{Pattern}{⊙}} \, \lottsym{ε}$}
        }
        \and
        \inferrule*[right=\lottref{P}{label}]{
        }{
          \text{$iⱼ \, \lottsym{\mathcolor{Pattern}{:}} \, \lottsym{ℕ} \, \lottsym{\mathcolor{Pattern}{⊙}} \, iⱼ \, \lottsym{:} \, \lottsym{ℕ}$}
        }
      }{
        \text{$iⱼ \, \lottsym{\mathcolor{Pattern}{:}} \, iⱼ \, \lottsym{:} \, \lottsym{ℕ} \, \lottsym{\mathcolor{Pattern}{⊙}} \, iⱼ \, \lottsym{:} \, \lottsym{ℕ}$}
      }
      \and
      \inferrule*[right=\lottref{P}{label}]{
      }{
        \text{$iₖ \, \lottsym{\mathcolor{Pattern}{:}} \, \lottsym{ℕ} \, \lottsym{\mathcolor{Pattern}{⊙}} \, iₖ \, \lottsym{:} \, \lottsym{ℕ}$}
      }
    }{
      \text{$iⱼ \lottsym{,} \, iₖ \, \lottsym{\mathcolor{Pattern}{:}} \, iⱼ \, \lottsym{:} \, \lottsym{ℕ} \lottsym{,} \, iₖ \, \lottsym{:} \, \lottsym{ℕ} \, \lottsym{\mathcolor{Pattern}{⊙}} \, iⱼ \, \lottsym{:} \, \lottsym{ℕ} \lottsym{,} \, iₖ \, \lottsym{:} \, \lottsym{ℕ}$}
    }
  }{
    \text{$\lottsym{\{} \, iⱼ \lottsym{,} \, iₖ \, \lottsym{\}} \, \lottsym{\mathcolor{Pattern}{:}} \, \lottsym{\{} \, iⱼ \, \lottsym{:} \, \lottsym{ℕ} \lottsym{,} \, iₖ \, \lottsym{:} \, \lottsym{ℕ} \, \lottsym{\}} \, \lottsym{\mathcolor{Pattern}{⊙}} \, iⱼ \, \lottsym{:} \, \lottsym{ℕ} \lottsym{,} \, iₖ \, \lottsym{:} \, \lottsym{ℕ}$}
  }
\end{mathpar}
Finally, rule \lottref{T}{let} combines the two derivations, typing the body
under the environment extended with the parameter names:
\begin{mathpar}
  \footnotesize
  \renewcommand{\TirNameStyle}[1]{\scriptsize\textsc{#1}}
  \inferrule*[right=\lottref{T}{let}]{
    \fbox{\textbf{B}}
    \and
    \fbox{\textbf{P}}
    \and
    \inferrule*[right=\textsc{T-op}]{ }{
      \text{$\ifdefined\store \mathcolor{Store}{\lottsym{ε}} \lottsym{;} \, \fi iⱼ \, \lottsym{:} \, \lottsym{ℕ} \lottsym{,} \, iₖ \, \lottsym{:} \, \lottsym{ℕ} \, \lottsym{⊢} \, \lottsym{@} \lottsym{.} iⱼ \, \lottsym{+} \, \lottsym{@} \lottsym{.} iₖ \, \lottsym{:} \, \lottsym{ℕ}$}
    }
  }{
    \text{$\ifdefined\store \mathcolor{Store}{\lottsym{ε}} \lottsym{;} \, \fi \lottsym{ε} \, \lottsym{⊢} \, \lottkw{let} \, \lottsym{\{} \, iⱼ \lottsym{,} \, iₖ \, \lottsym{\}} \, \lottsym{=} \, \lottsym{\{} \, iₖ \, \lottsym{:} \, 3 \lottsym{,} \, iⱼ \, \lottsym{:} \, 5 \, \lottsym{\}} \, \lottkw{in} \, \lottsym{@} \lottsym{.} iⱼ \, \lottsym{+} \, \lottsym{@} \lottsym{.} iₖ \, \lottsym{:} \, \lottsym{ℕ}$}
  }
\end{mathpar}

\paragraph{\textbf{Concatenation with duplicate labels}}

Consider the following concatenation of two records that share the label $i$:
\[
  \{ \, i = 3 \, \} \, \lottsym{\doubleplus} \, \{ \, i = \{ \, \} \, \}
\]
The concatenation judgement assumes that the entries of its operands line up.
Subtyping therefore first adds an optional entry to the type of the left
operand through \lottref{S}{empty}:
\begin{mathpar}
  \footnotesize
  \renewcommand{\TirNameStyle}[1]{\scriptsize\textsc{#1}}
  \inferrule*[left=\fbox{\textbf{L}}\textrm{ =},right=\lottref{T}{sub}]{
    \inferrule*[right=\lottref{T}{record}]{ }{
      \text{$\ifdefined\store \mathcolor{Store}{\lottsym{ε}} \lottsym{;} \, \fi \lottsym{ε} \, \lottsym{⊢} \, \lottsym{\{} \, i \, \lottsym{=} \, 3 \, \lottsym{\}} \, \lottsym{:} \, \lottsym{\{} \, i \, \lottsym{▹} \, \lottsym{ℕ} \, \lottsym{\}}$}
    }
    \and
    \inferrule*[right=\lottref{S}{record}]{
      \inferrule*[right=\lottref{S}{ext}]{
        \inferrule*[right=\lottref{S}{empty}]{
          \text{${\lottsym{⌊} i \, \lottsym{?} \, \lottsym{\{} \, \lottsym{\}} \lottsym{⌋}}^{\lottsym{?}}$}
        }{
          \text{$\lottsym{ε} \, \lottsym{\mathcolor{Subtyping}{⩽}} \, i \, \lottsym{?} \, \lottsym{\{} \, \lottsym{\}}$}
        }
        \and
        \cdots
      }{
        \text{$i \, \lottsym{▹} \, \lottsym{ℕ} \, \lottsym{\mathcolor{Subtyping}{⩽}} \, i \, \lottsym{▹} \, \lottsym{ℕ} \lottsym{,} \, i \, \lottsym{?} \, \lottsym{\{} \, \lottsym{\}}$}
      }
    }{
      \text{$\lottsym{\{} \, i \, \lottsym{▹} \, \lottsym{ℕ} \, \lottsym{\}} \, \lottsym{\mathcolor{Subtyping}{⩽}} \, \lottsym{\{} \, i \, \lottsym{▹} \, \lottsym{ℕ} \lottsym{,} \, i \, \lottsym{?} \, \lottsym{\{} \, \lottsym{\}} \, \lottsym{\}}$}
    }
  }{
    \text{$\ifdefined\store \mathcolor{Store}{\lottsym{ε}} \lottsym{;} \, \fi \lottsym{ε} \, \lottsym{⊢} \, \lottsym{\{} \, i \, \lottsym{=} \, 3 \, \lottsym{\}} \, \lottsym{:} \, \lottsym{\{} \, i \, \lottsym{▹} \, \lottsym{ℕ} \lottsym{,} \, i \, \lottsym{?} \, \lottsym{\{} \, \lottsym{\}} \, \lottsym{\}}$}
  }
\end{mathpar}
Concatenation then preserves the strong entry of the left operand, while the
entry of the right operand overwrites the optional one:
\begin{mathpar}
  \footnotesize
  \renewcommand{\TirNameStyle}[1]{\scriptsize\textsc{#1}}
  \inferrule*[left=\fbox{\textbf{C}}\textrm{ =},right=\lottref{C}{ext}]{
    \inferrule*[right=\lottref{C}{strongL}]{
      \inferrule*[right=\lottref{C}{empty}]{
      }{
        \text{$\lottsym{ε} \, \lottsym{\mathcolor{Concat}{⊕}} \, \lottsym{ε} \, \lottsym{\mathcolor{Concat}{↬}} \, \lottsym{ε}$}
      }
    }{
      \text{$i \, \lottsym{▹} \, \lottsym{ℕ} \, \lottsym{\mathcolor{Concat}{⊕}} \, \lottsym{ε} \, \lottsym{\mathcolor{Concat}{↬}} \, i \, \lottsym{▹} \, \lottsym{ℕ}$}
    }
    \and
    \inferrule*[right=\lottref{C}{later}]{
    }{
      \text{$\lottsym{?} \, \lottsym{\mathcolor{Concat}{⊕}} \, \lottsym{▹} \, \lottsym{\mathcolor{Concat}{↬}} \, \lottsym{▹}$}
    }
  }{
    \text{$i \, \lottsym{▹} \, \lottsym{ℕ} \lottsym{,} \, i \, \lottsym{?} \, \lottsym{\{} \, \lottsym{\}} \, \lottsym{\mathcolor{Concat}{⊕}} \, i \, \lottsym{▹} \, \lottsym{\{} \, \lottsym{\}} \, \lottsym{\mathcolor{Concat}{↬}} \, i \, \lottsym{▹} \, \lottsym{ℕ} \lottsym{,} \, i \, \lottsym{▹} \, \lottsym{\{} \, \lottsym{\}}$}
  }
\end{mathpar}
Finally, rule \lottref{T}{concat} combines the two derivations, yielding a
record with two entries labelled $i$:
\begin{mathpar}
  \footnotesize
  \renewcommand{\TirNameStyle}[1]{\scriptsize\textsc{#1}}
  \inferrule*[right=\lottref{T}{concat}]{
    \fbox{\textbf{L}}
    \and
    \inferrule*[right=\lottref{T}{record}]{ }{
      \text{$\ifdefined\store \mathcolor{Store}{\lottsym{ε}} \lottsym{;} \, \fi \lottsym{ε} \, \lottsym{⊢} \, \lottsym{\{} \, i \, \lottsym{=} \, \lottsym{\{} \, \lottsym{\}} \, \lottsym{\}} \, \lottsym{:} \, \lottsym{\{} \, i \, \lottsym{▹} \, \lottsym{\{} \, \lottsym{\}} \, \lottsym{\}}$}
    }
    \and
    \fbox{\textbf{C}}
  }{
    \text{$\ifdefined\store \mathcolor{Store}{\lottsym{ε}} \lottsym{;} \, \fi \lottsym{ε} \, \lottsym{⊢} \, \lottsym{\{} \, i \, \lottsym{=} \, 3 \, \lottsym{\}} \, \lottsym{⧺} \, \lottsym{\{} \, i \, \lottsym{=} \, \lottsym{\{} \, \lottsym{\}} \, \lottsym{\}} \, \lottsym{:} \, \lottsym{\{} \, i \, \lottsym{▹} \, \lottsym{ℕ} \lottsym{,} \, i \, \lottsym{▹} \, \lottsym{\{} \, \lottsym{\}} \, \lottsym{\}}$}
  }
\end{mathpar}

\section{Complete Syntax and Semantics of \lang}

\def\store{\relax}
\allowdisplaybreaks

\begin{tabularx}{\linewidth}{r c >{\raggedright}X >{\kern-1em}r}
  $e$ & $\Coloneqq$ & \renewcommand{\lottnonterminal}[3]{#3}$n\mid\text{$\lottsym{λ} \, \lottsym{\{} \, c \, \lottsym{\}} \lottsym{.} \, e$}\mid\text{$\lottsym{⟨} o₀ \lottsym{;} \, o₁ \, \lottsym{|} \, \lottsym{λ} \, \lottsym{\{} \, c \, \lottsym{\}} \lottsym{.} \, e \lottsym{⟩}$}\mid\lottnonterminalfilterout{nat,op,lam,closure,mproj,diff,let,loc,ref,deref,set}{%
\lottnonterminal{Term}{
\lottalias{e}
}{
\lottproduction{nat}{n}
\lottproductionsep
\lottproduction{op}{e₀ \, b \, e₁}
\lottproductionsep
\lottproduction{lam}{\lottsym{λ} \, \lottsym{\{} \, c \, \lottsym{\}} \lottsym{.} \, e}
\lottproductionsep
\lottproduction{closure}{\lottsym{⟨} r₀ \lottsym{;} \, r₁ \, \lottsym{|} \, \lottsym{λ} \, \lottsym{\{} \, c \, \lottsym{\}} \lottsym{.} \, e \lottsym{⟩}}
\lottproductionsep
\lottproduction{app}{e₀ \, e₁}
\lottproductionsep
\lottproduction{force}{\lottkw{force} \, e}
\lottproductionsep
\lottproduction{box}{e₀ \, \lottsym{▶} \, e₁}
\lottproductionsep
\lottproduction{env}{\lottsym{@}}
\lottproductionsep
\lottproduction{record}{\lottsym{\{} \, r \, \lottsym{\}}}
\lottproductionsep
\lottproduction{concat}{e₀ \, \lottsym{⧺} \, e₁}
\lottproductionsep
\lottproduction{proj}{e \lottsym{.} i}
\lottproductionsep
\lottproduction{mproj}{e \lottsym{.} \lottsym{\{} \, \lottsepbycomprehension{{i}_{i}} \, \lottsym{\}}}
\lottproductionsep
\lottproduction{diff}{e \lottsym{\textbackslash} \lottsym{\{} \, \lottsepbycomprehension{{i}_{i}} \, \lottsym{\}}}
\lottproductionsep
\lottproduction{let}{\lottkw{let} \, t \, \lottsym{=} \, e₀ \, \lottkw{in} \, e₁}
\lottproductionsep
\lottproduction{loc}{l}
\lottproductionsep
\lottproduction{ref}{\lottkw{ref} \, e}
\lottproductionsep
\lottproduction{deref}{\lottsym{!} e}
\lottproductionsep
\lottproduction{set}{e₀ \, \lottsym{≔} \, e₁}
}
}
  \mid\text{$e \lottsym{.} \lottsym{\{} \, \lottsepbycomprehension{{i}_{j}} \, \lottsym{\}}$}\mid e \lottsym{\textbackslash} \lottsym{\{} \, \lottsepbycomprehension{{i}_{j}} \, \lottsym{\}}\lottnonterminalfilterout{nat,op,lam,closure,app,force,box,env,record,concat,proj,mproj,diff}{%
\lottnonterminal{Term}{
\lottalias{e}
}{
\lottproduction{nat}{n}
\lottproductionsep
\lottproduction{op}{e₀ \, b \, e₁}
\lottproductionsep
\lottproduction{lam}{\lottsym{λ} \, \lottsym{\{} \, c \, \lottsym{\}} \lottsym{.} \, e}
\lottproductionsep
\lottproduction{closure}{\lottsym{⟨} r₀ \lottsym{;} \, r₁ \, \lottsym{|} \, \lottsym{λ} \, \lottsym{\{} \, c \, \lottsym{\}} \lottsym{.} \, e \lottsym{⟩}}
\lottproductionsep
\lottproduction{app}{e₀ \, e₁}
\lottproductionsep
\lottproduction{force}{\lottkw{force} \, e}
\lottproductionsep
\lottproduction{box}{e₀ \, \lottsym{▶} \, e₁}
\lottproductionsep
\lottproduction{env}{\lottsym{@}}
\lottproductionsep
\lottproduction{record}{\lottsym{\{} \, r \, \lottsym{\}}}
\lottproductionsep
\lottproduction{concat}{e₀ \, \lottsym{⧺} \, e₁}
\lottproductionsep
\lottproduction{proj}{e \lottsym{.} i}
\lottproductionsep
\lottproduction{mproj}{e \lottsym{.} \lottsym{\{} \, \lottsepbycomprehension{{i}_{i}} \, \lottsym{\}}}
\lottproductionsep
\lottproduction{diff}{e \lottsym{\textbackslash} \lottsym{\{} \, \lottsepbycomprehension{{i}_{i}} \, \lottsym{\}}}
\lottproductionsep
\lottproduction{let}{\lottkw{let} \, t \, \lottsym{=} \, e₀ \, \lottkw{in} \, e₁}
\lottproductionsep
\lottproduction{loc}{l}
\lottproductionsep
\lottproduction{ref}{\lottkw{ref} \, e}
\lottproductionsep
\lottproduction{deref}{\lottsym{!} e}
\lottproductionsep
\lottproduction{set}{e₀ \, \lottsym{≔} \, e₁}
}
}$& \textit{term}\\
\lottnonterminal{Record Pattern}{
\lottalias{c}
}{
\lottproduction{empty}{\lottsym{ε}}
\lottproductionsep
\lottproduction{ext}{c \lottsym{,} \, i \, \lottsym{◇} \, t}
}
\\
\lottnonterminal{Pattern}{
\lottalias{t}
}{
\lottproduction{label}{i}
\lottproductionsep
\lottproduction{record}{\lottsym{\{} \, c \, \lottsym{\}}}
}
\\
\lottnonterminal{Record}{
\lottalias{r}
}{
\lottproduction{empty}{\lottsym{ε}}
\lottproductionsep
\lottproduction{ext}{r \lottsym{,} \, i \, \rstrength \, e}
\lottproductionsep
\lottproduction{ext?}{r \lottsym{,} \, i \, \lottsym{?}}
}
\\
\lottnonterminal{Required Strength}{
\lottalias{\rstrength}
}{
\lottproduction{strong}{\lottsym{=}}
\lottproductionsep
\lottproduction{weak}{\lottsym{:}}
}
\\
  \\
\lottnonterminal{Value}{
\lottalias{v}
}{
\lottproduction{nat}{n}
\lottproductionsep
\lottproduction{closure}{\lottsym{⟨} o₀ \lottsym{;} \, o₁ \, \lottsym{|} \, \lottsym{λ} \, \lottsym{\{} \, c \, \lottsym{\}} \lottsym{.} \, e \lottsym{⟩}}
\lottproductionsep
\lottproduction{record}{\lottsym{\{} \, o \, \lottsym{\}}}
\lottproductionsep
\lottproduction{loc}{l}
}
\\
\lottnonterminal{Record Value}{
\lottalias{o}
}{
\lottproduction{empty}{\lottsym{ε}}
\lottproductionsep
\lottproduction{ext}{o \lottsym{,} \, i \, \rstrength \, v}
\lottproductionsep
\lottproduction{ext?}{o \lottsym{,} \, i \, \lottsym{?}}
}
\\
  \\
\lottnonterminal{Type}{
\lottalias{A}
\lottaliassep
\lottalias{B}
}{
\lottproduction{var}{a}
\lottproductionsep
\lottproduction{nat}{\lottsym{ℕ}}
\lottproductionsep
\lottproduction{arr}{\lottsym{\{} \, ρ \, \lottsym{\}} \, \lottsym{→} \, A}
\lottproductionsep
\lottproduction{record}{\lottsym{\{} \, ρ \, \lottsym{\}}}
\lottproductionsep
\lottproduction{quant}{\lottsym{∀} \, a \lottsym{.} \, A}
\lottproductionsep
\lottproduction{ref}{\lottkw{Ref} \, A}
}
\\
\lottnonterminal{Record Type}{
\lottalias{ρ}
}{
\lottproduction{empty}{\lottsym{ε}}
\lottproductionsep
\lottproduction{ext}{ρ \lottsym{,} \, i \, \strength \, A}
}
\\
\lottnonterminal{Strength}{
\lottalias{\strength}
}{
\lottproduction{strong'}{\lottsym{▹}}
\lottproductionsep
\lottproduction{weak}{\lottsym{:}}
\lottproductionsep
\lottproduction{optional}{\lottsym{?}}
}
\\
  \\
\lottnonterminal{Monotype}{
\lottalias{τ}
}{
\lottproduction{var}{a}
\lottproductionsep
\lottproduction{nat}{\lottsym{ℕ}}
\lottproductionsep
\lottproduction{arr}{\lottsym{\{} \, ϱ \, \lottsym{\}} \, \lottsym{→} \, τ}
\lottproductionsep
\lottproduction{record}{\lottsym{\{} \, ϱ \, \lottsym{\}}}
\lottproductionsep
\lottproduction{ref}{\lottkw{Ref} \, τ}
}
\\
\lottnonterminal{Record Monotype}{
\lottalias{ϱ}
}{
\lottproduction{empty}{\lottsym{ε}}
\lottproductionsep
\lottproduction{ext}{ϱ \lottsym{,} \, i \, \strength \, τ}
}
\\
\end{tabularx}

\medskip

\lottjudgementlabel{T}{\lottjudgementfilterout{op}{\lottjudgementnarrower[sub=0.3,lam=0.4,closure=0.6,app=0.4,box=0.4,concat=0.4,let=0.4,set=0.5]{0.2}{%
\lottjudgement{Typing}{\ifdefined\store \mathcolor{Store}{ρₛ} \lottsym{;} \, \fi ρ \, \lottsym{⊢} \, e \, \lottsym{:} \, A}{
\lottinferencerule{sub}{
\lotthypothesis{\ifdefined\store \mathcolor{Store}{ρₛ} \lottsym{;} \, \fi ρ \, \lottsym{⊢} \, e \, \lottsym{:} \, A}\\
\lotthypothesis{A \, \lottsym{\mathcolor{Subtyping}{⩽}} \, B}
}{\ifdefined\store \mathcolor{Store}{ρₛ} \lottsym{;} \, \fi ρ \, \lottsym{⊢} \, e \, \lottsym{:} \, B}
\lottinferencerulesep
\lottinferencerule{nat}{
}{\ifdefined\store \mathcolor{Store}{ρₛ} \lottsym{;} \, \fi ρ \, \lottsym{⊢} \, n \, \lottsym{:} \, \lottsym{ℕ}}
\lottinferencerulesep
\lottinferencerule{op}{
\lotthypothesis{\ifdefined\store \mathcolor{Store}{ρₛ} \lottsym{;} \, \fi ρ \, \lottsym{⊢} \, e₀ \, \lottsym{:} \, \lottsym{ℕ}}\\
\lotthypothesis{\ifdefined\store \mathcolor{Store}{ρₛ} \lottsym{;} \, \fi ρ \, \lottsym{⊢} \, e₁ \, \lottsym{:} \, \lottsym{ℕ}}
}{\ifdefined\store \mathcolor{Store}{ρₛ} \lottsym{;} \, \fi ρ \, \lottsym{⊢} \, e₀ \, b \, e₁ \, \lottsym{:} \, \lottsym{ℕ}}
\lottinferencerulesep
\lottinferencerule{lam}{\ifdefined\store\overline{a_j} \, \lottsym{∉} \, \lottkw{ftv} \lottsym{(} \mathcolor{Store}{ρₛ} \lottsym{)}\\\fi\overline{a_j} \, \lottsym{∉} \, \lottkw{ftv} \lottsym{(} ρ₀ \lottsym{)}\\
\lotthypothesis{c \, \lottsym{\mathcolor{Pattern}{:}} \, ρ₁ \, \lottsym{\mathcolor{Pattern}{⊙}} \, ρ₂}\\
\lotthypothesis{\ifdefined\store \mathcolor{Store}{ρₛ} \lottsym{;} \, \fi ρ₀ \lottsym{,} \, ρ₂ \, \lottsym{⊢} \, e \, \lottsym{:} \, A}
}{\ifdefined\store \mathcolor{Store}{ρₛ} \lottsym{;} \, \fi ρ₀ \, \lottsym{⊢} \, \lottsym{λ} \, \lottsym{\{} \, c \, \lottsym{\}} \lottsym{.} \, e \, \lottsym{:} \, \lottsym{∀} \, \lottsepbycomprehension{{a}_{j}} \lottsym{.} \, \lottsym{\{} \, ρ₁ \, \lottsym{\}} \, \lottsym{→} \, A}
\lottinferencerulesep
\lottinferencerule{closure}{\ifdefined\store\overline{a_j} \, \lottsym{∉} \, \lottkw{ftv} \lottsym{(} \mathcolor{Store}{ρₛ} \lottsym{)}\\\fi
\lotthypothesis{\ifdefined\store \mathcolor{Store}{ρₛ} \lottsym{;} \, \fi \lottsym{ε} \, \lottsym{⊢} \, \lottsym{\{} \, o₀ \, \lottsym{\}} \, \lottsym{:} \, \lottsym{\{} \, ρ₁ \, \lottsym{\}}}\\
\lotthypothesis{\lottsym{⌊} ρ₁ \lottsym{⌋}}\\\overline{a_j} \, \lottsym{∉} \, \lottkw{ftv} \lottsym{(} ρ₁ \lottsym{)}\\
\lotthypothesis{\ifdefined\store \mathcolor{Store}{ρₛ} \lottsym{;} \, \fi \lottsym{ε} \, \lottsym{⊢} \, \lottsym{\{} \, o₁ \, \lottsym{\}} \, \lottsym{:} \, \lottsym{∀} \, \lottsepbycomprehension{{a}_{j}} \lottsym{.} \, \lottsym{\{} \, ρ₂ \, \lottsym{\}}}\\
\lotthypothesis{c \, \lottsym{\mathcolor{Pattern}{:}} \, ρ₃ \, \lottsym{\mathcolor{Pattern}{⊙}} \, ρ₄}\\
\lotthypothesis{ρ₃ \, \lottsym{\mathcolor{App}{●}} \, ρ₂ \, \lottsym{\mathcolor{App}{↠}} \, ρ₅}\\
\lotthypothesis{\ifdefined\store \mathcolor{Store}{ρₛ} \lottsym{;} \, \fi ρ₁ \lottsym{,} \, ρ₄ \, \lottsym{⊢} \, e \, \lottsym{:} \, A}
}{\ifdefined\store \mathcolor{Store}{ρₛ} \lottsym{;} \, \fi ρ₀ \, \lottsym{⊢} \, \lottsym{⟨} o₀ \lottsym{;} \, o₁ \, \lottsym{|} \, \lottsym{λ} \, \lottsym{\{} \, c \, \lottsym{\}} \lottsym{.} \, e \lottsym{⟩} \, \lottsym{:} \, \lottsym{∀} \, \lottsepbycomprehension{{a}_{j}} \lottsym{.} \, \lottsym{\{} \, ρ₅ \, \lottsym{\}} \, \lottsym{→} \, A}
\lottinferencerulesep
\lottinferencerule{app}{
\lotthypothesis{\ifdefined\store \mathcolor{Store}{ρₛ} \lottsym{;} \, \fi ρ₀ \, \lottsym{⊢} \, e₀ \, \lottsym{:} \, \lottsym{∀} \, \lottsepbycomprehension{{a}_{j}} \lottsym{.} \, \lottsym{\{} \, ρ₁ \, \lottsym{\}} \, \lottsym{→} \, A}\\
\lotthypothesis{\ifdefined\store \mathcolor{Store}{ρₛ} \lottsym{;} \, \fi ρ₀ \, \lottsym{⊢} \, e₁ \, \lottsym{:} \, \lottsym{∀} \, \lottsepbycomprehension{{a}_{j}} \lottsym{.} \, \lottsym{\{} \, ρ₂ \, \lottsym{\}}}\\
\lotthypothesis{ρ₁ \, \lottsym{\mathcolor{App}{●}} \, ρ₂ \, \lottsym{\mathcolor{App}{↠}} \, ρ₃}
}{\ifdefined\store \mathcolor{Store}{ρₛ} \lottsym{;} \, \fi ρ₀ \, \lottsym{⊢} \, e₀ \, e₁ \, \lottsym{:} \, \lottsym{∀} \, \lottsepbycomprehension{{a}_{j}} \lottsym{.} \, \lottsym{\{} \, ρ₃ \, \lottsym{\}} \, \lottsym{→} \, A}
\lottinferencerulesep
\lottinferencerule{force}{
\lotthypothesis{\ifdefined\store \mathcolor{Store}{ρₛ} \lottsym{;} \, \fi ρ \, \lottsym{⊢} \, e \, \lottsym{:} \, \lottsym{\{} \, \lottsym{\}} \, \lottsym{→} \, A}
}{\ifdefined\store \mathcolor{Store}{ρₛ} \lottsym{;} \, \fi ρ \, \lottsym{⊢} \, \lottkw{force} \, e \, \lottsym{:} \, A}
\lottinferencerulesep
\lottinferencerule{box}{
\lotthypothesis{\ifdefined\store \mathcolor{Store}{ρₛ} \lottsym{;} \, \fi ρ₀ \, \lottsym{⊢} \, e₀ \, \lottsym{:} \, \lottsym{\{} \, ρ₁ \, \lottsym{\}}}\\
\lotthypothesis{\lottsym{⌊} ρ₁ \lottsym{⌋}}\\
\lotthypothesis{\ifdefined\store \mathcolor{Store}{ρₛ} \lottsym{;} \, \fi ρ₁ \, \lottsym{⊢} \, e₁ \, \lottsym{:} \, A}
}{\ifdefined\store \mathcolor{Store}{ρₛ} \lottsym{;} \, \fi ρ₀ \, \lottsym{⊢} \, e₀ \, \lottsym{▶} \, e₁ \, \lottsym{:} \, A}
\lottinferencerulesep
\lottinferencerule{env}{
}{\ifdefined\store \mathcolor{Store}{ρₛ} \lottsym{;} \, \fi ρ \, \lottsym{⊢} \, \lottsym{@} \, \lottsym{:} \, \lottsym{\{} \, ρ \, \lottsym{\}}}
\lottinferencerulesep
\lottinferencerule{record}{
\lotthypothesis{\ifdefined\store \mathcolor{Store}{ρₛ} \lottsym{;} \, \fi ρ₀ \, \lottsym{⊢} \, r \, \lottsym{:} \, ρ₁}
}{\ifdefined\store \mathcolor{Store}{ρₛ} \lottsym{;} \, \fi ρ₀ \, \lottsym{⊢} \, \lottsym{\{} \, r \, \lottsym{\}} \, \lottsym{:} \, \lottsym{\{} \, ρ₁ \, \lottsym{\}}}
\lottinferencerulesep
\lottinferencerule{concat}{
\lotthypothesis{\ifdefined\store \mathcolor{Store}{ρₛ} \lottsym{;} \, \fi ρ₀ \, \lottsym{⊢} \, e₀ \, \lottsym{:} \, \lottsym{\{} \, ρ₁ \, \lottsym{\}}}\\
\lotthypothesis{\ifdefined\store \mathcolor{Store}{ρₛ} \lottsym{;} \, \fi ρ₀ \, \lottsym{⊢} \, e₁ \, \lottsym{:} \, \lottsym{\{} \, ρ₂ \, \lottsym{\}}}\\
\lotthypothesis{ρ₁ \, \lottsym{\mathcolor{Concat}{⊕}} \, ρ₂ \, \lottsym{\mathcolor{Concat}{↬}} \, ρ₃}
}{\ifdefined\store \mathcolor{Store}{ρₛ} \lottsym{;} \, \fi ρ₀ \, \lottsym{⊢} \, e₀ \, \lottsym{⧺} \, e₁ \, \lottsym{:} \, \lottsym{\{} \, ρ₃ \, \lottsym{\}}}
\lottinferencerulesep
\lottinferencerule{proj}{
\lotthypothesis{\ifdefined\store \mathcolor{Store}{ρₛ} \lottsym{;} \, \fi ρ₀ \, \lottsym{⊢} \, e \, \lottsym{:} \, \lottsym{\{} \, ρ₁ \lottsym{,} \, i \, \rstrength \, A \, \lottsym{\}}}
}{\ifdefined\store \mathcolor{Store}{ρₛ} \lottsym{;} \, \fi ρ₀ \, \lottsym{⊢} \, e \lottsym{.} i \, \lottsym{:} \, A}
\lottinferencerulesep
\lottinferencerule{mproj}{
\lotthypothesis{\ifdefined\store \mathcolor{Store}{ρₛ} \lottsym{;} \, \fi ρ₀ \, \lottsym{⊢} \, e \, \lottsym{:} \, \lottsym{\{} \, ρ₁ \lottsym{,} \, \lottsepbycomprehension{{i}_{j} \, {\rstrength}_{j} \, {A}_{j}} \, \lottsym{\}}}
}{\ifdefined\store \mathcolor{Store}{ρₛ} \lottsym{;} \, \fi ρ₀ \, \lottsym{⊢} \, e \lottsym{.} \lottsym{\{} \, \lottsepbycomprehension{{i}_{j}} \, \lottsym{\}} \, \lottsym{:} \, \lottsym{\{} \, \lottsepbycomprehension{{i}_{j} \, {\rstrength}_{j} \, {A}_{j}} \, \lottsym{\}}}
\lottinferencerulesep
\lottinferencerule{diff}{
\lotthypothesis{\ifdefined\store \mathcolor{Store}{ρₛ} \lottsym{;} \, \fi ρ₀ \, \lottsym{⊢} \, e \, \lottsym{:} \, \lottsym{\{} \, ρ₁ \lottsym{,} \, \lottsepbycomprehension{{i}_{j} \, {\rstrength}_{j} \, {A}_{j}} \, \lottsym{\}}}
}{\ifdefined\store \mathcolor{Store}{ρₛ} \lottsym{;} \, \fi ρ₀ \, \lottsym{⊢} \, e \lottsym{\textbackslash} \lottsym{\{} \, \lottsepbycomprehension{{i}_{j}} \, \lottsym{\}} \, \lottsym{:} \, \lottsym{\{} \, ρ₁ \, \lottsym{\}}}
\lottinferencerulesep
\lottinferencerule{let}{
\lotthypothesis{\ifdefined\store \mathcolor{Store}{ρₛ} \lottsym{;} \, \fi ρ₀ \, \lottsym{⊢} \, e₀ \, \lottsym{:} \, A}\\
\lotthypothesis{t \, \lottsym{\mathcolor{Pattern}{:}} \, A \, \lottsym{\mathcolor{Pattern}{⊙}} \, ρ₁}\\
\lotthypothesis{\ifdefined\store \mathcolor{Store}{ρₛ} \lottsym{;} \, \fi ρ₀ \lottsym{,} \, ρ₁ \, \lottsym{⊢} \, e₁ \, \lottsym{:} \, B}
}{\ifdefined\store \mathcolor{Store}{ρₛ} \lottsym{;} \, \fi ρ₀ \, \lottsym{⊢} \, \lottkw{let} \, t \, \lottsym{=} \, e₀ \, \lottkw{in} \, e₁ \, \lottsym{:} \, B}
\lottinferencerulesep
\lottinferencerule{loc}{
}{\ifdefined\store \mathcolor{Store}{ρₛ₀ \lottsym{,} \, l \, \lottsym{:} \, A \lottsym{,} \, ρₛ₁} \lottsym{;} \, \fi ρ \, \lottsym{⊢} \, l \, \lottsym{:} \, \lottkw{Ref} \, A}
\lottinferencerulesep
\lottinferencerule{ref}{
\lotthypothesis{\ifdefined\store \mathcolor{Store}{ρₛ} \lottsym{;} \, \fi ρ \, \lottsym{⊢} \, e \, \lottsym{:} \, A}
}{\ifdefined\store \mathcolor{Store}{ρₛ} \lottsym{;} \, \fi ρ \, \lottsym{⊢} \, \lottkw{ref} \, e \, \lottsym{:} \, \lottkw{Ref} \, A}
\lottinferencerulesep
\lottinferencerule{deref}{
\lotthypothesis{\ifdefined\store \mathcolor{Store}{ρₛ} \lottsym{;} \, \fi ρ \, \lottsym{⊢} \, e \, \lottsym{:} \, \lottkw{Ref} \, A}
}{\ifdefined\store \mathcolor{Store}{ρₛ} \lottsym{;} \, \fi ρ \, \lottsym{⊢} \, \lottsym{!} e \, \lottsym{:} \, A}
\lottinferencerulesep
\lottinferencerule{set}{
\lotthypothesis{\ifdefined\store \mathcolor{Store}{ρₛ} \lottsym{;} \, \fi ρ \, \lottsym{⊢} \, e₀ \, \lottsym{:} \, \lottkw{Ref} \, A}\\
\lotthypothesis{\ifdefined\store \mathcolor{Store}{ρₛ} \lottsym{;} \, \fi ρ \, \lottsym{⊢} \, e₁ \, \lottsym{:} \, A}
}{\ifdefined\store \mathcolor{Store}{ρₛ} \lottsym{;} \, \fi ρ \, \lottsym{⊢} \, e₀ \, \lottsym{≔} \, e₁ \, \lottsym{:} \, \lottsym{\{} \, \lottsym{\}}}
}
}}}

\medskip

\lottjudgementforcename{Record Typing}{\lottjudgementlabel{T}{%
\lottjudgement{Typing}{\ifdefined\store \mathcolor{Store}{ρₛ} \lottsym{;} \, \fi ρ₀ \, \lottsym{⊢} \, r \, \lottsym{:} \, ρ₁}{
\lottinferencerule{empty}{
}{\ifdefined\store \mathcolor{Store}{ρₛ} \lottsym{;} \, \fi ρ \, \lottsym{⊢} \, \lottsym{ε} \, \lottsym{:} \, \lottsym{ε}}
\lottinferencerulesep
\lottinferencerule{ext}{
\lotthypothesis{\ifdefined\store \mathcolor{Store}{ρₛ} \lottsym{;} \, \fi ρ₀ \, \lottsym{⊢} \, r \, \lottsym{:} \, ρ₁}\\
\lotthypothesis{\ifdefined\store \mathcolor{Store}{ρₛ} \lottsym{;} \, \fi ρ₀ \, \lottsym{⊢} \, e \, \lottsym{:} \, A}
}{\ifdefined\store \mathcolor{Store}{ρₛ} \lottsym{;} \, \fi ρ₀ \, \lottsym{⊢} \, r \lottsym{,} \, i \, \rstrength \, e \, \lottsym{:} \, ρ₁ \lottsym{,} \, i \, \rstrength \, A}
\lottinferencerulesep
\lottinferencerule{ext?}{
\lotthypothesis{\ifdefined\store \mathcolor{Store}{ρₛ} \lottsym{;} \, \fi ρ₀ \, \lottsym{⊢} \, r \, \lottsym{:} \, ρ₁}
}{\ifdefined\store \mathcolor{Store}{ρₛ} \lottsym{;} \, \fi ρ₀ \, \lottsym{⊢} \, r \lottsym{,} \, i \, \lottsym{?} \, \lottsym{:} \, ρ₁ \lottsym{,} \, i \, \lottsym{?} \, A}
}
}}

\medskip

\lottjudgementforcename{Subtyping}{\lottjudgementlabel{S}{%
\lottjudgement{Left Subtyping}{A \, \lottsym{\mathcolor{Subtyping}{⩽}} \, B}{
\lottinferencerule{var}{
}{a \, \lottsym{\mathcolor{Subtyping}{⩽}} \, a}
\lottinferencerulesep
\lottinferencerule{nat}{
}{\lottsym{ℕ} \, \lottsym{\mathcolor{Subtyping}{⩽}} \, \lottsym{ℕ}}
\lottinferencerulesep
\lottinferencerule{arr}{
\lotthypothesis{ρ₁ \, \lottsym{\mathcolor{Subtyping}{⩽}} \, ρ₀}\\
\lotthypothesis{A \, \lottsym{\mathcolor{Subtyping}{⩽}} \, B}
}{\lottsym{\{} \, ρ₀ \, \lottsym{\}} \, \lottsym{→} \, A \, \lottsym{\mathcolor{Subtyping}{⩽}} \, \lottsym{\{} \, ρ₁ \, \lottsym{\}} \, \lottsym{→} \, B}
\lottinferencerulesep
\lottinferencerule{record}{
\lotthypothesis{ρ₀ \, \lottsym{\mathcolor{Subtyping}{⩽}} \, ρ₁}
}{\lottsym{\{} \, ρ₀ \, \lottsym{\}} \, \lottsym{\mathcolor{Subtyping}{⩽}} \, \lottsym{\{} \, ρ₁ \, \lottsym{\}}}
\lottinferencerulesep
\lottinferencerule{quantL}{
\lotthypothesis{A\left[τ/a\right] \, \lottsym{\mathcolor{Subtyping}{⩽}} \, B}
}{\lottsym{∀} \, a \lottsym{.} \, A \, \lottsym{\mathcolor{Subtyping}{⩽}} \, B}
\lottinferencerulesep
\lottinferencerule{quantR}{a \, \lottsym{∉} \, \lottkw{ftv} \lottsym{(} A \lottsym{)}\\
\lotthypothesis{A \, \lottsym{\mathcolor{Subtyping}{⩽}} \, B}
}{A \, \lottsym{\mathcolor{Subtyping}{⩽}} \, \lottsym{∀} \, a \lottsym{.} \, B}
\lottinferencerulesep
\lottinferencerule{ref}{
}{\lottkw{Ref} \, A \, \lottsym{\mathcolor{Subtyping}{⩽}} \, \lottkw{Ref} \, A}
}
}}

\medskip

\lottjudgementforcename{Record Subtyping}{\lottjudgementlabel{S}{\lottjudgementnarrower{1}{%
\lottjudgement{Left Subtyping}{ρ₀ \, \lottsym{\mathcolor{Subtyping}{⩽}} \, ρ₁}{
\lottinferencerule{empty}{
\lotthypothesis{{\lottsym{⌊} ρ \lottsym{⌋}}^{\lottsym{?}}}
}{\lottsym{ε} \, \lottsym{\mathcolor{Subtyping}{⩽}} \, ρ}
\lottinferencerulesep
\lottinferencerule{ext}{
\lotthypothesis{ρ₀ \, \lottsym{\mathcolor{Subtyping}{⩽}} \, ρ₁ \lottsym{,} \, ρ₂}\\
\lotthypothesis{i \, \lottsym{∉} \, ρ₁}\\
\lotthypothesis{\strength₀ \, \lottsym{\mathcolor{Subtyping}{⩽}} \, \strength₁}\\
\lotthypothesis{A \, \lottsym{\mathcolor{Subtyping}{⩽}} \, B}
}{i \, \strength₀ \, A \lottsym{,} \, ρ₀ \, \lottsym{\mathcolor{Subtyping}{⩽}} \, ρ₁ \lottsym{,} \, i \, \strength₁ \, B \lottsym{,} \, ρ₂}
}
}}}

\medskip

\lottjudgementforcename{Strength Subtyping}{\lottjudgementlabel{S}{%
\lottjudgement{Subtyping}{\strength₀ \, \lottsym{\mathcolor{Subtyping}{⩽}} \, \strength₁}{
\lottinferencerule{refl}{
}{\strength \, \lottsym{\mathcolor{Subtyping}{⩽}} \, \strength}
\lottinferencerulesep
\lottinferencerule{lt}{
}{\lottsym{:} \, \lottsym{\mathcolor{Subtyping}{⩽}} \, \lottsym{?}}
}
}}

\medskip

\columnratio{0.525}

\begin{paracol}{2}
  \lottjudgementforcename{Record Pattern Separation}{\lottjudgementlabel{P}{%
\lottjudgement{Sep}{c \, \lottsym{\mathcolor{Pattern}{:}} \, ρ₀ \, \lottsym{\mathcolor{Pattern}{⊙}} \, ρ₁}{
\lottinferencerule{empty}{
}{\lottsym{ε} \, \lottsym{\mathcolor{Pattern}{:}} \, \lottsym{ε} \, \lottsym{\mathcolor{Pattern}{⊙}} \, \lottsym{ε}}
\lottinferencerulesep
\lottinferencerule{ext}{
\lotthypothesis{c \, \lottsym{\mathcolor{Pattern}{:}} \, ρ₀ \, \lottsym{\mathcolor{Pattern}{⊙}} \, ρ₁}\\
\lotthypothesis{t \, \lottsym{\mathcolor{Pattern}{:}} \, A \, \lottsym{\mathcolor{Pattern}{⊙}} \, ρ₂}
}{\lottsym{(} c \lottsym{,} \, i \, \lottsym{◇} \, t \lottsym{)} \, \lottsym{\mathcolor{Pattern}{:}} \, \lottsym{(} ρ₀ \lottsym{,} \, i \, \lottsym{:} \, A \lottsym{)} \, \lottsym{\mathcolor{Pattern}{⊙}} \, ρ₁ \lottsym{,} \, ρ₂}
}
}}

  \switchcolumn

  \lottjudgementforcename{Pattern Separation}{\lottjudgementlabel{P}{%
\lottjudgement{Sep}{t \, \lottsym{\mathcolor{Pattern}{:}} \, A \, \lottsym{\mathcolor{Pattern}{⊙}} \, ρ}{
\lottinferencerule{label}{
}{i \, \lottsym{\mathcolor{Pattern}{:}} \, A \, \lottsym{\mathcolor{Pattern}{⊙}} \, \lottsym{(} i \, \lottsym{:} \, A \lottsym{)}}
\lottinferencerulesep
\lottinferencerule{record}{
\lotthypothesis{c \, \lottsym{\mathcolor{Pattern}{:}} \, ρ₀ \, \lottsym{\mathcolor{Pattern}{⊙}} \, ρ₁}
}{\lottsym{\{} \, c \, \lottsym{\}} \, \lottsym{\mathcolor{Pattern}{:}} \, \lottsym{\{} \, ρ₀ \, \lottsym{\}} \, \lottsym{\mathcolor{Pattern}{⊙}} \, ρ₁}
}
}}
\end{paracol}

\medskip

\lottjudgementforcename{Record Type App}{\lottjudgementlabel{A}{%
\lottjudgement{App}{ρ₀ \, \lottsym{\mathcolor{App}{●}} \, ρ₁ \, \lottsym{\mathcolor{App}{↠}} \, ρ₂}{
\lottinferencerule{empty}{
}{\lottsym{ε} \, \lottsym{\mathcolor{App}{●}} \, \lottsym{ε} \, \lottsym{\mathcolor{App}{↠}} \, \lottsym{ε}}
\lottinferencerulesep
\lottinferencerule{weakStrong}{
\lotthypothesis{ρ₀ \, \lottsym{\mathcolor{App}{●}} \, ρ₁ \, \lottsym{\mathcolor{App}{↠}} \, ρ₂}
}{ρ₀ \lottsym{,} \, i \, \lottsym{:} \, A \, \lottsym{\mathcolor{App}{●}} \, ρ₁ \lottsym{,} \, i \, \lottsym{▹} \, A \, \lottsym{\mathcolor{App}{↠}} \, ρ₂}
\lottinferencerulesep
\lottinferencerule{optionalStrong}{
\lotthypothesis{ρ₀ \, \lottsym{\mathcolor{App}{●}} \, ρ₁ \, \lottsym{\mathcolor{App}{↠}} \, ρ₂}
}{ρ₀ \lottsym{,} \, i \, \lottsym{?} \, A \, \lottsym{\mathcolor{App}{●}} \, ρ₁ \lottsym{,} \, i \, \lottsym{▹} \, A \, \lottsym{\mathcolor{App}{↠}} \, ρ₂}
\lottinferencerulesep
\lottinferencerule{ext}{
\lotthypothesis{ρ₀ \, \lottsym{\mathcolor{App}{●}} \, ρ₁ \, \lottsym{\mathcolor{App}{↠}} \, ρ₂}\\
\lotthypothesis{\strength₀ \, \lottsym{\mathcolor{App}{●}} \, \strength₁ \, \lottsym{\mathcolor{App}{↠}} \, \strength₂}
}{ρ₀ \lottsym{,} \, i \, \strength₀ \, A \, \lottsym{\mathcolor{App}{●}} \, ρ₁ \lottsym{,} \, i \, \strength₁ \, A \, \lottsym{\mathcolor{App}{↠}} \, ρ₂ \lottsym{,} \, i \, \strength₂ \, A}
}
}}

\medskip

\lottjudgementforcename{Strength App}{\lottjudgementlabel{A}{%
\lottjudgement{App}{\strength₀ \, \lottsym{\mathcolor{App}{●}} \, \strength₁ \, \lottsym{\mathcolor{App}{↠}} \, \strength₂}{
\lottinferencerule{later}{
}{\lottsym{:} \, \lottsym{\mathcolor{App}{●}} \, \lottsym{?} \, \lottsym{\mathcolor{App}{↠}} \, \lottsym{:}}
\lottinferencerulesep
\lottinferencerule{set}{
}{\lottsym{:} \, \lottsym{\mathcolor{App}{●}} \, \lottsym{:} \, \lottsym{\mathcolor{App}{↠}} \, \lottsym{?}}
\lottinferencerulesep
\lottinferencerule{leave}{
}{\lottsym{?} \, \lottsym{\mathcolor{App}{●}} \, \lottsym{?} \, \lottsym{\mathcolor{App}{↠}} \, \lottsym{?}}
\lottinferencerulesep
\lottinferencerule{replace}{
}{\lottsym{?} \, \lottsym{\mathcolor{App}{●}} \, \lottsym{:} \, \lottsym{\mathcolor{App}{↠}} \, \lottsym{?}}
}
}}

\medskip

\lottjudgementforcename{Record Type Concat}{\lottjudgementlabel{C}{%
\lottjudgement{Concat}{ρ₀ \, \lottsym{\mathcolor{Concat}{⊕}} \, ρ₁ \, \lottsym{\mathcolor{Concat}{↬}} \, ρ₂}{
\lottinferencerule{empty}{
}{\lottsym{ε} \, \lottsym{\mathcolor{Concat}{⊕}} \, \lottsym{ε} \, \lottsym{\mathcolor{Concat}{↬}} \, \lottsym{ε}}
\lottinferencerulesep
\lottinferencerule{strongL}{
\lotthypothesis{ρ₀ \, \lottsym{\mathcolor{Concat}{⊕}} \, ρ₁ \, \lottsym{\mathcolor{Concat}{↬}} \, ρ₂}
}{ρ₀ \lottsym{,} \, i \, \lottsym{▹} \, A \, \lottsym{\mathcolor{Concat}{⊕}} \, ρ₁ \, \lottsym{\mathcolor{Concat}{↬}} \, ρ₂ \lottsym{,} \, i \, \lottsym{▹} \, A}
\lottinferencerulesep
\lottinferencerule{ext}{
\lotthypothesis{ρ₀ \, \lottsym{\mathcolor{Concat}{⊕}} \, ρ₁ \, \lottsym{\mathcolor{Concat}{↬}} \, ρ₂}\\
\lotthypothesis{\strength₀ \, \lottsym{\mathcolor{Concat}{⊕}} \, \strength₁ \, \lottsym{\mathcolor{Concat}{↬}} \, \strength₂}
}{ρ₀ \lottsym{,} \, i \, \strength₀ \, A \, \lottsym{\mathcolor{Concat}{⊕}} \, ρ₁ \lottsym{,} \, i \, \strength₁ \, A \, \lottsym{\mathcolor{Concat}{↬}} \, ρ₂ \lottsym{,} \, i \, \strength₂ \, A}
}
}}

\medskip

\lottjudgementforcename{Strength Concat}{\lottjudgementlabel{C}{%
\lottjudgement{Concat}{\strength₀ \, \lottsym{\mathcolor{Concat}{⊕}} \, \strength₁ \, \lottsym{\mathcolor{Concat}{↬}} \, \strength₂}{
\lottinferencerule{strongR}{
}{\lottsym{:} \, \lottsym{\mathcolor{Concat}{⊕}} \, \lottsym{▹} \, \lottsym{\mathcolor{Concat}{↬}} \, \lottsym{▹}}
\lottinferencerulesep
\lottinferencerule{later}{
}{\lottsym{?} \, \lottsym{\mathcolor{Concat}{⊕}} \, \strength \, \lottsym{\mathcolor{Concat}{↬}} \, \strength}
\lottinferencerulesep
\lottinferencerule{leave}{
}{\lottsym{:} \, \lottsym{\mathcolor{Concat}{⊕}} \, \lottsym{?} \, \lottsym{\mathcolor{Concat}{↬}} \, \lottsym{:}}
\lottinferencerulesep
\lottinferencerule{replace}{
}{\lottsym{:} \, \lottsym{\mathcolor{Concat}{⊕}} \, \lottsym{:} \, \lottsym{\mathcolor{Concat}{↬}} \, \lottsym{:}}
}
}}

\medskip

\lottjudgementlabel{U}{%
\lottjudgement{Uniform}{{\lottsym{⌊} ρ \lottsym{⌋}}^{\strength}}{
\lottinferencerule{empty}{
}{{\lottsym{⌊} \lottsym{ε} \lottsym{⌋}}^{\strength}}
\lottinferencerulesep
\lottinferencerule{ext}{
\lotthypothesis{{\lottsym{⌊} ρ \lottsym{⌋}}^{\strength}}
}{{\lottsym{⌊} ρ \lottsym{,} \, i \, \strength \, A \lottsym{⌋}}^{\strength}}
}
}

\medskip

\lottjudgementforceexample{o \, \lottsym{⊢} \, e \, \lottsym{\longrightarrow} \, e'}{\lottjudgementlabel{E}{\lottjudgementfilterout{opl,opr,op}{%
\lottjudgement{Small Step}{r \, \lottsym{⊢} \, e \ifdefined\store \lottsym{;} \, \mathcolor{Store}{rₛ} \fi \, \lottsym{\longrightarrow} \, e' \ifdefined\store \lottsym{;} \, \mathcolor{Store}{rₛ'} \fi}{
\lottinferencerule{opl}{
\lotthypothesis{o \, \lottsym{⊢} \, e₀ \ifdefined\store \lottsym{;} \, \mathcolor{Store}{oₛ} \fi \, \lottsym{\longrightarrow} \, e₀' \ifdefined\store \lottsym{;} \, \mathcolor{Store}{oₛ'} \fi}
}{o \, \lottsym{⊢} \, e₀ \, b \, e₁ \ifdefined\store \lottsym{;} \, \mathcolor{Store}{oₛ} \fi \, \lottsym{\longrightarrow} \, e₀' \, b \, e₁ \ifdefined\store \lottsym{;} \, \mathcolor{Store}{oₛ'} \fi}
\lottinferencerulesep
\lottinferencerule{opr}{
\lotthypothesis{o \, \lottsym{⊢} \, e \ifdefined\store \lottsym{;} \, \mathcolor{Store}{oₛ} \fi \, \lottsym{\longrightarrow} \, e' \ifdefined\store \lottsym{;} \, \mathcolor{Store}{oₛ'} \fi}
}{o \, \lottsym{⊢} \, v \, b \, e \ifdefined\store \lottsym{;} \, \mathcolor{Store}{oₛ} \fi \, \lottsym{\longrightarrow} \, v \, b \, e' \ifdefined\store \lottsym{;} \, \mathcolor{Store}{oₛ'} \fi}
\lottinferencerulesep
\lottinferencerule{op}{
\lotthypothesis{n₀ \, b \, n₁ \, \lottsym{=} \, n₂}
}{o \, \lottsym{⊢} \, n₀ \, b \, n₁ \ifdefined\store \lottsym{;} \, \mathcolor{Store}{oₛ} \fi \, \lottsym{\longrightarrow} \, n₂ \ifdefined\store \lottsym{;} \, \mathcolor{Store}{oₛ} \fi}
\lottinferencerulesep
\lottinferencerule{lam}{
}{o \, \lottsym{⊢} \, \lottsym{λ} \, \lottsym{\{} \, c \, \lottsym{\}} \lottsym{.} \, e \ifdefined\store \lottsym{;} \, \mathcolor{Store}{oₛ} \fi \, \lottsym{\longrightarrow} \, \lottsym{⟨} o \lottsym{;} \, \lottsym{ε} \, \lottsym{|} \, \lottsym{λ} \, \lottsym{\{} \, c \, \lottsym{\}} \lottsym{.} \, e \lottsym{⟩} \ifdefined\store \lottsym{;} \, \mathcolor{Store}{oₛ} \fi}
\lottinferencerulesep
\lottinferencerule{appl}{
\lotthypothesis{o \, \lottsym{⊢} \, e₀ \ifdefined\store \lottsym{;} \, \mathcolor{Store}{oₛ} \fi \, \lottsym{\longrightarrow} \, e₀' \ifdefined\store \lottsym{;} \, \mathcolor{Store}{oₛ'} \fi}
}{o \, \lottsym{⊢} \, e₀ \, e₁ \ifdefined\store \lottsym{;} \, \mathcolor{Store}{oₛ} \fi \, \lottsym{\longrightarrow} \, e₀' \, e₁ \ifdefined\store \lottsym{;} \, \mathcolor{Store}{oₛ'} \fi}
\lottinferencerulesep
\lottinferencerule{appr}{
\lotthypothesis{o \, \lottsym{⊢} \, e \ifdefined\store \lottsym{;} \, \mathcolor{Store}{oₛ} \fi \, \lottsym{\longrightarrow} \, e' \ifdefined\store \lottsym{;} \, \mathcolor{Store}{oₛ'} \fi}
}{o \, \lottsym{⊢} \, v \, e \ifdefined\store \lottsym{;} \, \mathcolor{Store}{oₛ} \fi \, \lottsym{\longrightarrow} \, v \, e' \ifdefined\store \lottsym{;} \, \mathcolor{Store}{oₛ'} \fi}
\lottinferencerulesep
\lottinferencerule{app}{
}{o₀ \, \lottsym{⊢} \, \lottsym{⟨} o₁ \lottsym{;} \, o₂ \, \lottsym{|} \, \lottsym{λ} \, \lottsym{\{} \, c \, \lottsym{\}} \lottsym{.} \, e \lottsym{⟩} \, \lottsym{\{} \, o₃ \, \lottsym{\}} \ifdefined\store \lottsym{;} \, \mathcolor{Store}{oₛ} \fi \, \lottsym{\longrightarrow} \, \lottsym{⟨} o₁ \lottsym{;} \, o₂ \, \lottsym{⧺} \, o₃ \, \lottsym{|} \, \lottsym{λ} \, \lottsym{\{} \, c \, \lottsym{\}} \lottsym{.} \, e \lottsym{⟩} \ifdefined\store \lottsym{;} \, \mathcolor{Store}{oₛ} \fi}
\lottinferencerulesep
\lottinferencerule{force}{
\lotthypothesis{o \, \lottsym{⊢} \, e \ifdefined\store \lottsym{;} \, \mathcolor{Store}{oₛ} \fi \, \lottsym{\longrightarrow} \, e' \ifdefined\store \lottsym{;} \, \mathcolor{Store}{oₛ'} \fi}
}{o \, \lottsym{⊢} \, \lottkw{force} \, e \ifdefined\store \lottsym{;} \, \mathcolor{Store}{oₛ} \fi \, \lottsym{\longrightarrow} \, \lottkw{force} \, e' \ifdefined\store \lottsym{;} \, \mathcolor{Store}{oₛ'} \fi}
\lottinferencerulesep
\lottinferencerule{forceClosure}{
\lotthypothesis{\lottkw{expand} \lottsym{(} c \lottsym{;} \, o₂ \lottsym{)} \, \lottsym{=} \, o₃}
}{o₀ \, \lottsym{⊢} \, \lottkw{force} \, \lottsym{⟨} o₁ \lottsym{;} \, o₂ \, \lottsym{|} \, \lottsym{λ} \, \lottsym{\{} \, c \, \lottsym{\}} \lottsym{.} \, e \lottsym{⟩} \ifdefined\store \lottsym{;} \, \mathcolor{Store}{oₛ} \fi \, \lottsym{\longrightarrow} \, \lottsym{\{} \, o₁ \lottsym{,} \, o₃ \, \lottsym{\}} \, \lottsym{▶} \, e \ifdefined\store \lottsym{;} \, \mathcolor{Store}{oₛ} \fi}
\lottinferencerulesep
\lottinferencerule{boxl}{
\lotthypothesis{o \, \lottsym{⊢} \, e₀ \ifdefined\store \lottsym{;} \, \mathcolor{Store}{oₛ} \fi \, \lottsym{\longrightarrow} \, e₀' \ifdefined\store \lottsym{;} \, \mathcolor{Store}{oₛ'} \fi}
}{o \, \lottsym{⊢} \, e₀ \, \lottsym{▶} \, e₁ \ifdefined\store \lottsym{;} \, \mathcolor{Store}{oₛ} \fi \, \lottsym{\longrightarrow} \, e₀' \, \lottsym{▶} \, e₁ \ifdefined\store \lottsym{;} \, \mathcolor{Store}{oₛ'} \fi}
\lottinferencerulesep
\lottinferencerule{boxr}{
\lotthypothesis{o₁ \, \lottsym{⊢} \, e \ifdefined\store \lottsym{;} \, \mathcolor{Store}{oₛ} \fi \, \lottsym{\longrightarrow} \, e' \ifdefined\store \lottsym{;} \, \mathcolor{Store}{oₛ'} \fi}
}{o₀ \, \lottsym{⊢} \, \lottsym{\{} \, o₁ \, \lottsym{\}} \, \lottsym{▶} \, e \ifdefined\store \lottsym{;} \, \mathcolor{Store}{oₛ} \fi \, \lottsym{\longrightarrow} \, \lottsym{\{} \, o₁ \, \lottsym{\}} \, \lottsym{▶} \, e' \ifdefined\store \lottsym{;} \, \mathcolor{Store}{oₛ'} \fi}
\lottinferencerulesep
\lottinferencerule{box}{
}{o₀ \, \lottsym{⊢} \, \lottsym{\{} \, o₁ \, \lottsym{\}} \, \lottsym{▶} \, v \ifdefined\store \lottsym{;} \, \mathcolor{Store}{oₛ} \fi \, \lottsym{\longrightarrow} \, v \ifdefined\store \lottsym{;} \, \mathcolor{Store}{oₛ} \fi}
\lottinferencerulesep
\lottinferencerule{env}{
}{o \, \lottsym{⊢} \, \lottsym{@} \ifdefined\store \lottsym{;} \, \mathcolor{Store}{oₛ} \fi \, \lottsym{\longrightarrow} \, \lottsym{\{} \, o \, \lottsym{\}} \ifdefined\store \lottsym{;} \, \mathcolor{Store}{oₛ} \fi}
\lottinferencerulesep
\lottinferencerule{extl}{
\lotthypothesis{o \, \lottsym{⊢} \, \lottsym{\{} \, r \, \lottsym{\}} \ifdefined\store \lottsym{;} \, \mathcolor{Store}{oₛ} \fi \, \lottsym{\longrightarrow} \, \lottsym{\{} \, r' \, \lottsym{\}} \ifdefined\store \lottsym{;} \, \mathcolor{Store}{oₛ'} \fi}
}{o \, \lottsym{⊢} \, \lottsym{\{} \, r \lottsym{,} \, i \, \rstrength \, e \, \lottsym{\}} \ifdefined\store \lottsym{;} \, \mathcolor{Store}{oₛ} \fi \, \lottsym{\longrightarrow} \, \lottsym{\{} \, r' \lottsym{,} \, i \, \rstrength \, e \, \lottsym{\}} \ifdefined\store \lottsym{;} \, \mathcolor{Store}{oₛ'} \fi}
\lottinferencerulesep
\lottinferencerule{extl?}{
\lotthypothesis{o \, \lottsym{⊢} \, \lottsym{\{} \, r \, \lottsym{\}} \ifdefined\store \lottsym{;} \, \mathcolor{Store}{oₛ} \fi \, \lottsym{\longrightarrow} \, \lottsym{\{} \, r' \, \lottsym{\}} \ifdefined\store \lottsym{;} \, \mathcolor{Store}{oₛ'} \fi}
}{o \, \lottsym{⊢} \, \lottsym{\{} \, r \lottsym{,} \, i \, \lottsym{?} \, \lottsym{\}} \ifdefined\store \lottsym{;} \, \mathcolor{Store}{oₛ} \fi \, \lottsym{\longrightarrow} \, \lottsym{\{} \, r' \lottsym{,} \, i \, \lottsym{?} \, \lottsym{\}} \ifdefined\store \lottsym{;} \, \mathcolor{Store}{oₛ'} \fi}
\lottinferencerulesep
\lottinferencerule{extr}{
\lotthypothesis{o₀ \, \lottsym{⊢} \, e \ifdefined\store \lottsym{;} \, \mathcolor{Store}{oₛ} \fi \, \lottsym{\longrightarrow} \, e' \ifdefined\store \lottsym{;} \, \mathcolor{Store}{oₛ'} \fi}
}{o₀ \, \lottsym{⊢} \, \lottsym{\{} \, o₁ \lottsym{,} \, i \, \rstrength \, e \, \lottsym{\}} \ifdefined\store \lottsym{;} \, \mathcolor{Store}{oₛ} \fi \, \lottsym{\longrightarrow} \, \lottsym{\{} \, o₁ \lottsym{,} \, i \, \rstrength \, e' \, \lottsym{\}} \ifdefined\store \lottsym{;} \, \mathcolor{Store}{oₛ'} \fi}
\lottinferencerulesep
\lottinferencerule{concatl}{
\lotthypothesis{o \, \lottsym{⊢} \, e₀ \ifdefined\store \lottsym{;} \, \mathcolor{Store}{oₛ} \fi \, \lottsym{\longrightarrow} \, e₀' \ifdefined\store \lottsym{;} \, \mathcolor{Store}{oₛ'} \fi}
}{o \, \lottsym{⊢} \, e₀ \, \lottsym{⧺} \, e₁ \ifdefined\store \lottsym{;} \, \mathcolor{Store}{oₛ} \fi \, \lottsym{\longrightarrow} \, e₀' \, \lottsym{⧺} \, e₁ \ifdefined\store \lottsym{;} \, \mathcolor{Store}{oₛ'} \fi}
\lottinferencerulesep
\lottinferencerule{concatr}{
\lotthypothesis{o \, \lottsym{⊢} \, e \ifdefined\store \lottsym{;} \, \mathcolor{Store}{oₛ} \fi \, \lottsym{\longrightarrow} \, e' \ifdefined\store \lottsym{;} \, \mathcolor{Store}{oₛ'} \fi}
}{o \, \lottsym{⊢} \, v \, \lottsym{⧺} \, e \ifdefined\store \lottsym{;} \, \mathcolor{Store}{oₛ} \fi \, \lottsym{\longrightarrow} \, v \, \lottsym{⧺} \, e' \ifdefined\store \lottsym{;} \, \mathcolor{Store}{oₛ'} \fi}
\lottinferencerulesep
\lottinferencerule{concat}{
}{o₀ \, \lottsym{⊢} \, \lottsym{\{} \, o₁ \, \lottsym{\}} \, \lottsym{⧺} \, \lottsym{\{} \, o₂ \, \lottsym{\}} \ifdefined\store \lottsym{;} \, \mathcolor{Store}{oₛ} \fi \, \lottsym{\longrightarrow} \, \lottsym{\{} \, o₁ \, \lottsym{⧺} \, o₂ \, \lottsym{\}} \ifdefined\store \lottsym{;} \, \mathcolor{Store}{oₛ} \fi}
\lottinferencerulesep
\lottinferencerule{proj}{
\lotthypothesis{o \, \lottsym{⊢} \, e \ifdefined\store \lottsym{;} \, \mathcolor{Store}{oₛ} \fi \, \lottsym{\longrightarrow} \, e' \ifdefined\store \lottsym{;} \, \mathcolor{Store}{oₛ'} \fi}
}{o \, \lottsym{⊢} \, e \lottsym{.} i \ifdefined\store \lottsym{;} \, \mathcolor{Store}{oₛ} \fi \, \lottsym{\longrightarrow} \, e' \lottsym{.} i \ifdefined\store \lottsym{;} \, \mathcolor{Store}{oₛ'} \fi}
\lottinferencerulesep
\lottinferencerule{projRecord}{
\lotthypothesis{\lottkw{proj} \lottsym{(} o₁ \lottsym{;} \, i \lottsym{)} \, \lottsym{=} \, v}
}{o₀ \, \lottsym{⊢} \, \lottsym{\{} \, o₁ \, \lottsym{\}} \lottsym{.} i \ifdefined\store \lottsym{;} \, \mathcolor{Store}{oₛ} \fi \, \lottsym{\longrightarrow} \, v \ifdefined\store \lottsym{;} \, \mathcolor{Store}{oₛ} \fi}
\lottinferencerulesep
\lottinferencerule{mproj}{
\lotthypothesis{o \, \lottsym{⊢} \, e \ifdefined\store \lottsym{;} \, \mathcolor{Store}{oₛ} \fi \, \lottsym{\longrightarrow} \, e' \ifdefined\store \lottsym{;} \, \mathcolor{Store}{oₛ'} \fi}
}{o \, \lottsym{⊢} \, e \lottsym{.} \lottsym{\{} \, \lottsepbycomprehension{{i}_{j}} \, \lottsym{\}} \ifdefined\store \lottsym{;} \, \mathcolor{Store}{oₛ} \fi \, \lottsym{\longrightarrow} \, e' \lottsym{.} \lottsym{\{} \, \lottsepbycomprehension{{i}_{j}} \, \lottsym{\}} \ifdefined\store \lottsym{;} \, \mathcolor{Store}{oₛ'} \fi}
\lottinferencerulesep
\lottinferencerule{mprojRecord}{
\lotthypothesis{\lottkw{mproj} \lottsym{(} o₁ \lottsym{;} \, \lottsepbycomprehension{{i}_{j}} \lottsym{)} \, \lottsym{=} \, o₂}
}{o₀ \, \lottsym{⊢} \, \lottsym{\{} \, o₁ \, \lottsym{\}} \lottsym{.} \lottsym{\{} \, \lottsepbycomprehension{{i}_{j}} \, \lottsym{\}} \ifdefined\store \lottsym{;} \, \mathcolor{Store}{oₛ} \fi \, \lottsym{\longrightarrow} \, \lottsym{\{} \, o₂ \, \lottsym{\}} \ifdefined\store \lottsym{;} \, \mathcolor{Store}{oₛ} \fi}
\lottinferencerulesep
\lottinferencerule{diff}{
\lotthypothesis{o \, \lottsym{⊢} \, e \ifdefined\store \lottsym{;} \, \mathcolor{Store}{oₛ} \fi \, \lottsym{\longrightarrow} \, e' \ifdefined\store \lottsym{;} \, \mathcolor{Store}{oₛ'} \fi}
}{o \, \lottsym{⊢} \, e \lottsym{\textbackslash} \lottsym{\{} \, \lottsepbycomprehension{{i}_{j}} \, \lottsym{\}} \ifdefined\store \lottsym{;} \, \mathcolor{Store}{oₛ} \fi \, \lottsym{\longrightarrow} \, e' \lottsym{\textbackslash} \lottsym{\{} \, \lottsepbycomprehension{{i}_{j}} \, \lottsym{\}} \ifdefined\store \lottsym{;} \, \mathcolor{Store}{oₛ'} \fi}
\lottinferencerulesep
\lottinferencerule{diffRecord}{
\lotthypothesis{\lottkw{diff} \lottsym{(} o₁ \lottsym{;} \, \lottsepbycomprehension{{i}_{j}} \lottsym{)} \, \lottsym{=} \, o₂}
}{o₀ \, \lottsym{⊢} \, \lottsym{\{} \, o₁ \, \lottsym{\}} \lottsym{\textbackslash} \lottsym{\{} \, \lottsepbycomprehension{{i}_{j}} \, \lottsym{\}} \ifdefined\store \lottsym{;} \, \mathcolor{Store}{oₛ} \fi \, \lottsym{\longrightarrow} \, \lottsym{\{} \, o₂ \, \lottsym{\}} \ifdefined\store \lottsym{;} \, \mathcolor{Store}{oₛ} \fi}
\lottinferencerulesep
\lottinferencerule{let}{
\lotthypothesis{o \, \lottsym{⊢} \, e₀ \ifdefined\store \lottsym{;} \, \mathcolor{Store}{oₛ} \fi \, \lottsym{\longrightarrow} \, e₀' \ifdefined\store \lottsym{;} \, \mathcolor{Store}{oₛ'} \fi}
}{o \, \lottsym{⊢} \, \lottkw{let} \, t \, \lottsym{=} \, e₀ \, \lottkw{in} \, e₁ \ifdefined\store \lottsym{;} \, \mathcolor{Store}{oₛ} \fi \, \lottsym{\longrightarrow} \, \lottkw{let} \, t \, \lottsym{=} \, e₀' \, \lottkw{in} \, e₁ \ifdefined\store \lottsym{;} \, \mathcolor{Store}{oₛ'} \fi}
\lottinferencerulesep
\lottinferencerule{letv}{
\lotthypothesis{\lottkw{expand} \lottsym{(} t \lottsym{;} \, v \lottsym{)} \, \lottsym{=} \, o₁}
}{o₀ \, \lottsym{⊢} \, \lottkw{let} \, t \, \lottsym{=} \, v \, \lottkw{in} \, e \ifdefined\store \lottsym{;} \, \mathcolor{Store}{oₛ} \fi \, \lottsym{\longrightarrow} \, \lottsym{\{} \, o₀ \lottsym{,} \, o₁ \, \lottsym{\}} \, \lottsym{▶} \, e \ifdefined\store \lottsym{;} \, \mathcolor{Store}{oₛ} \fi}
\lottinferencerulesep
\lottinferencerule{ref}{
\lotthypothesis{o \, \lottsym{⊢} \, e \ifdefined\store \lottsym{;} \, \mathcolor{Store}{oₛ} \fi \, \lottsym{\longrightarrow} \, e' \ifdefined\store \lottsym{;} \, \mathcolor{Store}{oₛ'} \fi}
}{o \, \lottsym{⊢} \, \lottkw{ref} \, e \ifdefined\store \lottsym{;} \, \mathcolor{Store}{oₛ} \fi \, \lottsym{\longrightarrow} \, \lottkw{ref} \, e' \ifdefined\store \lottsym{;} \, \mathcolor{Store}{oₛ'} \fi}
\lottinferencerulesep
\lottinferencerule{refv}{
\lotthypothesis{l \, \lottsym{∉} \, oₛ}
}{o \, \lottsym{⊢} \, \lottkw{ref} \, v \ifdefined\store \lottsym{;} \, \mathcolor{Store}{oₛ} \fi \, \lottsym{\longrightarrow} \, l \ifdefined\store \lottsym{;} \, \mathcolor{Store}{oₛ \lottsym{,} \, l \, \lottsym{:} \, v} \fi}
\lottinferencerulesep
\lottinferencerule{deref}{
\lotthypothesis{o \, \lottsym{⊢} \, e \ifdefined\store \lottsym{;} \, \mathcolor{Store}{oₛ} \fi \, \lottsym{\longrightarrow} \, e' \ifdefined\store \lottsym{;} \, \mathcolor{Store}{oₛ'} \fi}
}{o \, \lottsym{⊢} \, \lottsym{!} e \ifdefined\store \lottsym{;} \, \mathcolor{Store}{oₛ} \fi \, \lottsym{\longrightarrow} \, \lottsym{!} e' \ifdefined\store \lottsym{;} \, \mathcolor{Store}{oₛ'} \fi}
\lottinferencerulesep
\lottinferencerule{derefLoc}{
}{o \, \lottsym{⊢} \, \lottsym{!} l \ifdefined\store \lottsym{;} \, \mathcolor{Store}{oₛ} \fi \, \lottsym{\longrightarrow} \, \lottsym{\{} \, oₛ \, \lottsym{\}} \lottsym{.} l \ifdefined\store \lottsym{;} \, \mathcolor{Store}{oₛ} \fi}
\lottinferencerulesep
\lottinferencerule{setl}{
\lotthypothesis{o \, \lottsym{⊢} \, e₀ \ifdefined\store \lottsym{;} \, \mathcolor{Store}{oₛ} \fi \, \lottsym{\longrightarrow} \, e₀' \ifdefined\store \lottsym{;} \, \mathcolor{Store}{oₛ'} \fi}
}{o \, \lottsym{⊢} \, e₀ \, \lottsym{≔} \, e₁ \ifdefined\store \lottsym{;} \, \mathcolor{Store}{oₛ} \fi \, \lottsym{\longrightarrow} \, e₀' \, \lottsym{≔} \, e₁ \ifdefined\store \lottsym{;} \, \mathcolor{Store}{oₛ'} \fi}
\lottinferencerulesep
\lottinferencerule{setr}{
\lotthypothesis{o \, \lottsym{⊢} \, e \ifdefined\store \lottsym{;} \, \mathcolor{Store}{oₛ} \fi \, \lottsym{\longrightarrow} \, e' \ifdefined\store \lottsym{;} \, \mathcolor{Store}{oₛ'} \fi}
}{o \, \lottsym{⊢} \, v \, \lottsym{≔} \, e \ifdefined\store \lottsym{;} \, \mathcolor{Store}{oₛ} \fi \, \lottsym{\longrightarrow} \, v \, \lottsym{≔} \, e' \ifdefined\store \lottsym{;} \, \mathcolor{Store}{oₛ'} \fi}
\lottinferencerulesep
\lottinferencerule{set}{
}{o \, \lottsym{⊢} \, l \, \lottsym{≔} \, v \ifdefined\store \lottsym{;} \, \mathcolor{Store}{oₛ} \fi \, \lottsym{\longrightarrow} \, \lottsym{\{} \, \lottsym{\}} \ifdefined\store \lottsym{;} \, \mathcolor{Store}{oₛ \, \lottsym{⧺} \, l \, \lottsym{:} \, v} \fi}
}
}}}

\medskip

\lottjudgementforceexample{o \, \lottsym{⊢} \, e \lottsym{;} \, \mathcolor{Store}{oₛ} \, \lottsym{⇒} \, v \lottsym{;} \, \mathcolor{Store}{oₛ'}}{\lottjudgementlabel{B}{\lottjudgementfilterout{op}{\lottjudgementnarrower[force=0.5,box=0.5,ext=0.5,deref=0.5,set=0.5]{0.3}{%
\lottjudgement{Big Step}{r \, \lottsym{⊢} \, e \ifdefined\store \lottsym{;} \, \mathcolor{Store}{rₛ} \fi \, \lottsym{⇒} \, e' \ifdefined\store \lottsym{;} \, \mathcolor{Store}{rₛ'} \fi}{
\lottinferencerule{nat}{
}{o \, \lottsym{⊢} \, n \ifdefined\store \lottsym{;} \, \mathcolor{Store}{oₛ} \fi \, \lottsym{⇒} \, n \ifdefined\store \lottsym{;} \, \mathcolor{Store}{oₛ} \fi}
\lottinferencerulesep
\lottinferencerule{op}{
\lotthypothesis{o \, \lottsym{⊢} \, e₀ \ifdefined\store \lottsym{;} \, \mathcolor{Store}{oₛ₀} \fi \, \lottsym{⇒} \, n₀ \ifdefined\store \lottsym{;} \, \mathcolor{Store}{oₛ₁} \fi}\\
\lotthypothesis{o \, \lottsym{⊢} \, e₁ \ifdefined\store \lottsym{;} \, \mathcolor{Store}{oₛ₁} \fi \, \lottsym{⇒} \, n₁ \ifdefined\store \lottsym{;} \, \mathcolor{Store}{oₛ₂} \fi}\\
\lotthypothesis{n₀ \, b \, n₁ \, \lottsym{=} \, n}
}{o \, \lottsym{⊢} \, e₀ \, b \, e₁ \ifdefined\store \lottsym{;} \, \mathcolor{Store}{oₛ₀} \fi \, \lottsym{⇒} \, n \ifdefined\store \lottsym{;} \, \mathcolor{Store}{oₛ₂} \fi}
\lottinferencerulesep
\lottinferencerule{lam}{
}{o \, \lottsym{⊢} \, \lottsym{λ} \, \lottsym{\{} \, c \, \lottsym{\}} \lottsym{.} \, e \ifdefined\store \lottsym{;} \, \mathcolor{Store}{oₛ} \fi \, \lottsym{⇒} \, \lottsym{⟨} o \lottsym{;} \, \lottsym{ε} \, \lottsym{|} \, \lottsym{λ} \, \lottsym{\{} \, c \, \lottsym{\}} \lottsym{.} \, e \lottsym{⟩} \ifdefined\store \lottsym{;} \, \mathcolor{Store}{oₛ} \fi}
\lottinferencerulesep
\lottinferencerule{closure}{
}{o₀ \, \lottsym{⊢} \, \lottsym{⟨} o₁ \lottsym{;} \, o₂ \, \lottsym{|} \, \lottsym{λ} \, \lottsym{\{} \, c \, \lottsym{\}} \lottsym{.} \, e \lottsym{⟩} \ifdefined\store \lottsym{;} \, \mathcolor{Store}{oₛ} \fi \, \lottsym{⇒} \, \lottsym{⟨} o₁ \lottsym{;} \, o₂ \, \lottsym{|} \, \lottsym{λ} \, \lottsym{\{} \, c \, \lottsym{\}} \lottsym{.} \, e \lottsym{⟩} \ifdefined\store \lottsym{;} \, \mathcolor{Store}{oₛ} \fi}
\lottinferencerulesep
\lottinferencerule{app}{
\lotthypothesis{o₀ \, \lottsym{⊢} \, e₀ \ifdefined\store \lottsym{;} \, \mathcolor{Store}{oₛ₀} \fi \, \lottsym{⇒} \, \lottsym{⟨} o₁ \lottsym{;} \, o₂ \, \lottsym{|} \, \lottsym{λ} \, \lottsym{\{} \, c \, \lottsym{\}} \lottsym{.} \, e₀' \lottsym{⟩} \ifdefined\store \lottsym{;} \, \mathcolor{Store}{oₛ₁} \fi}\\
\lotthypothesis{o₀ \, \lottsym{⊢} \, e₁ \ifdefined\store \lottsym{;} \, \mathcolor{Store}{oₛ₁} \fi \, \lottsym{⇒} \, \lottsym{\{} \, o₃ \, \lottsym{\}} \ifdefined\store \lottsym{;} \, \mathcolor{Store}{oₛ₂} \fi}
}{o₀ \, \lottsym{⊢} \, e₀ \, e₁ \ifdefined\store \lottsym{;} \, \mathcolor{Store}{oₛ₀} \fi \, \lottsym{⇒} \, \lottsym{⟨} o₁ \lottsym{;} \, o₂ \, \lottsym{⧺} \, o₃ \, \lottsym{|} \, \lottsym{λ} \, \lottsym{\{} \, c \, \lottsym{\}} \lottsym{.} \, e₀' \lottsym{⟩} \ifdefined\store \lottsym{;} \, \mathcolor{Store}{oₛ₂} \fi}
\lottinferencerulesep
\lottinferencerule{force}{
\lotthypothesis{o₀ \, \lottsym{⊢} \, e \ifdefined\store \lottsym{;} \, \mathcolor{Store}{oₛ₀} \fi \, \lottsym{⇒} \, \lottsym{⟨} o₁ \lottsym{;} \, o₂ \, \lottsym{|} \, \lottsym{λ} \, \lottsym{\{} \, c \, \lottsym{\}} \lottsym{.} \, e' \lottsym{⟩} \ifdefined\store \lottsym{;} \, \mathcolor{Store}{oₛ₁} \fi}\\
\lotthypothesis{\lottkw{expand} \lottsym{(} c \lottsym{;} \, o₂ \lottsym{)} \, \lottsym{=} \, o₃}\\
\lotthypothesis{o₁ \lottsym{,} \, o₃ \, \lottsym{⊢} \, e' \ifdefined\store \lottsym{;} \, \mathcolor{Store}{oₛ₁} \fi \, \lottsym{⇒} \, v \ifdefined\store \lottsym{;} \, \mathcolor{Store}{oₛ₂} \fi}
}{o₀ \, \lottsym{⊢} \, \lottkw{force} \, e \ifdefined\store \lottsym{;} \, \mathcolor{Store}{oₛ₀} \fi \, \lottsym{⇒} \, v \ifdefined\store \lottsym{;} \, \mathcolor{Store}{oₛ₂} \fi}
\lottinferencerulesep
\lottinferencerule{box}{
\lotthypothesis{o₀ \, \lottsym{⊢} \, e₀ \ifdefined\store \lottsym{;} \, \mathcolor{Store}{oₛ₀} \fi \, \lottsym{⇒} \, \lottsym{\{} \, o₁ \, \lottsym{\}} \ifdefined\store \lottsym{;} \, \mathcolor{Store}{oₛ₁} \fi}\\
\lotthypothesis{o₁ \, \lottsym{⊢} \, e₁ \ifdefined\store \lottsym{;} \, \mathcolor{Store}{oₛ₁} \fi \, \lottsym{⇒} \, v₁ \ifdefined\store \lottsym{;} \, \mathcolor{Store}{oₛ₂} \fi}
}{o₀ \, \lottsym{⊢} \, e₀ \, \lottsym{▶} \, e₁ \ifdefined\store \lottsym{;} \, \mathcolor{Store}{oₛ₀} \fi \, \lottsym{⇒} \, v₁ \ifdefined\store \lottsym{;} \, \mathcolor{Store}{oₛ₂} \fi}
\lottinferencerulesep
\lottinferencerule{env}{
}{o \, \lottsym{⊢} \, \lottsym{@} \ifdefined\store \lottsym{;} \, \mathcolor{Store}{oₛ} \fi \, \lottsym{⇒} \, \lottsym{\{} \, o \, \lottsym{\}} \ifdefined\store \lottsym{;} \, \mathcolor{Store}{oₛ} \fi}
\lottinferencerulesep
\lottinferencerule{empty}{
}{o \, \lottsym{⊢} \, \lottsym{\{} \, \lottsym{\}} \ifdefined\store \lottsym{;} \, \mathcolor{Store}{oₛ} \fi \, \lottsym{⇒} \, \lottsym{\{} \, \lottsym{\}} \ifdefined\store \lottsym{;} \, \mathcolor{Store}{oₛ} \fi}
\lottinferencerulesep
\lottinferencerule{ext}{
\lotthypothesis{o₀ \, \lottsym{⊢} \, \lottsym{\{} \, r \, \lottsym{\}} \ifdefined\store \lottsym{;} \, \mathcolor{Store}{oₛ₀} \fi \, \lottsym{⇒} \, \lottsym{\{} \, o₁ \, \lottsym{\}} \ifdefined\store \lottsym{;} \, \mathcolor{Store}{oₛ₁} \fi}\\
\lotthypothesis{o₀ \, \lottsym{⊢} \, e \ifdefined\store \lottsym{;} \, \mathcolor{Store}{oₛ₁} \fi \, \lottsym{⇒} \, v \ifdefined\store \lottsym{;} \, \mathcolor{Store}{oₛ₂} \fi}
}{o₀ \, \lottsym{⊢} \, \lottsym{\{} \, r \lottsym{,} \, i \, \rstrength \, e \, \lottsym{\}} \ifdefined\store \lottsym{;} \, \mathcolor{Store}{oₛ₀} \fi \, \lottsym{⇒} \, \lottsym{\{} \, o₁ \lottsym{,} \, i \, \rstrength \, v \, \lottsym{\}} \ifdefined\store \lottsym{;} \, \mathcolor{Store}{oₛ₂} \fi}
\lottinferencerulesep
\lottinferencerule{ext?}{
\lotthypothesis{o₀ \, \lottsym{⊢} \, \lottsym{\{} \, r \, \lottsym{\}} \ifdefined\store \lottsym{;} \, \mathcolor{Store}{oₛ} \fi \, \lottsym{⇒} \, \lottsym{\{} \, o₁ \, \lottsym{\}} \ifdefined\store \lottsym{;} \, \mathcolor{Store}{oₛ'} \fi}
}{o₀ \, \lottsym{⊢} \, \lottsym{\{} \, r \lottsym{,} \, i \, \lottsym{?} \, \lottsym{\}} \ifdefined\store \lottsym{;} \, \mathcolor{Store}{oₛ} \fi \, \lottsym{⇒} \, \lottsym{\{} \, o₁ \lottsym{,} \, i \, \lottsym{?} \, \lottsym{\}} \ifdefined\store \lottsym{;} \, \mathcolor{Store}{oₛ'} \fi}
\lottinferencerulesep
\lottinferencerule{concat}{
\lotthypothesis{o₀ \, \lottsym{⊢} \, e₀ \ifdefined\store \lottsym{;} \, \mathcolor{Store}{oₛ₀} \fi \, \lottsym{⇒} \, \lottsym{\{} \, o₁ \, \lottsym{\}} \ifdefined\store \lottsym{;} \, \mathcolor{Store}{oₛ₁} \fi}\\
\lotthypothesis{o₀ \, \lottsym{⊢} \, e₁ \ifdefined\store \lottsym{;} \, \mathcolor{Store}{oₛ₁} \fi \, \lottsym{⇒} \, \lottsym{\{} \, o₂ \, \lottsym{\}} \ifdefined\store \lottsym{;} \, \mathcolor{Store}{oₛ₂} \fi}
}{o₀ \, \lottsym{⊢} \, e₀ \, \lottsym{⧺} \, e₁ \ifdefined\store \lottsym{;} \, \mathcolor{Store}{oₛ₀} \fi \, \lottsym{⇒} \, \lottsym{\{} \, o₁ \, \lottsym{⧺} \, o₂ \, \lottsym{\}} \ifdefined\store \lottsym{;} \, \mathcolor{Store}{oₛ₂} \fi}
\lottinferencerulesep
\lottinferencerule{proj}{
\lotthypothesis{o₀ \, \lottsym{⊢} \, e \ifdefined\store \lottsym{;} \, \mathcolor{Store}{oₛ} \fi \, \lottsym{⇒} \, \lottsym{\{} \, o₁ \, \lottsym{\}} \ifdefined\store \lottsym{;} \, \mathcolor{Store}{oₛ'} \fi}\\
\lotthypothesis{\lottkw{proj} \lottsym{(} o₁ \lottsym{;} \, i \lottsym{)} \, \lottsym{=} \, v}
}{o₀ \, \lottsym{⊢} \, e \lottsym{.} i \ifdefined\store \lottsym{;} \, \mathcolor{Store}{oₛ} \fi \, \lottsym{⇒} \, v \ifdefined\store \lottsym{;} \, \mathcolor{Store}{oₛ'} \fi}
\lottinferencerulesep
\lottinferencerule{mproj}{
\lotthypothesis{o₀ \, \lottsym{⊢} \, e \ifdefined\store \lottsym{;} \, \mathcolor{Store}{oₛ} \fi \, \lottsym{⇒} \, \lottsym{\{} \, o₁ \, \lottsym{\}} \ifdefined\store \lottsym{;} \, \mathcolor{Store}{oₛ'} \fi}\\
\lotthypothesis{\lottkw{mproj} \lottsym{(} o₁ \lottsym{;} \, \lottsepbycomprehension{{i}_{j}} \lottsym{)} \, \lottsym{=} \, o₂}
}{o₀ \, \lottsym{⊢} \, e \lottsym{.} \lottsym{\{} \, \lottsepbycomprehension{{i}_{j}} \, \lottsym{\}} \ifdefined\store \lottsym{;} \, \mathcolor{Store}{oₛ} \fi \, \lottsym{⇒} \, \lottsym{\{} \, o₂ \, \lottsym{\}} \ifdefined\store \lottsym{;} \, \mathcolor{Store}{oₛ'} \fi}
\lottinferencerulesep
\lottinferencerule{diff}{
\lotthypothesis{o₀ \, \lottsym{⊢} \, e \ifdefined\store \lottsym{;} \, \mathcolor{Store}{oₛ} \fi \, \lottsym{⇒} \, \lottsym{\{} \, o₁ \, \lottsym{\}} \ifdefined\store \lottsym{;} \, \mathcolor{Store}{oₛ'} \fi}\\
\lotthypothesis{\lottkw{diff} \lottsym{(} o₁ \lottsym{;} \, \lottsepbycomprehension{{i}_{j}} \lottsym{)} \, \lottsym{=} \, o₂}
}{o₀ \, \lottsym{⊢} \, e \lottsym{\textbackslash} \lottsym{\{} \, \lottsepbycomprehension{{i}_{j}} \, \lottsym{\}} \ifdefined\store \lottsym{;} \, \mathcolor{Store}{oₛ} \fi \, \lottsym{⇒} \, \lottsym{\{} \, o₂ \, \lottsym{\}} \ifdefined\store \lottsym{;} \, \mathcolor{Store}{oₛ'} \fi}
\lottinferencerulesep
\lottinferencerule{let}{
\lotthypothesis{o₀ \, \lottsym{⊢} \, e₀ \ifdefined\store \lottsym{;} \, \mathcolor{Store}{oₛ₀} \fi \, \lottsym{⇒} \, v₀ \ifdefined\store \lottsym{;} \, \mathcolor{Store}{oₛ₁} \fi}\\
\lotthypothesis{\lottkw{expand} \lottsym{(} t \lottsym{;} \, v₀ \lottsym{)} \, \lottsym{=} \, o₁}\\
\lotthypothesis{o₀ \lottsym{,} \, o₁ \, \lottsym{⊢} \, e₁ \ifdefined\store \lottsym{;} \, \mathcolor{Store}{oₛ₁} \fi \, \lottsym{⇒} \, v₁ \ifdefined\store \lottsym{;} \, \mathcolor{Store}{oₛ₂} \fi}
}{o₀ \, \lottsym{⊢} \, \lottkw{let} \, t \, \lottsym{=} \, e₀ \, \lottkw{in} \, e₁ \ifdefined\store \lottsym{;} \, \mathcolor{Store}{oₛ₀} \fi \, \lottsym{⇒} \, v₁ \ifdefined\store \lottsym{;} \, \mathcolor{Store}{oₛ₂} \fi}
\lottinferencerulesep
\lottinferencerule{loc}{
}{o \, \lottsym{⊢} \, l \ifdefined\store \lottsym{;} \, \mathcolor{Store}{oₛ} \fi \, \lottsym{⇒} \, l \ifdefined\store \lottsym{;} \, \mathcolor{Store}{oₛ} \fi}
\lottinferencerulesep
\lottinferencerule{ref}{
\lotthypothesis{o \, \lottsym{⊢} \, e \ifdefined\store \lottsym{;} \, \mathcolor{Store}{oₛ} \fi \, \lottsym{⇒} \, v \ifdefined\store \lottsym{;} \, \mathcolor{Store}{oₛ'} \fi}\\
\lotthypothesis{l \, \lottsym{∉} \, oₛ'}
}{o \, \lottsym{⊢} \, \lottkw{ref} \, e \ifdefined\store \lottsym{;} \, \mathcolor{Store}{oₛ} \fi \, \lottsym{⇒} \, l \ifdefined\store \lottsym{;} \, \mathcolor{Store}{oₛ' \lottsym{,} \, l \, \lottsym{:} \, v} \fi}
\lottinferencerulesep
\lottinferencerule{deref}{
\lotthypothesis{o \, \lottsym{⊢} \, e \ifdefined\store \lottsym{;} \, \mathcolor{Store}{oₛ} \fi \, \lottsym{⇒} \, l \ifdefined\store \lottsym{;} \, \mathcolor{Store}{oₛ'} \fi}\\
\lotthypothesis{\lottkw{proj} \lottsym{(} oₛ' \lottsym{;} \, l \lottsym{)} \, \lottsym{=} \, v}
}{o \, \lottsym{⊢} \, \lottsym{!} e \ifdefined\store \lottsym{;} \, \mathcolor{Store}{oₛ} \fi \, \lottsym{⇒} \, v \ifdefined\store \lottsym{;} \, \mathcolor{Store}{oₛ'} \fi}
\lottinferencerulesep
\lottinferencerule{set}{
\lotthypothesis{o \, \lottsym{⊢} \, e₀ \ifdefined\store \lottsym{;} \, \mathcolor{Store}{oₛ₀} \fi \, \lottsym{⇒} \, l \ifdefined\store \lottsym{;} \, \mathcolor{Store}{oₛ₁} \fi}\\
\lotthypothesis{o \, \lottsym{⊢} \, e₁ \ifdefined\store \lottsym{;} \, \mathcolor{Store}{oₛ₁} \fi \, \lottsym{⇒} \, v \ifdefined\store \lottsym{;} \, \mathcolor{Store}{oₛ₂} \fi}
}{o \, \lottsym{⊢} \, e₀ \, \lottsym{≔} \, e₁ \ifdefined\store \lottsym{;} \, \mathcolor{Store}{oₛ₀} \fi \, \lottsym{⇒} \, \lottsym{\{} \, \lottsym{\}} \ifdefined\store \lottsym{;} \, \mathcolor{Store}{oₛ₂ \, \lottsym{⧺} \, l \, \lottsym{:} \, v} \fi}
}
}}}}

\newcommand{\expand}[2]{\lottkw{expand} \lottsym{(} #1 \lottsym{;} \, #2 \lottsym{)}}
\newcommand{\proj}[2]{\lottkw{proj} \lottsym{(} #1 \lottsym{;} \, #2 \lottsym{)}}
\newcommand{\mproj}[2]{\lottkw{mproj} \lottsym{(} #1 \lottsym{;} \, #2 \lottsym{)}}
\newcommand{\diff}[2]{\lottkw{diff} \lottsym{(} #1 \lottsym{;} \, #2 \lottsym{)}}
\begin{align*}
  & \boxed{\expand{c}{o}}
  & \expand{\lottsym{ε}}{\lottsym{ε}} &= \lottsym{ε} \\
  && \expand{c \lottsym{,} \, i \, \lottsym{◇} \, t}{o₀ \lottsym{,} \, i \, \rstrength \, v \lottsym{,} \, o₁} &= \expand{c}{o₀ \lottsym{,} \, o₁} \lottsym{,} \, \expand{t}{v} & \tag{if $i \, \lottsym{∉} \, o₁$} \\
  \\
  & \boxed{\expand{t}{v}}
  & \expand{i}{v} &= i \, \lottsym{:} \, v \\
  && \expand{\lottsym{\{} \, c \, \lottsym{\}}}{\lottsym{\{} \, o \, \lottsym{\}}} &= \expand{c}{o} \\
  \\
  & \boxed{\proj{o}{i}}
  & \proj{o₀ \lottsym{,} \, i \, \rstrength \, v \lottsym{,} \, o₁}{i} &= v & \tag{if $i \, \lottsym{∉} \, o₁$} \\
  \\
  & \boxed{\mproj{o}{\lottsepbycomprehension{i}}}
  & \mproj{o}{\lottsym{ε}} &= \lottsym{ε} \\
  && \mproj{o₀ \lottsym{,} \, i \, \rstrength \, v \lottsym{,} \, o₁}{\lottsepbycomprehension{i'} \lottsym{,} \, i} &= \mproj{o₀ \lottsym{,} \, o₁}{\lottsepbycomprehension{i'}} \lottsym{,} \, i \, \rstrength \, v & \tag{if $i \, \lottsym{∉} \, o₁$} \\
  \\
  & \boxed{\diff{o}{\lottsepbycomprehension{i}}}
  & \diff{o}{\lottsym{ε}} &= o \\
  && \diff{o₀ \lottsym{,} \, i \, \rstrength \, v \lottsym{,} \, o₁}{\lottsepbycomprehension{i'} \lottsym{,} \, i} &= \diff{o₀ \lottsym{,} \, o₁}{\lottsepbycomprehension{i'}} & \tag{if $i \, \lottsym{∉} \, o₁$} \\
  \\
  & \boxed{o₀ \, \lottsym{\doubleplus} \, o₁}
  & \lottsym{ε} \, \lottsym{\doubleplus} \, o &= o \\
  && i \, \lottsym{=} \, v \lottsym{,} \, o₀ \, \lottsym{\doubleplus} \, o₁ &= i \, \lottsym{=} \, v \lottsym{,} \, \lottsym{(} o₀ \, \lottsym{\doubleplus} \, o₁ \lottsym{)} \\
  && i \, \lottsym{:} \, v₀ \lottsym{,} \, o₀ \, \lottsym{\doubleplus} \, o₁ \lottsym{,} \, i \, \rstrength \, v₁ \lottsym{,} \, o₂ &= i \, \rstrength \, v₁ \lottsym{,} \, \lottsym{(} o₀ \, \lottsym{\doubleplus} \, o₁ \lottsym{,} \, o₂ \lottsym{)} & \tag{if $i \, \lottsym{∉} \, o₁$} \\
  && i \, \lottsym{:} \, v \lottsym{,} \, o₀ \, \lottsym{\doubleplus} \, o₁ \lottsym{,} \, i \, \lottsym{?} \lottsym{,} \, o₂ &= i \, \lottsym{:} \, v \lottsym{,} \, \lottsym{(} o₀ \, \lottsym{\doubleplus} \, o₁ \lottsym{,} \, o₂ \lottsym{)} & \tag{if $i \, \lottsym{∉} \, o₁$} \\
  && i \, \lottsym{:} \, v \lottsym{,} \, o₀ \, \lottsym{\doubleplus} \, o₁ &= i \, \lottsym{:} \, v \lottsym{,} \, \lottsym{(} o₀ \, \lottsym{\doubleplus} \, o₁ \lottsym{)} & \tag{if $i \, \lottsym{∉} \, o₁$} \\
  && i \, \lottsym{?} \lottsym{,} \, o₀ \, \lottsym{\doubleplus} \, o₁ \lottsym{,} \, i \, \rstrength \, v \lottsym{,} \, o₂ &= i \, \rstrength \, v \lottsym{,} \, \lottsym{(} o₀ \, \lottsym{\doubleplus} \, o₁ \lottsym{,} \, o₂ \lottsym{)} & \tag{if $i \, \lottsym{∉} \, o₁$} \\
  && i \, \lottsym{?} \lottsym{,} \, o₀ \, \lottsym{\doubleplus} \, o₁ \lottsym{,} \, i \, \lottsym{?} \lottsym{,} \, o₂ &= i \, \lottsym{?} \lottsym{,} \, \lottsym{(} o₀ \, \lottsym{\doubleplus} \, o₁ \lottsym{,} \, o₂ \lottsym{)} & \tag{if $i \, \lottsym{∉} \, o₁$} \\
  && i \, \lottsym{?} \lottsym{,} \, o₀ \, \lottsym{\doubleplus} \, o₁ &= i \, \lottsym{?} \lottsym{,} \, \lottsym{(} o₀ \, \lottsym{\doubleplus} \, o₁ \lottsym{)} & \tag{if $i \, \lottsym{∉} \, o₁$} \\
\end{align*}

\section{\lang Syntax-Directed Typing}

The full collection of syntax-directed typing rules are as follows:

\medskip

\lottjudgementforcename{Any Instantiation}{\lottjudgementlabel{ST}{%
\lottjudgement{Instantiate Any}{\ifdefined\store \mathcolor{Store}{ρₛ} \lottsym{;} \, \fi ρ \, \lottsym{⊢}_s \, e \, \lottsym{::} \, A}{
\lottinferencerule{instantiate}{
\lotthypothesis{\ifdefined\store \mathcolor{Store}{ρₛ} \lottsym{;} \, \fi ρ \, \lottsym{⊢}_s \, e \, \lottsym{:} \, \lottsym{∀} \, \lottsepbycomprehension{{a}_{j}} \lottsym{.} \, {a}_{k}}
}{\ifdefined\store \mathcolor{Store}{ρₛ} \lottsym{;} \, \fi ρ \, \lottsym{⊢}_s \, e \, \lottsym{::} \, τ}
\lottinferencerulesep
\lottinferencerule{ignore}{
\lotthypothesis{\ifdefined\store \mathcolor{Store}{ρₛ} \lottsym{;} \, \fi ρ \, \lottsym{⊢}_s \, e \, \lottsym{:} \, A}\\\lottsym{∀} a \, k, A \, \lottsym{≠} \, \lottsym{∀} \, \lottsepbycomprehension{{a}_{j}} \lottsym{.} \, a_k
}{\ifdefined\store \mathcolor{Store}{ρₛ} \lottsym{;} \, \fi ρ \, \lottsym{⊢}_s \, e \, \lottsym{::} \, A}
}
}}

\medskip

\lottjudgementforcename{Syntax-Directed Typing}{\lottjudgementlabel{ST}{\lottjudgementfilterout{op}{\lottjudgementnarrower[lam=1,concat=0.65,mproj=1,diff=1]{0.5}{%
\lottjudgement{Syntax Directed}{\ifdefined\store \mathcolor{Store}{ρₛ} \lottsym{;} \, \fi ρ \, \lottsym{⊢}_s \, e \, \lottsym{:} \, A}{
\lottinferencerule{nat}{
}{\ifdefined\store \mathcolor{Store}{ρₛ} \lottsym{;} \, \fi ρ \, \lottsym{⊢}_s \, n \, \lottsym{:} \, \lottsym{ℕ}}
\lottinferencerulesep
\lottinferencerule{op}{
\lotthypothesis{\ifdefined\store \mathcolor{Store}{ρₛ} \lottsym{;} \, \fi ρ \, \lottsym{⊢}_s \, e₀ \, \lottsym{:} \, A}\\
\lotthypothesis{A \, \lottsym{\mathcolor{Subtyping}{⩽}} \, \lottsym{ℕ}}\\
\lotthypothesis{\ifdefined\store \mathcolor{Store}{ρₛ} \lottsym{;} \, \fi ρ \, \lottsym{⊢}_s \, e₁ \, \lottsym{:} \, B}\\
\lotthypothesis{B \, \lottsym{\mathcolor{Subtyping}{⩽}} \, \lottsym{ℕ}}
}{\ifdefined\store \mathcolor{Store}{ρₛ} \lottsym{;} \, \fi ρ \, \lottsym{⊢}_s \, e₀ \, b \, e₁ \, \lottsym{:} \, \lottsym{ℕ}}
\lottinferencerulesep
\lottinferencerule{lam}{\ifdefined\store\overline{a_j} \, \lottsym{∉} \, \lottkw{ftv} \lottsym{(} \mathcolor{Store}{ρₛ} \lottsym{)}\\\fi\overline{a_j} \, \lottsym{∉} \, \lottkw{ftv} \lottsym{(} ρ₀ \lottsym{)}\\
\lotthypothesis{c \, \lottsym{\mathcolor{Pattern}{:}} \, ρ₁ \, \lottsym{\mathcolor{Pattern}{⊙}} \, ρ₂}\\
\lotthypothesis{\ifdefined\store \mathcolor{Store}{ρₛ} \lottsym{;} \, \fi ρ₀ \lottsym{,} \, ρ₂ \, \lottsym{⊢}_s \, e \, \lottsym{:} \, A}
}{\ifdefined\store \mathcolor{Store}{ρₛ} \lottsym{;} \, \fi ρ₀ \, \lottsym{⊢}_s \, \lottsym{λ} \, \lottsym{\{} \, c \, \lottsym{\}} \lottsym{.} \, e \, \lottsym{:} \, \lottsym{∀} \, \lottsepbycomprehension{{a}_{j}} \lottsym{.} \, \lottsym{\{} \, ρ₁ \, \lottsym{\}} \, \lottsym{→} \, A}
\lottinferencerulesep
\lottinferencerule{closure}{\ifdefined\store\overline{a_j} \, \lottsym{∉} \, \lottkw{ftv} \lottsym{(} \mathcolor{Store}{ρₛ} \lottsym{)}\\\fi
\lotthypothesis{\ifdefined\store \mathcolor{Store}{ρₛ} \lottsym{;} \, \fi \lottsym{ε} \, \lottsym{⊢}_s \, \lottsym{\{} \, o₀ \, \lottsym{\}} \, \lottsym{:} \, A}\\
\lotthypothesis{A \, \lottsym{\mathcolor{Subtyping}{⩽}} \, \lottsym{\{} \, ρ₁ \, \lottsym{\}}}\\
\lotthypothesis{\lottsym{⌊} ρ₁ \lottsym{⌋}}\\\overline{a_j} \, \lottsym{∉} \, \lottkw{ftv} \lottsym{(} ρ₁ \lottsym{)}\\
\lotthypothesis{\ifdefined\store \mathcolor{Store}{ρₛ} \lottsym{;} \, \fi \lottsym{ε} \, \lottsym{⊢}_s \, \lottsym{\{} \, o₁ \, \lottsym{\}} \, \lottsym{:} \, B}\\
\lotthypothesis{B \, \lottsym{\mathcolor{Subtyping}{⩽}} \, \lottsym{∀} \, \lottsepbycomprehension{{a}_{j}} \lottsym{.} \, \lottsym{\{} \, ρ₂ \, \lottsym{\}}}\\
\lotthypothesis{c \, \lottsym{\mathcolor{Pattern}{:}} \, ρ₃ \, \lottsym{\mathcolor{Pattern}{⊙}} \, ρ₄}\\
\lotthypothesis{ρ₃ \, \lottsym{\mathcolor{App}{●}} \, ρ₂ \, \lottsym{\mathcolor{App}{↠}} \, ρ₅}\\
\lotthypothesis{\ifdefined\store \mathcolor{Store}{ρₛ} \lottsym{;} \, \fi ρ₁ \lottsym{,} \, ρ₄ \, \lottsym{⊢}_s \, e \, \lottsym{:} \, C}
}{\ifdefined\store \mathcolor{Store}{ρₛ} \lottsym{;} \, \fi ρ₀ \, \lottsym{⊢}_s \, \lottsym{⟨} o₀ \lottsym{;} \, o₁ \, \lottsym{|} \, \lottsym{λ} \, \lottsym{\{} \, c \, \lottsym{\}} \lottsym{.} \, e \lottsym{⟩} \, \lottsym{:} \, \lottsym{∀} \, \lottsepbycomprehension{{a}_{j}} \lottsym{.} \, \lottsym{\{} \, ρ₅ \, \lottsym{\}} \, \lottsym{→} \, C}
\lottinferencerulesep
\lottinferencerule{app}{\overline{a₂_l} \, \lottsym{∉} \, \lottkw{ftv} \lottsym{(} ρ₁ \lottsym{)}\\\overline{a₂_l} \, \lottsym{∉} \, \lottkw{ftv} \lottsym{(} A \lottsym{)}\\\overline{a₂_l} \, \lottsym{∉} \, \lottkw{ftv} \lottsym{(} ρ₂ \lottsym{)}\\
\lotthypothesis{\ifdefined\store \mathcolor{Store}{ρₛ} \lottsym{;} \, \fi ρ₀ \, \lottsym{⊢}_s \, e₀ \, \lottsym{:} \, \lottsym{∀} \, \lottsepbycomprehension{{a₀}_{j}} \lottsym{.} \, \lottsym{\{} \, ρ₁ \, \lottsym{\}} \, \lottsym{→} \, A}\\
\lotthypothesis{\ifdefined\store \mathcolor{Store}{ρₛ} \lottsym{;} \, \fi ρ₀ \, \lottsym{⊢}_s \, e₁ \, \lottsym{:} \, \lottsym{∀} \, \lottsepbycomprehension{{a₁}_{k}} \lottsym{.} \, \lottsym{\{} \, ρ₂ \, \lottsym{\}}}\\
\lotthypothesis{ρ₁\left[\lottsepbycomprehension{{τ₀}_j/{a₀}_{j}}\right] \, \lottsym{\mathcolor{App}{●}}_s \, ρ₂\left[\lottsepbycomprehension{{τ₁}_j/{a₁}_{k}}\right] \, \lottsym{\mathcolor{App}{↠}} \, ρ₃}
}{\ifdefined\store \mathcolor{Store}{ρₛ} \lottsym{;} \, \fi ρ₀ \, \lottsym{⊢}_s \, e₀ \, e₁ \, \lottsym{:} \, \lottsym{∀} \, \lottsepbycomprehension{{a₂}_{l}} \lottsym{.} \, \lottsym{\{} \, ρ₃ \, \lottsym{\}} \, \lottsym{→} \, A\left[\lottsepbycomprehension{{τ₀}_j/{a₀}_{j}}\right]}
\lottinferencerulesep
\lottinferencerule{appAnyL}{\overline{a₂_l} \, \lottsym{∉} \, \lottkw{ftv} \lottsym{(} ρ₂ \lottsym{)}\\
\lotthypothesis{\ifdefined\store \mathcolor{Store}{ρₛ} \lottsym{;} \, \fi ρ₀ \, \lottsym{⊢}_s \, e₀ \, \lottsym{:} \, \lottsym{∀} \, \lottsepbycomprehension{{a₀}_{j}} \lottsym{.} \, {a}_{m}}\\
\lotthypothesis{\ifdefined\store \mathcolor{Store}{ρₛ} \lottsym{;} \, \fi ρ₀ \, \lottsym{⊢}_s \, e₁ \, \lottsym{:} \, \lottsym{∀} \, \lottsepbycomprehension{{a₁}_{k}} \lottsym{.} \, \lottsym{\{} \, ρ₂ \, \lottsym{\}}}\\
\lotthypothesis{ϱ \, \lottsym{\mathcolor{App}{●}}_s \, ρ₂\left[\lottsepbycomprehension{{τ₁}_j/{a₁}_{k}}\right] \, \lottsym{\mathcolor{App}{↠}} \, ρ₃}
}{\ifdefined\store \mathcolor{Store}{ρₛ} \lottsym{;} \, \fi ρ₀ \, \lottsym{⊢}_s \, e₀ \, e₁ \, \lottsym{:} \, \lottsym{∀} \, \lottsepbycomprehension{{a₂}_{l}} \lottsym{.} \, \lottsym{\{} \, ρ₃ \, \lottsym{\}} \, \lottsym{→} \, τ₀}
\lottinferencerulesep
\lottinferencerule{appAnyR}{\overline{a₂_l} \, \lottsym{∉} \, \lottkw{ftv} \lottsym{(} ρ₁ \lottsym{)}\\\overline{a₂_l} \, \lottsym{∉} \, \lottkw{ftv} \lottsym{(} A \lottsym{)}\\
\lotthypothesis{\ifdefined\store \mathcolor{Store}{ρₛ} \lottsym{;} \, \fi ρ₀ \, \lottsym{⊢}_s \, e₀ \, \lottsym{:} \, \lottsym{∀} \, \lottsepbycomprehension{{a₀}_{j}} \lottsym{.} \, \lottsym{\{} \, ρ₁ \, \lottsym{\}} \, \lottsym{→} \, A}\\
\lotthypothesis{\ifdefined\store \mathcolor{Store}{ρₛ} \lottsym{;} \, \fi ρ₀ \, \lottsym{⊢}_s \, e₁ \, \lottsym{:} \, \lottsym{∀} \, \lottsepbycomprehension{{a₁}_{k}} \lottsym{.} \, {a}_{m}}\\
\lotthypothesis{ρ₁\left[\lottsepbycomprehension{{τ}_j/{a₀}_{j}}\right] \, \lottsym{\mathcolor{App}{●}}_s \, ϱ \, \lottsym{\mathcolor{App}{↠}} \, ρ₃}
}{\ifdefined\store \mathcolor{Store}{ρₛ} \lottsym{;} \, \fi ρ₀ \, \lottsym{⊢}_s \, e₀ \, e₁ \, \lottsym{:} \, \lottsym{∀} \, \lottsepbycomprehension{{a₂}_{l}} \lottsym{.} \, \lottsym{\{} \, ρ₃ \, \lottsym{\}} \, \lottsym{→} \, A\left[\lottsepbycomprehension{{τ}_j/{a₀}_{j}}\right]}
\lottinferencerulesep
\lottinferencerule{appAny}{
\lotthypothesis{\ifdefined\store \mathcolor{Store}{ρₛ} \lottsym{;} \, \fi ρ₀ \, \lottsym{⊢}_s \, e₀ \, \lottsym{:} \, \lottsym{∀} \, \lottsepbycomprehension{{a₀}_{j}} \lottsym{.} \, {a}_{m₀}}\\
\lotthypothesis{\ifdefined\store \mathcolor{Store}{ρₛ} \lottsym{;} \, \fi ρ₀ \, \lottsym{⊢}_s \, e₁ \, \lottsym{:} \, \lottsym{∀} \, \lottsepbycomprehension{{a₁}_{k}} \lottsym{.} \, {a}_{m₁}}\\
\lotthypothesis{ϱ₁ \, \lottsym{\mathcolor{App}{●}}_s \, ϱ₂ \, \lottsym{\mathcolor{App}{↠}} \, ρ₃}
}{\ifdefined\store \mathcolor{Store}{ρₛ} \lottsym{;} \, \fi ρ₀ \, \lottsym{⊢}_s \, e₀ \, e₁ \, \lottsym{:} \, \lottsym{∀} \, \lottsepbycomprehension{{a₂}_{l}} \lottsym{.} \, \lottsym{\{} \, ρ₃ \, \lottsym{\}} \, \lottsym{→} \, τ}
\lottinferencerulesep
\lottinferencerule{force}{
\lotthypothesis{\ifdefined\store \mathcolor{Store}{ρₛ} \lottsym{;} \, \fi ρ₀ \, \lottsym{⊢}_s \, e \, \lottsym{::} \, \lottsym{∀} \, \lottsepbycomprehension{{a}_{j}} \lottsym{.} \, \lottsym{\{} \, ρ₁ \, \lottsym{\}} \, \lottsym{→} \, A}\\
\lotthypothesis{{\lottsym{⌊} ρ₁ \lottsym{⌋}}^{\lottsym{?}}}
}{\ifdefined\store \mathcolor{Store}{ρₛ} \lottsym{;} \, \fi ρ₀ \, \lottsym{⊢}_s \, \lottkw{force} \, e \, \lottsym{:} \, A\left[\lottsepbycomprehension{{τ}_j/{a}_{j}}\right]}
\lottinferencerulesep
\lottinferencerule{box}{
\lotthypothesis{\ifdefined\store \mathcolor{Store}{ρₛ} \lottsym{;} \, \fi ρ₀ \, \lottsym{⊢}_s \, e₀ \, \lottsym{::} \, \lottsym{∀} \, \lottsepbycomprehension{{a}_{j}} \lottsym{.} \, \lottsym{\{} \, ρ₁ \, \lottsym{\}}}\\
\lotthypothesis{\lottsym{⌊} ρ₁ \lottsym{⌋}}\\
\lotthypothesis{\ifdefined\store \mathcolor{Store}{ρₛ} \lottsym{;} \, \fi ρ₁\left[\lottsepbycomprehension{{τ}_j/{a}_{j}}\right] \, \lottsym{⊢}_s \, e₁ \, \lottsym{:} \, A}
}{\ifdefined\store \mathcolor{Store}{ρₛ} \lottsym{;} \, \fi ρ₀ \, \lottsym{⊢}_s \, e₀ \, \lottsym{▶} \, e₁ \, \lottsym{:} \, A}
\lottinferencerulesep
\lottinferencerule{env}{
}{\ifdefined\store \mathcolor{Store}{ρₛ} \lottsym{;} \, \fi ρ \, \lottsym{⊢}_s \, \lottsym{@} \, \lottsym{:} \, \lottsym{\{} \, ρ \, \lottsym{\}}}
\lottinferencerulesep
\lottinferencerule{record}{
\lotthypothesis{\ifdefined\store \mathcolor{Store}{ρₛ} \lottsym{;} \, \fi ρ₀ \, \lottsym{⊢}_s \, r \, \lottsym{:} \, ρ₁}
}{\ifdefined\store \mathcolor{Store}{ρₛ} \lottsym{;} \, \fi ρ₀ \, \lottsym{⊢}_s \, \lottsym{\{} \, r \, \lottsym{\}} \, \lottsym{:} \, \lottsym{\{} \, ρ₁ \, \lottsym{\}}}
\lottinferencerulesep
\lottinferencerule{concat}{
\lotthypothesis{\ifdefined\store \mathcolor{Store}{ρₛ} \lottsym{;} \, \fi ρ₀ \, \lottsym{⊢}_s \, e₀ \, \lottsym{::} \, \lottsym{∀} \, \lottsepbycomprehension{{a₁}_{j}} \lottsym{.} \, \lottsym{\{} \, ρ₁ \, \lottsym{\}}}\\
\lotthypothesis{\ifdefined\store \mathcolor{Store}{ρₛ} \lottsym{;} \, \fi ρ₀ \, \lottsym{⊢}_s \, e₁ \, \lottsym{::} \, \lottsym{∀} \, \lottsepbycomprehension{{a₂}_{k}} \lottsym{.} \, \lottsym{\{} \, ρ₂ \, \lottsym{\}}}\\
\lotthypothesis{ρ₁\left[\lottsepbycomprehension{{τ₁}_j/{a₁}_{j}}\right] \, \lottsym{\mathcolor{Concat}{⊕}}_s \, ρ₂\left[\lottsepbycomprehension{{τ₂}_j/{a₂}_{k}}\right] \, \lottsym{\mathcolor{Concat}{↬}} \, ρ₃}
}{\ifdefined\store \mathcolor{Store}{ρₛ} \lottsym{;} \, \fi ρ₀ \, \lottsym{⊢}_s \, e₀ \, \lottsym{⧺} \, e₁ \, \lottsym{:} \, \lottsym{\{} \, ρ₃ \, \lottsym{\}}}
\lottinferencerulesep
\lottinferencerule{proj}{
\lotthypothesis{\ifdefined\store \mathcolor{Store}{ρₛ} \lottsym{;} \, \fi ρ₀ \, \lottsym{⊢}_s \, e \, \lottsym{::} \, \lottsym{∀} \, \lottsepbycomprehension{{a}_{j}} \lottsym{.} \, \lottsym{\{} \, ρ₁ \lottsym{,} \, i \, \rstrength \, A \lottsym{,} \, ρ₂ \, \lottsym{\}}}\\
\lotthypothesis{i \, \lottsym{∉} \, ρ₂}
}{\ifdefined\store \mathcolor{Store}{ρₛ} \lottsym{;} \, \fi ρ₀ \, \lottsym{⊢}_s \, e \lottsym{.} i \, \lottsym{:} \, A\left[\lottsepbycomprehension{{τ}_j/{a}_{j}}\right]}
\lottinferencerulesep
\lottinferencerule{mproj}{
\lotthypothesis{\ifdefined\store \mathcolor{Store}{ρₛ} \lottsym{;} \, \fi ρ₀ \, \lottsym{⊢}_s \, e \, \lottsym{::} \, \lottsym{∀} \, \lottsepbycomprehension{{a}_{j}} \lottsym{.} \, \lottsym{\{} \, ρ₁ \lottsym{,} \, \lottsepbycomprehensionpatcollection{{i₀}_{k} \, {\rstrength₀}_{k} \, {A}_{k} \lottsym{,} \, {ρ₂}_{k}}{k}{[:n]} \, \lottsym{\}}}\\
\lotthypothesis{\lottjudgementcomprehensionpatcollection{\lottjudgementcomprehensionpatcollection{{i₀}_{k} \, \lottsym{∉} \, {ρ₂}_{l}}{l}{[:k+1]}}{k}{[:n]}}\\
\lotthypothesis{\lottsepbycomprehensionpatcollection{{i₀}_{k} \, {\rstrength₀}_{k} \, {A}_{k}}{k}{[:n]} \, \lottkw{permutes} \, \lottsepbycomprehensionpatcollection{{i₁}_{k} \, {\rstrength₁}_{k} \, {B}_{k}}{k}{[:n]}}
}{\ifdefined\store \mathcolor{Store}{ρₛ} \lottsym{;} \, \fi ρ₀ \, \lottsym{⊢}_s \, e \lottsym{.} \lottsym{\{} \, \lottsepbycomprehensionpatcollection{{i₁}_{k}}{k}{[:n]} \, \lottsym{\}} \, \lottsym{:} \, \lottsym{\{} \, \lottsepbycomprehensionpatcollection{{i₁}_{k} \, {\rstrength₁}_{k} \, {B}_{k}\left[\lottsepbycomprehension{{τ}_j/{a}_{j}}\right]}{k}{[:n]} \, \lottsym{\}}}
\lottinferencerulesep
\lottinferencerule{diff}{
\lotthypothesis{\ifdefined\store \mathcolor{Store}{ρₛ} \lottsym{;} \, \fi ρ₀ \, \lottsym{⊢}_s \, e \, \lottsym{::} \, \lottsym{∀} \, \lottsepbycomprehension{{a}_{j}} \lottsym{.} \, \lottsym{\{} \, ρ₁ \lottsym{,} \, \lottsepbycomprehensionpatcollection{{i₀}_{k} \, {\rstrength₀}_{k} \, {A}_{k} \lottsym{,} \, {ρ₂}_{k}}{k}{[:n]} \, \lottsym{\}}}\\
\lotthypothesis{\lottjudgementcomprehensionpatcollection{\lottjudgementcomprehensionpatcollection{{i₀}_{k} \, \lottsym{∉} \, {ρ₂}_{l}}{l}{[:k+1]}}{k}{[:n]}}\\
\lotthypothesis{\lottsepbycomprehensionpatcollection{{i₀}_{k} \, {\rstrength₀}_{k} \, {A}_{k}}{k}{[:n]} \, \lottkw{permutes} \, \lottsepbycomprehensionpatcollection{{i₁}_{k} \, {\rstrength₁}_{k} \, {B}_{k}}{k}{[:n]}}
}{\ifdefined\store \mathcolor{Store}{ρₛ} \lottsym{;} \, \fi ρ₀ \, \lottsym{⊢}_s \, e \lottsym{\textbackslash} \lottsym{\{} \, \lottsepbycomprehensionpatcollection{{i₁}_{k}}{k}{[:n]} \, \lottsym{\}} \, \lottsym{:} \, \lottsym{\{} \, ρ₁\left[\lottsepbycomprehension{{τ}_j/{a}_{j}}\right] \lottsym{,} \, \lottsepbycomprehensionpatcollection{{ρ₂}_{k}\left[\lottsepbycomprehension{{τ}_j/{a}_{j}}\right]}{k}{[:n]} \, \lottsym{\}}}
\lottinferencerulesep
\lottinferencerule{let}{
\lotthypothesis{\ifdefined\store \mathcolor{Store}{ρₛ} \lottsym{;} \, \fi ρ₀ \, \lottsym{⊢}_s \, e₀ \, \lottsym{:} \, A}\\
\lotthypothesis{t \, \lottsym{\mathcolor{Pattern}{:}}_s \, A \, \lottsym{\mathcolor{Pattern}{⊙}} \, ρ₁}\\
\lotthypothesis{\ifdefined\store \mathcolor{Store}{ρₛ} \lottsym{;} \, \fi ρ₀ \lottsym{,} \, ρ₁ \, \lottsym{⊢}_s \, e₁ \, \lottsym{:} \, B}
}{\ifdefined\store \mathcolor{Store}{ρₛ} \lottsym{;} \, \fi ρ₀ \, \lottsym{⊢}_s \, \lottkw{let} \, t \, \lottsym{=} \, e₀ \, \lottkw{in} \, e₁ \, \lottsym{:} \, B}
\lottinferencerulesep
\lottinferencerule{loc}{
}{\ifdefined\store \mathcolor{Store}{ρₛ₀ \lottsym{,} \, l \, \lottsym{:} \, A \lottsym{,} \, ρₛ₁} \lottsym{;} \, \fi ρ \, \lottsym{⊢}_s \, l \, \lottsym{:} \, \lottkw{Ref} \, A}
\lottinferencerulesep
\lottinferencerule{ref}{
\lotthypothesis{\ifdefined\store \mathcolor{Store}{ρₛ} \lottsym{;} \, \fi ρ \, \lottsym{⊢}_s \, e \, \lottsym{:} \, A}\\
\lotthypothesis{A \, \lottsym{\mathcolor{Subtyping}{⩽}} \, B}
}{\ifdefined\store \mathcolor{Store}{ρₛ} \lottsym{;} \, \fi ρ \, \lottsym{⊢}_s \, \lottkw{ref} \, e \, \lottsym{:} \, \lottkw{Ref} \, B}
\lottinferencerulesep
\lottinferencerule{deref}{
\lotthypothesis{\ifdefined\store \mathcolor{Store}{ρₛ} \lottsym{;} \, \fi ρ \, \lottsym{⊢}_s \, e \, \lottsym{::} \, \lottsym{∀} \, \lottsepbycomprehension{{a}_{j}} \lottsym{.} \, \lottkw{Ref} \, A}
}{\ifdefined\store \mathcolor{Store}{ρₛ} \lottsym{;} \, \fi ρ \, \lottsym{⊢}_s \, \lottsym{!} e \, \lottsym{:} \, A\left[\lottsepbycomprehension{{τ}_j/{a}_{j}}\right]}
\lottinferencerulesep
\lottinferencerule{set}{
\lotthypothesis{\ifdefined\store \mathcolor{Store}{ρₛ} \lottsym{;} \, \fi ρ \, \lottsym{⊢}_s \, e₀ \, \lottsym{::} \, \lottsym{∀} \, \lottsepbycomprehension{{a}_{j}} \lottsym{.} \, \lottkw{Ref} \, A}\\
\lotthypothesis{\ifdefined\store \mathcolor{Store}{ρₛ} \lottsym{;} \, \fi ρ \, \lottsym{⊢}_s \, e₁ \, \lottsym{:} \, B}\\
\lotthypothesis{B \, \lottsym{\mathcolor{Subtyping}{⩽}} \, A\left[\lottsepbycomprehension{{τ}_j/{a}_{j}}\right]}
}{\ifdefined\store \mathcolor{Store}{ρₛ} \lottsym{;} \, \fi ρ \, \lottsym{⊢}_s \, e₀ \, \lottsym{≔} \, e₁ \, \lottsym{:} \, \lottsym{\{} \, \lottsym{\}}}
}
}}}}

\medskip

\lottjudgementforcename{Record Syntax-Directed Typing}{\lottjudgementlabel{ST}{%
\lottjudgement{Syntax Directed}{\ifdefined\store \mathcolor{Store}{ρₛ} \lottsym{;} \, \fi ρ₀ \, \lottsym{⊢}_s \, r \, \lottsym{:} \, ρ₁}{
\lottinferencerule{empty}{
}{\ifdefined\store \mathcolor{Store}{ρₛ} \lottsym{;} \, \fi ρ \, \lottsym{⊢}_s \, \lottsym{ε} \, \lottsym{:} \, \lottsym{ε}}
\lottinferencerulesep
\lottinferencerule{ext}{
\lotthypothesis{\ifdefined\store \mathcolor{Store}{ρₛ} \lottsym{;} \, \fi ρ₀ \, \lottsym{⊢}_s \, r \, \lottsym{:} \, ρ₁}\\
\lotthypothesis{\ifdefined\store \mathcolor{Store}{ρₛ} \lottsym{;} \, \fi ρ₀ \, \lottsym{⊢}_s \, e \, \lottsym{:} \, A}
}{\ifdefined\store \mathcolor{Store}{ρₛ} \lottsym{;} \, \fi ρ₀ \, \lottsym{⊢}_s \, r \lottsym{,} \, i \, \rstrength \, e \, \lottsym{:} \, ρ₁ \lottsym{,} \, i \, \rstrength \, A}
\lottinferencerulesep
\lottinferencerule{ext?}{
\lotthypothesis{\ifdefined\store \mathcolor{Store}{ρₛ} \lottsym{;} \, \fi ρ₀ \, \lottsym{⊢}_s \, r \, \lottsym{:} \, ρ₁}
}{\ifdefined\store \mathcolor{Store}{ρₛ} \lottsym{;} \, \fi ρ₀ \, \lottsym{⊢}_s \, r \lottsym{,} \, i \, \lottsym{?} \, \lottsym{:} \, ρ₁ \lottsym{,} \, i \, \lottsym{?} \, A}
}
}}

Note that application is covered by four rules, \lottref{ST}{app},
\lottref{ST}{appAnyL}, \lottref{ST}{appAnyR}, and \lottref{ST}{appAny}. This
does not compromise syntax-directedness: the rules differ only in whether the
type inferred for the function and for the argument is an ``any'' type
($\lottsym{∀} \, \lottsepbycomprehension{{a}_{j}} \lottsym{.} \, {a}_{k}$) or not, and these shapes are mutually exclusive, so
at most one rule applies to a given application. The four rules could
equivalently be presented as a single \lottref{ST}{app} whose premise
dispatches on the shapes of the two types through an auxiliary judgement,
much like record type application does on entry strengths; we keep them
separate to match the mechanization.

\lottjudgementforcename{Syntax-Directed Record Pattern Separation}{\lottjudgementlabel{SP}{%
\lottjudgement{With Subtyping}{c \, \lottsym{\mathcolor{Pattern}{:}}_s \, ρ₀ \, \lottsym{\mathcolor{Pattern}{⊙}} \, ρ₁}{
\lottinferencerule{empty}{
}{\lottsym{ε} \, \lottsym{\mathcolor{Pattern}{:}}_s \, \lottsym{ε} \, \lottsym{\mathcolor{Pattern}{⊙}} \, \lottsym{ε}}
\lottinferencerulesep
\lottinferencerule{ext}{
\lotthypothesis{t \, \lottsym{\mathcolor{Pattern}{:}}_s \, A \, \lottsym{\mathcolor{Pattern}{⊙}} \, ρ₁}\\
\lotthypothesis{c \, \lottsym{\mathcolor{Pattern}{:}}_s \, ρ₀₀ \lottsym{,} \, ρ₀₁ \, \lottsym{\mathcolor{Pattern}{⊙}} \, ρ₂}\\
\lotthypothesis{i \, \lottsym{∉} \, ρ₀₀}
}{\lottsym{(} i \, \lottsym{◇} \, t \lottsym{,} \, c \lottsym{)} \, \lottsym{\mathcolor{Pattern}{:}}_s \, \lottsym{(} ρ₀₀ \lottsym{,} \, i \, \lottsym{:} \, A \lottsym{,} \, ρ₀₁ \lottsym{)} \, \lottsym{\mathcolor{Pattern}{⊙}} \, ρ₁ \lottsym{,} \, ρ₂}
}
}}

\medskip

\lottjudgementforcename{Syntax-Directed Pattern Separation}{\lottjudgementlabel{SP}{%
\lottjudgement{With Subtyping}{t \, \lottsym{\mathcolor{Pattern}{:}}_s \, A \, \lottsym{\mathcolor{Pattern}{⊙}} \, ρ}{
\lottinferencerule{label}{
}{i \, \lottsym{\mathcolor{Pattern}{:}}_s \, A \, \lottsym{\mathcolor{Pattern}{⊙}} \, \lottsym{(} i \, \lottsym{:} \, A \lottsym{)}}
\lottinferencerulesep
\lottinferencerule{record}{
\lotthypothesis{c \, \lottsym{\mathcolor{Pattern}{:}}_s \, ρ₀\left[\lottsepbycomprehension{{τ}_j/{a}_{j}}\right] \, \lottsym{\mathcolor{Pattern}{⊙}} \, ρ₁}
}{\lottsym{\{} \, c \, \lottsym{\}} \, \lottsym{\mathcolor{Pattern}{:}}_s \, \lottsym{∀} \, \lottsepbycomprehension{{a}_{j}} \lottsym{.} \, \lottsym{\{} \, ρ₀ \, \lottsym{\}} \, \lottsym{\mathcolor{Pattern}{⊙}} \, ρ₁}
\lottinferencerulesep
\lottinferencerule{recordAny}{
\lotthypothesis{c \, \lottsym{\mathcolor{Pattern}{:}}_s \, ϱ \, \lottsym{\mathcolor{Pattern}{⊙}} \, ρ}
}{\lottsym{\{} \, c \, \lottsym{\}} \, \lottsym{\mathcolor{Pattern}{:}}_s \, \lottsym{∀} \, \lottsepbycomprehension{{a}_{j}} \lottsym{.} \, {a}_{k} \, \lottsym{\mathcolor{Pattern}{⊙}} \, ρ}
}
}}

\medskip

\lottjudgementforcename{Syntax-Directed Record Type App}{\lottjudgementlabel{SA}{%
\lottjudgement{With Subtyping}{ρ₀ \, \lottsym{\mathcolor{App}{●}}_s \, ρ₁ \, \lottsym{\mathcolor{App}{↠}} \, ρ₂}{
\lottinferencerule{empty}{
}{\lottsym{ε} \, \lottsym{\mathcolor{App}{●}}_s \, \lottsym{ε} \, \lottsym{\mathcolor{App}{↠}} \, \lottsym{ε}}
\lottinferencerulesep
\lottinferencerule{weakStrong}{
\lotthypothesis{ρ₀ \, \lottsym{\mathcolor{App}{●}}_s \, ρ₁₀ \lottsym{,} \, ρ₁₁ \, \lottsym{\mathcolor{App}{↠}} \, ρ₂}\\
\lotthypothesis{i \, \lottsym{∉} \, ρ₁₀}\\
\lotthypothesis{B \, \lottsym{\mathcolor{Subtyping}{⩽}} \, A}
}{i \, \lottsym{:} \, A \lottsym{,} \, ρ₀ \, \lottsym{\mathcolor{App}{●}}_s \, ρ₁₀ \lottsym{,} \, i \, \lottsym{▹} \, B \lottsym{,} \, ρ₁₁ \, \lottsym{\mathcolor{App}{↠}} \, ρ₂}
\lottinferencerulesep
\lottinferencerule{optionalStrong}{
\lotthypothesis{ρ₀ \, \lottsym{\mathcolor{App}{●}}_s \, ρ₁₀ \lottsym{,} \, ρ₁₁ \, \lottsym{\mathcolor{App}{↠}} \, ρ₂}\\
\lotthypothesis{i \, \lottsym{∉} \, ρ₁₀}\\
\lotthypothesis{B \, \lottsym{\mathcolor{Subtyping}{⩽}} \, A}
}{i \, \lottsym{?} \, A \lottsym{,} \, ρ₀ \, \lottsym{\mathcolor{App}{●}}_s \, ρ₁₀ \lottsym{,} \, i \, \lottsym{▹} \, B \lottsym{,} \, ρ₁₁ \, \lottsym{\mathcolor{App}{↠}} \, ρ₂}
\lottinferencerulesep
\lottinferencerule{ext}{
\lotthypothesis{ρ₀ \, \lottsym{\mathcolor{App}{●}}_s \, ρ₁₀ \lottsym{,} \, ρ₁₁ \, \lottsym{\mathcolor{App}{↠}} \, ρ₂}\\
\lotthypothesis{i \, \lottsym{∉} \, ρ₁₀}\\
\lotthypothesis{\strength₀ \, \lottsym{\mathcolor{App}{●}}_s \, \strength₁ \, \lottsym{\mathcolor{App}{↠}} \, \strength₂}\\
\lotthypothesis{B \, \lottsym{\mathcolor{Subtyping}{⩽}} \, A}
}{i \, \strength₀ \, A \lottsym{,} \, ρ₀ \, \lottsym{\mathcolor{App}{●}}_s \, ρ₁₀ \lottsym{,} \, i \, \strength₁ \, B \lottsym{,} \, ρ₁₁ \, \lottsym{\mathcolor{App}{↠}} \, i \, \strength₂ \, A \lottsym{,} \, ρ₂}
\lottinferencerulesep
\lottinferencerule{weakExt?}{
\lotthypothesis{ρ₀ \, \lottsym{\mathcolor{App}{●}}_s \, ρ₁ \, \lottsym{\mathcolor{App}{↠}} \, ρ₂}\\
\lotthypothesis{i \, \lottsym{∉} \, ρ₁}
}{i \, \lottsym{:} \, A \lottsym{,} \, ρ₀ \, \lottsym{\mathcolor{App}{●}}_s \, ρ₁ \, \lottsym{\mathcolor{App}{↠}} \, i \, \lottsym{:} \, A \lottsym{,} \, ρ₂}
\lottinferencerulesep
\lottinferencerule{optionalExt?}{
\lotthypothesis{ρ₀ \, \lottsym{\mathcolor{App}{●}}_s \, ρ₁ \, \lottsym{\mathcolor{App}{↠}} \, ρ₂}\\
\lotthypothesis{i \, \lottsym{∉} \, ρ₁}
}{i \, \lottsym{?} \, A \lottsym{,} \, ρ₀ \, \lottsym{\mathcolor{App}{●}}_s \, ρ₁ \, \lottsym{\mathcolor{App}{↠}} \, i \, \lottsym{?} \, A \lottsym{,} \, ρ₂}
}
}}

\medskip

\lottjudgementforcename{Syntax-Directed Strength App}{\lottjudgementlabel{SA}{%
\lottjudgement{With Subtyping}{\strength₀ \, \lottsym{\mathcolor{App}{●}}_s \, \strength₁ \, \lottsym{\mathcolor{App}{↠}} \, \strength₂}{
\lottinferencerule{later}{
}{\lottsym{:} \, \lottsym{\mathcolor{App}{●}}_s \, \lottsym{?} \, \lottsym{\mathcolor{App}{↠}} \, \lottsym{:}}
\lottinferencerulesep
\lottinferencerule{ignore}{
}{\lottsym{:} \, \lottsym{\mathcolor{App}{●}}_s \, \lottsym{:} \, \lottsym{\mathcolor{App}{↠}} \, \lottsym{:}}
\lottinferencerulesep
\lottinferencerule{set}{
}{\lottsym{:} \, \lottsym{\mathcolor{App}{●}}_s \, \lottsym{:} \, \lottsym{\mathcolor{App}{↠}} \, \lottsym{?}}
\lottinferencerulesep
\lottinferencerule{replace}{
}{\lottsym{?} \, \lottsym{\mathcolor{App}{●}}_s \, \lottsym{:} \, \lottsym{\mathcolor{App}{↠}} \, \lottsym{?}}
\lottinferencerulesep
\lottinferencerule{leave}{
}{\lottsym{?} \, \lottsym{\mathcolor{App}{●}}_s \, \lottsym{?} \, \lottsym{\mathcolor{App}{↠}} \, \lottsym{?}}
}
}}

\medskip

\lottjudgementforcename{Syntax-Directed Record Type Concat}{\lottjudgementlabel{SC}{%
\lottjudgement{With Subtyping}{ρ₀ \, \lottsym{\mathcolor{Concat}{⊕}}_s \, ρ₁ \, \lottsym{\mathcolor{Concat}{↬}} \, ρ₂}{
\lottinferencerule{empty}{
}{\lottsym{ε} \, \lottsym{\mathcolor{Concat}{⊕}}_s \, \lottsym{ε} \, \lottsym{\mathcolor{Concat}{↬}} \, \lottsym{ε}}
\lottinferencerulesep
\lottinferencerule{strongL}{
\lotthypothesis{ρ₀ \, \lottsym{\mathcolor{Concat}{⊕}}_s \, ρ₁ \, \lottsym{\mathcolor{Concat}{↬}} \, ρ₂}
}{i \, \lottsym{▹} \, A \lottsym{,} \, ρ₀ \, \lottsym{\mathcolor{Concat}{⊕}}_s \, ρ₁ \, \lottsym{\mathcolor{Concat}{↬}} \, i \, \lottsym{▹} \, A \lottsym{,} \, ρ₂}
\lottinferencerulesep
\lottinferencerule{ext}{
\lotthypothesis{ρ₀ \, \lottsym{\mathcolor{Concat}{⊕}}_s \, ρ₁₀ \lottsym{,} \, ρ₁₁ \, \lottsym{\mathcolor{Concat}{↬}} \, ρ₂}\\
\lotthypothesis{i \, \lottsym{∉} \, ρ₁₀}\\
\lotthypothesis{\strength₀ \, \lottsym{\mathcolor{Concat}{⊕}}_s \, \strength₁ \, \lottsym{\mathcolor{Concat}{↬}} \, \strength₂}\\
\lotthypothesis{A \, \lottsym{\mathcolor{Subtyping}{⩽}} \, C}\\
\lotthypothesis{B \, \lottsym{\mathcolor{Subtyping}{⩽}} \, C}
}{i \, \strength₀ \, A \lottsym{,} \, ρ₀ \, \lottsym{\mathcolor{Concat}{⊕}}_s \, ρ₁₀ \lottsym{,} \, i \, \strength₁ \, B \lottsym{,} \, ρ₁₁ \, \lottsym{\mathcolor{Concat}{↬}} \, i \, \strength₂ \, C \lottsym{,} \, ρ₂}
\lottinferencerulesep
\lottinferencerule{weakExtl?}{
\lotthypothesis{ρ₀ \, \lottsym{\mathcolor{Concat}{⊕}}_s \, ρ₁ \, \lottsym{\mathcolor{Concat}{↬}} \, ρ₂}\\
\lotthypothesis{i \, \lottsym{∉} \, ρ₁}
}{i \, \lottsym{:} \, A \lottsym{,} \, ρ₀ \, \lottsym{\mathcolor{Concat}{⊕}}_s \, ρ₁ \, \lottsym{\mathcolor{Concat}{↬}} \, i \, \lottsym{:} \, A \lottsym{,} \, ρ₂}
\lottinferencerulesep
\lottinferencerule{optionalExtl?}{
\lotthypothesis{ρ₀ \, \lottsym{\mathcolor{Concat}{⊕}}_s \, ρ₁ \, \lottsym{\mathcolor{Concat}{↬}} \, ρ₂}\\
\lotthypothesis{i \, \lottsym{∉} \, ρ₁}\\
\lotthypothesis{\strength \, \lottsym{\mathcolor{Subtyping}{⩽}} \, \lottsym{?}}
}{i \, \strength \, A \lottsym{,} \, ρ₀ \, \lottsym{\mathcolor{Concat}{⊕}}_s \, ρ₁ \, \lottsym{\mathcolor{Concat}{↬}} \, i \, \lottsym{?} \, A \lottsym{,} \, ρ₂}
\lottinferencerulesep
\lottinferencerule{strongExtr?}{
\lotthypothesis{\lottsym{ε} \, \lottsym{\mathcolor{Concat}{⊕}}_s \, ρ₁ \, \lottsym{\mathcolor{Concat}{↬}} \, ρ₂}
}{\lottsym{ε} \, \lottsym{\mathcolor{Concat}{⊕}}_s \, i \, \lottsym{▹} \, A \lottsym{,} \, ρ₁ \, \lottsym{\mathcolor{Concat}{↬}} \, i \, \lottsym{▹} \, A \lottsym{,} \, ρ₂}
\lottinferencerulesep
\lottinferencerule{weakExtr?}{
\lotthypothesis{\lottsym{ε} \, \lottsym{\mathcolor{Concat}{⊕}}_s \, ρ₁ \, \lottsym{\mathcolor{Concat}{↬}} \, ρ₂}
}{\lottsym{ε} \, \lottsym{\mathcolor{Concat}{⊕}}_s \, i \, \lottsym{:} \, A \lottsym{,} \, ρ₁ \, \lottsym{\mathcolor{Concat}{↬}} \, i \, \lottsym{:} \, A \lottsym{,} \, ρ₂}
\lottinferencerulesep
\lottinferencerule{optionalExtr?}{
\lotthypothesis{\lottsym{ε} \, \lottsym{\mathcolor{Concat}{⊕}}_s \, ρ₁ \, \lottsym{\mathcolor{Concat}{↬}} \, ρ₂}\\
\lotthypothesis{\strength \, \lottsym{\mathcolor{Subtyping}{⩽}} \, \lottsym{?}}
}{\lottsym{ε} \, \lottsym{\mathcolor{Concat}{⊕}}_s \, i \, \strength \, A \lottsym{,} \, ρ₁ \, \lottsym{\mathcolor{Concat}{↬}} \, i \, \lottsym{?} \, A \lottsym{,} \, ρ₂}
}
}}

\medskip

\lottjudgementforcename{Syntax-Directed Strength Concat}{\lottjudgementlabel{SC}{%
\lottjudgement{With Subtyping}{\strength₀ \, \lottsym{\mathcolor{Concat}{⊕}}_s \, \strength₁ \, \lottsym{\mathcolor{Concat}{↬}} \, \strength₂}{
\lottinferencerule{strongR}{
}{\lottsym{:} \, \lottsym{\mathcolor{Concat}{⊕}}_s \, \lottsym{▹} \, \lottsym{\mathcolor{Concat}{↬}} \, \lottsym{▹}}
\lottinferencerulesep
\lottinferencerule{later}{
}{\lottsym{?} \, \lottsym{\mathcolor{Concat}{⊕}}_s \, \strength \, \lottsym{\mathcolor{Concat}{↬}} \, \strength}
\lottinferencerulesep
\lottinferencerule{leave}{
}{\lottsym{:} \, \lottsym{\mathcolor{Concat}{⊕}}_s \, \lottsym{?} \, \lottsym{\mathcolor{Concat}{↬}} \, \lottsym{:}}
\lottinferencerulesep
\lottinferencerule{replace}{
}{\lottsym{:} \, \lottsym{\mathcolor{Concat}{⊕}}_s \, \lottsym{:} \, \lottsym{\mathcolor{Concat}{↬}} \, \lottsym{:}}
\lottinferencerulesep
\lottinferencerule{forget}{
}{\lottsym{:} \, \lottsym{\mathcolor{Concat}{⊕}}_s \, \lottsym{?} \, \lottsym{\mathcolor{Concat}{↬}} \, \lottsym{?}}
}
}}

\section{Lambda Calculus Syntax and Semantics}

{\small
\begin{tabularx}{\linewidth}{r c l r X r c l r}
\lottnonterminal{Type}{
\mathcolor{STLC}{\lottalias{τ}}
}{
\lottproduction{arr}{\mathcolor{STLC}{ \mathcolor{STLC}{ τ₀ } \, \lottsym{→} \, \mathcolor{STLC}{ τ₁ } }}
\lottproductionsep
\lottproduction{nat}{\mathcolor{STLC}{ \lottsym{ℕ} }}
}
     &&\lottnonterminalforcename{environment}{%
\lottnonterminal{Environment}{
\mathcolor{STLC}{\lottalias{Γ}}
}{
\lottproduction{empty}{\mathcolor{STLC}{ \lottsym{ε} }}
\lottproductionsep
\lottproduction{ext}{\mathcolor{STLC}{ \mathcolor{STLC}{ Γ } \lottsym{,} \, \mathcolor{STLC}{ x } \, \lottsym{:} \, \mathcolor{STLC}{ τ } }}
}
}\\
\lottnonterminal{Term}{
\mathcolor{STLC}{\lottalias{e}}
}{
\lottproduction{var}{\mathcolor{STLC}{ \mathcolor{STLC}{ x } }}
\lottproductionsep
\lottproduction{lam}{\mathcolor{STLC}{ \lottsym{λ} \, \mathcolor{STLC}{ x } \lottsym{.} \, \mathcolor{STLC}{ e } }}
\lottproductionsep
\lottproduction{app}{\mathcolor{STLC}{ \mathcolor{STLC}{ e₀ } \, \mathcolor{STLC}{ e₁ } }}
\lottproductionsep
\lottproduction{nat}{\mathcolor{STLC}{ n }}
}
     &&\lottnonterminalforcename{translation environment}{%
\lottnonterminal{Trans Environment}{
\lottalias{ι}
}{
\lottproduction{empty}{\lottsym{ε}}
\lottproductionsep
\lottproduction{ext}{ι \lottsym{,} \, \mathcolor{STLC}{ x } \, \lottsym{∼} \, i}
}
}\\
\end{tabularx}
}

\medskip

\lottjudgementforcename{STLC Typing}{\lottjudgementlabel{LT}{%
\lottjudgement{Typing}{\mathcolor{STLC}{ Γ } \, \lottsym{⊢} \, \mathcolor{STLC}{ e } \, \lottsym{:} \, \mathcolor{STLC}{ τ }}{
\lottinferencerule{var}{
\lotthypothesis{\mathcolor{STLC}{ x } \, \lottsym{:} \, \mathcolor{STLC}{ τ } \, \lottsym{∈} \, \mathcolor{STLC}{ Γ }}
}{\mathcolor{STLC}{ Γ } \, \lottsym{⊢} \, \mathcolor{STLC}{ \mathcolor{STLC}{ x } } \, \lottsym{:} \, \mathcolor{STLC}{ τ }}
\lottinferencerulesep
\lottinferencerule{lam}{
\lotthypothesis{\mathcolor{STLC}{ \mathcolor{STLC}{ Γ } \lottsym{,} \, \mathcolor{STLC}{ x } \, \lottsym{:} \, \mathcolor{STLC}{ τ₀ } } \, \lottsym{⊢} \, \mathcolor{STLC}{ \mathcolor{STLC}{ e } } \, \lottsym{:} \, \mathcolor{STLC}{ τ₁ }}
}{\mathcolor{STLC}{ Γ } \, \lottsym{⊢} \, \mathcolor{STLC}{ \lottsym{λ} \, \mathcolor{STLC}{ x } \lottsym{.} \, \mathcolor{STLC}{ e } } \, \lottsym{:} \, \mathcolor{STLC}{ \mathcolor{STLC}{ τ₀ } \, \lottsym{→} \, \mathcolor{STLC}{ τ₁ } }}
\lottinferencerulesep
\lottinferencerule{app}{
\lotthypothesis{\mathcolor{STLC}{ Γ } \, \lottsym{⊢} \, \mathcolor{STLC}{ e₀ } \, \lottsym{:} \, \mathcolor{STLC}{ \mathcolor{STLC}{ τ₀ } \, \lottsym{→} \, \mathcolor{STLC}{ τ₁ } }}\\
\lotthypothesis{\mathcolor{STLC}{ Γ } \, \lottsym{⊢} \, \mathcolor{STLC}{ e₁ } \, \lottsym{:} \, \mathcolor{STLC}{ τ₀ }}
}{\mathcolor{STLC}{ Γ } \, \lottsym{⊢} \, \mathcolor{STLC}{ \mathcolor{STLC}{ e₀ } \, \mathcolor{STLC}{ e₁ } } \, \lottsym{:} \, \mathcolor{STLC}{ τ₁ }}
\lottinferencerulesep
\lottinferencerule{nat}{
}{\mathcolor{STLC}{ Γ } \, \lottsym{⊢} \, \mathcolor{STLC}{ n } \, \lottsym{:} \, \mathcolor{STLC}{ \lottsym{ℕ} }}
}
}}

\lottjudgementforcename{STLC Small-Step}{\lottjudgementlabel{LE}{%
\lottjudgement{Reduction}{\mathcolor{STLC}{ e } \, \lottsym{\longrightarrow} \, \mathcolor{STLC}{ e' }}{
\lottinferencerule{appl}{
\lotthypothesis{\mathcolor{STLC}{ e₀ } \, \lottsym{\longrightarrow} \, \mathcolor{STLC}{ e₀' }}
}{\mathcolor{STLC}{ \mathcolor{STLC}{ e₀ } \, \mathcolor{STLC}{ e₁ } } \, \lottsym{\longrightarrow} \, \mathcolor{STLC}{ \mathcolor{STLC}{ e₀' } \, \mathcolor{STLC}{ e₁ } }}
\lottinferencerulesep
\lottinferencerule{appr}{
\lotthypothesis{\mathcolor{STLC}{ e } \, \lottsym{\longrightarrow} \, \mathcolor{STLC}{ e' }}
}{\mathcolor{STLC}{ \mathcolor{STLC}{ \mathcolor{STLC}{ v } } \, \mathcolor{STLC}{ e } } \, \lottsym{\longrightarrow} \, \mathcolor{STLC}{ \mathcolor{STLC}{ \mathcolor{STLC}{ v } } \, \mathcolor{STLC}{ e' } }}
\lottinferencerulesep
\lottinferencerule{lamApp}{
}{\mathcolor{STLC}{ \mathcolor{STLC}{ \lottsym{(} \mathcolor{STLC}{ \lottsym{λ} \, \mathcolor{STLC}{ x } \lottsym{.} \, \mathcolor{STLC}{ e } } \lottsym{)} } \, \mathcolor{STLC}{ \mathcolor{STLC}{ v } } } \, \lottsym{\longrightarrow} \, \mathcolor{STLC}{ \mathcolor{STLC}{ e }\left[\mathcolor{STLC}{ \mathcolor{STLC}{ v } }/\mathcolor{STLC}{ x }\right] }}
}
}}

\section{Lambda Calculus Term Approximation}

\begin{lottgrammar}
  \lottnonterminalforcename{approximation environment}{%
\lottnonterminal{Approx Environment}{
\lottalias{δ}
}{
\lottproduction{empty}{\lottsym{ε}}
\lottproductionsep
\lottproduction{varIdExt}{δ \lottsym{,} \, \mathcolor{STLC}{ x } \, \lottsym{≈} \, i}
\lottproductionsep
\lottproduction{valueExt}{δ \lottsym{,} \, i \, \lottsym{:} \, v}
}
}
\end{lottgrammar}

\medskip

\lottjudgementforcename{Term Approximation}{\lottjudgementlabel{A}{%
\lottjudgement{Approx}{δ \, \lottsym{⊢} \, \mathcolor{STLC}{ eₛ } \, \lottsym{≈} \, e}{
\lottinferencerule{varBound}{
\lotthypothesis{i \, \lottsym{:} \, v \, \lottsym{∈} \, δ}\\
\lotthypothesis{\lottsym{ε} \, \lottsym{⊢} \, \mathcolor{STLC}{ \mathcolor{STLC}{ vₛ } } \, \lottsym{≈} \, v}
}{δ \, \lottsym{⊢} \, \mathcolor{STLC}{ \mathcolor{STLC}{ vₛ } } \, \lottsym{≈} \, \lottsym{@} \lottsym{.} i}
\lottinferencerulesep
\lottinferencerule{varFree}{
\lotthypothesis{\mathcolor{STLC}{ x } \, \lottsym{≈} \, i \, \lottsym{∈} \, δ}
}{δ \, \lottsym{⊢} \, \mathcolor{STLC}{ \mathcolor{STLC}{ x } } \, \lottsym{≈} \, \lottsym{@} \lottsym{.} i}
\lottinferencerulesep
\lottinferencerule{lam}{
\lotthypothesis{δ \lottsym{,} \, \mathcolor{STLC}{ x } \, \lottsym{≈} \, i \, \lottsym{⊢} \, \mathcolor{STLC}{ \mathcolor{STLC}{ eₛ } } \, \lottsym{≈} \, e}
}{δ \, \lottsym{⊢} \, \mathcolor{STLC}{ \lottsym{λ} \, \mathcolor{STLC}{ x } \lottsym{.} \, \mathcolor{STLC}{ eₛ } } \, \lottsym{≈} \, \lottsym{λ} \, \lottsym{\{} \, \_ \, \lottsym{◇} \, i \, \lottsym{\}} \lottsym{.} \, e}
\lottinferencerulesep
\lottinferencerule{closure}{
\lotthypothesis{\lottsym{‖} o \lottsym{‖} \lottsym{,} \, \mathcolor{STLC}{ x } \, \lottsym{≈} \, i \, \lottsym{⊢} \, \mathcolor{STLC}{ \mathcolor{STLC}{ eₛ } } \, \lottsym{≈} \, e}
}{δ \, \lottsym{⊢} \, \mathcolor{STLC}{ \lottsym{λ} \, \mathcolor{STLC}{ x } \lottsym{.} \, \mathcolor{STLC}{ eₛ } } \, \lottsym{≈} \, \lottsym{⟨} o \lottsym{;} \, \lottsym{ε} \, \lottsym{|} \, \lottsym{λ} \, \lottsym{\{} \, \_ \, \lottsym{◇} \, i \, \lottsym{\}} \lottsym{.} \, e \lottsym{⟩}}
\lottinferencerulesep
\lottinferencerule{closureApp}{
\lotthypothesis{\lottsym{‖} o \lottsym{‖} \lottsym{,} \, \mathcolor{STLC}{ x } \, \lottsym{≈} \, i \, \lottsym{⊢} \, \mathcolor{STLC}{ \mathcolor{STLC}{ eₛ } } \, \lottsym{≈} \, e}\\
\lotthypothesis{\lottsym{ε} \, \lottsym{⊢} \, \mathcolor{STLC}{ \mathcolor{STLC}{ vₛ } } \, \lottsym{≈} \, v}
}{δ \, \lottsym{⊢} \, \mathcolor{STLC}{ \mathcolor{STLC}{ \lottsym{(} \mathcolor{STLC}{ \lottsym{λ} \, \mathcolor{STLC}{ x } \lottsym{.} \, \mathcolor{STLC}{ eₛ } } \lottsym{)} } \, \mathcolor{STLC}{ \mathcolor{STLC}{ vₛ } } } \, \lottsym{≈} \, \lottkw{force} \, \lottsym{⟨} o \lottsym{;} \, \_ \, \lottsym{:} \, v \, \lottsym{|} \, \lottsym{λ} \, \lottsym{\{} \, \_ \, \lottsym{◇} \, i \, \lottsym{\}} \lottsym{.} \, e \lottsym{⟩}}
\lottinferencerulesep
\lottinferencerule{app}{
\lotthypothesis{δ \, \lottsym{⊢} \, \mathcolor{STLC}{ eₛ₀ } \, \lottsym{≈} \, e₀}\\
\lotthypothesis{δ \, \lottsym{⊢} \, \mathcolor{STLC}{ eₛ₁ } \, \lottsym{≈} \, e₁}
}{δ \, \lottsym{⊢} \, \mathcolor{STLC}{ \mathcolor{STLC}{ eₛ₀ } \, \mathcolor{STLC}{ eₛ₁ } } \, \lottsym{≈} \, \lottkw{force} \, e₀ \, \lottsym{\{} \, \_ \, \lottsym{:} \, e₁ \, \lottsym{\}}}
\lottinferencerulesep
\lottinferencerule{box}{
\lotthypothesis{\lottsym{‖} o \lottsym{‖} \, \lottsym{⊢} \, \mathcolor{STLC}{ eₛ } \, \lottsym{≈} \, e}
}{δ \, \lottsym{⊢} \, \mathcolor{STLC}{ eₛ } \, \lottsym{≈} \, \lottsym{\{} \, o \, \lottsym{\}} \, \lottsym{▶} \, e}
\lottinferencerulesep
\lottinferencerule{nat}{
}{δ \, \lottsym{⊢} \, \mathcolor{STLC}{ n } \, \lottsym{≈} \, n}
}
}}

\medskip

\begin{paracol}{2}
  \begin{align*}
    & \boxed{\lottsym{‖} o \lottsym{‖}} \hspace{-15pt}
    & \lottsym{‖} ε \lottsym{‖} &= \text{$\lottsym{ε}$} \\
    && \lottsym{‖} o \lottsym{,} \, i \, \lottsym{:} \, v \lottsym{‖} &= \text{$\lottsym{‖} o \lottsym{‖} \lottsym{,} \, i \, \lottsym{:} \, v$}
  \end{align*}

  \switchcolumn

  \begin{align*}
    & \boxed{\lottsym{‖} ι \lottsym{‖}} \hspace{-15pt}
    & \text{$\lottsym{‖} \lottsym{ε} \lottsym{‖}$} &= \text{$\lottsym{ε}$} \\
    && \text{$\lottsym{‖} ι \lottsym{,} \, \mathcolor{STLC}{ x } \, \lottsym{∼} \, i \lottsym{‖}$} &= \text{$\lottsym{‖} ι \lottsym{‖} \lottsym{,} \, \mathcolor{STLC}{ x } \, \lottsym{≈} \, i$}
  \end{align*}
\end{paracol}

\endgroup

\end{document}